\documentclass[12pt,preprint]{aastex}

\usepackage{graphicx}
\usepackage{comment}
\usepackage{amsmath}
\usepackage{epstopdf}
\usepackage{longtable,lscape}

\def\beq{\begin{equation}} 
\def\eeq{\end{equation}} 

\def\fun#1#2{\lower3.6pt\vbox{\baselineskip0pt\lineskip.9pt
  \ialign{$\mathsurround=0pt#1\hfil##\hfil$\crcr#2\crcr\sim\crcr}}}

\begin{document} 

\title{The Third \textit{Swift} Burst Alert Telescope Gamma-Ray Burst Catalog}

\author{Amy Lien$^{1,2,}$\altaffilmark{*}, Takanori Sakamoto$^{3,}$\altaffilmark{*}, Scott~D.~Barthelmy$^4$, Wayne~H.~Baumgartner$^{1,2}$, John~K.~Cannizzo$^{1,2}$, Kevin Chen$^5$, Nicholas~R.~Collins$^{1,2}$, Jay R. Cummings$^4$, Neil Gehrels$^4$, Hans A. Krimm$^{1,6}$, Craig. B. Markwardt$^4$, David~M.~Palmer$^7$, Michael Stamatikos$^8$, Eleonora Troja$^{1,9}$, T. N. Ukwatta$^{7}$}
\affil{$^1$Center for Research and Exploration in Space Science and Technology (CRESST) and NASA Goddard Space Flight Center, Greenbelt, MD 20771, USA}
\affil{$^2$Department of Physics, University of Maryland, Baltimore County, 1000 Hilltop Circle, Baltimore, MD 21250, USA}
\affil{$^3$Department of Physics and Mathematics, College of Science and Engineering, Aoyama Gakuin University, 5-10-1 Fuchinobe, Chuo-ku, Sagamihara-shi, Kanagawa 252-5258, Japan}  
\affil{$^4$NASA Goddard Space Flight Center, Greenbelt, MD 20771, USA}
\affil{$^5$Department of Physics, University of California, Berkeley, 366 LeConte Hall MC 7300, Berkeley, CA, 9472}
\affil{$^6$Universities Space Research Association, 10211 Wincopin Circle, Suite 500, Columbia, MD 21044, USA}
\affil{$^7$Space and Remote Sensing (ISR-2), Los Alamos National Laboratory, Los Alamos, NM 87544, USA}
\affil{$^8$Department of Physics, Department of Astronomy \& Center for Cosmology and AstroParticle Physics, Ohio State University, Columbus, OH 43210 USA}
\affil{$^9$Department of Physics and Department of Astronomy, University of Maryland, College Park, MD 20742, USA}
\affil{* These authors contributed equally to this work.}

\begin{abstract}

To date, the Burst Alert Telescope (BAT) onboard {\it Swift} has detected $\sim$ 1000 gamma-ray bursts (GRBs), of which $\sim$ 360 GRBs have redshift measurements, ranging from $z=0.03$ to $z=9.38$. We present the analyses of the BAT-detected GRBs for the past $\sim$ 11 years up through GRB151027B. We report summaries of both the temporal and spectral analyses of the GRB characteristics using event data (i.e., data for each photon within approximately 250 s before and 950 s after the BAT trigger time), and discuss the instrumental sensitivity and selection effects of GRB detections. We also explore the GRB properties with redshift when possible. The result summaries and data products are available at http://swift.gsfc.nasa.gov/results/batgrbcat/index.html. In addition, we perform searches for GRB emissions before or after the event data using the BAT survey data. 
We estimate the false detection rate to be only one false detection in this sample.
There are 15 ultra-long GRBs ($\sim 2\%$ of the BAT GRBs) in this search with confirmed emission beyond $\sim$ 1000 s of event data,
and only two GRBs (GRB100316D and GRB101024A) with detections in the survey
data prior to the starting of event data.

\end{abstract}

\noindent

\section{Introduction}

\label{sect:intro}

Gamma-ray bursts (GRBs) are one of the most energetic explosions in the universe, and are important in many aspects of astrophysics and cosmology.
The remarkable amount of energy released by GRBs in such a short time scale provides a unique opportunity to study physics in an extreme environment, 
and also challenges the physical models of the progenitors. Both the observational evidence and theoretical studies connect
long GRBs (bursts with duration longer than $\sim 2$ s) with the death of massive stars \citep[see e.g.,][and references therein]{Woosley06, Fryer07, Gehrels12, Kumar15}. On the other hand, 
the origin of short bursts (durations $\lesssim 2$ s) remains mysterious. Current studies suggest that short GRBs are likely related to compact-object mergers and thus they are one of the candidate sources of gravitational waves \citep[see e.g.,][and reference therein]{Eichler89, Nakar07, Berger14}, one of which has recently been detected directly by LIGO \citep{Abbott16}.
Moreover, GRBs are one of the few astrophysical objects that can be directly detected out to very high redshift ($z \gtrsim 8$) due to their extraordinary brightness, 
and thus they provide a valuable tool to study the early universe.


{\it Swift}, a multi-wavelength telescope dedicated to GRB studies, was launched on Nov. 20, 2004 \citep{Gehrels04}. Over the past $\sim 11$ years, the Burst Alert Telescope (BAT) onboard {\it Swift} has detected $\sim 1000$ GRBs. The unique ability of {\it Swift} to observe a large portion of the sky, promptly localize the burst, and rapidly downlink and circulate the detection notice has enabled fast multi-wavelength follow-up observations, and vastly enhanced the scientific outcome. 

The BAT is one of the three telescopes onboard {\it Swift}, and is capable of detecting GRBs and localizing a burst to within a few arcmin. When the BAT detects a GRB, {\it Swift} will slew to the GRB position and observe the burst with the X-ray Telescope (XRT) and the UV-Optical Telescope (UVOT) onboard {\it Swift}, which can further refine the localization to $\lesssim$ arcsec. 
The BAT is composed of a detector plane that has 32,768 CdZnTe (CZT) detectors, and a coded-aperture mask that has $\sim 52,000$ lead tiles. 
The coded-mask technique is useful in X-ray and gamma-ray astronomy to obtain a large field of view while maintaining imaging capability.
The basic idea is that each point source casts a unique shadow through the coded-aperture mask onto the detector plane, and thus 
one can re-construct the source image/position by deconvolving the illuminated pattern on the detector plane and the mask pattern.
The BAT has a field of view of 2.2 sr when it is $> 10\%$ coded, and an energy range of 14-150 keV for imaging or up to 350 keV with no position information.
Details of the BAT instruments can be found in \citet{Barthelmy05} and the first BAT GRB catalog \citep{Sakamoto08}.

The BAT adopts two main trigger methods for detecting GRBs: (1) the rate trigger criteria, which search for GRBs based on count rate increases in the light curves,
and (2) the image trigger criteria, which discover bursts based on images created with different time intervals ($\gtrsim$ minute).  In addition, sometimes a burst that was not triggered on-board can be recovered later by ground analysis. We refer to these events as ground-detected GRBs. These ground-detected bursts usually happen when the BAT is not capable of triggering bursts, such as during spacecraft slews or close to the South Atlantic Anomaly (SAA, which is an area that contains a high level of background high-energy particles). Moreover, a burst can occur at a location that is highly off-axis relative to the BAT detector plane and hence only generate weak signals.
The search and discovery of ground-detected GRBs are usually motivated by detections from other instruments, such as {\it Fermi}, INTEGRAL, and MAXI.

Several catalogs related to the BAT-detected GRBs have been published, including the first and second BAT GRB catalog
from the {\it Swift}/BAT team \citep[][which will hereafter be referred as the BAT1 and BAT2 catalog, respectively]{Sakamoto08, Sakamoto11}. 
Some catalogs with selected BAT GRBs for specific usages have also been presented,
such as the catalog composed by \citet{Salvaterra12} (also known as the BAT6 catalog) that selects bright bursts detected by BAT with optimal conditions for ground follow-up observations, in order to construct a GRB sample with redshift completeness.
Furthermore, many online GRB tables are available. Those that are related to the BAT data include (1) the Swift GRB table\footnote{\label{Swift_table} http://swift.gsfc.nasa.gov/archive/grb\_table/} compiled by J. D. Myers using information from GCN circulars,
(2) an online GRB catalog\footnote{http://grb.pa.msu.edu/grbcatalog} maintained by Tilan Ukwatta that contains GRBs from {\it Swift}, (3) the {\it Swift} Burst Analyser\footnote{http://www.swift.ac.uk/burst\_analyser/} maintained by Phil Evans, which includes plots of both the BAT and XRT light curves at selected energy bands \citep{Evans10,Evans09,Evans07},
(4) the ``Swiftgrb database''\footnote{http://heasarc.gsfc.nasa.gov/docs/swift/archive/grbsummary/} produced by Padgett et al., which is completed through December of 2012 and includes data product for BAT and XRT,
and (5) an online repository\footnote{http://butler.lab.asu.edu/swift/} generated by Nathaniel Butler that includes the XRT and BAT light curves, spectra, and GRB redshifts \citep{Butler07, Butler10}.
Moreover, the ``GRB Online Index (GRBOX)\footnote{http://www.astro.caltech.edu/grbox/grbox.php}'' maintained by Daniel Perley compiles a list of GRB with redshift measurements and information of follow-up observations. The webpage maintained by Jochen Greiner\footnote{http://www.mpe.mpg.de/~jcg/grbgen.html} also presents a comprehensive information of GRB localizations and redshifts.

In this catalog, we update the results in the BAT2 catalog to include the GRBs detected by BAT after 2009. We include all the bursts through
the 1000th {\it Swift} GRB, GRB151027B, which was officially announced by the {\it Swift} team\footnote{http://www.nasa.gov/feature/goddard/nasas-swift-spots-its-thousandth-gamma-ray-burst/}. 
This 1000th burst is counted based on the list in the Swift GRB table compiled by J. D. Myers$^{\ref{Swift_table}}$,
which is slightly different than the list we compile here in the third BAT GRB catalog.
The Swift GRB table lists the GRBs that were first reported by ${\it Swift}$ in the GCNs. 
In the third  BAT GRB catalog, we include all GRBs that were reported being seen by BAT (either triggered onboard or found by ground analyses, some of which may be motivated by detections from other instruments).
For those GRBs without an XRT/UVOT afterglow detection, we will mark them as ``questionable detections'' if the signal-to-noise ratio is lower than 7 (see Section~\ref{sect:event_standard} and \ref{sect:ground_GRBs} for details of how the signal-to-noise ratio is determined).

To make sure the analyses for the new and old bursts are consistent, we reanalyze all the bursts in the BAT2 catalog as well,
using the same up-to-date software.
The main GRB characteristics (e.g., burst durations, spectral fits) are acquired 
from analyses using the event data (sometime also called the event-by-event data), which record information of individual photons and usually cover ranges between $\sim$ 250 s 
before and $\sim 950$ s after the BAT trigger time.
For the event data analyses, we follow the general pipelines adopted in the BAT2 catalog.
We report some extra information in this catalog regarding the GRB observation status, such as the partial coding fraction and the trigger method (rate or image trigger).
We also include summaries and discussions of the BAT observational constraints (e.g., the Sun/Moon constraints, fractions of time when BAT is able to trigger a burst, changes in the number of active detectors).
Furthermore, in addition to studies using event data, we perform further searches for possible extended emission beyond the event data range using the BAT survey data. The survey data are binned in $\sim$ 5-min intervals, and cover time periods that do not have event data.


This paper is organized as follows: 
Section \ref{sect:BAT_status} reports the update of the BAT status related to GRB observations, which includes status of the in-orbit
calibration using the Crab observations, and a summary of the BAT observing time.
Section \ref{sect:event_data_analysis} presents the method of the data analysis for the event data.
Section \ref{sect:event_results} reports the results of the event data analyses and includes discussions of observed burst properties and 
rest-frame characteristics for those GRBs with redshift measurements.   
Section \ref{sect:survey_data_analysis} describes the pipeline for analyzing the survey data 
and also discusses the false-detection rate of the survey data in order to search for weak emissions beyond the event data range.
Section \ref{sect:survey_results} summarizes the results from the survey data search.
The overall summary is presented in Section~\ref{sect:summary}.

\section{Updates of the BAT status}
\label{sect:BAT_status}

\subsection{Status of the in-orbit calibrations}

Each year, the {\it Swift} team schedules special observations of the Crab Nebula to perform an on-orbit 
calibration of the BAT for both energy and position measurements.
During the calibration observations, the BAT observes the Crab nebula at different incident angles
to check that the measurements show consistent results.

In 2015, five observations with different incident angles were performed:
(1) on-axis,
(2) off-axis with $\theta = 30$ deg, $\phi = -90$ deg.
(3) off-axis with $\theta = 30$ deg, $\phi = 90$ deg,
(4) off-axis with $\theta = 45$ deg, $\phi = 0$ deg, and 
(5) off-axis with $\theta = 45$ deg, $\phi = -180$ deg.
$\theta$ is the polar angle measured from the BAT pointing direction, and $\phi$ is the azimuth angle. 

\begin{figure}[!h]
\begin{center}
\includegraphics[width=1.0\textwidth]{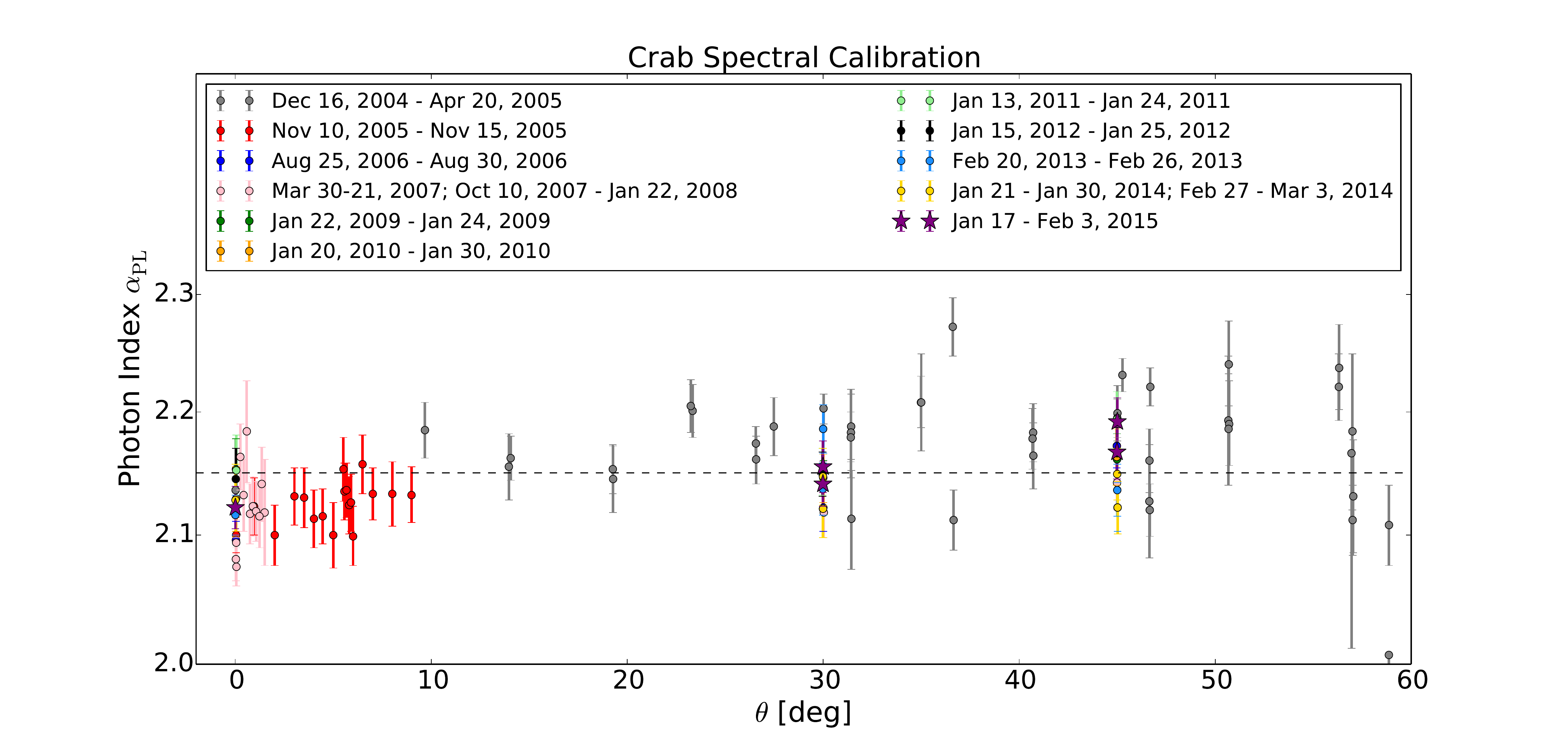}
\end{center}
\caption{
The best fit Crab photon index vs. $\theta$ angle. The dashed line marks the assumed Crab photon index of -2.15
\citep{Rothschild98, Jung89}.
}
\label{fig:Crab_alpha_theta}
\end{figure}

\begin{figure}[!h]
\begin{center}
\includegraphics[width=1.0\textwidth]{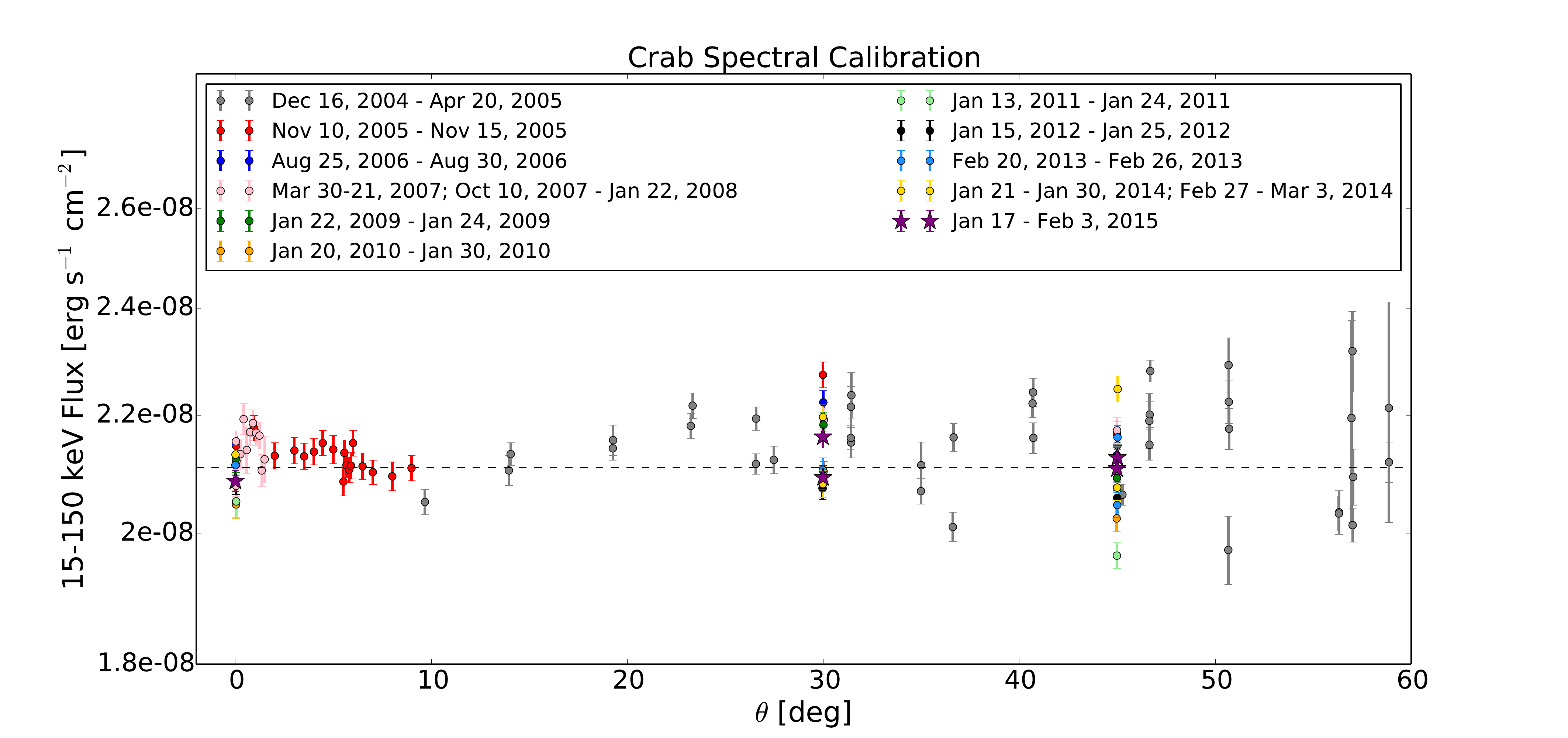}
\end{center}
\caption{
The best fit Crab flux in 15-150 keV vs. $\theta$ angle. The dashed line marks the assumed Crab flux of 
$2.11 \times 10^{-8} \ \rm erg \ cm^{-2} \ s^{-1}$ \citep{Rothschild98, Jung89}.
}
\label{fig:Crab_flux_theta}
\end{figure}

Figure \ref{fig:Crab_alpha_theta} and \ref{fig:Crab_flux_theta} show
the results of the spectral fits with different incident angles from year 2005 to 2015.
The Crab spectrum is fitted with a simple power-law model (see definition in Eq.~\ref{eq:PL}).
The photon indices of the simple power-law model $\alpha_{PL}$ are plotted in Fig.~\ref{fig:Crab_alpha_theta},
while the fluxes are presented in Fig.~\ref{fig:Crab_flux_theta}.
Results show that the photon index and the flux measured at different locations on the BAT detector plane
can vary by up to $\sim \pm 5\%$ and $\sim \pm 10\%$, respectively, from the canonical values of Crab photon index of -2.15
and flux of  $2.11 \times 10^{-8} \ \rm erg \ cm^{-2} \ s^{-1}$ \citep{Rothschild98, Jung89}.

\citet{Wilson-Hodge11} suggests that the flux of the Crab Nebula can fluctuate on a time scale of months to years,
with the value changes as much as $\sim10\%$ in the BAT energy range from 2008 to 2010. However, as seen in Fig.~\ref{fig:Crab_flux_theta}, the systematic
errors for the flux measurements at large incident angles can be as large as $\sim 10\%$. This is because of the unknown systematic
uncertainties included in the BAT energy response function. Thus, it can be difficult to place tight constraints for flux
variations less than $\sim 10\%$, if the spectral analysis is involved in the BAT data. 
Note that the BAT Crab light curve presented in \citet{Wilson-Hodge11} is based on
the survey data process which extracts the counts directly from the sky images.
The BAT calibration data base is last updated
in 2009 \citep{Sakamoto11}.

\begin{figure}[!h]
\begin{center}
\includegraphics[width=0.6\textwidth]{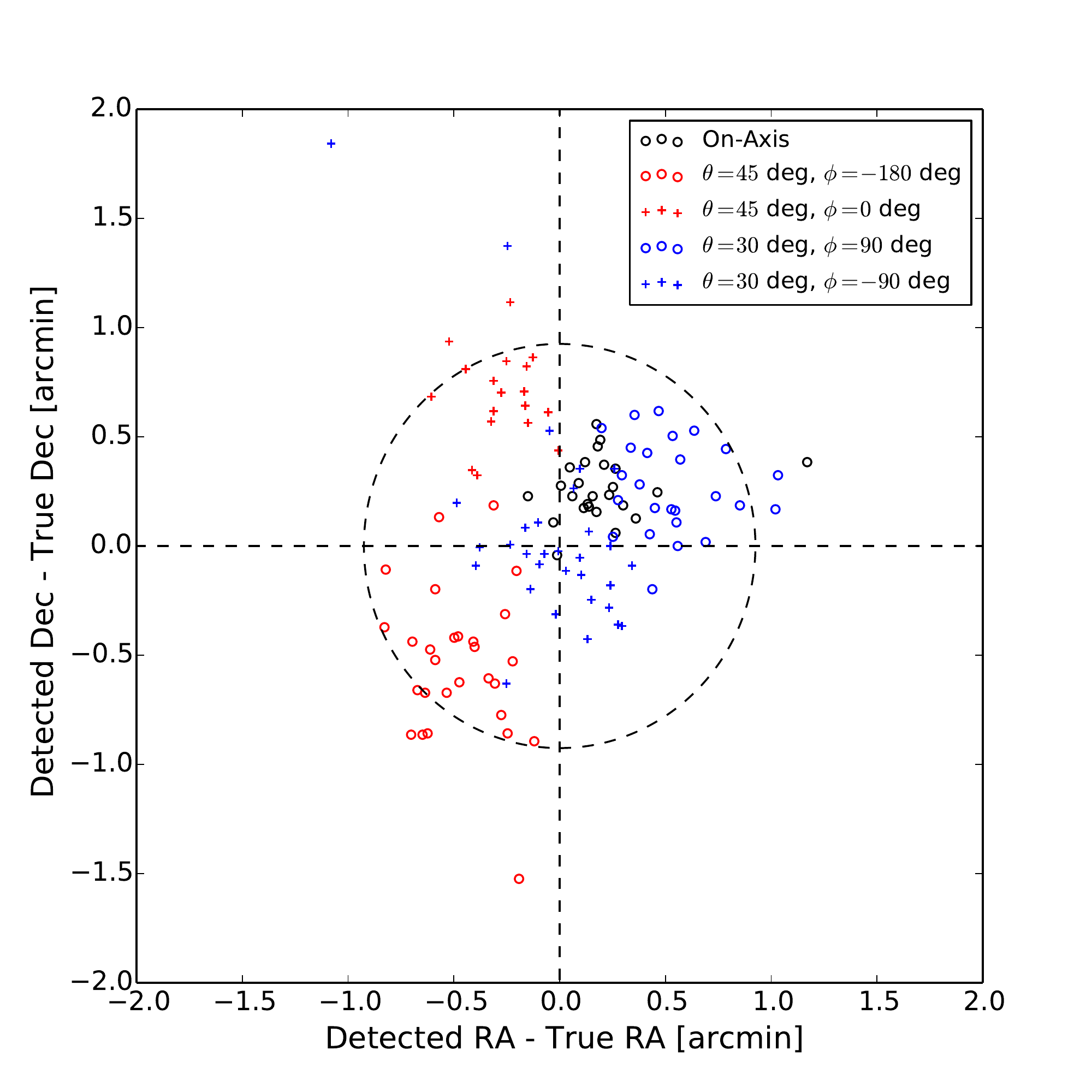}
\end{center}
\caption{
Crab position difference from ``true'' Crab position at ${\rm RA}=83.633$ deg and ${\rm DEC}=22.014$ deg. The dashed circle with a radius of 0.93 arcmin encloses 90\% of the measurements.
}
\label{fig:Crab_position}
\end{figure}

Figure \ref{fig:Crab_position} shows the differences between the true Crab position (${\rm RA}=83.633$, ${\rm DEC} = 22.014$) and the Crab positions calculated using observations
from the five different incident angles taken in 2015.
Results show that the Crab position measurements can change up to $\sim 2$ arcmin (with respect to the assumed ``true'' location)
when measuring with different incident locations. However, 90\% of the measurements are within 0.93 arcmin from the ``true'' location. 

\subsection{Summary of the BAT observing time}

\begin{figure}[!h]
\begin{center}
\includegraphics[width=1.0\textwidth]{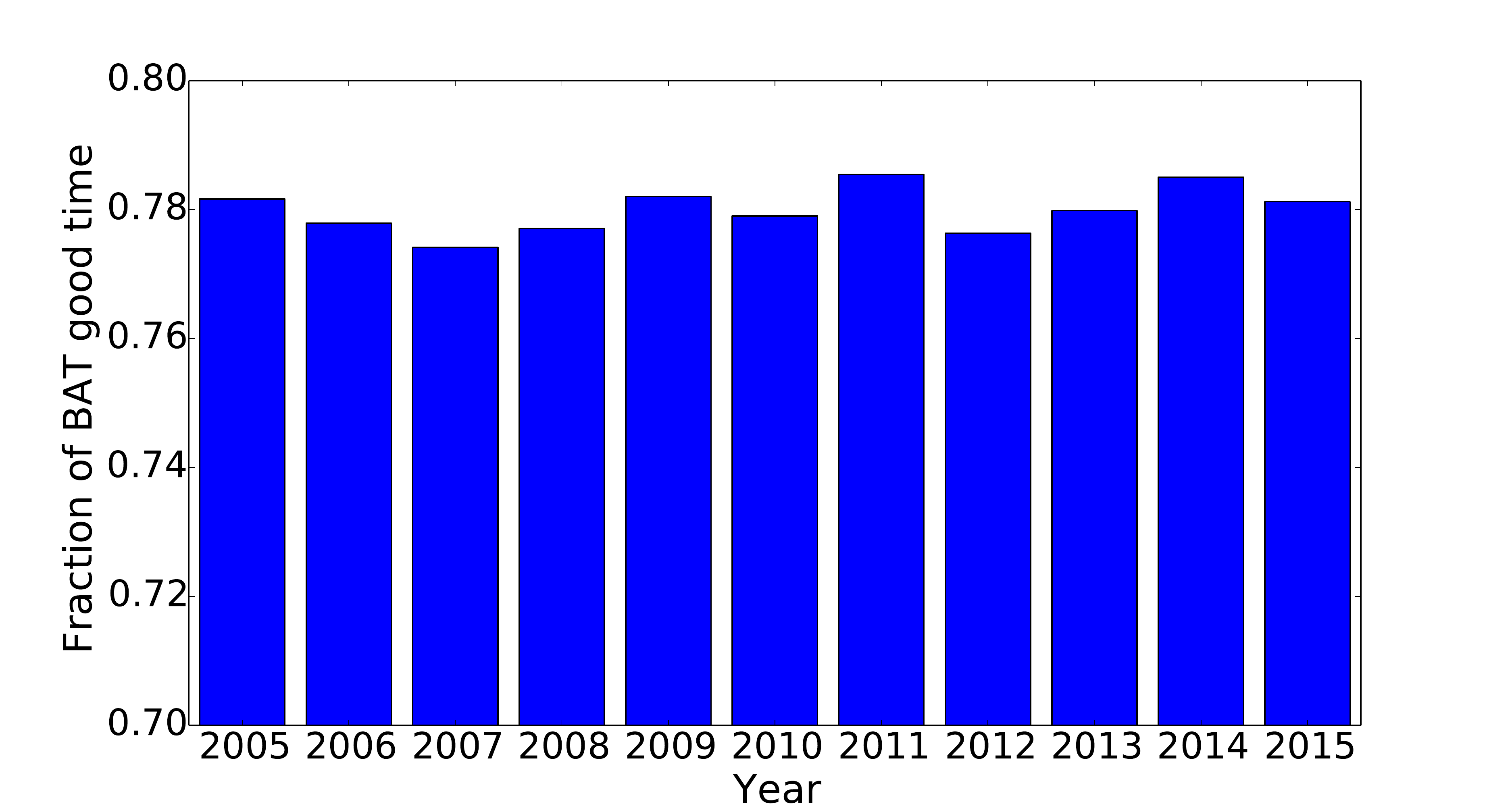}
\end{center}
\caption{
Fraction of the BAT observing time (i.e., the time when BAT is able to trigger bursts) as a function of year. The fraction remains very stable at $\sim 0.78$.
}
\label{fig:BAT_goodtime}
\end{figure}

\begin{figure}[!h]
\begin{center}
\includegraphics[width=1.0\textwidth]{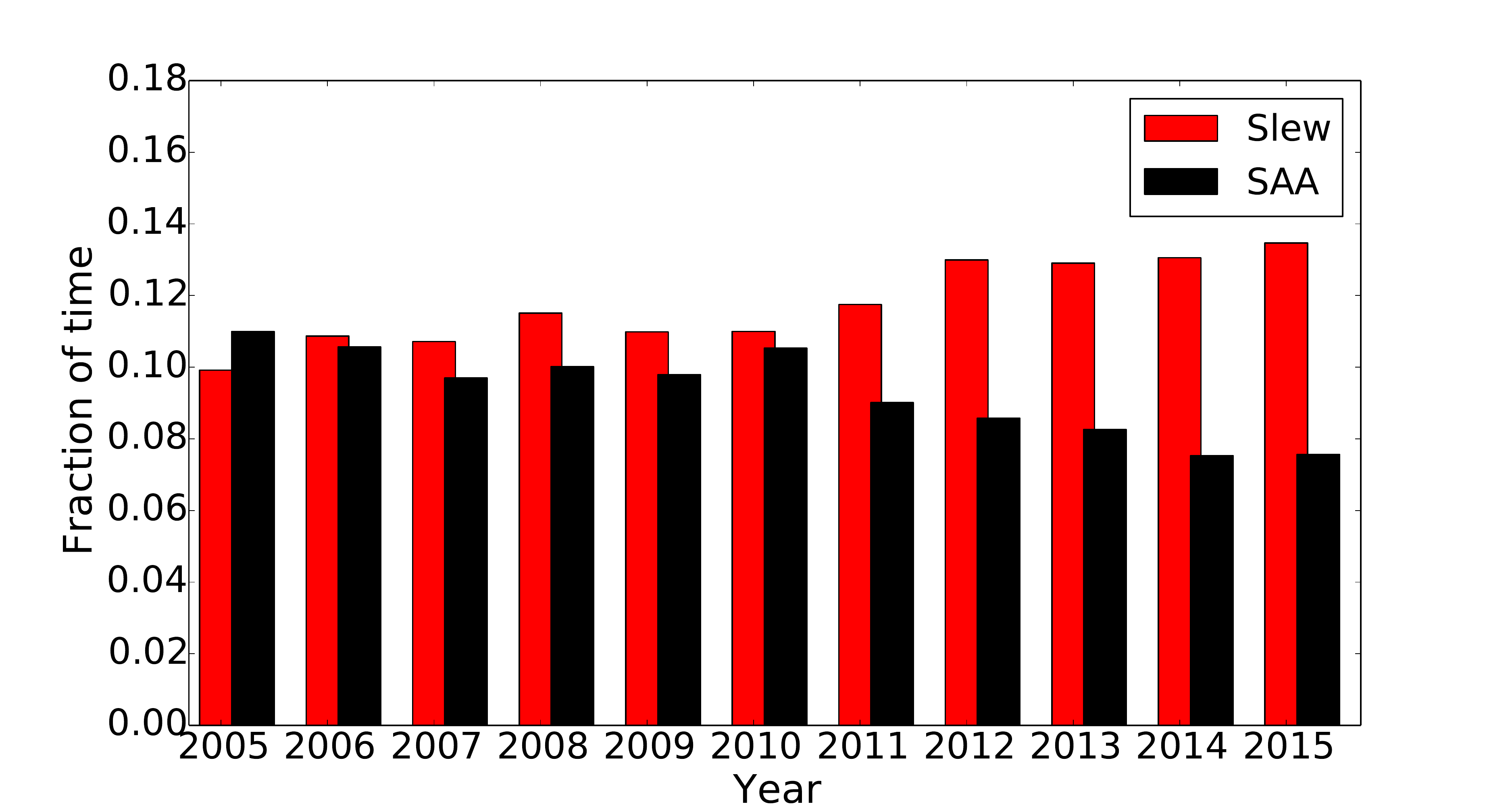}
\end{center}
\caption{
Fraction of the BAT slew time (red bars), and the time in SAA (black bars), as a function of year.
Note that this is the SAA time for BAT only, which determines the SAA time based on instant count rate and backlog in the ring buffer. 
XRT and UVOT adopts a more strict criteria for SAA using the sky location, and thus have a larger SAA time fraction in general.
}
\label{fig:BAT_slew_SAA_time}
\end{figure}

Based on the BAT log files, BAT spends $\sim 78\%$ of the time performing observations and searching for GRBs. Figure \ref{fig:BAT_goodtime} shows the fraction of the time when BAT was capable of triggering bursts from year 2005 to 2015. 
BAT is unable to trigger a burst mainly due to spacecraft slewing and when the spacecraft passes through the SAA. Fig.~\ref{fig:BAT_slew_SAA_time} shows the fraction of time during spacecraft slews and SAA. 
There are $\sim 12\%$ of spacecraft slew time. This fraction gradually increases with time as {\it Swift} observes more and more Target-of-Opportunity (ToO) targets.
Record shows that the spacecraft spends $\sim 9\%$ of the time in SAA. Note that this is the SAA time as defined for BAT operations, 
which is determined based on instant count rate and backlog in the ring buffer.
XRT and UVOT adopt a more strict criteria for SAA using the location, and thus have a larger SAA time fraction in general.
As shown in the figure, the fraction of the BAT SAA time decreases slightly from 2005 to 2015. 
This is because the solar activity increases during these years, which results in a slightly lower particle density in the SAA region. 
Since BAT uses the count rate to define the SAA time,
the fraction of SAA time decreases as fewer cosmic-ray particles appear in the SAA region.

\section{Data Analysis for BAT event data}
\label{sect:event_data_analysis}

\subsection{Standard analysis}
\label{sect:event_standard}

All the BAT event data used in this analysis are downloaded from HEASARC\footnote{http://heasarc.gsfc.nasa.gov/cgi-bin/W3Browse/swift.pl}.
We use the standard BAT software (HEASOFT 6.15\footnote{http://heasarc.nasa.gov/lheasoft/}) 
and the latest calibration database (CALDB\footnote{http://heasarc.gsfc.nasa.gov/docs/heasarc/caldb/swift/})
to perform analysis for event data.
Specifically, we use the script {\it bateconvert} for energy calibration, and
{\it batgrbproduct}\footnote{http://heasarc.gsfc.nasa.gov/ftools/caldb/help/batgrbproduct.html} to perform 
a series of standard analyses of the event data, which includes filtering out hot pixels of the detectors,
occultation time periods, refined-position analysis, duration estimation, making light curves 
with different time segments and bin sizes, and generating spectra.
We adopt the default options of {\it batgrbproduct}, except the minimum partial coding fraction (pcodethresh),
which is set to 0.05 instead of 0.0.

The burst durations estimated by {\it batgrbproduct} include $T_{100}$, $T_{90}$, and $T_{50}$,
which correspond to the durations that contain 100\%, 90\%, and 50\% of the burst emission, respectively.
Specifically, the start and end times of $T_{90}$ in this standard pipeline are defined as the times when
the fraction of photons in the accumulated light curve 
reaches 5\% and 95\%\footnote{http://heasarc.gsfc.nasa.gov/ftools/caldb/help/battblocks.html}.
Similarly, the start and end time of $T_{50}$ is defined as the times when the accumulated light curve reaches 25\% and 75\%.
These definitions for $T_{90}$ and $T_{50}$ are commonly adopted for quantifying burst durations by other teams, including 
BATSE, {\it BeppoSAX}, and {\it Fermi} \citep{Koshut96, Paciesas99, Frontera09, vonKienlin14}.
In this paper, we follow the convention and use the $T_{90} = 2$ s as the separation between the long and short GRBs categories \citep{Kouveliotou93}. 
Specifically, the bursts that are referred to as short GRBs in this paper have $T_{90} \leq 2$ s, without taking into account of the uncertainty in $T_{90}$.

The burst refined positions are generally also found by {\it batgrbproduct} by running {\it batcelldetect} on the image with the burst emission. 
This image will end before the spacecraft slews if $T_{100}$ lasts beyond the slew time. 
These refined positions are not used in any of the further analyses, such as calculating the mask-weighted light curve and spectrum.
However, we do use the signal-to-noise ratio reported with these refined positions to determine whether the burst could be a questionable detection.
For a few dozen bursts, the signal-to-noise ratio associated with the refined positions found by 
this auto pipeline is lower than 7 (the typical image-trigger threshold). However, most of these cases are due to a long quiescent period of the burst emission before the spacecraft slews.
We thus rerun the search for detections for these bursts using images created with the time interval and energy range determined by the flight software.
If the signal-to-noise ratio is still below 7 and there is not an accompanying an XRT afterglows, we will mark it as a ``tentative detection'' in Table \ref{tab:summary_general}.
The readers are advised to treat these bursts with special caution.

For spectral analysis, we use the commonly-adopted X-ray fitting package, XSPEC\footnote{http://heasarc.gsfc.nasa.gov/xanadu/xspec/}.
When the spectrum covers time periods including spacecraft slew time,
we generate multiple response files in this period in order to create an ``average'' response file for the whole period.
{\it Swift} slews at a rate of $\sim$ 1 deg per second \citep{Markwardt07} and hence
the telescope motion can be safely ignored within 5 seconds \citep{Sakamoto11_calibrate}.
Therefore, we create 
a response file for each five seconds during slew time,
and a response file for each time segment when the spacecraft is settled.
We create an average response file for the whole time period using the HEASARC tool {\it addrmf},
with weighting factors equal to the fraction of photon counts in the specific time periods of each response file.

Following the BAT2 catalog,
we fit the GRB spectra with two different models: simple power law (PL) and cutoff power law (CPL).
The simple power-law model is described by the following equation,
\beq
\label{eq:PL}
f(E) = K^{\rm PL}_{50} \bigg(\frac{E}{50 \ \rm keV} \bigg)^{\alpha^{\rm PL}},
\eeq
where f(E) is the photon flux at energy E, $\alpha^{\rm PL}$ is the PL index, and ${K^{PL}_{50} }$ is the normalization factor at 50 keV, with units of photons $\rm cm^{-2} \ s^{-1} \ keV^{-1}$.
The cutoff power-law model is expressed as,
\beq
\label{eq:CPL}
f(E) = K^{\rm CPL}_{50} \bigg(\frac{E}{50 \ \rm keV} \bigg)^{\alpha^{\rm CPL}} \exp \bigg(\frac{-E(2+\alpha^{\rm CPL})}{E_{\rm peak}}\bigg),
\eeq
where $\alpha^{\rm CPL}$ is the CPL index, ${K^{CPL}_{50} }$ is the normalization factor at 50 keV, with units of photons $\rm cm^{-2} \ s^{-1} \ keV^{-1}$,
and $E_{\rm peak}$ is the peak energy in the $\nu F_\nu$ (i.e., $E^2 f(E)$) spectrum, where $F_\nu = E f(E)$ is the energy flux density.

The BAT spectra produced using the mask-weighting techniques have Gaussian statistics \citep{Markwardt07}. Therefore, we use the ``fit'' command
in XSPEC with the default ``statistic chi'' option, which finds a fit with the maximum likelihood for Gaussian data (in other words, finds a fit with a minimum $\chi^2$; 
see detailed descriptions in the Appendix B of the XSPEC manual\footnote{\label{note1}http://heasarc.gsfc.nasa.gov/xanadu/xspec/manual/XSappendixStatistics.html})

We use the ``error'' command in XSPEC to estimate the 90\% confidence region for each parameter. This command changes the assigned parameter values
until it finds a value that gives a fit with statistics differing by a number $\Delta$ from the best-fit statistics. When the data follow a Gaussian distribution, 
$\Delta$ follows a $\chi^2$ distribution. 
In our case, we use the default $\Delta = 2.706$, which 
is equivalent to the 90\% confidence region in the $\chi^2$ distribution. 
Based on the XSPEC manual, the ``error'' command is one of the more reliable and recommended methods for constraining uncertainties, 
and it is not as computationally expensive as the Monte Carlo techniques (see detailed descriptions in the Appendix B of the XSPEC manual\textsuperscript{\ref{note1}}).
However,  in order to use the ``error'' command to estimate the uncertainties of the implicit parameters, such as the fluxes, 
one would need to re-write the function to make flux one of the parameters (instead of the normalization factor K), and re-do the fit. 
Therefore, for the flux error estimation, we use the ``cflux'' and ``cpflux'' command in XSPEC,
which performs this conversion for energy flux and photon flux, respectively.
Ideally, if XSPEC does find the global minimum, all these fits should find the same solution. 
However, if XSPEC only finds a local minimum, fitting using the different forms of the same function 
can converge to different solutions. Thus, we cross check the fits that use different forms of the same functions and only accept the fits
if they find solutions that are consistent with each other (i.e., the fitted values have overlapping uncertainty ranges).

Note that in order to set up an automatic pipeline for all the bursts, we accept the solutions found by the ``fit'' command after about 100 iterations. 
We cannot rule out the possibility that this is a local minimum. 
Sometimes XSPEC finds better fits when going through the search with the ``error'' command. However, we always 
discard these new fits to make sure the uncertainties for all the parameters are estimate based on the same best fit,
and to prevent XSPEC from going into an infinite loop if it keeps finding new fits when constraining parameter errors.  

As mentioned in \citet{Sakamoto11}, most of the GRB spectra in the BAT energy range can be well-fitted by the simple PL model,
and show no significant improvement in their fit when changing to the CPL model. 
We adopt the same criteria as the one in \citet{Sakamoto11} and determine when CPL is a better fit
when $\Delta \chi^2 \equiv \chi^2_{\rm PL} - \chi^2_{\rm CPL} > 6$ (and when there is no problem in the CPL fit; see the criteria for acceptable fits below).

In the following list, we summarize the criteria we use to decide whether a fit is acceptable:
\begin{itemize}
\item All the parameters (normalization factor, photon index, photon and energy fluxes for different energy bands, and $E_{\rm peak}$ for the CPL fit) and their errors are constrained.
\item The parameters (photon index and $E_{\rm peak}$) found by fitting different forms of the same functions are consistent with each other (i.e., the 1-sigma uncertainty regions overlap).
We compare the fits that are used to constrain the photon and energy fluxes in different energy bands with the original fit for constraining the normalization factor and photon index.
\item The normalization factors and fluxes for different energy bands are not consistent with zero (i.e., the lower limit found is greater than zero).
\item The lower limit is not larger than the upper limit (this should always be satisfied in principle, but we place this criteria regardless just in case).
\item The ``probability of the null hypothesis'' from the resulting XSPEC fit (estimated based on the $\chi^2$ distribution) needs to be larger than 0.1. That is, the null hypothesis needs to be consistent with the data
within 90\% of confidence range.
\item If $\Delta \chi^2 \equiv \chi^2_{\rm PL} - \chi^2_{\rm CPL} > 6$ and the fit satisfies all the criteria above, we adopt the CPL fit in the following discussions and mark these bursts as better-fitted by the CPL model in the summary tables (Appendix \ref{sect:appendix_tables}). 
\end{itemize}
Sect~\ref{sect:unaccept_spec} contains further discussions regarding the bursts with unacceptable spectra under these criteria. Note that spectral fits from both the simple PL and CPL are presented for all GRBs 
in the summary tables (Appendix \ref{sect:appendix_tables}) regardless of whether or not they satisfy these criteria. The names of GRBs with their best-fit models are given in separate lists in Section~\ref{sect:appendix_tables}.

We perform spectral analyses for the following four types of spectra: (1) Time-averaged spectra, which are the spectra created using the $T_{100}$ duration\footnote{There are ten GRBs with unconstrained $T_{100}$ due to the weakness of the burst. In such cases, we use the time interval determined by the flight software to create the time-averaged spectrum.}, 
(2) The 1-s peak spectra, which cover the 1-s peak time selected by {\it battblocks}, 
(3) The time-resolved spectra, which are a series of spectra for each burst created based on the sub-time-segments within $T_{100}$ that are selected by {\it battblocks}. Ideally,
these sub-time-segments pick out the sub-structure in the light curve variations, but the selections are not always perfect.
(4) The 20-ms peak spectra, which cover the 20-ms peak time selected by {\it battblocks} using the 4-ms binned light curve. This is not one of the standard products included by {\it batgrbproduct},
however, we create this additional peak time and spectral analyses for those extremely short GRBs in particular. Due to the extremely short time interval, we generate a spectrum with only
10 energy bins (equally spaced in the log-scale from 14.0 to 149.99 keV), following the pipeline in the BAT2 catalog.

\subsection{Ground-detected GRBs}
\label{sect:ground_GRBs}

There are
81
GRBs found in ground analysis, including
25
discovered during spacecraft slews. That is, these GRBs did not pass the on-board trigger criteria, 
but were identified by miscellaneous ground processing, such as searching in the BAT data for those GRBs triggered by other spacecrafts, and/or searching for possible GRBs during spacecraft slews (since BAT cannot trigger during this time). 
Most of these bursts are found in either the ``failed event data'' or the ``slew event data''.
The ``failed event data'' are $\sim 10$-second-long event data that are downlinked when a burst passes the first-stage detection threshold (i.e., the rate-trigger criteria),
but failed to pass the second-stage trigger criteria (i.e., the image-trigger criteria; see \citet{Barthelmy05} for more information of the two-stage trigger criteria for BAT).
The ``slew event data'' are event data collected during spacecraft slews.
GRB150407A, GRB140909A, GRB110604A, GRB070125, and GRB060123 were only found manually in images created by the flight software, and no event data are available.

When at least some event data exist for a ground-detected burst, we re-analyze the burst using standard BAT data analysis scripts, {\it bateconvert} and {\it batgrbproduct}, as mentioned in Section~\ref{sect:event_standard}.
To perform the analysis, {\it batgrbproduct} requires some information from the prompt data collected through the Tracking and Data Relay Satellite System (TDRSS),
which includes the burst observation ID, trigger number, trigger time,
the RA and DEC of the burst found in the onboard analysis, background duration used for the trigger, and information of whether this is a rate or image trigger. 
For bursts triggered onboard, the TDRSS data are downlinked to the ground within seconds to minutes of the trigger time. 
However, only limited data are transfered due to the downlink bandwidth.
The complete data are downlinked to ground stations $\sim$ hours later. 
Since ground-detected GRBs are not triggered onboard, they do not have TDRSS data. We thus manually create fake TDRSS messages that contain the relevant information
required by {\it batgrbproduct}.

Bursts that have ``failed event data'' are assigned with unique trigger numbers because they have passed the rate trigger criterion.
We use these trigger numbers in the fake TDRSS messages. 
Bursts found using the ``slew event data'' do not have a unique trigger number. Hence, we use the observation ID corresponding to the
relevant slew event data as the trigger number in the fake TDRSS message.
For the burst position, we use the best position reported in a GCN circular, which was found by previous manual analysis using the BAT data, or
from the follow-up XRT/UVOT observations if the afterglow was detected (for simplicity, we do not use position from ground-based follow-up). 
The burst position is the only important 
information in the TDRSS data that is used for the actual analysis. Other information, such as the background duration, are only used in the summary 
report produced by {\it batgrbproduct}, and thus we use arbitrary numbers in the fake TDRSS messages.

Because the ``failed event data'' are much shorter than regular event data, 
it is common that the burst duration lasts longer than the $\sim 3 - 10$ s event data range. In such cases,
we use the un-maskweighted count rate data (the so-called ``quad-rate data''\footnote{Descriptions for the quad-rate data can be found at http://swift.gsfc.nasa.gov/archive/archiveguide1/node1.html} to be specific, which is the raw count rate binned in 1.6 s).

Many of the ground-detected bursts have very low partial coding fractions, we follow the same guidelines described in Section \ref{sect:low_pcode}
for analysis of these bursts. There are
15
ground-detected bursts that were outside of the BAT calibrated field-of-view (i.e., the region where the FLUX mask\footnote{\label{note_mask} See Section \ref{sect:low_pcode} for explanation of the FLUX and DETECTION masks.}
is applicable)
in the whole event data range.
These bursts require using DETECTION mask$^{\ref{note_mask}}$ for finding burst durations and refined positions, and the spectral analyses are unavailable.
There are
2
ground-detected bursts requiring DETECTION mask for finding the burst refined positions,
the rest of the analyses are done using the FLUX mask.

Similar to the onboard triggered bursts, any ground-detected GRBs without an XRT afterglow and with signal-to-noise ratio less than 7 is marked as a "tentative burst".
However, the limited event data and the lack of flight trigger information make it difficult to determine the time interval
and energy range that enclose the maximum burst emission to estimate the signal-to-noise ratio. 
A trial-and-error approach might find a higher signal-to-noise ratio,
but it also increases the expected number of false detections, which are hard to quantify in a manual process.
We therefore estimate the signal-to-noise ratio from the image created in 15-350 keV in the time interval of $T_{100}$ range if possible, or the whole event-data range if $T_{100}$
extends beyond that. If the $T_{100}$ range includes some spacecraft-slewing periods, we create a mosaic image with 
small time steps (usually $\sim 0.5$ s).
Moreover, we note that for many ground-detected bursts, the lack of XRT afterglows might be due to delayed observations,
since the bursts was discovered on the ground after full data downlinks, and manually submitting a ToO observation request.
It is hard to make a robust conclusion on this issue, however, since the information is lost forever. 

\subsection{GRBs with low partial coding fraction}
\label{sect:low_pcode}

There are some bursts that are detected at the very edge of the BAT field of view, and thus have 
a very low partial coding fraction. We examine every burst with a partial coding fraction lower
than 10\%. If the standard analysis method fails to perform part or all of the analysis due to the 
small partial coding fraction, we redo the analysis using the ``DETECTION'' mask aperture
setting. In comparison to the default setting of the ``FLUX'' mask aperture, the DETECTION 
mask aperture is the full aperture that includes the largest solid
angle and most illuminated detectors. However, it also includes regions with shadows from 
the mounting brackets, and thus will reduce the accuracy of flux measurements and 
is only recommended to use for finding bursts \citep{Markwardt07}.
We only use DETECTION mask for finding burst refined position and estimating burst duration
if necessary.  We only perform spectral analysis for the part of the light curve that is available 
with the FLUX mask setting.

There are
62
bursts with partial-coding fractions lower than 10\% 
(including the ground-detected GRBs).
We examine and assign these bursts
into three groups
with different analysis approaches:
\begin{enumerate}
\item FLUX mask is okay: the {\it batgrbproduct} pipeline completes the analysis with the default FLUX mask setting, and there no need to 
perform further analysis. There are
28
bursts in this category.
\item DETECTION mask needed only for finding refined position: the pipeline successfully performs most of the analysis except finding the refined position. 
We redo the search for the refined position with the DETECTION mask setting. There are
13
bursts requiring such analysis.
\item DETECTION mask needed for finding burst duration and refined position: 
There are
21
bursts with either part or all of the burst durations outside of the BAT calibrated field of view (i.e., the region included in the 
FLUX mask). For these bursts, we use the DETECTION mask for estimating the burst durations and refined positions. The spectral analysis
is only available for the time period when the burst is in the BAT calibrated field of view.
\end{enumerate}

\subsection{Manual examinations of the analysis results: re-analyzing bursts with problems and/or adding comments for special bursts}
Occasionally, the standard analysis can fail or generate erroneous results for several reasons. For example, peculiar background behavior, such as rapid background rise when the telescope enters the South Atlantic Anomaly (SAA), can cause incomplete background subtraction and result in wrong estimations of burst durations. Additionally, if some bright X-ray sources appear in the same field of view as the burst, the background subtraction can be incorrect since the mask-weighting technique assumes the burst to be the brightest source in the BAT field of view.
Thus, the light curve might be contaminated by these bright sources. 
Furthermore, sometimes there are gaps in the event data due to downlink problems. 
Therefore, we examine the result of each GRB by eye and add comments for those bursts that require special treatment. For problems that appear in more than one GRB, we make the comments in standard format, in order to enable automatic searches afterwards. We also mark the short GRBs with extended emission. However, we do not adopt any quantifiable criteria for determining short GRBs with extended emission. We simply follow the discussions from previous GCN circulars and eye inspections from the light curves. 


The adopted comments in standard format include:
\begin{itemize}
\item The event data are not available.
\item The event data are only available for part of the burst duration.
\item battblocks failed because of the weak nature of the burst.
\item GRB found by the ground process (failed event data).
\item GRB found by the ground process (slew event data).
\item DETECTION mask with pcodethresh = 0.0 is used for finding the refined position.
\item DETECTION mask with pcodethresh = 0.0 is used for everything except spectral analysis.
\item Refined positions found by mosaic image (DETECTION mask with pcodethresh = 0.01, time bin = XX s, and energy band = XX-XX keV).
\item Spectral analysis is not available.
\item Spectral analysis is only available for part of the burst duration.
\item The detector plane histogram data are used for the spectral analysis.
\item T100, T90, and T50 are lower limits.
\item T100, T90, and T50 might be lower limits.
\item Only part of the event data are used in order to have a more reasonable estimation of burst durations.
\item Burst durations are found using quad-rate data from ${\rm T0}_{\rm GCN}-\rm XX$ s to ${\rm T0}_{\rm GCN}+{\rm XX}$ s.
\item Burst durations are found using FRED-model fitting.
\item Tentative detection.
\item Short GRB with extended emission. 
\item Maybe short GRB with extended emission.
\item Refined position calculated with time interval and energy band determined by the flight software (T0 to T0+XX s; XX-XX keV).
\item Spectral analysis failed, likely because the burst is too weak.
\item Obvious data gap within the burst duration.
\end{itemize}

In the following sub-subsections, we summarize further discussion of the common problems.

\subsubsubsection{GRBs without event data or event data range shorter than the burst duration}
There are only two bursts, GRB041219A and GRB071112C, which were triggered on-board and have no event data due to downlink problems.
For GRB071112C, the burst duration is found by applying {\it battblocks} on the quad-rate data. The spectral analysis is not available, because the closest survey data bin lasts from 
$\sim T0-120$ s to $\sim T0+10$ s, which includes more background period than the burst duration. 
For GRB041219A, all the analyses are not available due to the lack of event data, rate data, and survey data, because this burst occurred at the very beginning of the mission. 

There are
77
bursts that have burst durations that last longer than the event data. For the
35
ground-detected bursts that last longer than the available $\sim 10$ s of event data, 
we apply {\it battblocks} on quad-rate data (un-maskweighted) to estimate burst durations.
However, for the bursts triggered on-board, we add the comment ``T100, T90, and T50 are lower limits'', instead of using 
the quad-rate data for quantifying the burst durations. This is because the on-board triggers have much longer event data. 
Thus, the bursts that extend beyond the event data ranges are usually those with fairly long durations, and quantifying 
the durations using un-maskweighted rate data becomes more inaccurate due to changes of the background levels.
The spectral analyses for these bursts with incomplete event data are only available for the part of the burst emission that occurs within the event data range.

\subsubsubsection{GRBs with unusual background changes or background problems}
Unusual background changes are most likely to occur when the spacecraft is entering or leaving the SAA. During these times, the background count rates can increase/decrease by $\sim \rm few \times 10^4 \rm \ count \ s^{-1}$ within a few hundreds of seconds, and cause problems in background subtraction and burst-duration estimations. 
We examine the burst light curves individually. When the burst duration seems to be incorrect, we re-do the analysis with a modified light curve that excludes the peculiar background period (when possible, i.e., when the problematic background period is far enough from the burst emission) to check if the burst duration changes significantly by excluding the problematic part of the data. 
It was necessary to recalculate the burst durations of
53
GRBs using only part of the event data.

\subsubsubsection{GRBs with bright X-ray sources in the field of view}
\label{sect:bright_src}

Other bright X-ray sources in the field of view that have similar or higher signal-to-noise ratios as the GRB
can cause problems in background subtraction and give incorrect estimations of the GRB counts.
We thus list the bursts with bright X-ray sources in their field of view in Table \ref{tab:brigh_sources} in Appendix~\ref{sect:appendix_tables}. 
The analysis results for these bursts need to be treated with caution,
particular the reliability of potential weak emissions in the light curves, and suspicious bumps or dips in the spectra.
Extra manual analyses to remove the bright sources might be needed to obtain more reliable results.

We adopt the following criterion for selecting bursts with bright X-ray sources in their field of view:
(1) If the burst has signal-to-noise ratio ${\rm SNR}_{\rm GRB} \geq 10.0$, the X-ray sources in the field of view need to have signal-to-noise ratios ${\rm SNR}_{\rm source} \geq (0.9 \times {\rm SNR}_{\rm GRB})$, 
and
(2) if the burst has signal-to-noise ratio ${\rm SNR}_{\rm GRB} < 10.0$, the X-ray sources in the field of view need to have signal-to-noise ratios ${\rm SNR}_{\rm source} \geq ({\rm SNR}_{\rm GRB} - 1.0)$.
There are
315
bursts that satisfy this criterion,
indicating that for a large fraction of the BAT-detected bursts ($\sim 31\%$),
extra caution is needed when determining the reality of weak emissions in the light curves.

\subsubsubsection{GRBs with data gap}
For the bursts triggered onboard, there are only 12 GRBs that have data gaps in the T100 range. Most of the data gaps are around one or two seconds. The only one with a large data gap of $58$ s is GRB080319B because it happened shortly after the ``A'' burst and had some problem in data collection. The 12 GRBs are: GRB151027A, GRB131002A, GRB130907A, GRB111209A, GRB111022B, GRB110709B, GRB090516, GRB081017, GRB080928, GRB080319B, GRB060526, and GRB041224.

\subsubsubsection{Tentative detections}
There are some events with marginal BAT detections ($< 7 \sigma$; see Section~\ref{sect:event_standard} and \ref{sect:ground_GRBs} for details of how the detection significance is estimated). 
We mark these as ``tentative detections'' in Table \ref{tab:summary_general} if there was no XRT afterglow detected.
There are 16 bursts that are marked as tentative detections.
These bursts should be treated with caution, as some of them might be due to noise fluctuation even though they have a given GRB name.
However, we note that 10 out of 16 tentative detections are lacking XRT information due to observational constraints, and thus are difficult to determine the true nature
of the event with the BAT detections alone. Information from other instruments (e.g., {\it Fermi}, Konus-Wind, etc) might provide further clues to 
identify the burst nature. We leave the judgement to the readers because it differs on a case-by-case basis.

\subsubsection{Comparison with the BAT2 Catalog}
We reprocess the main plots presented in the BAT2 catalog. All the figures are very similar to those in the BAT2 catalog.
Therefore, we do not notice any significant differences between the data analyses in the BAT2 catalog and the current one.
However, we do introduce some new criteria for selecting acceptable spectral fits in this catalog.

\section{Results for the BAT event data analyses}
\label{sect:event_results}

The analysis results are available through the following interfaces:
\begin{enumerate}
\item Tables in Appendix~\ref{sect:appendix_tables}, which list all the numbers from the analyses, including GRB names, trigger IDs, trigger times, burst durations, spectral fits, energy and photon fluxes, redshift (if available)...etc (see detailed descriptions in Appendix~\ref{sect:appendix_tables}).
\item Webpages that summarize the GRB light curves and spectral analyses in plots, and also include special comments for the burst if available. There is one webpage for each burst, with an index for all the pages at 
http://swift.gsfc.nasa.gov/results/batgrbcat/index.html
\item The data products from the analyses, including the light curve fits files, spectra created for different time ranges ($T_{100}$, 1-s peak, 20-ms peak, and the time-resolved spectra). The corresponding response files generated by averaging through the slew time interval is also included. The data products can be found via the GRB index page http://swift.gsfc.nasa.gov/results/batgrbcat/index.html.
\end{enumerate}

There are three bursts that triggered BAT twice (GRB111209A, GRB110709B, GRB140716A). We merged event data
from triggers of GRB111209A and GRB110709B, and only listed one of the trigger numbers in the trigger ID column in the summary tables.
However, GRB140716A is a ground detected burst, and there is a large data gap between the two triggers. Therefore, we listed both triggers
independently in all the summary tables as GRB140716A-1 and GRB140716A-2, which correspond to different trigger IDs.

\subsection{BAT GRB demographics}

The BAT has detected
1006
bursts to date (until 
GRB151027B), including
925
GRBs triggered on-board and
81
bursts found by ground analyses (within which
25
events are found during spacecraft slews). 
As mentioned in Section \ref{sect:intro}, although GRB151027B is the officially announced 1000th GRB detected by {\it Swift}, there
are 
1006
in our list due to a slightly different criteria of counting the BAT-detected GRBs.
Table \ref{tab:GRBnum_vs_ndet} lists the average number of GRB detected each year and the number 
of active detectors of BAT. Both the total number of GRB detections (i.e., including ground-detected GRBs)
and the number of GRB triggered onboard are included in the table\footnote{The numbers for 2015 is not listed in the table
because not all GRBs in 2015 are included in this catalog}.
The averaged number of active BAT detectors per year is gradually decreasing because some detectors get noisier and are thus turned off. 
Results show that the number of GRB detections per year remains similar throughout the ten years of {\it Swift} observations,
despite the continuously decreasing number of active BAT detectors.
Table \ref{tab:GRB_category} summarizes the number of bursts
in the commonly adopted categories (long, short, short GRBs with extended emission, etc). 
In this paper, the term ``ultra-long burst'' refers to GRBs with the observed durations longer than 1000 s (the usual BAT event data range),
and we only considered duration measured using the BAT emission.

\begin{table}[h!]
\caption{\label{tab:GRBnum_vs_ndet}
Summary of yearly GRB detection and the averaged number of BAT active detectors from 2004 to 2014.
}
\begin{center}
\begin{tabular}{|c|c|c|c|}
\hline
Year & Number of detections & Number of detections & Average number of active \\
         & (with ground-detected GRBs) & (no ground-detected GRBs) & BAT detectors \\
\hline\hline
2005 & 88 & 86 & 29413 \\
\hline
2006 & 102 &  100 & 26997 \\
\hline
2007 & 87 & 80 & 27147 \\
\hline
2008 & 105 & 96 & 26478 \\
\hline
2009 & 91 & 81 & 24387 \\
\hline
2010 & 85 & 72 & 24050 \\
\hline
2011 & 82 & 75 & 22817 \\
\hline
2012 & 92 & 89 & 23017 \\
\hline
2013 & 96 & 85 & 22053 \\
\hline
2014 & 94 & 84 & 20413 \\
\hline
\end{tabular}
\end{center}
\end{table}

\begin{table}[h!]
\caption{\label{tab:GRB_category}
Summary of number of GRBs in each category.
}
\begin{center}
\begin{tabular}{|c|c|}
\hline
GRB category & Number of bursts (percentage)\\
\hline\hline
Long & 850 (84.49\%) \\
\hline
Short & 90 (8.95\%) \\
\hline
Short with Extended Emission & 12 (1.19\%) \\
\hline
Ultra long ($T_{90} \gtrsim 1000$ s) & 16 (1.59\%)\footnotemark \\
\hline
Bursts with un-constrained durations & 66 (6.56\%) \\
\hline
\end{tabular}
\end{center}
\end{table}
\footnotetext{From the false-detection rate estimation, we expect one false detection in this sample (see Sect.~{\ref{sect:survey_data_analysis}} for more details). Also, the ultra-long GRB130925A \citep{Evans14, Piro14} detected by BAT is not in the list of GRBs with confirmed detection in survey data (Table \ref{tab:survey_GRB_duration}), because the currently existing survey data product required for the analysis ends before this burst.} 

\begin{figure}[!h]
\begin{center}
\includegraphics[width=1.0\textwidth]{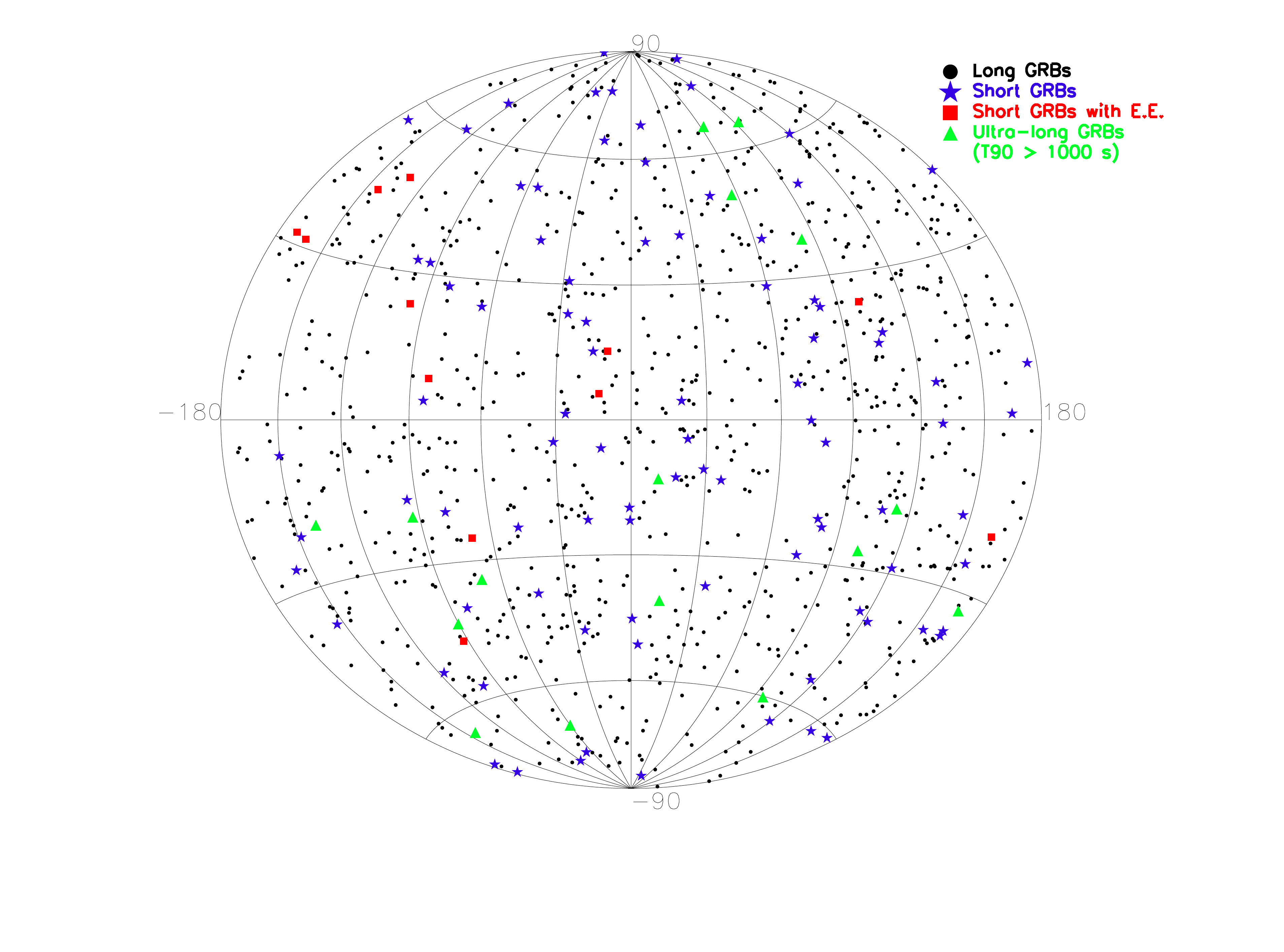}
\end{center}
\caption{
All-sky map in Galactic coordinates for BAT-detected GRBs. Short GRBs are marked as blue stars;
short GRBs with extended emission (E. E.) are plotted as red square; GRBs with burst duration longer 
than 1000 s (i.e., longer than the event data range) are shown as green triangles. 
}
\label{fig:all_sky_map}
\end{figure}

The sky distribution (in Galactic coordinate) of all the BAT-detected GRBs are plotted in Fig.~\ref{fig:all_sky_map},
with  blue stars representing short GRBs, red square showing the short GRBs with extended emission (E. E.),
and green triangle marking the bursts with duration longer than 1000 s (i.e., longer than the event data range).

\subsection{Burst durations}

There are
990
GRBs that have available burst durations.
However, there are 
9
bursts that do not have available errors associated with 
the $T_{90}$,
because the burst durations are determined 
by the FRED-model fit.
For the 16
GRBs without $T_{90}$ measurements,
10
of them are missing burst durations because {\it battblocks} failed to find the burst durations
due to the weak nature of the bursts,
and the rest of the 6 GRBs are those without event data. 
In addition, there are
52
bursts that have incomplete GRB durations, i.e., the reported burst durations 
are lower limits, mostly because the burst durations last longer than
the available event data time range.
Two of these
bursts, GRB101225A and GRB060218, 
have unusually long burst durations without obvious structure in the light curve,
and thus are also in the list of 
GRBs for which {\it battblocks} failed to find the burst durations.

\begin{figure}[!h]
\begin{center}
\includegraphics[width=0.8\textwidth]{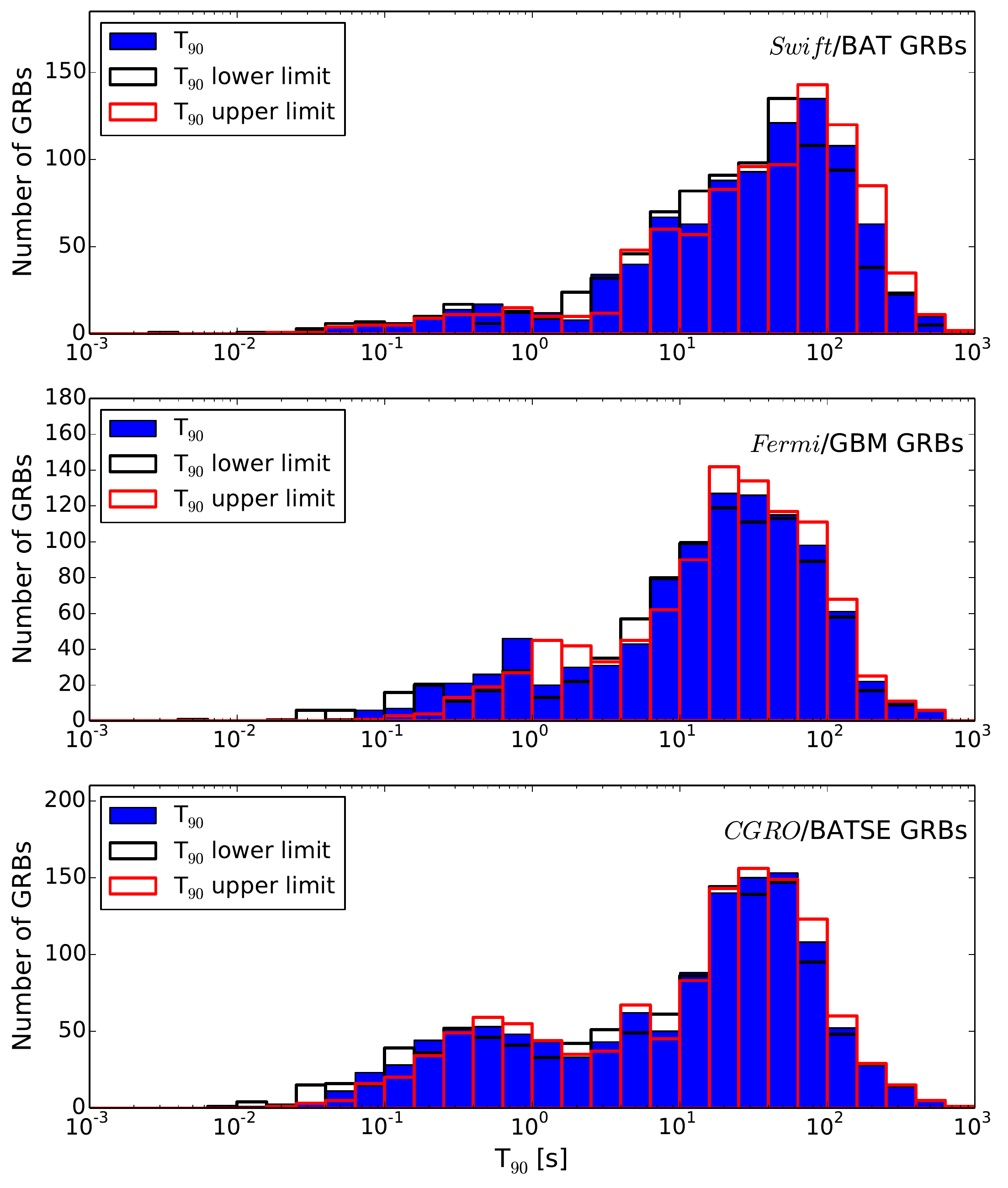}
\end{center}
\caption{
$T_{90}$ distribution for {\it Swift}/BAT (top panel), {\it Fermi}/GBM (middle panel), and {\it CGRO}/BATSE (bottom panel).
For the BAT GRBs, only bursts with successfully determined are included in the plot.
$T_{90}$ for {\it Fermi}/GBM bursts are obtained from the {\it Fermi} GBM burst catalog
\citep{Gruber14, vonKienlin14}.
$T_{90}$ for {\it CGRO}/BATSE bursts are from The Fourth Gamma-ray Bursts Catalog \citep{Paciesas99}.
Distributions using the upper and lower bounds of the $T_{90}$ uncertainty range are also plotted for comparison.
The bin size of this plot is 0.2 in log scale.
}
\label{fig:T90}
\end{figure}

The upper panel of Fig.~\ref{fig:T90} shows the $T_{90}$ distribution of
940
bursts
that have burst durations successfully determined.
That is, we exclude bursts with incomplete $T_{90}$,
and/or bursts without $T_{90}$ that are found by {\it battblocks} or FRED-model fit.
There are
850
GRBs with $T_{90} > 2$ s (long GRBs), and 
90
GRBs with $T_{90} \leq 2$ s (short GRBs).
When folding in the appropriate lower/upper limit,
there are
17
long bursts with the $T_{90}$ lower limit shorter
than 2 second, and
5
short bursts with $T_{90}$ upper limit
longer than 2 second.
For comparison, the histogram of $T_{90}$ upper limits
and  $T_{90}$ lower limits are also plotted in the figure.
Results show that the uncertainties in $T_{90}$ measurements
do not have significant effect on the overall $T_{90}$ distribution. 

The $T_{90}$ distribution remains very similar to the one shown in the BAT2 catalog, 
and thus is still significantly different from the distributions of GRBs detected by other instruments, such as {\it Fermi} and BATSE,
as mentioned in the BAT2 catalog. The middle and bottom panels of Fig.~\ref{fig:T90} shows the $T_{90}$ distributions for GRBs detected by 
{\it Fermi} and BATSE for comparison. $T_{90}$ of the {\it Fermi} GRBs are obtained from the {\it Fermi} GBM burst catalog\footnote{http://heasarc.gsfc.nasa.gov/W3Browse/fermi/fermigbrst.html} 
\citep{Gruber14, vonKienlin14}, and
$T_{90}$ of the BATSE GRBs are from The Fourth Gamma-ray Bursts Catalog \citep{Paciesas99}.
Compared to the short GRB fraction in the {\it Swift}/BAT GRB sample ($\sim 9\%$), the fractions of short bursts are larger in both the {\it Fermi} GRB sample ($\sim 17\%$) and the BATSE sample ($\sim 26\%$).


\subsection{Spectral Analyses}

\subsubsection{Time-averaged spectra}

After applying the criteria described in Sect~\ref{sect:event_standard},
there are
877
bursts that have acceptable spectral fits in their time-averaged spectra
(spectra made by photons in the $T_{100}$ range), 
in which 
90
bursts are better fitted by the CPL model.

\subsubsubsection{GRB classification of spectral characteristics: short-hard bursts vs. long-soft bursts}

In addition to the short and long categories using the burst durations, previous studies from GRBs detected by
multiple instruments, including BATSE, {\it Fermi}, and {\it Swift}, have found that 
short bursts tend to be harder than long GRBs \citep[e.g.,][]{Kouveliotou93, Qin00, Ripa09, Sakamoto11, Qin13, vonKienlin14}.

\begin{figure}[!h]
\begin{center}
\includegraphics[width=0.8\textwidth]{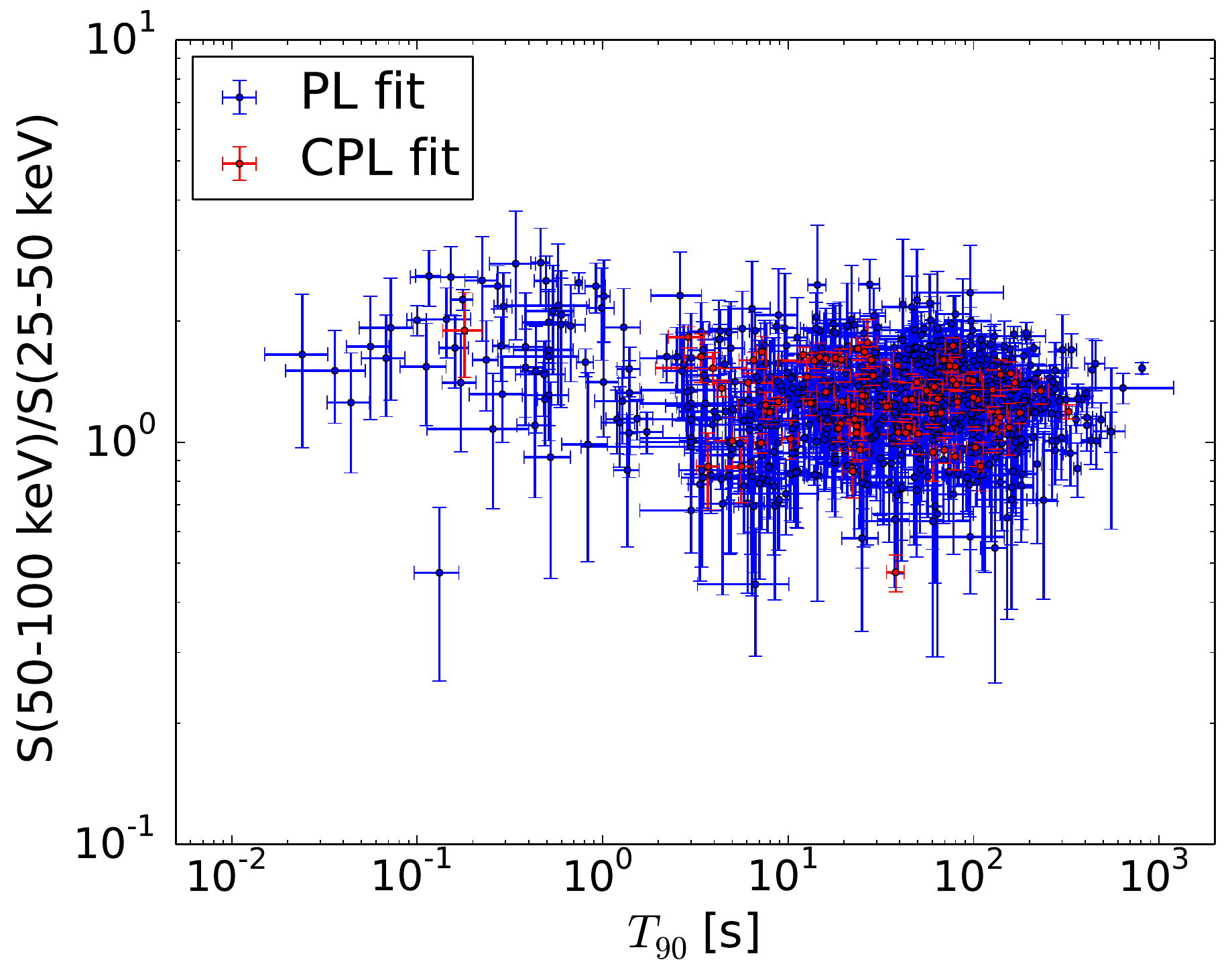}
\end{center}
\caption{
Hardness ratio (i.e., fluence in 50-100 keV over fluence in 25-50 keV) versus $T_{90}$.
The fluences are calculated using the best-fit models (either the simple PL or CPL).
The bursts that are better fitted by the CPL model are marked in red. Note that GRBs with
unconstrained durations, such as the ultra-long GRBs, are not included in this plot.
}
\label{fig:hardness_ratio_T90}
\end{figure}

Figure \ref{fig:hardness_ratio_T90} shows an updated version of the hardness ratio (i.e., the fluence in 50-100 keV divided by fluence in 25-50 keV)
versus $T_{90}$ to include all the new BAT-detected GRBs since the BAT2 catalog. 
The fluences are estimated from the better-fitted spectral model (see criterion described in Section~\ref{sect:event_standard}).
We only included bursts with acceptable spectral fits and with available values of $T_{90}$ and $T_{90}$ errors. In addition, 
we exclude bursts with incomplete $T_{90}$ (i.e., the burst durations are lower limits) and those bursts with $T_{90}$ consistent with zero (i.e.,
the lower limit of $T_{90}$ is equal or less than zero).
There are
815
bursts included in this plot.
There are
86
bursts that are better fitted by the CPL model in this plot (marked in red).

This plot is very similar to the one presented in the BAT2 catalog. The conventional
two GRB classes, short-hard bursts and long-soft bursts, can be roughly identified in this plot, though
the separation of the two groups is not very obvious.
The particularly soft short burst with the hardness ratio of 0.47 and $T_{90}$ of 0.132 s is 
GRB140622A. Despite the unusually soft spectrum, the fast fading XRT light curve of this burst
is consistent with the normal behavior of a short burst \citep{GCN16438,GCN16439}. Moreover, the redshift of $z \sim 0.96$ measured 
from the emission lines from the possible host galaxy \citep{GCN16437} suggests that this is not a Galactic source and is unlikely 
to be a soft gamma repeater \citep[e.g.,][]{Mereghetti08}.

\begin{figure}[!h]
\begin{center}
\includegraphics[width=0.8\textwidth]{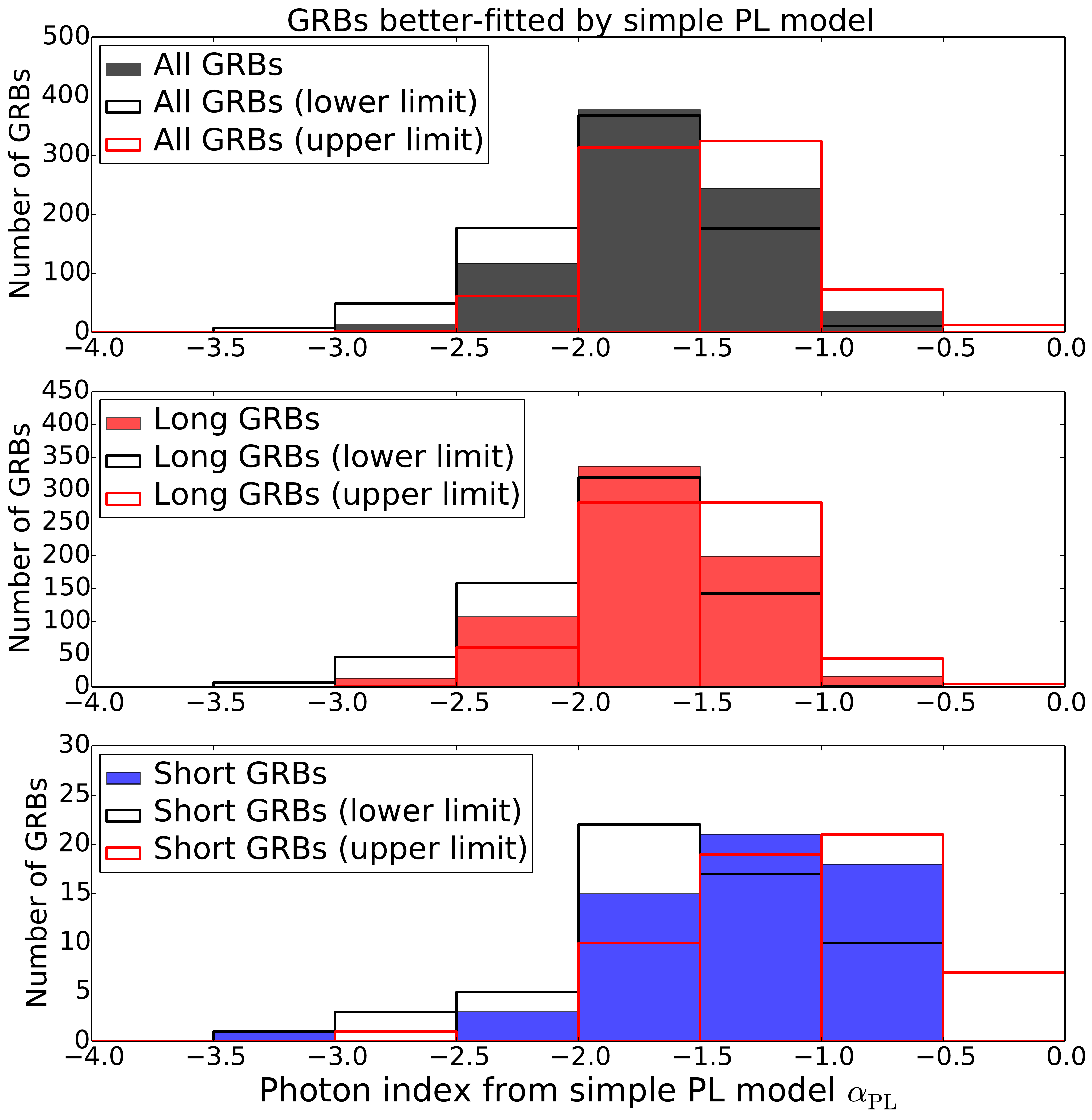}
\end{center}
\caption{
Distributions of the GRB power-law indices $\alpha_{PL}$ for those bursts that are better fitted by simple PL model.
Distributions using the upper and lower bounds of the $\alpha_{PL}$ value uncertainty range are also plotted for comparison.
}
\label{fig:alpha_PL}
\end{figure}

To further explore the difference in spectral hardness of the short and long bursts,
we plot the histogram of the photon index $\alpha$ for the bursts that are better fitted by the simple PL model
in Fig~\ref{fig:alpha_PL}. 
The upper panel shows the distribution for all
787
GRBs that have acceptable spectral fits and are better fitted by the simple PL model.
The middle and bottom panel show the distributions for long and short GRBs, respectively. We need the $T_{90}$ information
to distinguish short and long bursts. Thus, GRBs without available values of $T_{90}$ and/or $T_{90}$ errors are excluded 
from these two panels. Moreover, GRBs without complete burst durations (only lower limits reported) are also excluded.
The figure shows that the short bursts are slightly harder (i.e. higher $\alpha_{\rm PL}$) than long bursts, but 
the difference is not significant. 
There are
671
long GRBs, and
58
short GRBs in these two panels.

\subsubsubsection{BAT sensitivity on GRB detections}
\label{sect:BAT_sensitivity}

The BAT detector is a photon-counting instrument \citep{Barthelmy05}, and thus the sensitivity roughly
increases with $\sqrt{T}$ ($T$ is the exposure time), as the signal-to-noise ratio increases as $1/\sqrt{N}$ ($N$ is the number of photons) and the 
number of photons N increases as time $T$ (for BAT, the photon $N$ is dominated by background photons, and thus roughly increase linearly with $T$).

\begin{figure}[!h]
\begin{center}
\includegraphics[width=0.49\textwidth]{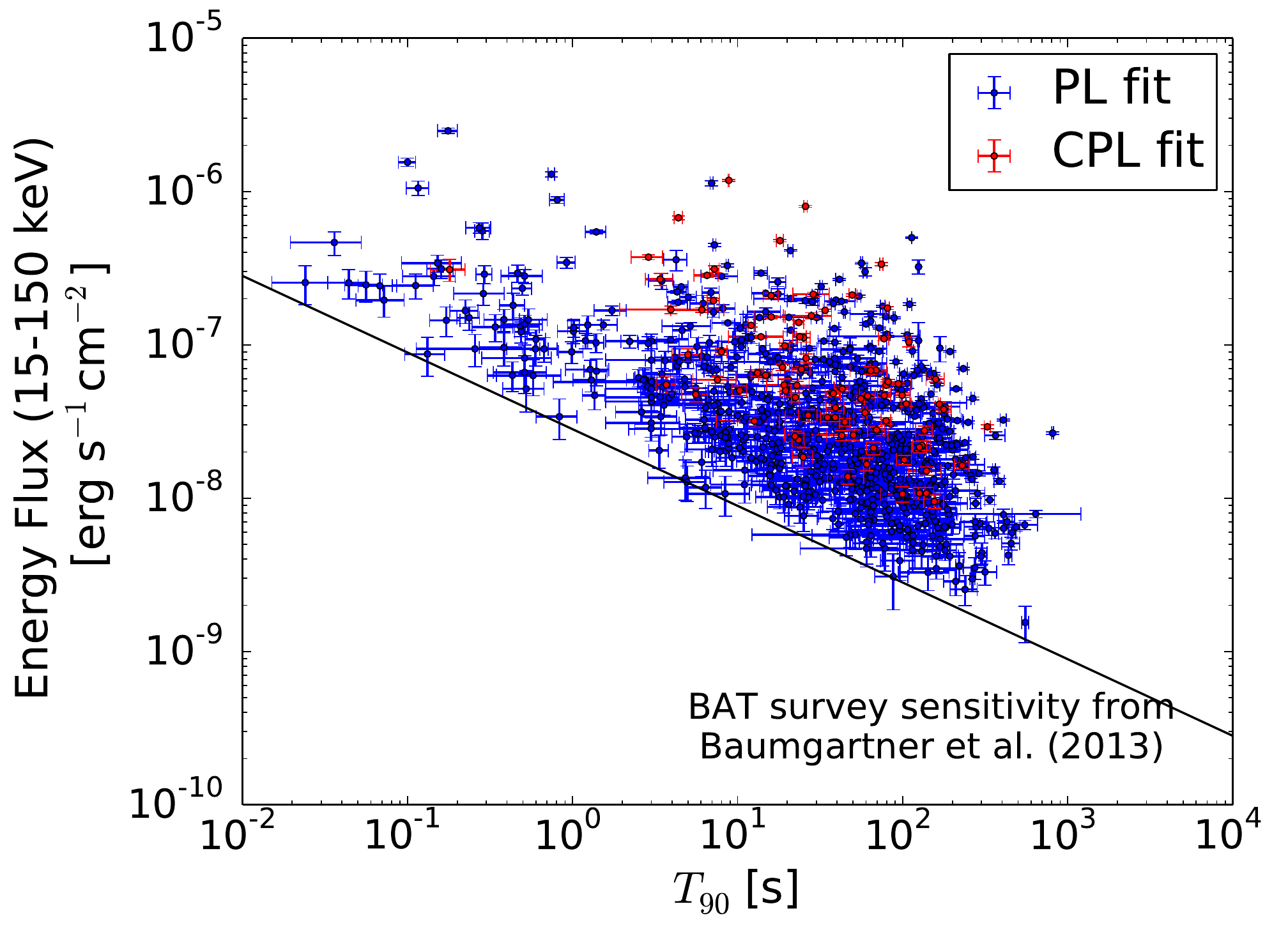}
\includegraphics[width=0.49\textwidth]{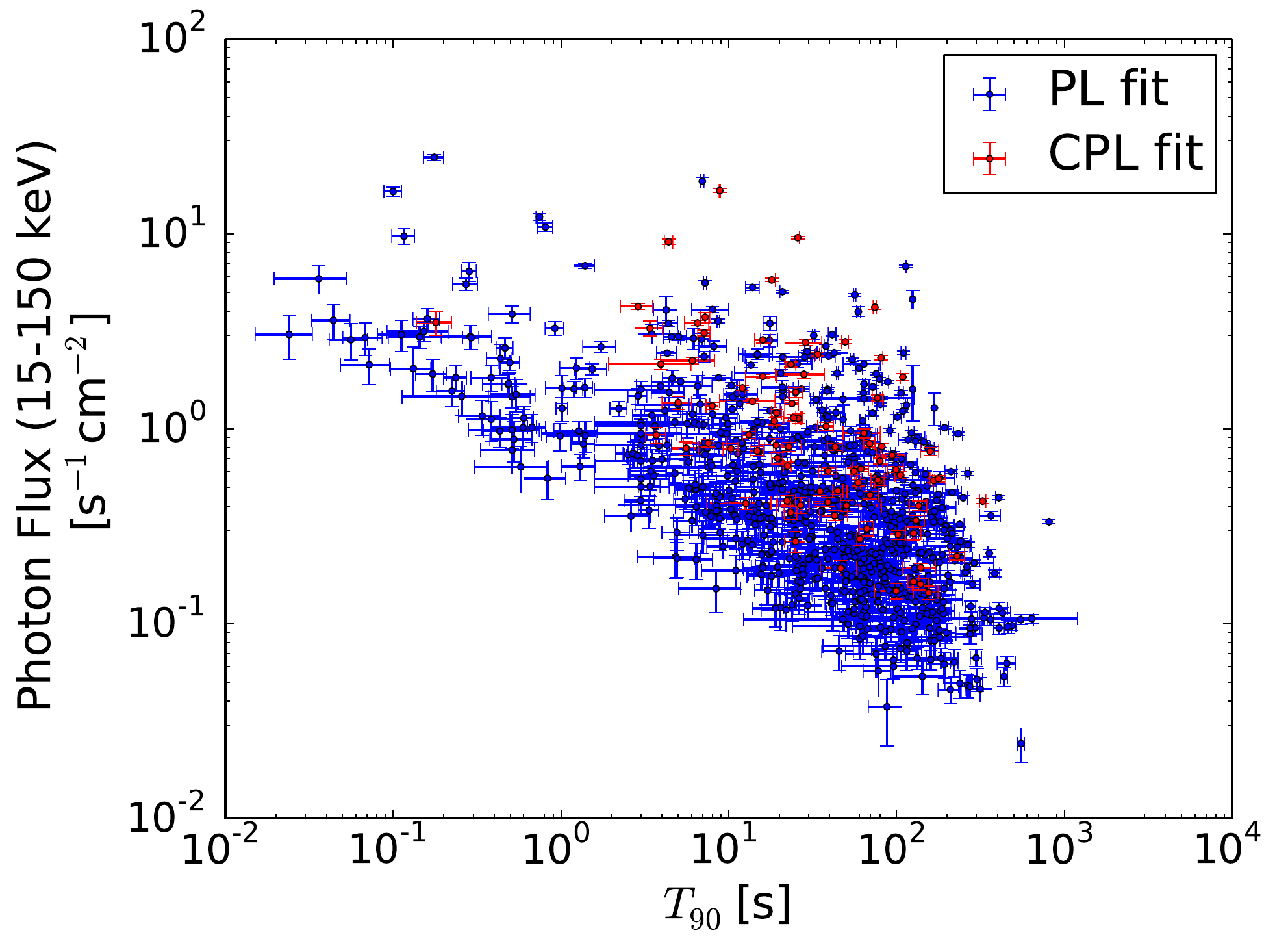}
\end{center}
\caption{
{\it Left panel:} Time-average energy flux (15-150 keV) vs. $T_{90}$. 
{\it Right panel:} Time-average photon flux (15-150 keV) vs. $T_{90}$.
For both plots, the fluxes are estimated by the best-fit model (either simple PL or CPL).
The bursts that are better fitted by the CPL model are marked in red.
Note that GRBs with
unconstrained durations, such as the ultra-long GRBs, are not included in this plot.
}
\label{fig:energy_photon_flux_T90}
\end{figure}

Figure \ref{fig:energy_photon_flux_T90} shows this effect by plotting the correlation between the energy/photon fluxes of the BAT-detected GRBs
versus the burst durations $T_{90}$\footnote{For historical reason, the time-averaged flux reported by the BAT team always refers to the one in the $T_{100}$ range instead of the $T_{90}$ range,
while the burst durations are reported in $T_{90}$. For most of the bursts, we do not expect the average flux using the $T_{100}$ range to differ significantly from
the one using the $T_{90}$ range, since the $T_{90}$ range includes the majority of the burst emission.}. 
The fluxes are estimated by the best-fit model (either the simple PL or the CPL).
The bursts that are better fitted by the CPL model are marked in red.
The figure shows a clear correlation of the minimum fluxes of the detected GRBs and the burst durations.
This correlation of the energy flux and $T_{90}$ is very similar to BAT sensitivity (as a function of exposure time) derived in \citet{Baumgartner13} (Eq. 9),
despite that in \citet{Baumgartner13} the sensitivity is derived for non-GRB sources with Crab-like spectra, and for a signal-to-noise ratio of $5 \sigma$ 
(instead of the $\sim 6.5$ to $7 \sigma$ threshold used for GRB detections).
However, note that this plot includes only the GRBs with ``acceptable spectral fits'' (as defined by criteria described in Section~\ref{sect:event_standard}).
Thus, this plots might exclude dim bursts that do not have data with low enough uncertainties to constraint the fits.

In addition, due to the complexity of the BAT trigger algorithm, this correlation between the minimum detectable fluxes with 
the burst durations should only be treated as an approximation.
For example,
the burst durations are not usually identical to the actual exposure time 
used by the trigger algorithm for detecting the burst, because the trigger algorithm might not correctly bracket the burst period. 
Thus, this correlation
does not necessary mean that GRBs with fluxes above this line
will be certainly detected, since the foreground period of the trigger algorithm needs 
to first correctly select the optimal period that maximizes the signal-to-noise ratio \citep{Lien14}.
Moreover, if the GRB flux decays significantly with time so that the average flux decreases faster than $T^{-1/2}$,
there would be no gain in the signal-to-noise ratio by increasing the exposure time.  

The figure also shows that bursts that are better fitted by the CPL model tend to have higher fluxes.
This is because a burst needs to be bright enough
to obtain a decent spectrum (i.e., with smaller uncertainties in each energy bin) that is capable of distinguishing the two models.

\begin{figure}[!h]
\begin{center}
\includegraphics[width=0.49\textwidth]{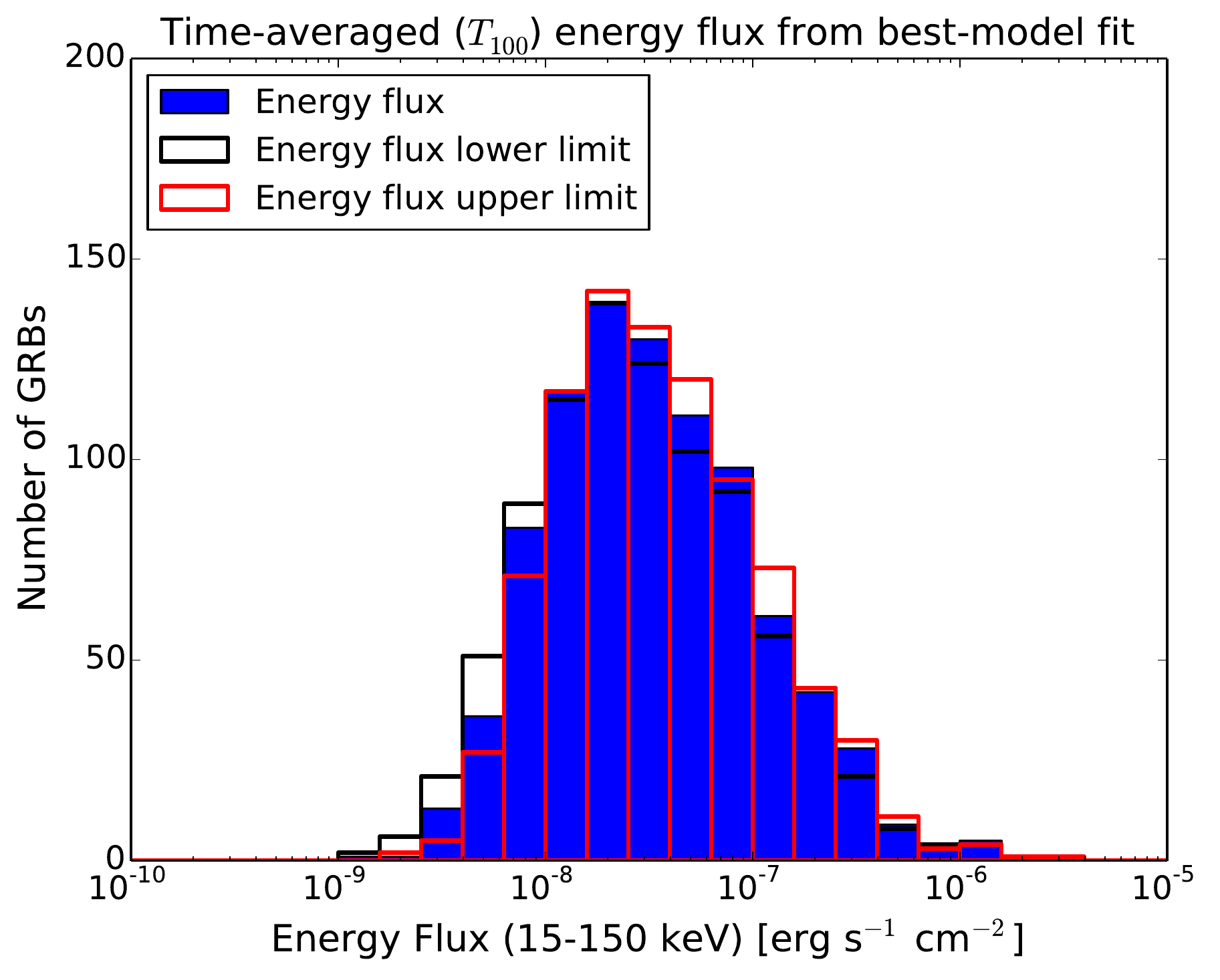}
\includegraphics[width=0.49\textwidth]{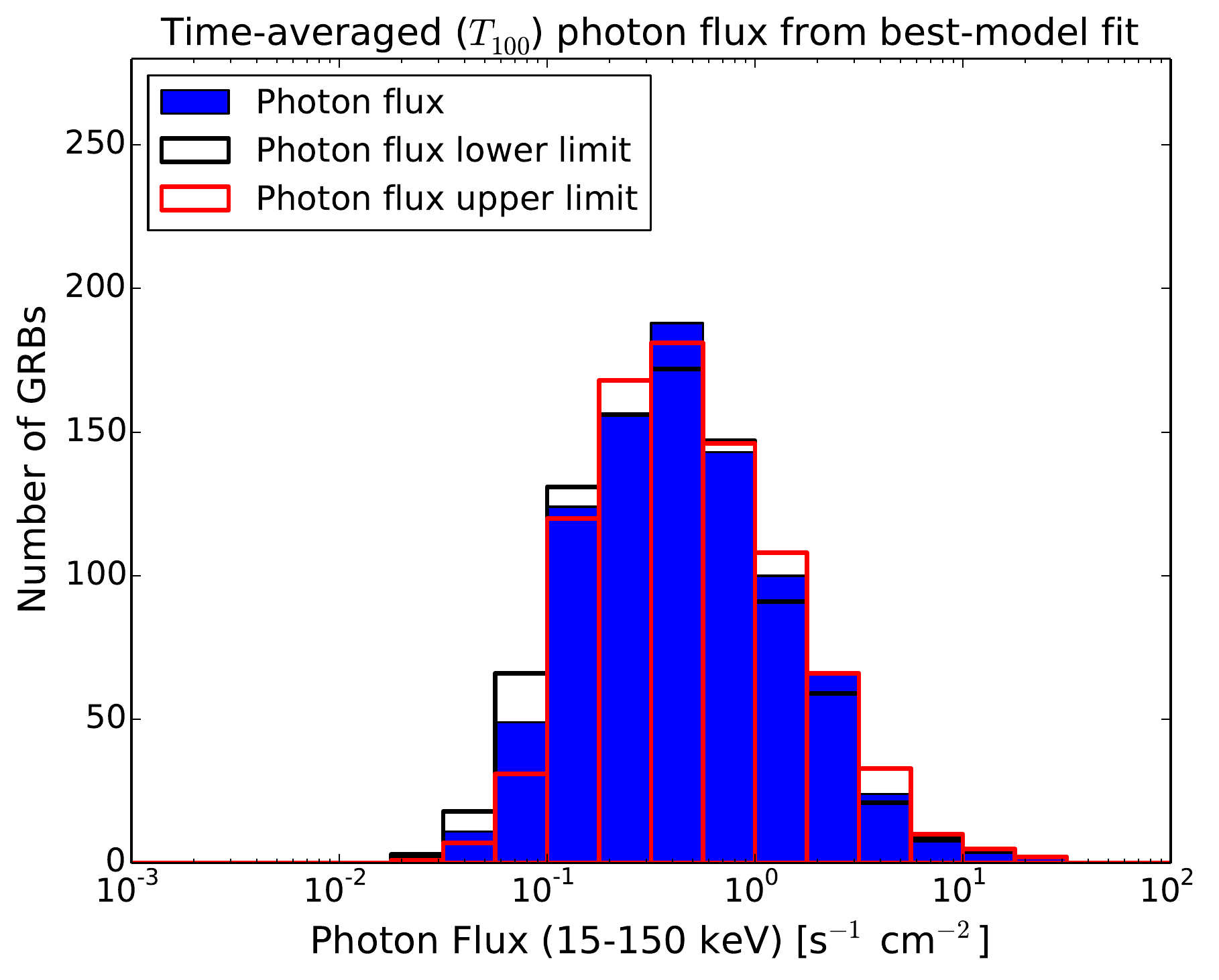}
\end{center}
\caption{
{\it Left Panel:} Distributions of the GRB energy flux (15-150 keV) estimated from best-fit model (either the simple PL or CPL).
{\it Right Panel:} Distributions of the GRB photon flux (15-150 keV) estimated from best-fit model (either the simple PL or CPL).
Distributions using the upper and lower bounds of the flux uncertainty range are also plotted for comparison.
}
\label{fig:energy_photon_flux_T100}
\end{figure}

Figure \ref{fig:energy_photon_flux_T100} shows the distribution of energy fluxes (left panel) and photon fluxes (right panel) for all the
877
bursts with acceptable spectral fits. The fluxes are estimated from 
the best-fit model (either the simple PL for CPL).
The distributions for both the energy and photon fluxes are roughly Gaussian, with 
an average of
$5.90 \times 10^{-8}$
 $\rm erg \ s^{-1} \ cm^{-2}$ 
and
0.75
 $\rm ph \ s^{-1} \ cm^{-2}$
for the energy and photon fluxes, respectively.
Again, because we only plots those bursts with acceptable spectral fits, 
the weak bursts with lower fluxes are likely to be removed from this sample.

\subsubsubsection{BAT selection effects on GRB spectral shapes}
\label{sect: BAT_selection_effect_spectrum}

Both the theoretical predictions from the synchrotron shock model \citep[][and reference therein]{Rees92, Preece98}
and empirical fits from observations with instruments that have wide-energy coverages (e.g., BATSE and {\it Fermi})
suggest that the GRB spectrum (photon flux as a function of photon energy) can be roughly described by a power-law decay at lower energy, followed by 
some steepening after the energy $E_{\rm peak}$, the peak energy in the $\nu F_\nu$ spectrum, where $F_\nu$ is the energy flux density.

\begin{figure}[!h]
\begin{center}
\includegraphics[width=0.8\textwidth]{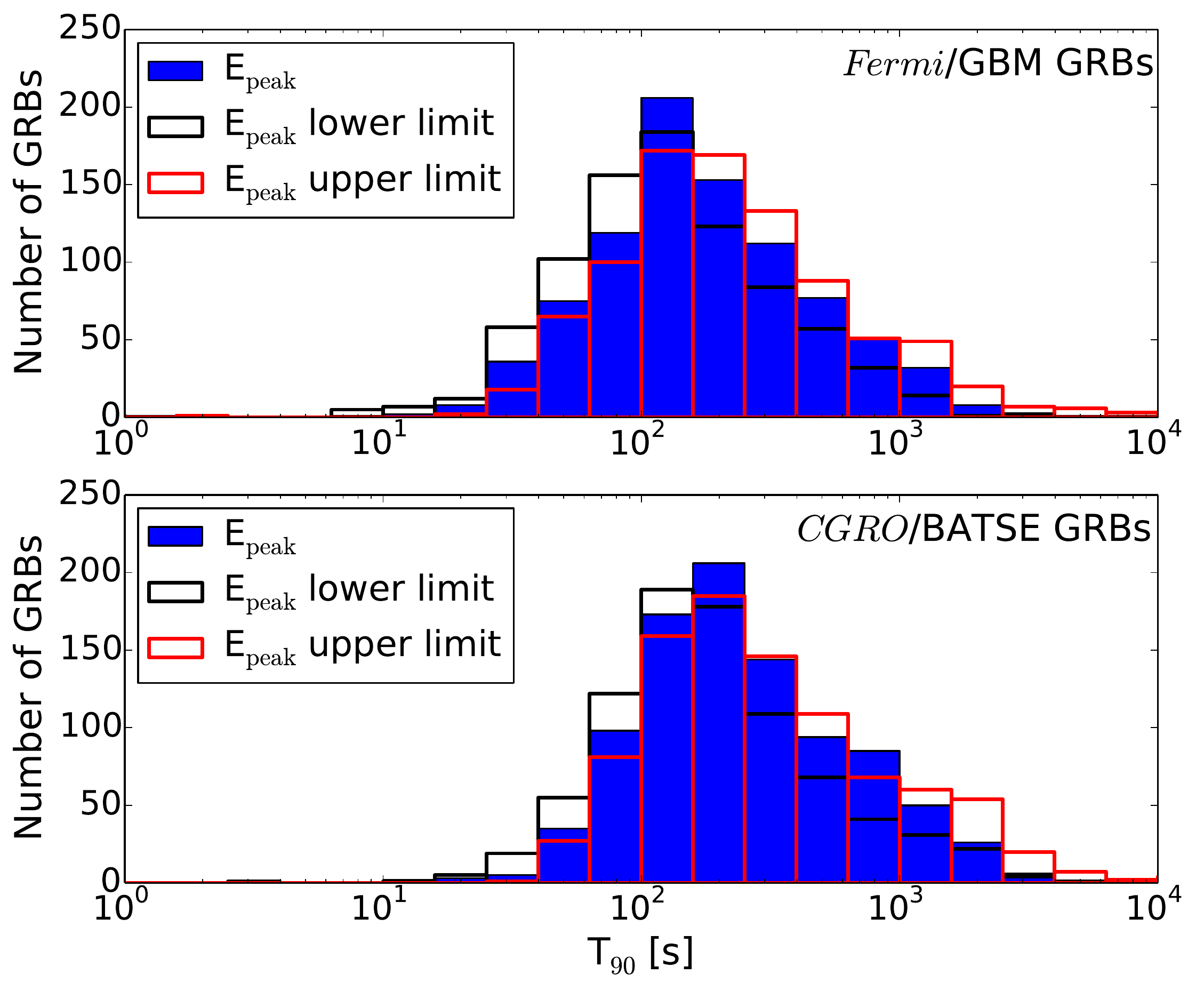}
\end{center}
\caption{
$E_{\rm peak}$ distributions for GRBs detected by {\it Fermi} \citep{Gruber14, vonKienlin14} and BATSE \citep{Goldstein13}.
Distributions using the upper and lower bounds of the $E_{\rm peak}$ uncertainty range are also plotted for comparison.
}
\label{fig:Epeak_other_missions}
\end{figure}

The BAT has a relatively narrow energy range. 
Therefore, it can be difficult to constrain $E_{\rm peak}$ with BAT data alone.
In fact, the $E_{\rm peak}$ distributions from the BATSE and {\it Fermi} GRB samples suggest that the $E_{\rm peak}$ distribution
peaks at around few hundreds keV (see Fig.~\ref{fig:Epeak_other_missions}). Moreover, even for those bursts with $E_{\rm peak}$ 
occurring within the BAT energy range, the narrow energy coverage also requires
a spectrum to have less uncertainty in order to be able to constrain $E_{\rm peak}$.

\begin{figure}[!h]
\begin{center}
\includegraphics[width=0.49\textwidth]{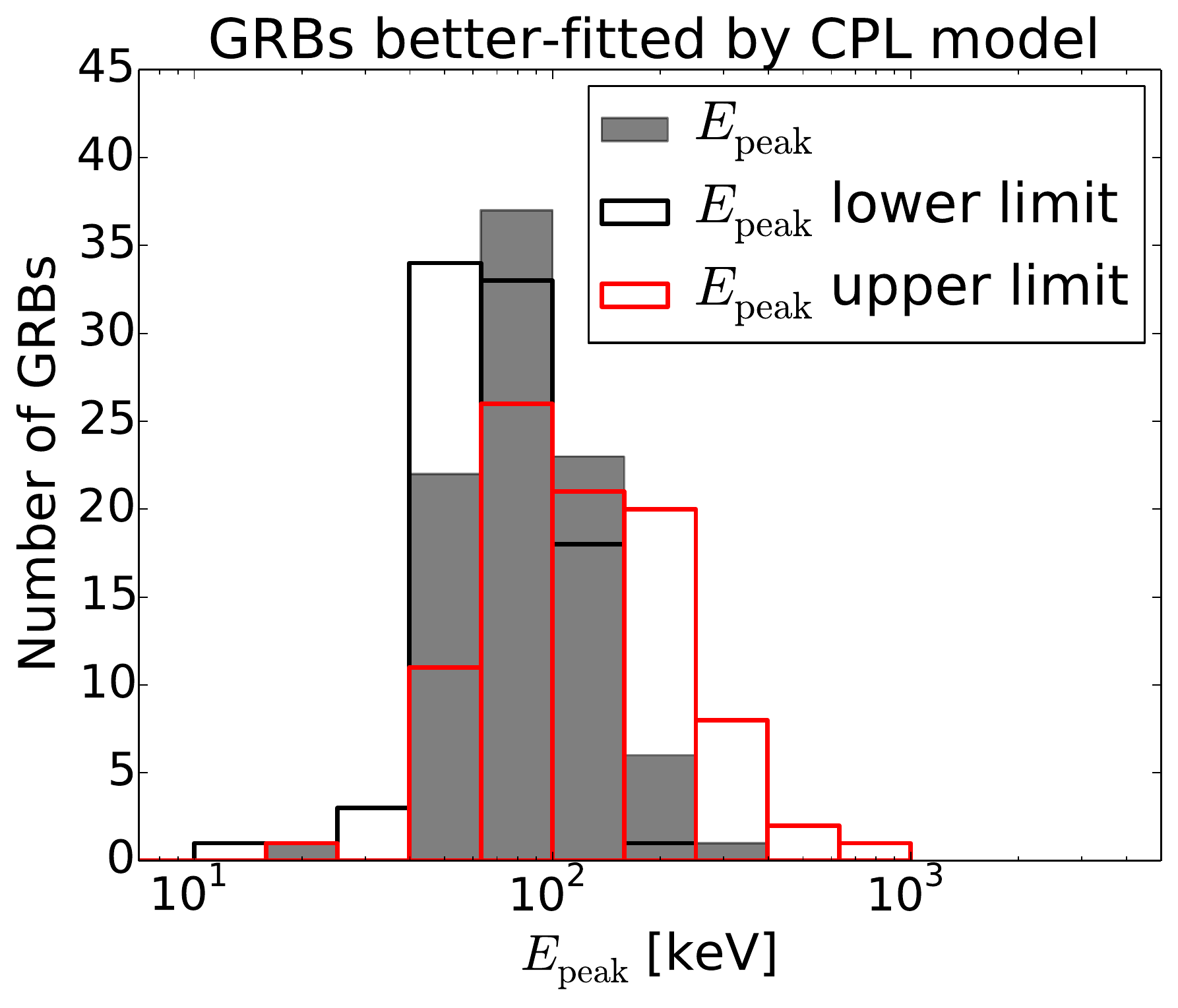}
\includegraphics[width=0.49\textwidth]{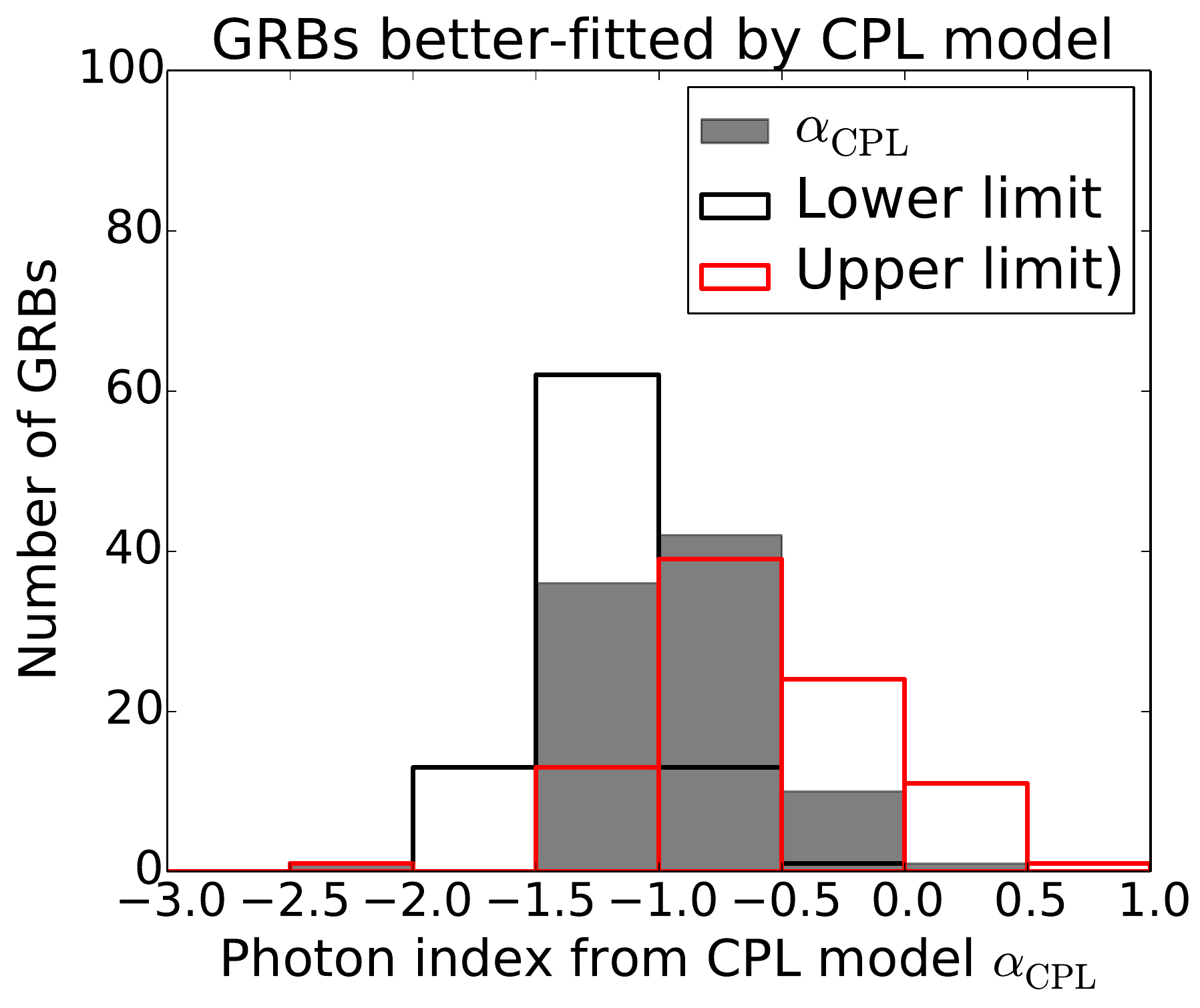}
\end{center}
\caption{
{\it Left panel}: $E_{\rm peak}$ distributions for GRBs that are better fitted by the CPL model.
{\it Right panel}: Photon index $\alpha_{\rm CPL}$ distributions for GRBs that are better fitted by the CPL model.
Distributions using the upper and lower bounds of the $E_{\rm peak}$/$\alpha_{\rm CPL}$  uncertainty range are also plotted for comparison.
}
\label{fig:Epeak}
\end{figure}

In the current BAT GRB sample, there are
90
bursts with acceptable spectra that are better fitted by the CPL model.
The $E_{\rm peak}$ and photon index $\alpha_{\rm CPL}$ distributions for these events are
plotted in the left and right panel of Fig.~\ref{fig:Epeak}, respectively.
The black and red lines show the distributions using the lower and upper limits.
As expected, all of the values of $E_{\rm peak}$ lie in the range of $\sim 10$ keV 
to $\sim 300$ keV, which is within the BAT energy range.
However, these might not be the only bursts in the BAT sample that have $E_{\rm peak}$ within the BAT energy range.
It is possible that some other bursts also have $E_{\rm peak}$ in this range, but are not bright enough to present good spectra
that can distinguish the two models \citep{Sakamoto08}.
The $E_{\rm peak}$ distribution peaks at $\sim 80$ keV (the center of the bin with the largest number of bursts),
which is similar to the one presented in the BAT2 catalog, but is
significantly different than the distributions from GRBs detected by other instruments, as shown in Fig.~\ref{fig:Epeak_other_missions} \citep{Sakamoto11, Goldstein13, Gruber14, vonKienlin14}.
The difference is likely due to instrumental biases with each instrument sensitives to a different energy range.

\begin{figure}[!h]
\begin{center}
\includegraphics[width=0.49\textwidth]{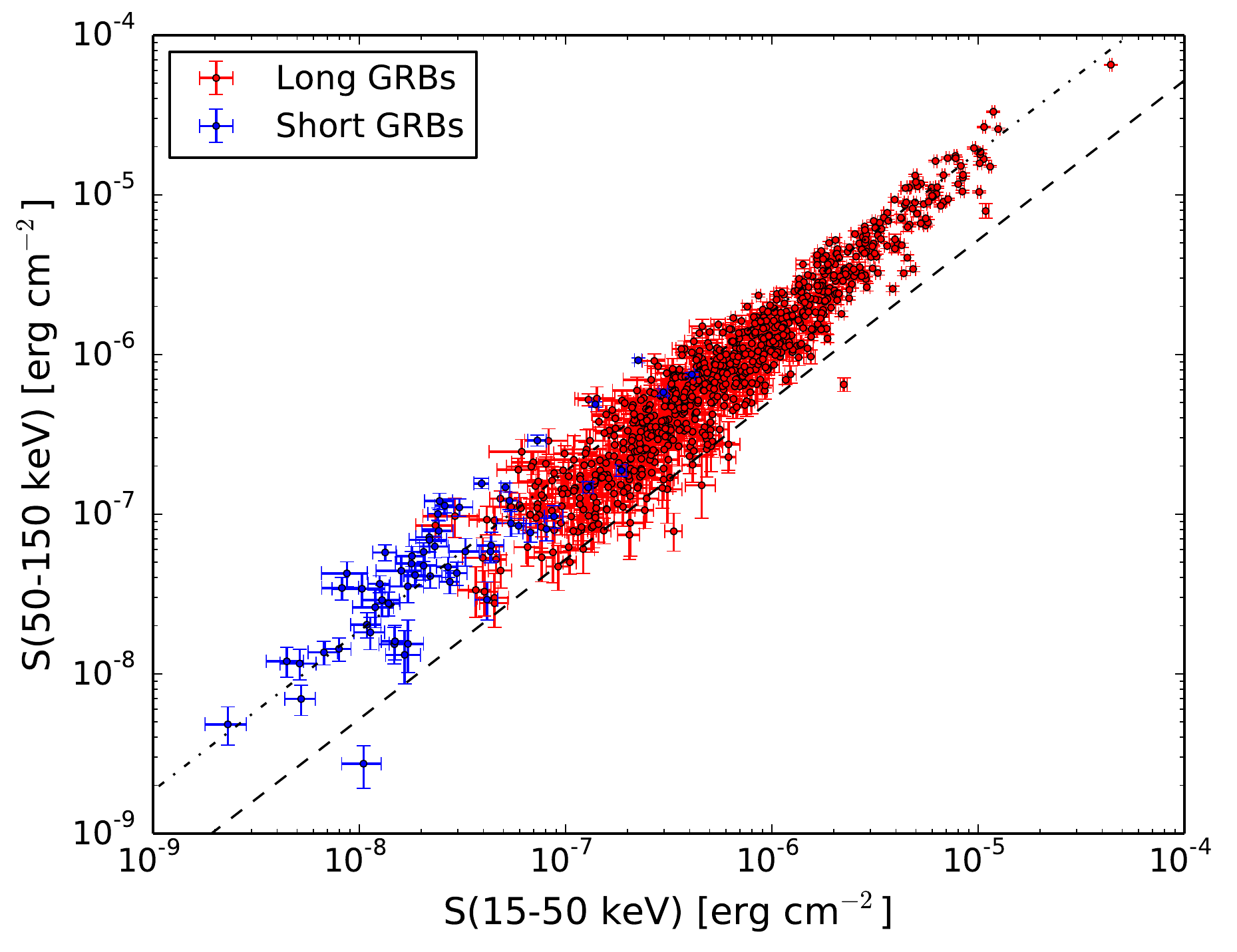}
\includegraphics[width=0.49\textwidth]{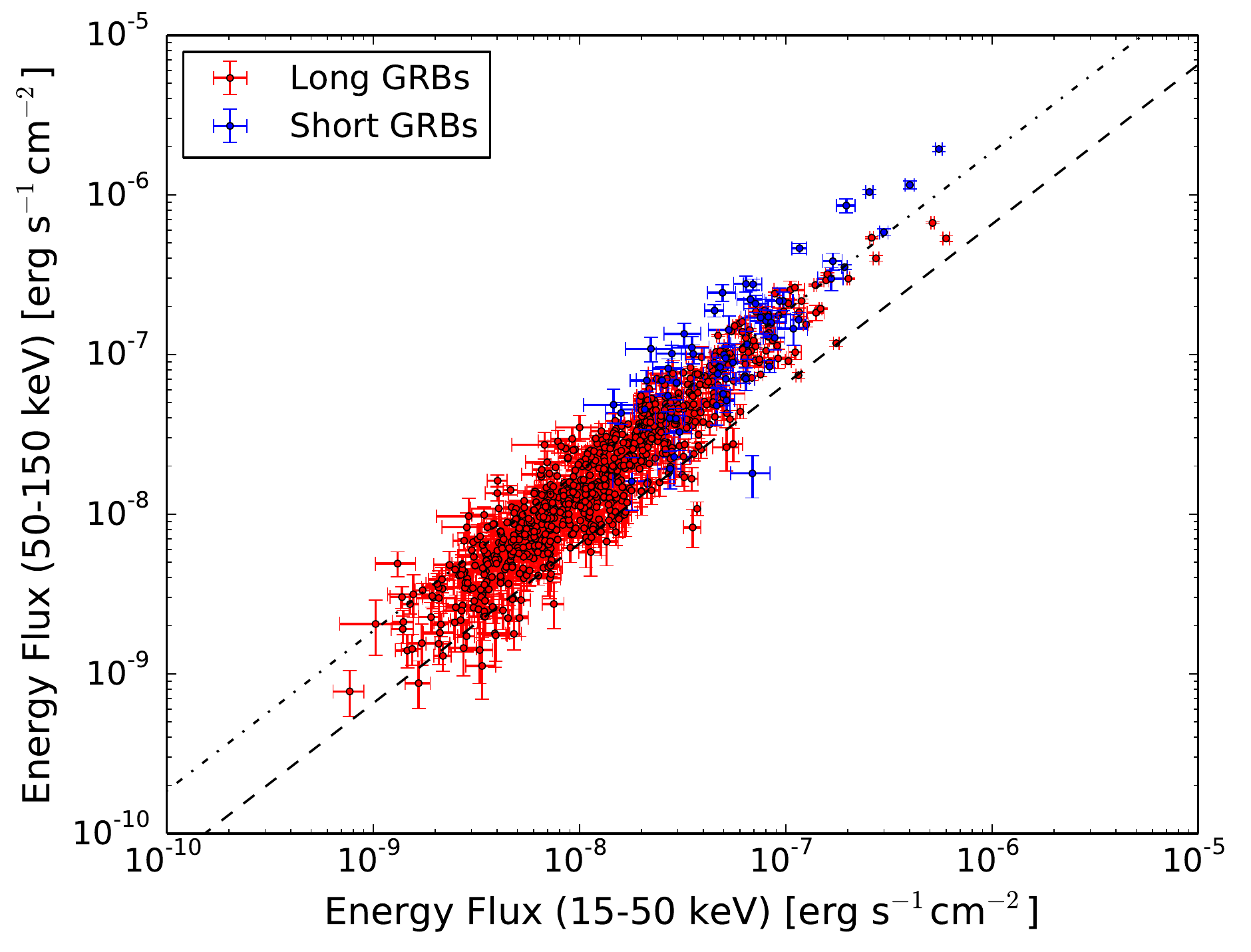}
\end{center}
\caption{
{\it Left Panel:} Fluence in the 50-150 keV range versus fluence in the 15-50 keV range. Fluences are estimated by the best-fit model (either the simple PL or CPL).
Red dot represents the GRBs with $T_{90} > 2$ s; blue points refers to GRBs with $T_{90} \leq 2$ s.
{\it Right Panel:} Similar plot as the left panel, but with energy flux instead of energy fluence.
The dash-dotted line and the dashed line traces the fluences calculated from the Band function with $E^{\rm obs}_{\rm peak} = 15$ and 150 keV, respectively.
Both lines assume a canonical values of $\alpha = -1$ and $\beta = -2.5$.
}
\label{fig:fluence_fluence_flux_flux}
\end{figure}

Both the BAT1 and BAT2 catalog have shown that most of the BAT-detected GRBs have spectral hardnesses
that are consistent with the spectral hardnesses calculated from a Band function fit with $E^{\rm obs}_{\rm peak}$ in the range of 
the BAT energy band.
Here we updated the same plot that shows the spectral hardness as in the BAT1 and BAT2 catalog with 
the new GRB detections, as shown in the left panel of Fig.~\ref{fig:fluence_fluence_flux_flux}. In addition, we also plot the spectral hardness in
flux instead of fluence in the right panel of Fig.~\ref{fig:fluence_fluence_flux_flux}.
As usual, we only include in these plots bursts with 
acceptable spectral fits and complete burst durations with valid numbers of $T_{90}$ and $T_{90}$ errors. There are
756
long bursts
(red dots) and
59
short bursts (blue dots) in these plots.
Similar to the BAT1 and BAT2 catalog, we plot the lines using the Band function with $E_{\rm peak}$ of 15 keV (dashed line)
and 150 keV (dash-dotted line), respectively. Both lines are calculated using canonical values of $\alpha = -1$, $\beta=-2.5$.
Each line traces the fluence ratios from the same $\alpha$, $\beta$, and $E_{\rm peak}$, with a range of normalizations.
Results show that $\sim 80\%$ of the bursts lie between the two lines (when including the errors of the burst fluences), 
indicating that most of the BAT-detected bursts have fluence ratios that are consistent with the ones from the Band function with $E_{\rm peak}$
lying within the BAT energy range. In other words, it is possible that these bursts have $E_{\rm peak}$ within the BAT energy range,
though most of the spectra do not have small enough uncertainties to constrain the $E_{\rm peak}$.

\begin{figure}[!h]
\begin{center}
\includegraphics[width=0.8\textwidth]{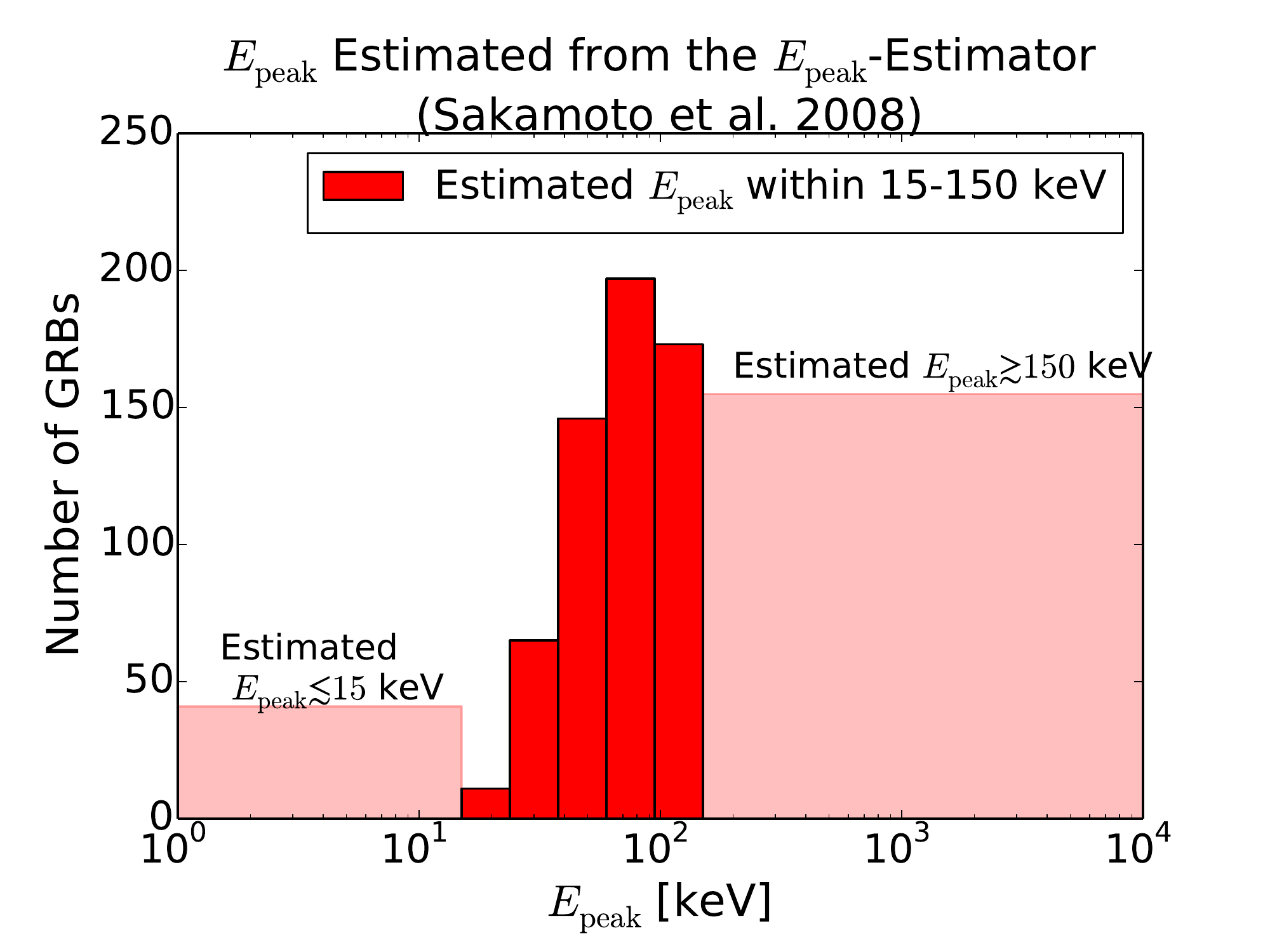}
\end{center}
\caption{
$E_{\rm peak}$ distributions for GRBs that are better fitted by the CPL model.
}
\label{fig:Epeak_estimator}
\end{figure}

Furthermore, \citet{Sakamoto08} studies the potential confusion between the simple PL, CPL, and Band function \citep{Band93} spectral fit
in the BAT observations using simulated spectra.
These authors found that most of the BAT-detected GRBs probably have $E_{\rm peak}$ within the BAT energy range,
and derived an equation to estimate $E_{\rm peak}$ (for the Band function) using the photon index from the simple PL fit.
Figure \ref{fig:Epeak_estimator} shows $E_{\rm peak}$ distribution from the $E_{\rm peak}$ estimator derived in \citet{Sakamoto08}.
Results show that majority ($\sim 78\%$) of the GRBs detected by BAT might have $E_{\rm peak}$ within the BAT energy range.
This fraction is very similar to the fraction ($\sim 80\%$) of the bursts that fall between the lines in Fig~\ref{fig:fluence_fluence_flux_flux},
which traces the fluence ratios with  $E^{\rm obs}_{\rm peak} = 15$ keV and $150$ keV, respectively.
Because the $E_{\rm peak}$ estimator only works for GRBs that have $E_{\rm peak}$ within $15-150$ keV, 
all the bursts estimated to have $E_{\rm peak}$ below or above this energy range are placed in single bins in light red.

\begin{figure}[!h]
\begin{center}
\includegraphics[width=0.8\textwidth]{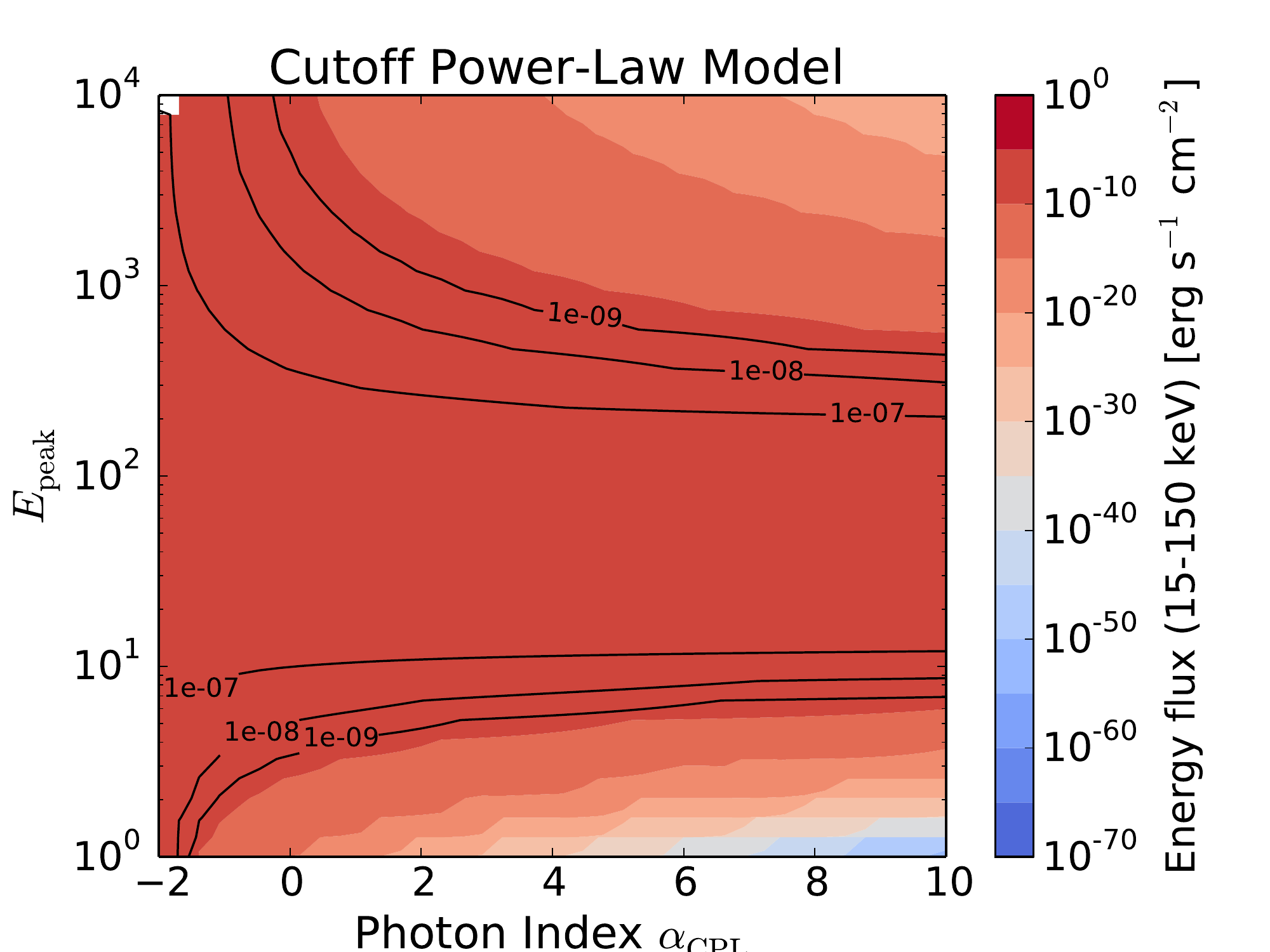}
\end{center}
\caption{
A contour plot that shows how the flux in $15-150$ keV changes as a function of $E_{\rm} peak$ and $\alpha_{\rm CPL}$ in the CPL model.
}
\label{fig:flux_contour}
\end{figure}

For bursts with similar total energies, 
it is reasonable to expect that BAT is most sensitive to those events that have $E_{\rm peak}$ within the instrument's energy range,
because in such cases most of the energy of these bursts are distributed in the detectable energy range.
However, it is also possible to have a burst with $E_{\rm peak}$ outside of the BAT energy being detectable 
because it has higher fluxes overall, and
it is not completely trivial how much the energy budget arrangement is sensitive to different 
spectral shapes.
To quantify the energy budget in the BAT energy range, we calculate the energy flux in $15-150$ keV for a range of $E_{\rm peak}$ and photon indices
in the CPL model, but with a fixed total flux (i.e., we assume a flux of $7.21 \times 10^{-7} \ \rm erg \ s^{-1} \ cm^{-2}$ from $1-10000$ keV, which is a relatively typical flux for GRB 
and it can be easily scaled up or down for bursts with different total flux/normalization).
Figure \ref{fig:flux_contour} shows the contour plot of how flux changes as a function of $E_{\rm peak}$ and photon indices.
We make the contour plot with the CPL model rather than the Band function simply because the Band function has one more parameter, which 
makes it difficult to present in a 2-dimensional plot. Moreover, since we are focusing on the flux in the narrow BAT energy range, 
the CPL model should be a good-enough approximation.
We show an extremely wide range of $\alpha_{\rm CPL}$ in this plot for demonstration. However, observations from 
BATSE, {\it Fermi}, and {\it Swift} suggest that the value never exceeds 1 \citep{Sakamoto11, Goldstein13, Gruber14, vonKienlin14}.
Moreover, theoretical predictions from the synchrotron shock model enforce that $\alpha_{\rm CPL} < -2/3$ \citep[][and references therein]{Rees92, Preece98}.
Therefore, one can see in the plot that in the reasonable range of $\alpha_{\rm CPL}$ from $\sim -2$ to $\sim 1$, GRBs with normal energy output are indeed most likely 
to be detected by BAT if they have $E_{\rm peak}$ in the BAT energy range. For GRBs with much higher $E_{\rm peak}$, they would need to be 
roughly one or two orders of magnitude brighter in order to be detected, and the consequences become more significant at larger $\alpha_{\rm CPL}$.

When comparing Fig.~\ref{fig:Epeak} and Fig.~\ref{fig:alpha_PL} for the spectral index distributions from the simple PL and CPL fits,
one would notice that in general the spectral indices from the CPL fits are higher than those from the simple PL fits.
This can be explained if most of the bursts actually have $E_{\rm peak}$ within the BAT energy range.
In such cases, the photon index from a simple PL fit will be an average of the true power law index before and after $E_{\rm peak}$,
and thus will be lower than the real first slope of the spectrum.

For those bursts that are better fitted by the CPL model, all except two have their lower limit of $\alpha_{\rm CPL}$ greater than -2/3. Therefore,
almost all the bursts are consistent with the synchrotron shock model.
The only two GRBs with the lower limits higher than -2/3 are GRB050219A and GRB130420B,
and the $\alpha_{\rm CPL}$ range for these two bursts are (-0.41, 0.18) and (-0.61, 0.44), respectively.
All the bursts with acceptable fits that are better fitted by the simple PL model also have spectral indices that are consistent with the the synchrotron shock model.
However, the comparison using the simple PL fit might not be physically meaningful, 
if the simple PL model does not represent the true underlying spectral shape due to the uncertainty in the data.



\subsubsection{Peak spectra}

For the 1-s peak spectra, there are only
728
bursts that have acceptable spectral fits
due to the smaller number of photon counts in the 1-s duration.
Within these bursts with acceptable spectral fits, there are 
68
GRBs that are better fitted by the CPL model.

There are
51
bursts with $T_{100}$ shorter than one second.
For these extremely short bursts, the 1-s peak flux is likely
to include some background intervals, and the 20-ms peak flux discussed below (and reported in Section \ref{sect:20ms_tables}) is probably better represent the true peak flux. 
Nonetheless, we still present the 1-s peak flux for GRBs with $T_{100} < 1$ s here for completeness,
and also because of the uncertainties in the burst duration measurements.

\subsubsubsection{Comparison with the time-averaged ($T_{100}$) spectra}

There are 
542
bursts that are better fitted by the simple PL model
for both the $T_{100}$ spectra and the 1-s peak spectra;
22
bursts that are better fitted by the CPL for 
both of the $T_{100}$ spectra and the 1-s peak spectra; 
42
bursts that are better fitted by the PL model for the $T_{100}$
spectra but change to the CPL fit for the 1-s peak spectra;
and
50
bursts that are better fitted by the CPL model for the $T_{100}$
but switch to the simple PL fit for the 1-s peak spectra.
Those bursts that switch between models for the two different spectral fits
usually either have $\Delta \chi^2 \sim 5$ (close to our threshold for adopting the CPL model
at  $\Delta \chi^2  = 6$) for one of the spectra, 
or the fits have fairly low null probability (close to our threshold of 0.1 for rejecting the fit).
Note that only one burst with $T_{100} < 1$ s (GRB081101) is better fitted by different models
for the time-averaged spectrum and the 1-s peak spectrum. To be specific, the time-averaged spectrum for this burst is better fitted by the CPL model, 
while the 1-s peak spectrum is better fitted by the simple PL model,
likely because the 1-s peak spectrum includes a larger interval than the true $T_{100}$ range and thus
is contaminated by some background photons.

\begin{figure}[!h]
\begin{center}
\includegraphics[width=0.6\textwidth]{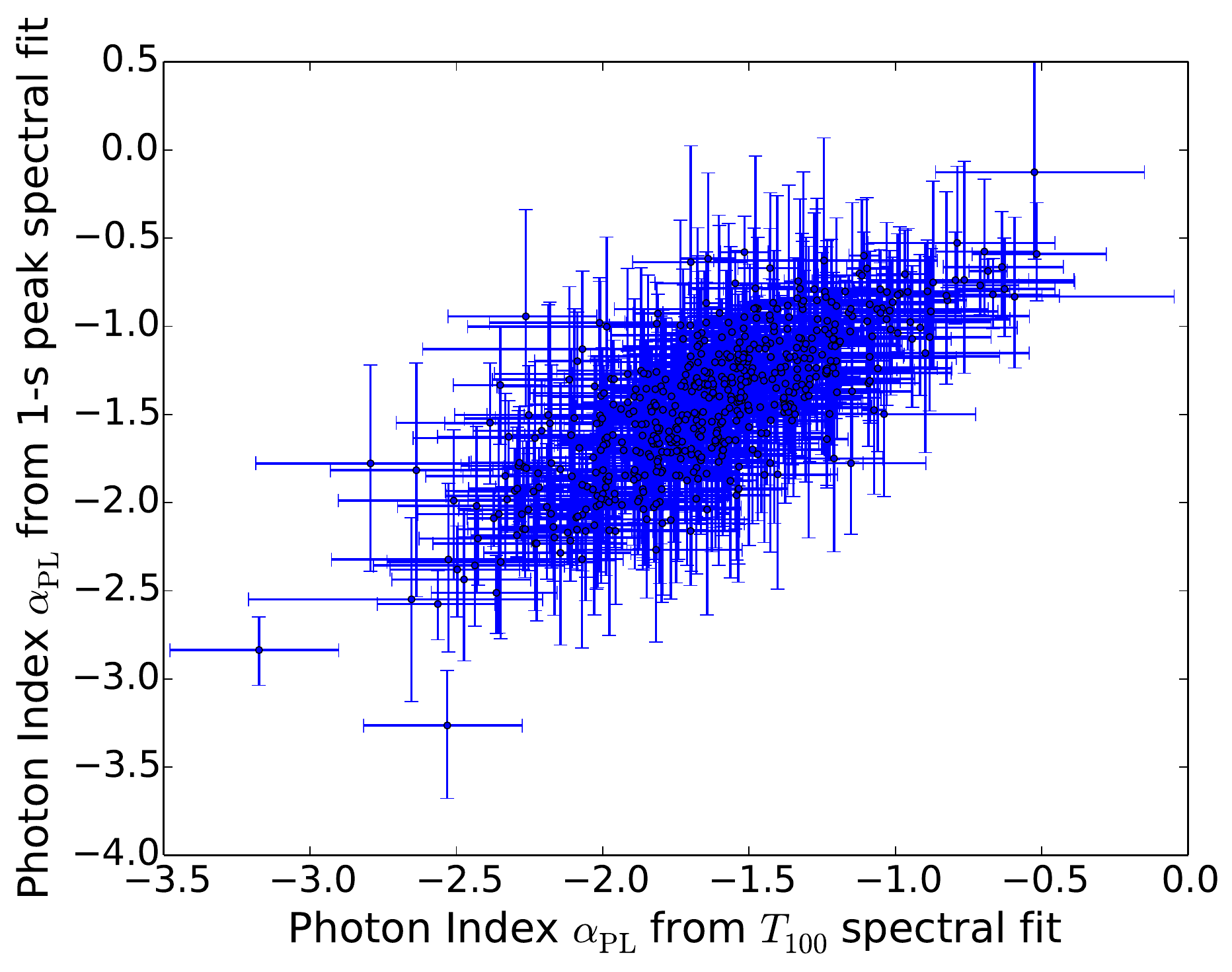}
\end{center}
\vspace{-15pt}
\caption{
Comparison between the photon indices $\alpha_{\rm PL}$ from the $T_{100}$ spectral fits 
and the 1-s peak spectral fits.
}
\label{fig:alpha_compare}
\end{figure}

Figure \ref{fig:alpha_compare} compares the photon indices $\alpha_{\rm PL}$ 
from the $T_{100}$ and 1-s peak spectral fits. The results
show that the fits from two different kinds of spectra gives similar
 $\alpha_{\rm PL}$. 
Several studies have shown that it is not uncommon that the spectral evolution 
follows the shape of the light curve \citep[e.g.,][]{Zhang07,Zhang11}. Therefore, one might expect
that the 1-s peak spectrum is harder (i.e., has larger $\alpha_{\rm PL}$) than
the time-averaged spectrum. Due to the large uncertainties in the values of $\alpha_{\rm PL}$,
it is difficult to determine whether such trend exists. However, we do notice this trend 
of spectral evolution for some brighter bursts with decent spectrum (see the ``Spectral Evolution'' plot in the 
individual burst webpage\footnote{http://swift.gsfc.nasa.gov/results/batgrbcat/index.html}).

\subsubsubsection{BAT detection limit with the 1-s flux}

\begin{figure}[!h]
\begin{center}
\includegraphics[width=0.49\textwidth]{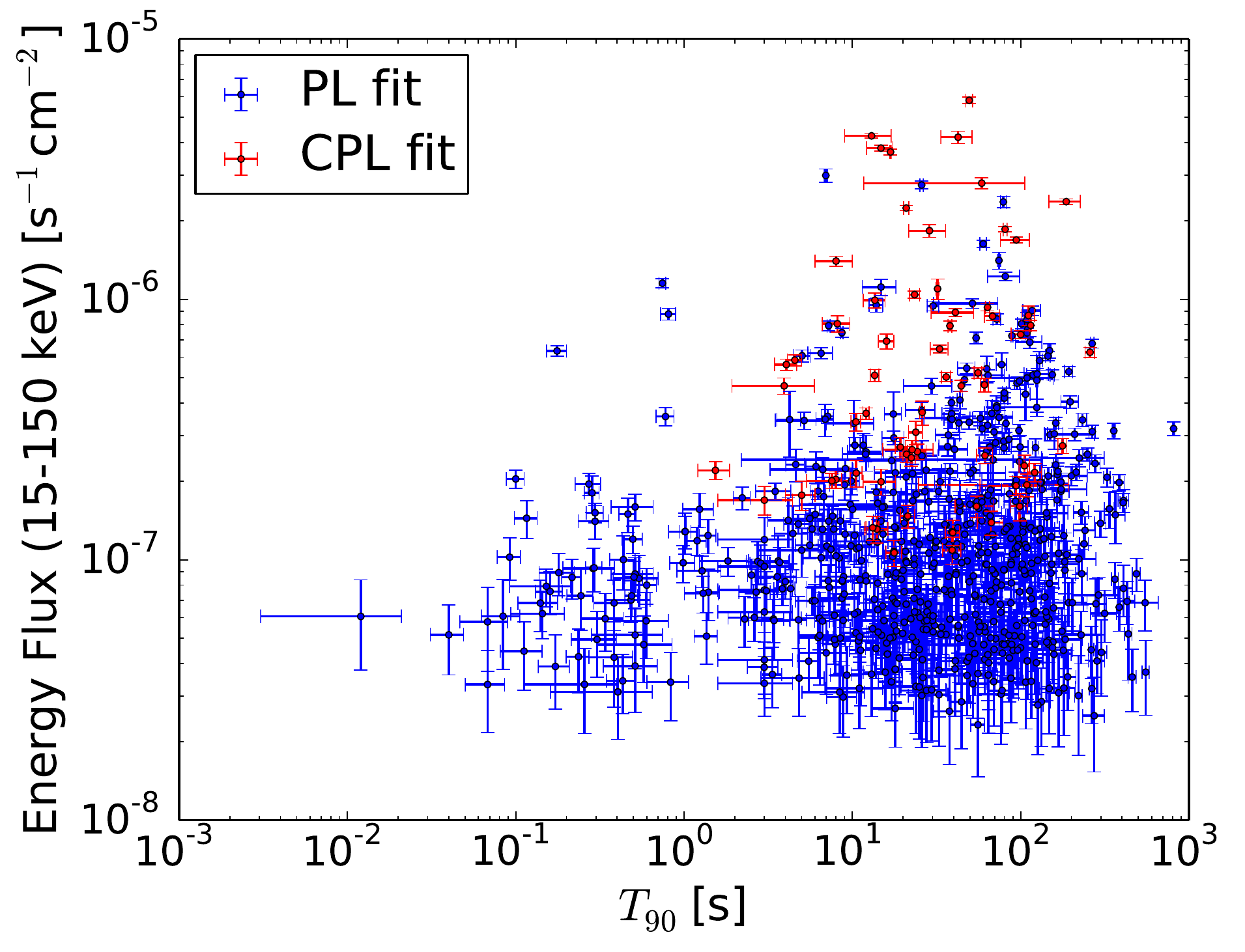}
\includegraphics[width=0.49\textwidth]{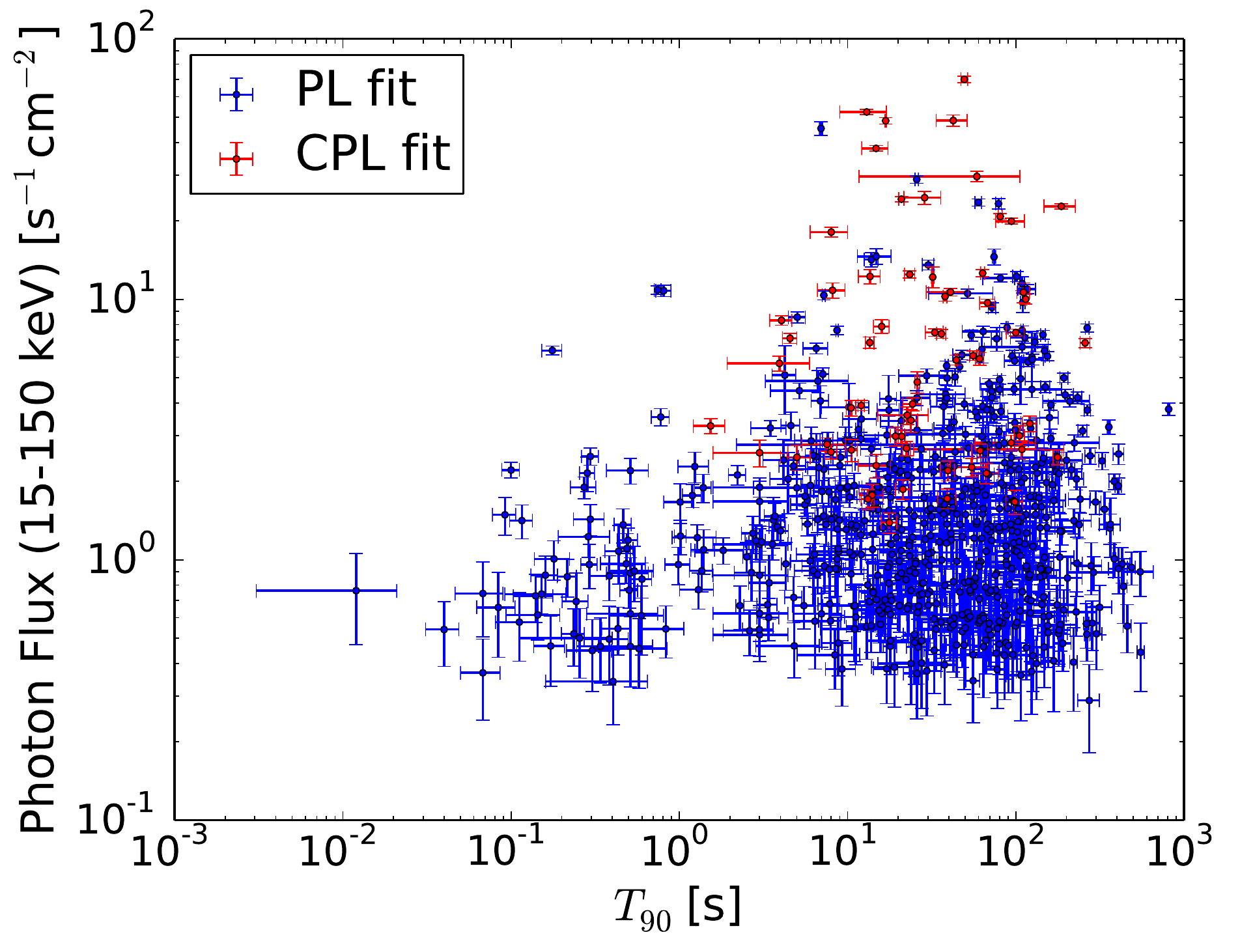}
\end{center}
\caption{
{\it Left panel:} 1-s peak energy flux (15-150 keV) vs. $T_{90}$. 
{\it Right panel:} 1-s peak photon flux (15-150 keV) vs. $T_{90}$.
For both plots, the fluxes are estimated by the best-fit model (either simple PL or CPL).
The bursts that are better fitted by the simple PL model are marked in blue, and 
GRBs that are better fitted by the CPL model are marked in red.
}
\label{fig:1speak_energy_photon_flux_T90}
\end{figure}

Similar to Fig.~\ref{fig:energy_photon_flux_T90}, we plot in Fig~\ref{fig:1speak_energy_photon_flux_T90} the 1-s peak energy flux (left panel) and the photon flux (right panel) as a function of $T_{90}$ 
to show the BAT detection limit with the 1-s flux.
As expected, there is no obvious correlation between the minimum 1-s peak energy/photon flux with respect to $T_{90}$.
Also, results show that the BAT sensitivity to 1-s flux is $\sim 3 \times 10^{-8} \ \rm erg \ s^{-1} \ cm^{-2}$ (or $\sim 0.3 \ \rm s^{-1} \ cm^{-2}$ for photon flux),
which are similar to the detection limit shown in Fig.~\ref{fig:energy_photon_flux_T90}.
Similar to Fig.~\ref{fig:energy_photon_flux_T90}, the bursts that are fitted-better by the CPL model have higher minimum fluxes.

\subsubsubsection{20-ms peak fluxes for short GRBs}

Because most of the short bursts are shorter than one second,
we also generate peak spectra for the 20-ms duration 
to have values that better represent the peak fluxes for short bursts.
As mentioned in Sect~\ref{sect:event_standard},
the 20-ms peak spectra are made with larger energy bins and only ten energy bands,
to have reasonable number of counts in each energy bin.
However, despite the larger energy bins used, there are only
226
bursts with acceptable spectral fits for the 20-ms spectral analyses, which is significant lower than
those from the 1-s peak spectral analyses and the time-averaged $T_{100}$ spectral analyses.
Only five GRBs (GRB140209A, GRB110715A, GRB101023A, GRB060117, and GRB050525A) have the 20-ms spectrum better fitted by the CPL model.

\begin{figure}[!h]
\begin{center}
\includegraphics[width=0.6\textwidth]{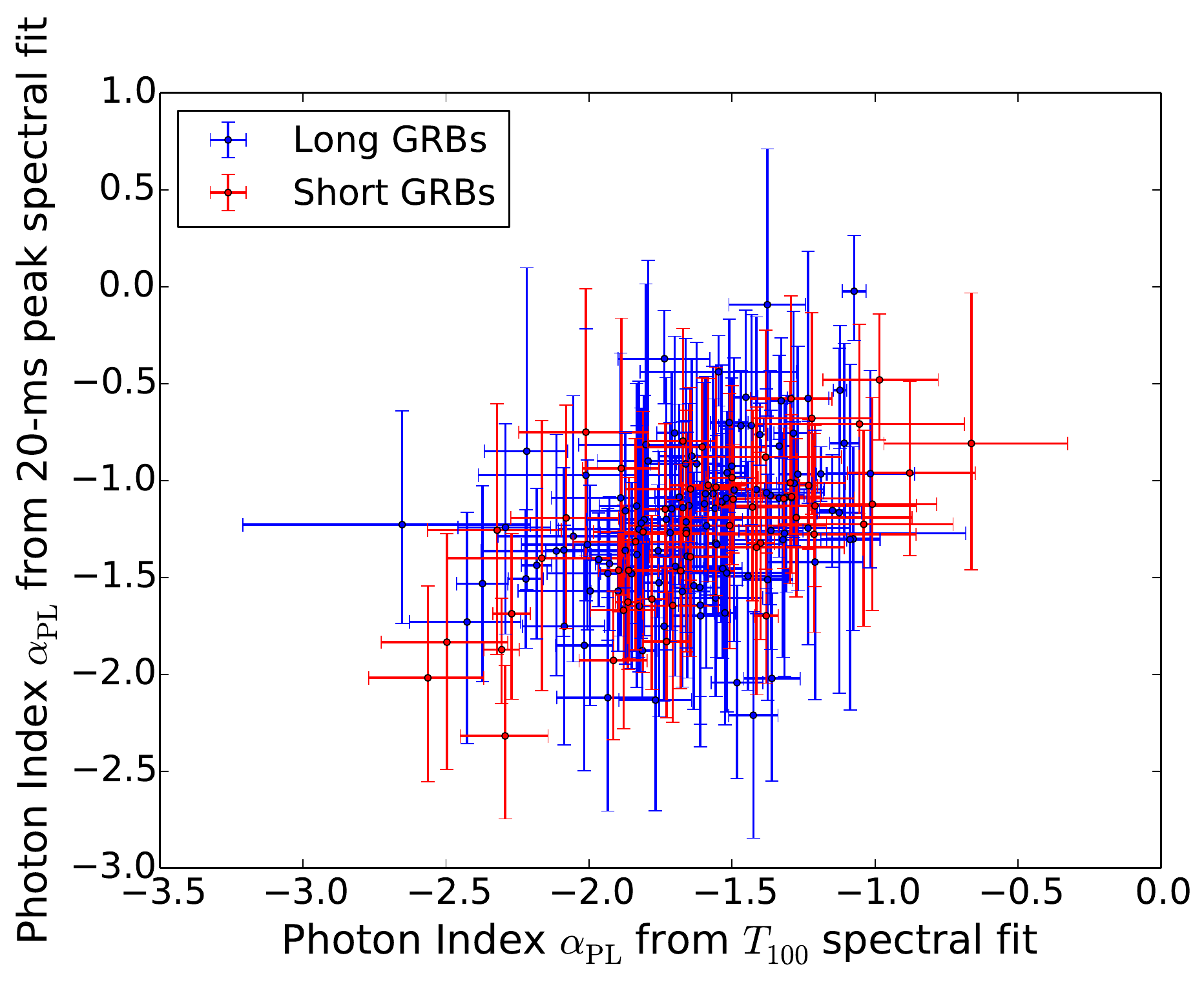}
\end{center}
\caption{
Comparison between the photon indices $\alpha_{\rm PL}$ from the $T_{100}$ spectral fits 
and the 20-ms peak spectral fits.
Long GRBs are shown in blue, and short GRBs are marked in red.
}
\label{fig:T100_alpha_20ms_alpha}
\end{figure}

Figure~\ref{fig:T100_alpha_20ms_alpha} 
shows the correlation between the photon indices $\alpha_{\rm PL}$
from the 20-ms peak spectral analyses and the time-averaged $T_{100}$ spectral analyses.
Comparing the the similar plot made for the 1-s peak spectral fits,
the correlation between the  $\alpha_{\rm PL}$ from the 20-ms peak and $T_{100}$ spectra are less significant.
Results also show that the long and short bursts follow similar trend.

\subsubsection{GRB that do not have acceptable spectral fits}
\label{sect:unaccept_spec}

In Sect~\ref{sect:event_standard}, we listed the criteria used for determining whether the results from spectral analysis are acceptable.
In order to obtain a more complete picture of the BAT-detected bursts, we include further discussions about those GRBs that have the spectral analysis results excluded from these criteria,
and hence are not included in the previous section. 

When we choose GRBs with acceptable spectra, we first select the bursts with $\Delta \chi^2 < 6$, 
and go through those CPL spectral fits to select the acceptable ones based on the criteria listed in Sect~\ref{sect:event_standard}.
We then go through the PL fits for the rest of the bursts, and choose those bursts with acceptable PL fits.
Therefore, for all the GRBs without acceptable spectra, there must be some problems in their PL fits.
However, even for the GRBs with acceptable PL fits, there might be some problems in their CPL fits (which is not adopted).
Hence, when sorting out the reasons for those unacceptable spectral fits, we only look through the PL fits.

We put the bursts with unacceptable spectral fits into three different categories:
\begin{enumerate}
\item GRBs with problematic spectral fits: These are GRBs with some problems in their fits, which includes GRBs with at least one unconstrained parameter (parameters here includes the photon index, normalization, energy and photon flux, the photon index for the fits used for finding energy and photon fluxes, and lower and upper limits for all these parameters), GRBs with fitted values outside of the uncertainty ranges (although this should not happen, we place this criteria here just to be safe), 
and bursts with inconsistent results from the original fits and those fits used to constrain the photon and energy fluxes. These are bursts that would need manual spectral analysis to figure out the exact causes of the problems, and whether a better fit could be found. In this category, there are 
37
GRBs for the time-averaged spectral fits,
109
bursts for the 1-s peak spectral fits,
and 
352
events for the 20-ms peak spectral fits.

\item GRBs with lower limits consistent with zero (but not in group one): These are bursts with at least one of the values of either the normalization, the photon flux, or the energy flux, consistent with zero, and thus only the upper limit of these parameters can be obtained. In this category, there are 
6
GRBs for the time-averaged spectral fits,
100
bursts for the 1-s peak spectral fits,
and 
418
events for the 20-ms peak spectral fits.

\item GRBs with spectral fits that have the null hypothesis probability $<$ 0.1 (but not in group one): These are fits that are likely to reject the null hypothesis\footnote{The null hypothesis here refers to the simple PL model, though for these events, the CPL fits either also have low probability for the null hypothesis, or have some other problems.}. However, we note that many of these fits that are inconsistent with the null hypothesis are likely due to systematic problems in data reduction rather than physical reasons. For example, other bright X-ray sources in the field-of-view can cause problems in background subtraction (as discussed in Sect~\ref{sect:bright_src}) and give  an incorrect estimation of the GRB source counts. Therefore, a careful examination of the potential data reduction problems must be carried out before seeking for alternative models. 
In this category, there are 
87
GRBs for the time-averaged spectral fits,
within which
43
bursts have X-ray sources with similar or higher signal-to-noise ratio in their field of view.
For the 1-s peak spectral fits, there are
77
bursts in this group,
and 
24
of these have X-ray sources with similar or higher signal-to-noise ratio in their field of view.
For the 20-ms peak spectral fits, there are
48
events in this group, and
28
of them have X-ray sources with similar or higher signal-to-noise ratio in their field of view.

\end{enumerate}

For the time-averaged spectral fits,
there are
one
burst, GRB061218, that belongs to both group 2 and group 3. 
For the 1-s peak and 20-ms peak spectral fits, there are more GRBs belonging to both group 2 and group 3
(7
bursts for the 1-s peak spectral fits,
and
38
events for the 20-ms peak spectral fits).

Also, note that since GRB140716A are listed as GRB140716A-1 and GRB140716A-2 in all the summary tables for the two separate triggers of this same burst (as described in Section \ref{sect:event_results}),
this burst sometime shows up in both lists of the acceptable spectral fits and unacceptable spectral fits. Specifically, GRB140716A-1 has acceptable spectral fits for the time-averaged spectrum and the 1-s peak spectrum, while GRB140716A-2 has acceptable spectral fits for the time-averaged spectrum but unacceptable fits for the 1-s peak spectrum. For the 20-ms peak spectrum, both GRB140716A-1 and GRB140716A-2 has unacceptable spectral fits, but they only count as one burst in the unacceptable spectral fit list of the 20-ms peak spectrum.

\subsection{Partial coding fraction}

\begin{figure}[!h]
\begin{center}
\includegraphics[width=0.8\textwidth]{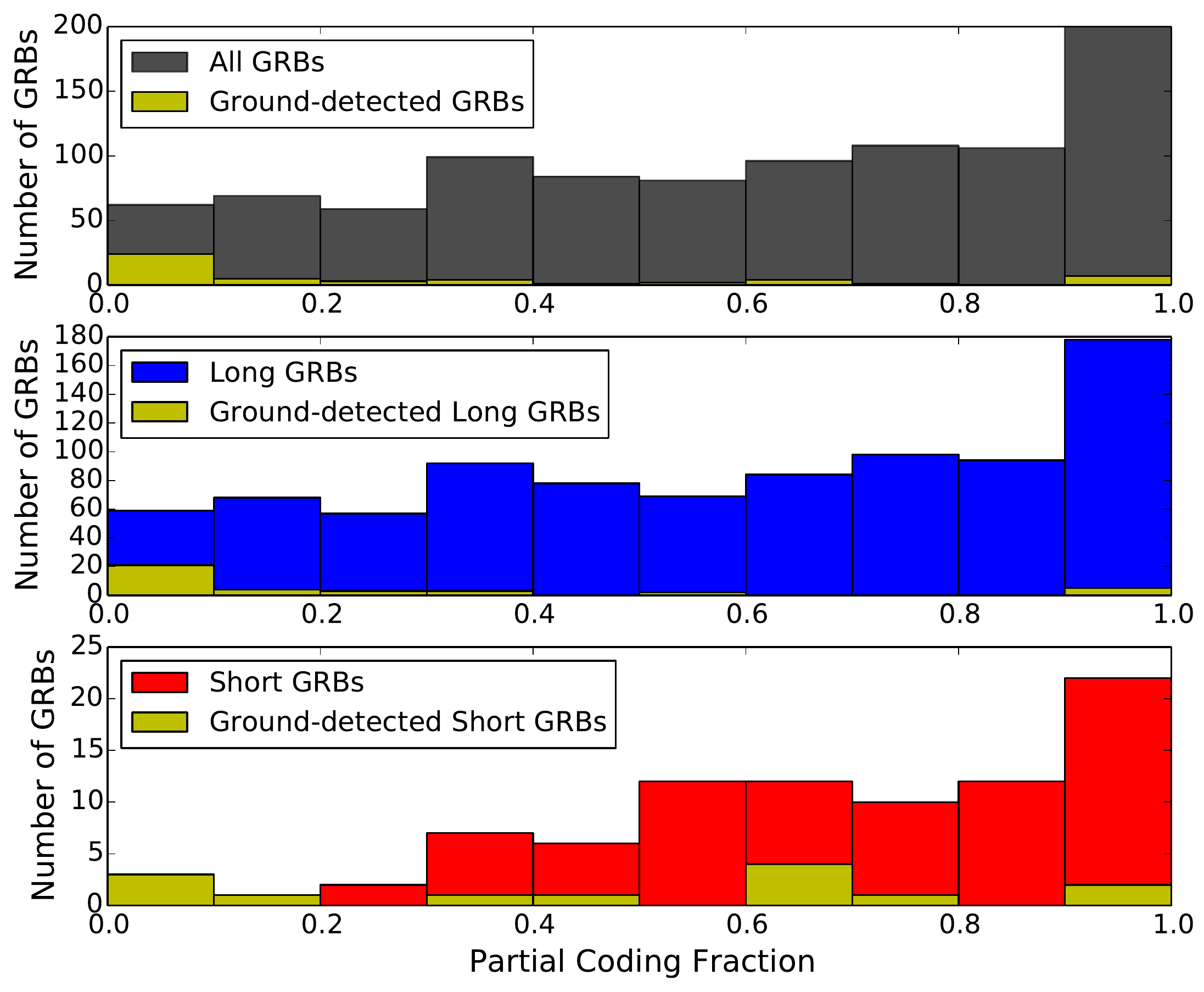}
\end{center}
\caption{
Distribution of the partial coding fractions for long, short, and ground-detected GRBs. The bursts found by ground-analyses during spacecraft slews 
are not included because the partial coding fraction changes constantly during slews.
}
\label{fig:pcode}
\end{figure}

Figure \ref{fig:pcode} shows the histogram of the partial coding fraction for long,
short, and ground-detected GRBs. The bursts found by ground-analyses during spacecraft slews 
are not included, because the partial coding fraction changes constantly for these events.
There are no short GRBs triggered on-board with partial coding fraction less than 0.2,
which is equivalent to an incident angle of $\sim 50^{\rm o}$. 

\subsection{Short GRBs with Extended Emission}
\label{sect:Short_with_EE}

Short GRBs with extended emissions have raised special interests among the GRB community,
partially due to their mixed characteristics between the short and long bursts \citep[e.g.,][]{Norris06, Gehrels06, Troja08, Kaneko15}.
We therefore include a special section of short GRBs with extended emission. We also present
both a more secure list of these bursts (Table~\ref{tab:sGRB_with_EE}), plus a list of possible short GRBs with extended emission (Table~\ref{tab:maybe_sGRB_with_EE}).

All of the short GRBs with extended emissions listed in Table~\ref{tab:sGRB_with_EE} that occurred before 2009 are included
in the BAT2 catalog. Note that we do not apply any quantitative criteria for selecting these short GRBs with extended emissions.
Most of the bursts before 2009 are adopted from the BAT2 catalog, which was classified by \citet{Norris10}. However,
the bursts after 2009 are chosen based on reports in GCN circulars. We also double check by eye inspection for all GRBs
to (1) make sure no other obvious GRBs with similar features are missed in previous GCN circulars,
and (2) those bursts reported as short GRBs with extended emissions do show such a feature.

\begin{table}[h!]
\caption{\label{tab:sGRB_with_EE}
A list of definite short GRBs with extended emission (E. E.). The times listed in the table are relative to the burst trigger time.
}
\begin{center}
\begin{tabular}{|c|c|c|c|c|c|}
\hline
GRB name & Short Pulse Start & Short Pulse End & E.E. End & Short Pulse  & E.E. \\
                    &   [s]  & (E.E. Start) [s] & [s]                      & $\alpha_{\rm PL}$        & $\alpha_{\rm PL}$\\
\hline
\hline
GRB150424A & -0.060 & 0.468 & 95.012 & $-0.78^{+0.06}_{-0.06}$ & $-2.10^{+0.46}_{-0.54}$ \\
\hline
GRB111121A & -0.336 & 1.000 & 138.264 & $-0.99^{+0.08}_{-0.08}$ & $-1.83^{+0.15}_{-0.15}$ \\
\hline
GRB090916 & -0.040 & 0.392 & 68.520 & $-1.58^{+0.27}_{-0.27}$ & $-1.37^{+0.38}_{-0.37}$ \\
\hline
GRB090715A & -0.200 & 0.800 & 48.936 & $-1.02^{+0.21}_{-0.20}$ & $-1.54^{+0.62}_{-0.62}$ \\
\hline
GRB090531B & 0.252 & 1.300 & 56.132 & $-0.99^{+0.16}_{-0.16}$ & $-1.69^{+0.27}_{-0.28}$ \\
\hline
GRB080503 & -2.192 & 0.600 & 221.808 & $-1.32^{+0.45}_{-0.43}$ & $-1.89^{+0.12}_{-0.12}$ \\
\hline
GRB071227 & -0.144 & 0.908 & 150.552 & $-1.01^{+0.24}_{-0.23}$ & $-2.23^{+0.41}_{-0.49}$ \\
\hline
GRB070714B & -0.792 & 1.976 & 74.640 & $-0.96^{+0.08}_{-0.08}$ & $-1.92^{+0.43}_{-0.49}$ \\
\hline
GRB061210 & -0.004 & 0.080 & 89.392 & $-0.69^{+0.12}_{-0.12}$ & $-1.80^{+0.34}_{-0.37}$ \\
\hline
GRB061006 & -23.288 & -22.000 & 137.720 & $-0.91^{+0.08}_{-0.07}$ & $-2.06^{+0.22}_{-0.23}$ \\
\hline
GRB051227 & -0.848 & 0.828 & 122.732 & $-0.94^{+0.25}_{-0.23}$ & $-1.48^{+0.27}_{-0.27}$ \\
\hline
GRB050724 & -0.024 & 0.416 & 107.140 & $-1.51^{+0.14}_{-0.14}$ & $-2.05^{+0.25}_{-0.26}$ \\
\hline
\hline
\end{tabular}
\end{center}
\end{table}

\begin{figure}[!h]
\begin{center}
\includegraphics[width=1.0\textwidth]{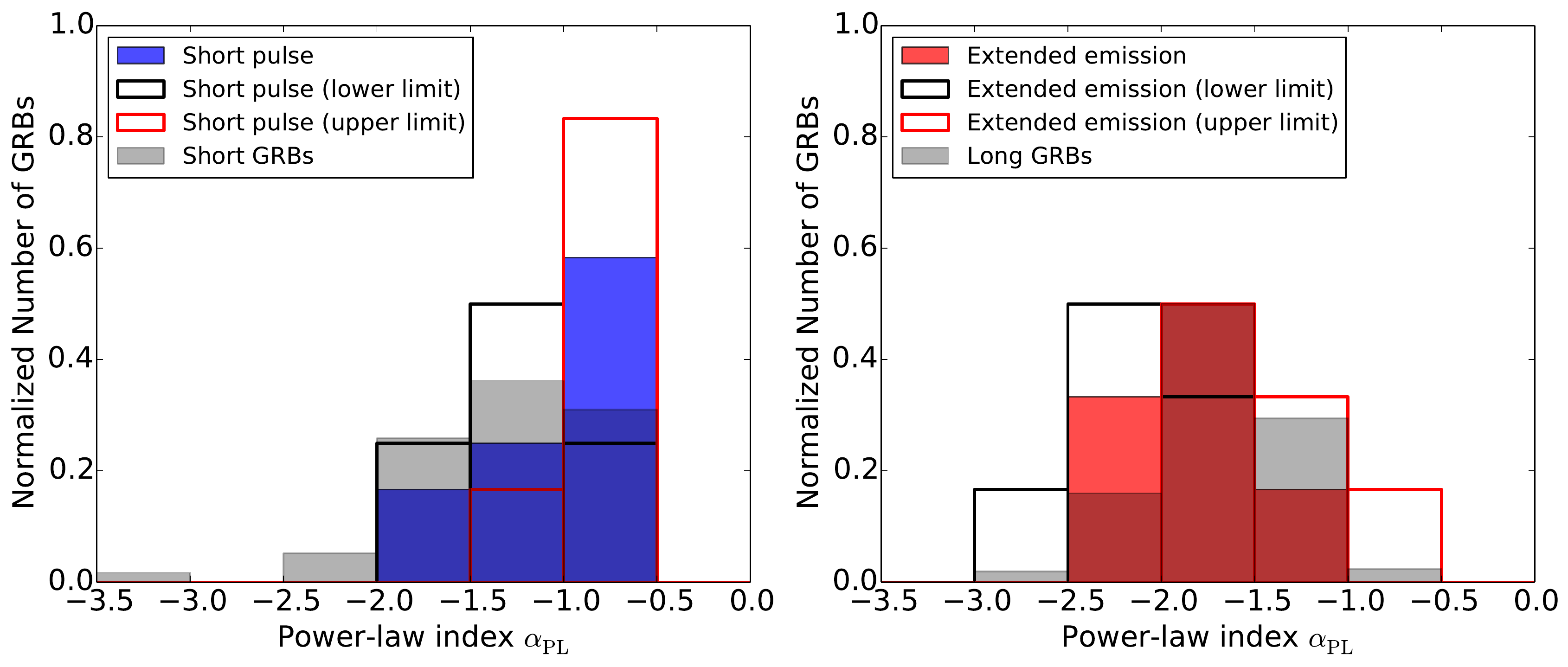}
\end{center}
\caption{
The distribution of the photon index from the simple PL fit $\alpha_{\rm PL}$ with spectra
from 
the short pulses (left panel) and extended emissions (right panel) of the short GRBs with extended emission (listed in Table~\ref{tab:sGRB_with_EE}).
The $\alpha_{\rm PL}$ distributions from short and long bursts are also plotted for comparison.
Furthermore, distributions using the upper and lower bounds of the $\alpha_{\rm PL}$ value uncertainty range are also plotted.
}
\label{fig:shortGRBs_with_EE}
\end{figure}

\begin{table}[h!]
\caption{\label{tab:maybe_sGRB_with_EE}
A list of possible short GRBs with extended emission.
The comments note the reasons for these bursts being in the ``possible'' category, with each number corresponding to the following descriptions:
(1) The short pulse is slightly longer than 2 s.
(2) The extended emission is not picked out by the auto-pipeline ({\it battblocks}).
(3) The extended emission is weak, and there are some significant fluctuations in the background and/or bright X-ray sources in the field of view, which may cause extra residuals when performing the mask-weighting.
(4) The bursts was found by ground analyses and there is not sufficient event data.
}
\begin{center}
\begin{tabular}{|c|c|c|}
\hline
GRB name & Trigger ID/Observation ID & Comment \\
\hline
\hline
GRB140302A & 589685 & 1 \\
\hline
GRB140209A & 586071 & 1, 3 \\
\hline
GRB140102A & 582760 & 1 \\
\hline
GRB130716A & 561974 & 3 \\
\hline
GRB130612A & 557976 & 1, 2, 3 \\
\hline
GRB110402A & 450545 & 1, 3 \\
\hline
GRB100816A & 431764 & 1, 2, 3 \\
\hline
GRB090831C & 361489 & 1, 3 \\
\hline
GRB090530 & 353567 & 1 \\
\hline
GRB090518 & 352420 & 1, 3 \\
\hline
GRB090510 & 351588 & 3 \\
\hline
GRB080303 & 304549 & 1 \\
\hline
GRB080123 & 301578 & 3 \\
\hline
GRB081211B & 00090053089 & 4 \\
\hline
GRB060614 & 214805 & 1 \\
\hline
\hline
\end{tabular}
\end{center}
\end{table}

The possible short GRBs with extended emission listed in Table~\ref{tab:maybe_sGRB_with_EE} are 
GRBs with similar structure as those listed in Table~\ref{tab:sGRB_with_EE}. They are selected because some literature (mostly the GCN circulars) mentioned 
potential extended emissions. However, these are not included in Table~\ref{tab:sGRB_with_EE}
because of at least one of the following reasons:
(1) The short pulse is slightly longer than 2 s.
(2) The extended emission is not picked out by the auto-pipeline ({\it battblocks}).
(3) The extended emission is weak, and there are some significant fluctuations in the background and/or bright X-ray sources in the field of view, which may cause extra residuals when performing the mask-weighting. 
(4) The bursts was found by ground analyses and there is not sufficient event data.
The corresponding comments for each burst are presented in the table.
Note that although GRB081211B is a ground-detected burst (during a spacecraft slew) with only $\sim 120$ s of event data,
several GCN circulars suggest that this burst is possibly a short GRB with extended emission, with the short spike detected by
Konus-Wind while the burst was outside of the BAT field of view (Golenetskii et al. 2008/GCN 8676; Perley et al. 2008/GCN 8914).
In fact, there is also a visible short spike at $\sim$ T0-150 s in the BAT raw light curve. However, due to the lack of event data at that time,
we cannot confirm whether this short pulse is related to the GRB.  

There are five bursts (GRB091117, GRB100724A, GRB100625A, GRB101219A, and GRB090621B) for which the GCN circular mentioned an indication of extended emission,
but further analysis suggests that the extended emissions of GRB091117, GRB100724A, GRB100625A, and GRB101219A are below $\sim 3 \sigma$ even when choosing an optimal time periods and optimal energy bands. GRB090621B 
shows no extended emission until $\sim$ T0+150 s, when a low-significant bump occurred and last $\sim 50$ s. Thus,
we conclude that these extended emissions are likely not real.


Figure \ref{fig:shortGRBs_with_EE} shows the results of spectral fits from the simple PL model ($\alpha_{\rm PL}$) for both the short pulses (left panel) and the extended emission (right panel), for the confirmed short GRBs with extended emissions (Table \ref{tab:sGRB_with_EE}). Because of the small sample of the short GRBs with extended emission, and all of these bursts have constrained $\alpha_{\rm PL}$ but not necessarily have constrained parameters in the CPL model, we only show the simple PL model fits in this figure. Moreover, we include all the bursts, even if the do not satisfied some of the criteria listed in Section \ref{sect:event_standard} (in fact, only a few simple PL fits here do not satisfied all the strict criteria in \ref{sect:event_standard}, such as the lower limit of one energy band is consistent with zero).
The $\alpha_{\rm PL}$ distributions for short and long GRBs are also plotted in gray bars in the left and right panel, respectively, to be compared with the distributions of short pulses and extended emissions.
As mentioned in previous studies \citep[e.g.,][]{Sakamoto11}, the spectra of short pulses are harder than the extended emission in general,
and resemble more of short GRBs, while the spectra for extended emission parts are softer and match better with the 
$\alpha_{\rm PL}$ distributions of long GRBs.

\subsection{GRBs with redshift measurements}

We include a list of GRBs with redshift measurements in this catalog (Table \ref{tab:redshift_list}).
The information in this list is
collected from and cross-checked between other online lists 
(e.g., GRBOX by Daniel Perley\footnote{http://www.astro.caltech.edu/grbox/grbox.php}, online list by Jochen Greiner\footnote{http://www.mpe.mpg.de/~jcg/grbgen.html}, online table by Nathaniel Butler\footnote{http://butler.lab.asu.edu/swift/bat\_spec\_table.html}), the Gamma-ray Coordinates Network (GCN) circulars \citep{Barthelmy95},
and papers. The redshift list with full references are included in Table \ref{tab:redshift_list}.

To date (till 
GRB151027B), there are
378
BAT-detected GRBs with redshift measurements
(within which 18 redshifts are marked as potential questionable measurements).
In Table~\ref{tab:redshift_list}, we mark the four common methods for redshift measurements: (1) absorption lines
measurement from the GRB afterglow spectra (noted by symbol ``ba''); 
(2) emission lines from the GRB host galaxy spectra (noted by symbol ``he'');
(3) photometric redshift from the GRB afterglow (noted by symbol ``bp'');
(4) photometric redshift from the GRB host galaxy (noted by symbol ``hp'').
Other less-common methods, such as the Lyman-alpha break, are described in short sentences in Table~\ref{tab:redshift_list}.
If we noticed that some questions were raised about the GRB redshifts (e.g., the potential host galaxy might not be related to the burst),
the redshift value in the table will be followed by a question mark,
and these redshift values are not included in the following summarized numbers and plots. 

There are
229
GRB spectroscopic redshifts from GRB afterglows,
96
spectroscopic redshifts measured from host galaxies,
17
photometric redshifts from GRB afterglows,
and
12
photometric redshifts from host galaxies.

\begin{figure}[!h]
\begin{center}
\includegraphics[width=0.8\textwidth]{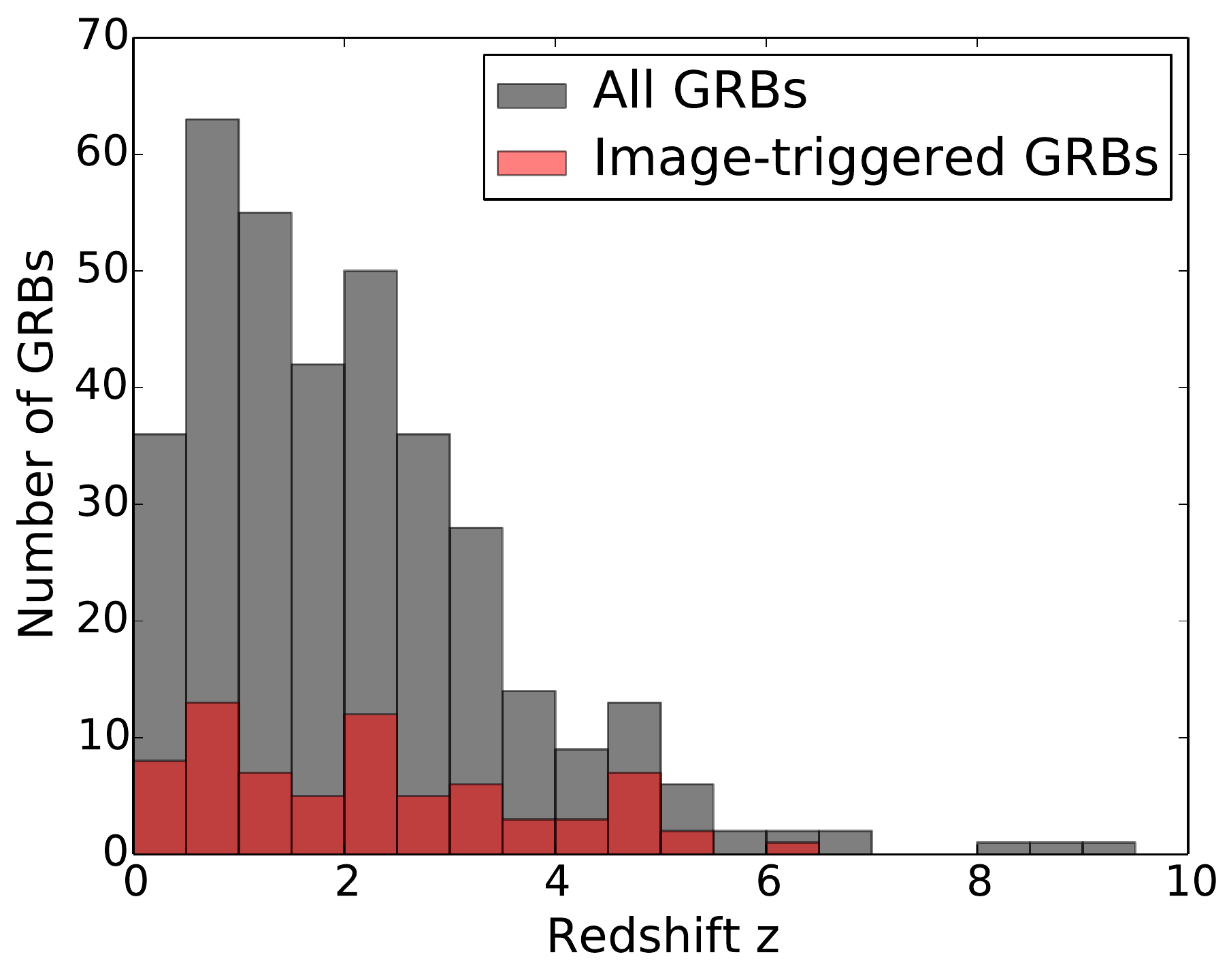}
\end{center}
\caption{
Redshift distribution for the BAT GRBs that have redshift measurements (gray bars).
Distribution for bursts that are detected by image trigger are shown in red bars.
}
\label{fig:redshift_with_image}
\end{figure}

The redshift distribution of the BAT-detected GRBs is shown in Fig.~\ref{fig:redshift_with_image}. The distribution of bursts found by the image trigger 
is plotted in red. Compared to GRBs detected by rate triggers, the image-triggered GRBs are more uniformly distributed 
throughout all redshifts. The image-triggered bursts compose of
20.0\% of the events with redshift measurements, 
which is very similar to the fraction of image-triggered GRBs out of all BAT-detected bursts
(17.5\%).


\subsubsection{Energy flux versus redshift: exploring potential selection effects in redshift measurements with the BAT trigger characteristic}

\begin{figure}[!h]
\begin{center}
\includegraphics[width=0.65\textwidth]{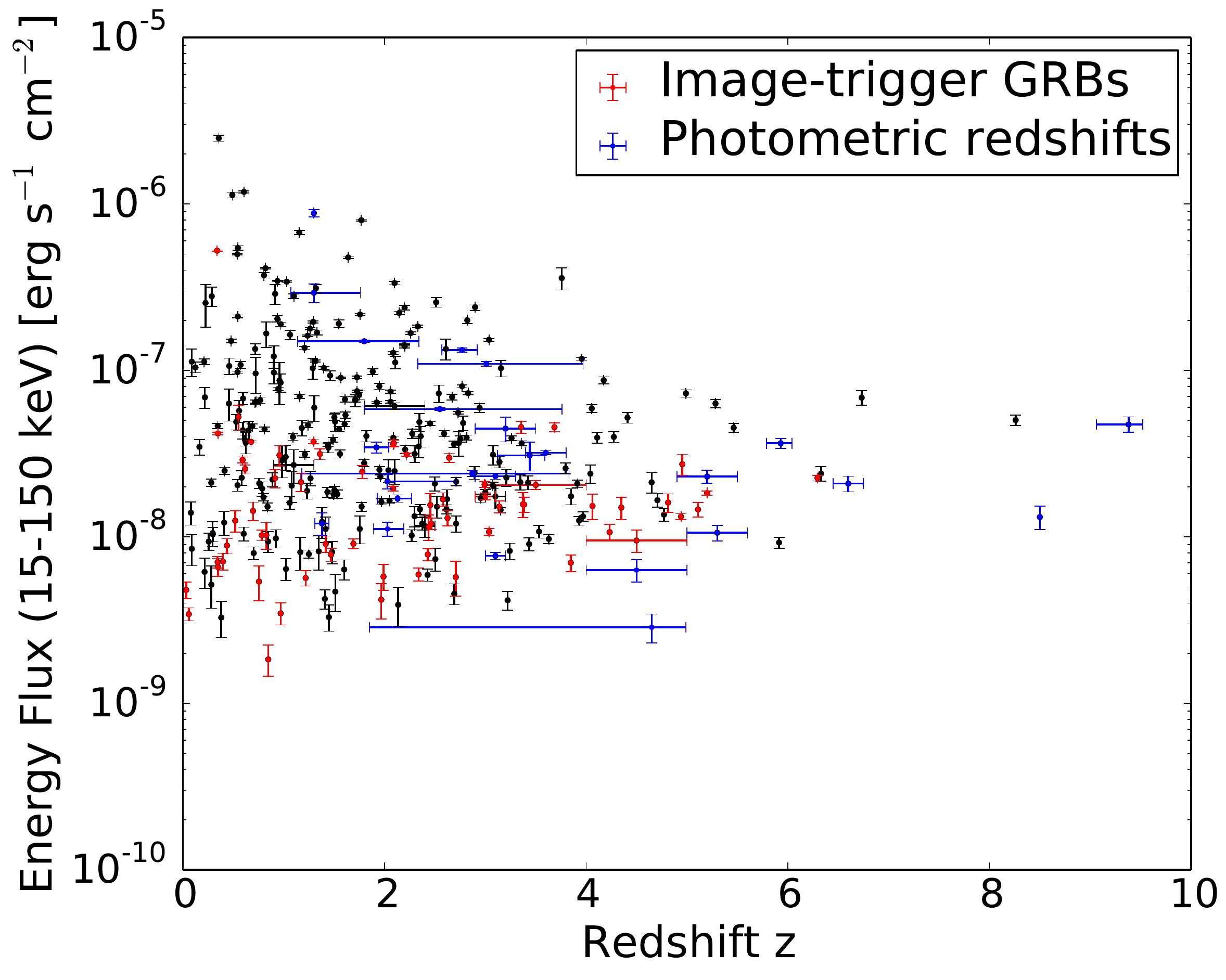}
\end{center}
\vspace{-15pt}
\caption{
Redshift versus energy flux (15-150 keV) for the BAT GRBs that have redshift measurements.
Bursts that are detected by the image trigger are marked in red; 
GRBs with photometric redshifts are marked in blue;
all other bursts are plotted in black.
The uncertainties for redshifts are shown when available. However, 
note that these uncertainties are adopted from different references and might 
refer to different confidence ranges (see table \ref{tab:redshift_list} for the original sources). 
}
\label{fig:energy_flux_vs_z}
\end{figure}

The energy fluxes in 15-150 keV as a function of redshift is plotted in Fig.~\ref{fig:energy_flux_vs_z}.
There are
35
bursts with redshift measurements excluded from this plot,
because those bursts do not have acceptable spectral fits.
The image-triggered bursts are marked in red, and 
the GRBs with photometric redshifts are marked in blue, with
the uncertainties shown when available.
Note that
these uncertainties are adopted from different sources (see Table \ref{tab:redshift_list}) and might 
refer to different confidence ranges. Thus, the values here are only 
presented as rough references.  
As expected, image triggers find bursts with lower fluxes in general throughout the redshift range.
The decline in the number of detections of high-flux bursts at higher redshift is likely due to the shrinking of the sample size of detected GRBs 
(it is less likely to detect bursts from the tail of the distribution).
There is also a slight decrease in the detections of low-flux bursts at higher redshift, though the effect is not obvious until higher redshift ($z \gtrsim 5$),
which might imply that the majority of the bursts (i.e., the center of the intrinsic flux distribution) does not lie far from the BAT sensitivity threshold.


\begin{figure}[!h]
\begin{center}
\includegraphics[width=0.65\textwidth]{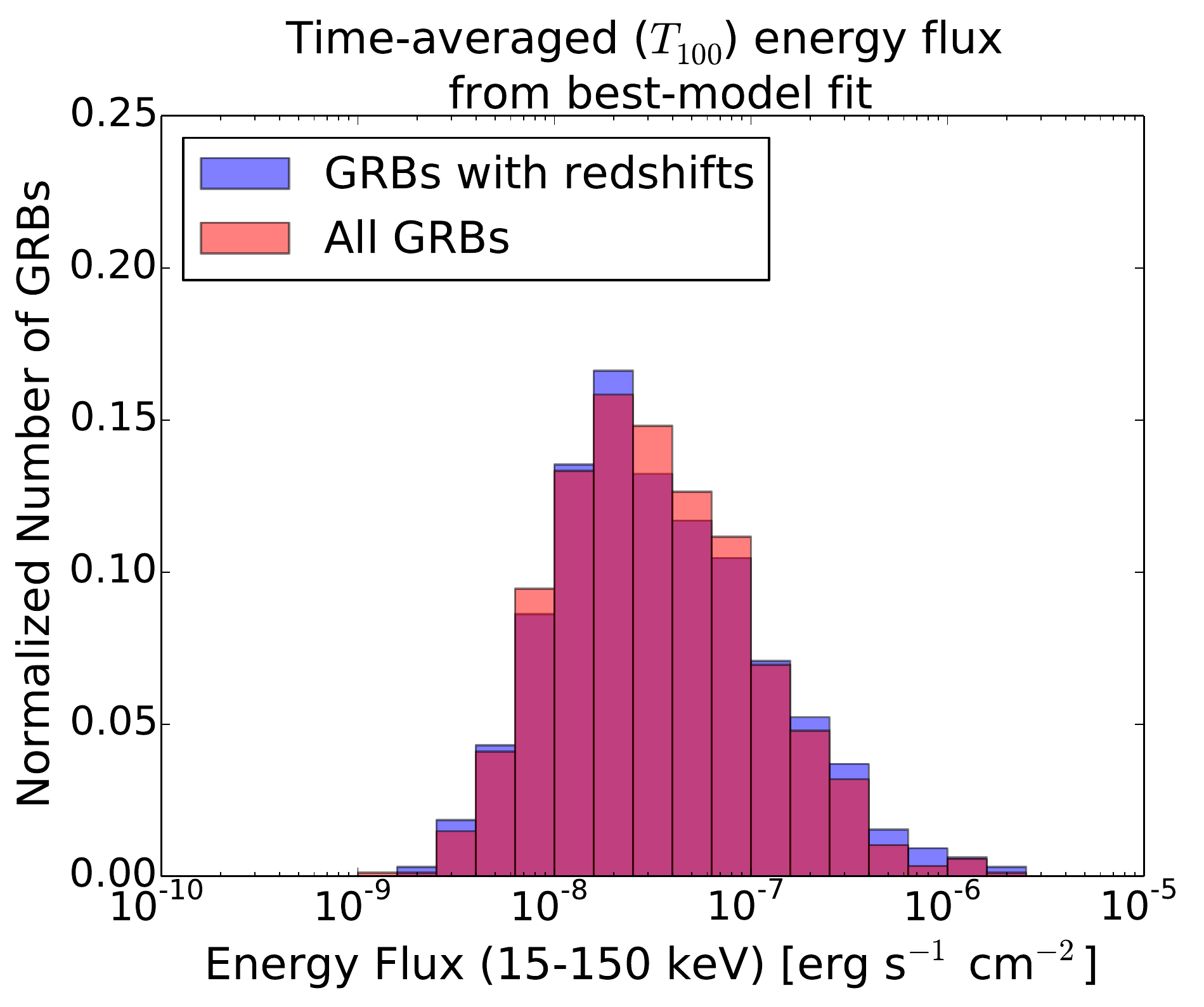}
\end{center}
\vspace{-15pt}
\caption{
Comparison of the distributions of the time-averaged ($T_{100}$) energy flux in 15-150 keV for the GRBs with redshifts and all GRBs (only bursts with acceptable spectral fits are included).
}
\label{fig:energy_flux_compare}
\end{figure}

Figure \ref{fig:energy_flux_compare} compares
the distributions of the time-averaged ($T_{100}$) energy flux in $15-150$ keV 
for the GRBs with redshifts and all GRBs
(but only those bursts with acceptable spectral fits are included).
The plot is shown in the normalized number of GRBs,
since there are $\sim 3$ times more bursts in the all-GRB sample
than in the GRB-with-redshift sample.
Results show that the two distributions are very similar,
which implies that the successful redshift measurements from 
the ground-based follow-up facilities have no correlation
with how bright the burst is in the BAT energy range. 

\subsubsection{Burst duration versus redshift}
\label{sect:T90_vs_z}

\subsubsubsection{The missing time-dilation effect and the observational biases}

Many studies have shown that the observed burst durations do not present a clear effect of time dilation for GRBs at higher redshift \citep[e.g.,][]{Sakamoto11,Kocevski13,Littlejohns14}.
One likely explanation is the ``tip-of-the-iceberg'' effect, where a larger fraction of the burst emission becomes hidden underneath the background noise, as the brightness of the GRB decreases at higher redshift. 
\citet{Kocevski13} demonstrates this effect with a single-pulse structure and concludes that the observed duration can miss up to $\sim 80\%$ of the true burst duration at high redshift $z \sim 5$.

\begin{figure}[!h]
\begin{center}
\includegraphics[width=0.55\textwidth]{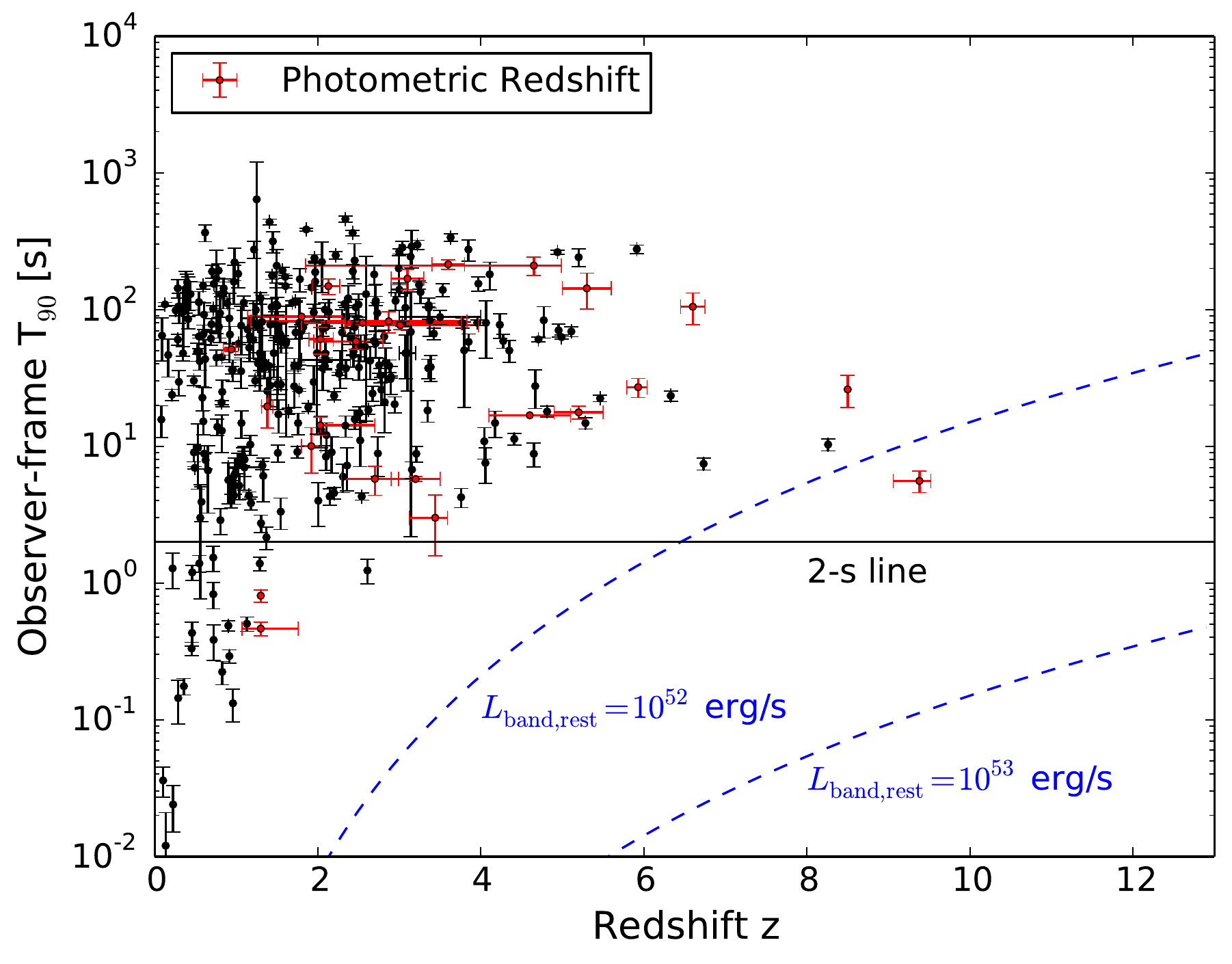}
\includegraphics[width=0.55\textwidth]{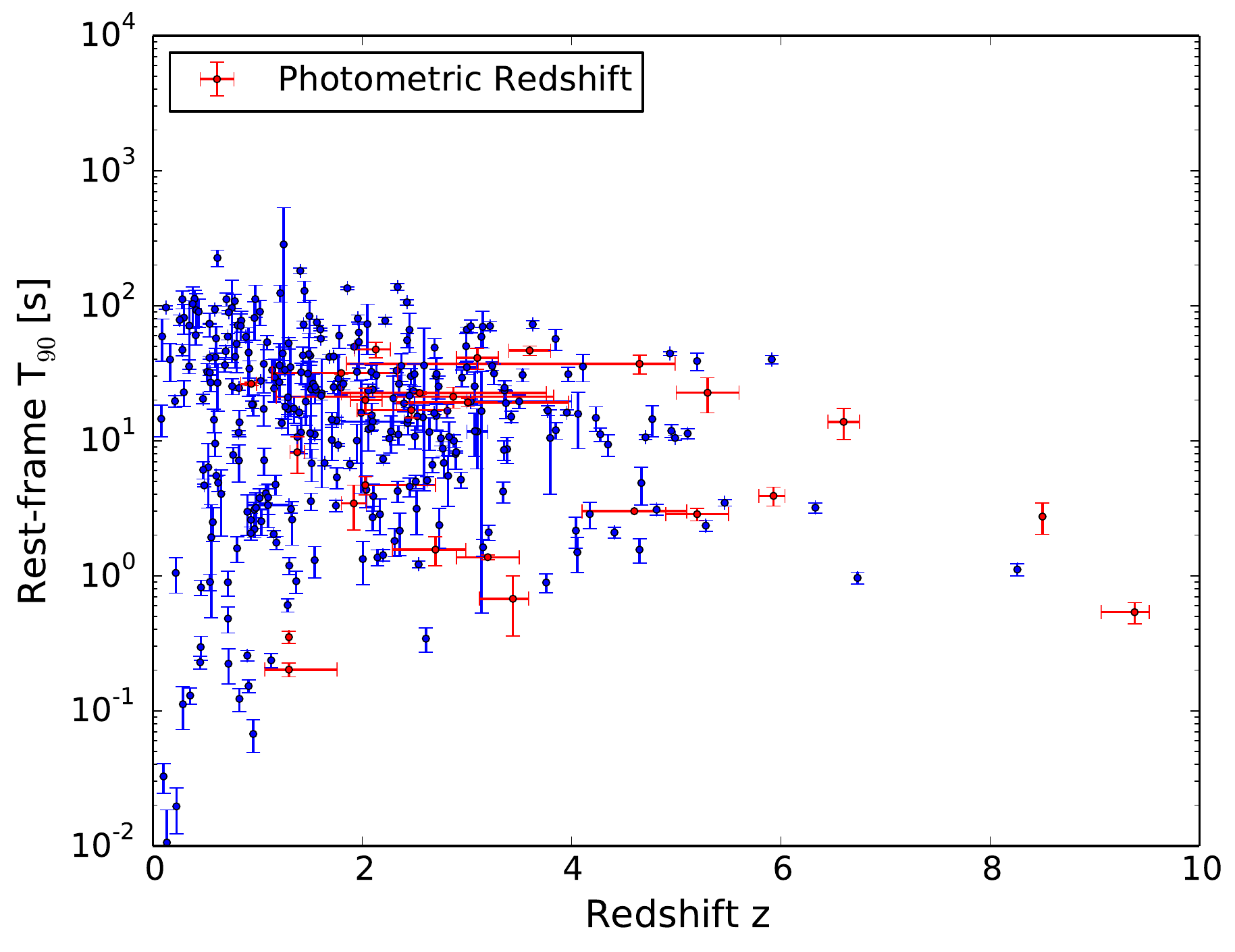}
\end{center}
\caption{
$T_{90}$ versus redshift $z$.
{\it Upper Panel:} the $T_{90}$ in the observer frame as a function of $z$. 
The blue-dotted line shows the expected correlation between the observed burst duration and redshift for bursts with luminosity of $L_{\rm band, rest} = 10^{52} \ \rm erg \ s^{-1}$ and $10^{53} \ \rm erg \ s^{-1}$, respectively.
{\it Bottom Panel:} the $T_{90}$ in the rest frame as a function of $z$.
Bursts with photometric redshifts are marked in red specifically due to the large uncertainties in their redshift measurements.
The uncertainties for redshifts are also shown when available. However, 
note that these uncertainties are adopted from different references and might 
refer to different confidence ranges (see Table \ref{tab:redshift_list} for the original sources).
Moreover, GRBs with unconstrained durations, such as the ultra-long GRBs, are not included in this plot. 
}
\label{fig:T90_obs_rest_vs_z}
\end{figure}

Figure \ref{fig:T90_obs_rest_vs_z} shows the comparison of the rest-frame $T_{90}$ and the observer-frame $T_{90}$
as a function of redshift $z$. The rest-frame $T_{90}$ is calculated simply by correcting the (1+z) time dilation effect (i.e., rest-frame $T_{90}$ = (observer-frame $T_{90}$)/(1+z)).
Indeed, there is no obvious trend from time dilation.
However, the observed $T_{90}$ seems to show some correlation between the minimum detectable $T_{90}$ with redshift. 
As discussed in Section~\ref{sect:BAT_sensitivity}, BAT can in general detect lower fluxes for bursts with longer burst durations. Therefore, one would generally expect that 
BAT can only detect longer bursts at higher redshift for GRBs with a specific intrinsic luminosity.
In other words, the correlation between the minimum detectable flux (at $5 \sigma$) in the observed energy band $F_{\rm band, obs}$ and the exposure time $T$  \citep[Eq. 9 in][]{Baumgartner13}: 
\beq
F_{\rm band, obs} = 1.18 \rm mCrab \bigg( \frac{\rm T}{\rm 1 \ Ms} \Big)^{-1/2} = 2.86 \times 10^{-11} \ \rm [erg \ s^{-1} \ cm^{-2}] \Big( \frac{\rm T}{\rm 1 \ Ms} \bigg)^{-1/2} 
\eeq
requires a longer exposure time to detect lower flux, and in the GRB case the maximum possible exposure time is determined by the burst duration.
The observed flux in the observed energy band is correlated with the luminosity in the corresponding redshifted bandpass as 
\beq
F_{\rm band, obs} = \frac{L_{\rm band, rest}}{4 \pi D^2_{\rm L}},
\eeq
where $D_{\rm L}$ is the luminosity distance and $L_{\rm band, rest}$ here
refers to the luminosity in an energy band that corresponds to the observed energy band, not the bolometric luminosity (see detailed descriptions in Appendix \ref{sect:lum_flux}).
Thus, for a specific luminosity, one can get the following correlation between the exposure time and the redshift,
\beq
 \bigg( \frac{\rm T}{\rm 1 \ Ms} \bigg)^{1/2} = 2.86 \times 10^{-11} \ \rm [erg \ s^{-1} \ cm^{-2}] \ \frac{4 \pi D^2_{\rm L}}{L_{\rm band, rest}}.
\eeq
For simplicity, we approximate the exposure time by the burst duration, i.e., assuming T = $T_{90}$,
and plot this relation
as the blue-dotted lines in Fig.~\ref{fig:T90_obs_rest_vs_z} (for $L_{\rm band, rest} = 10^{52} \ \rm erg \ s^{-1}$ and $10^{53} \ \rm erg \ s^{-1}$, respectively).
As expected, a larger luminosity corresponds to smaller burst duration throughout the redshift. 
As one can see in the plot, both of the blue-dashed lines lie below all but one of the detected burst with redshift measurements,
leaving a relatively large empty space between the blue dotted line with $L_{\rm band, rest} = 10^{52} \ \rm erg \ s^{-1}$ and the detected bursts with 
the shortest durations at each redshift.
Naively, one might expect that this indicates that there are no bursts with $L_{\rm band, rest} > 10^{52} \ \rm erg \ s^{-1}$,
otherwise BAT should have detected them even if they have durations that lies below the blue dotted line.
However, when we calculate the GRB luminosities in the following subsection (see Fig.~\ref{fig:lum_vs_z}),
more than one bursts have luminosities exceeding $10^{52} \ \rm erg \ s^{-1}$.
Therefore, we are only missing bright ``short'' bursts
that should have been detectable, and not all detectable bright bursts.
Another possibility to explain these missing bright short bursts in the plot would be the missing redshift measurements. 
In other words, we might have detected these bright short bursts, but do not have redshift measurements and thus 
they are not included in this plot.
Unfortunately, it remains difficult to distinguish these two possibilities due to the biases and incompleteness of redshift measurements.

\subsubsubsection{Long and short bursts with a rest-frame $T_{90}$}

\begin{figure}[!h]
\begin{center}
\includegraphics[width=0.6\textwidth]{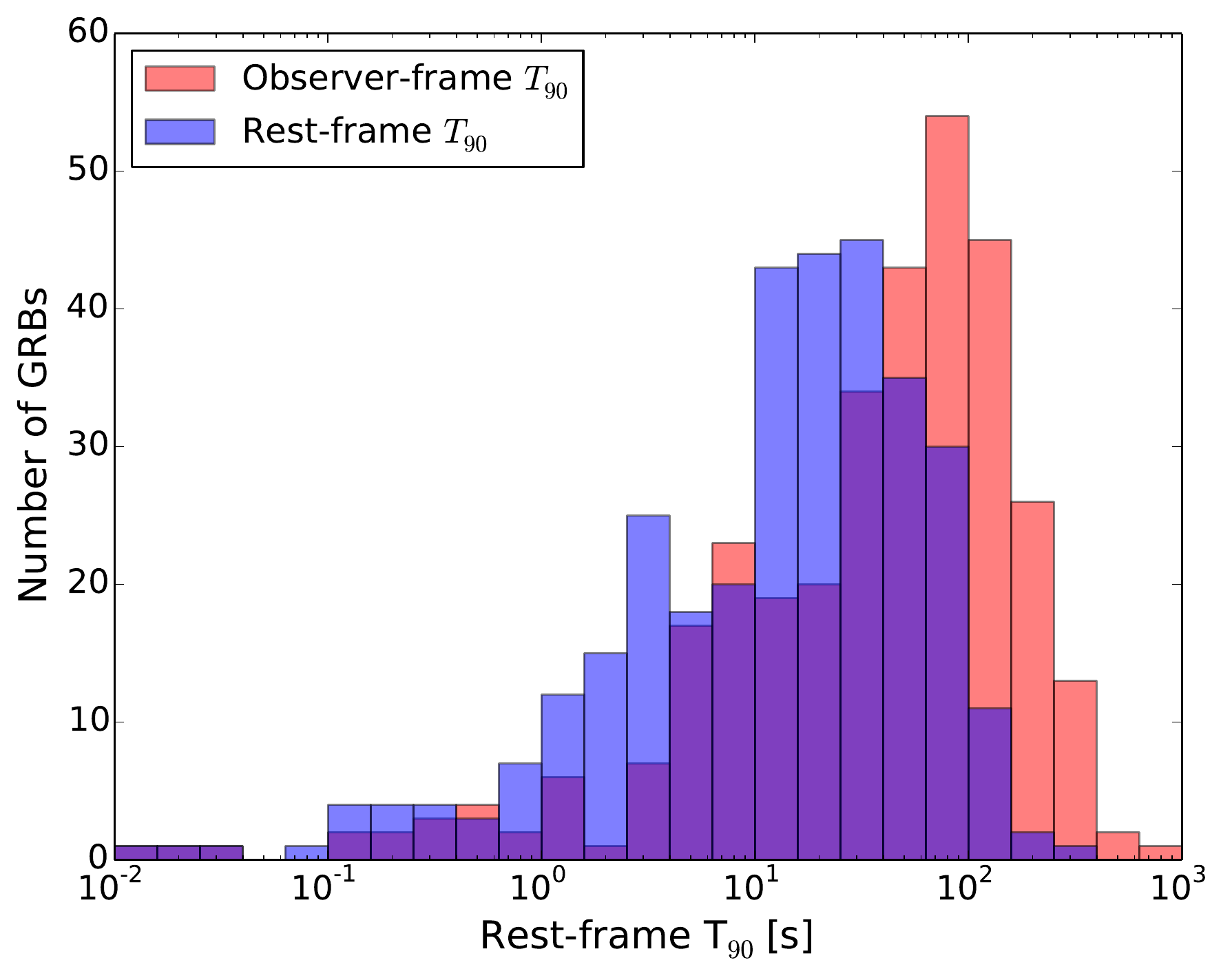}
\end{center}
\caption{
Comparison between the rest-frame $T_{90}$ distributions and the observer-frame $T_{90}$ distributions
for the BAT GRBs that have redshift measurements.
Note that GRBs with unconstrained durations, such as the ultra-long GRBs, are not included in this plot.
}
\label{fig:compare_T90}
\end{figure}

The GRB community has commonly used the observed burst duration to classify GRBs and to infer different 
physical origins, with the long GRBs related to deaths of massive starts, and short bursts linked to compact-object mergers.
However, the burst duration in the observed frame is affected by several biases, such as the tip-of-the-iceberg effect mention above and the time-dilation effect, 
and thus might not represent the true duration of a burst.
Although it is difficult to recover the intrinsic total burst duration from the tip-of-the-iceberg effect, we can easily correct the time-dilation effect and calculate the rest-frame $T{90}$ 
by dividing the observed $T_{90}$ by the (1+z) factor.
Figure \ref{fig:compare_T90} compares the rest-frame $T_{90}$ distribution and the observer-frame $T_{90}$ distribution.
For the observer-frame $T_{90}$ distribution, we only include GRBs with redshift measurements,
in order to have a fair comparison with the rest-frame $T_{90}$ distribution because 
there are more long bursts that have redshift measurements than short bursts.
Results show that the ``tail'' of short bursts in the distribution become slightly less significant 
in the rest-frame $T_{90}$ histogram. There are
21
bursts with the observer-frame $T_{90} > 2$ s,
but have the rest-frame $T_{90} \leq 2$ s. 
However, this rest-frame $T_{90}$ distribution is still unlikely to represent the distribution of the intrinsic burst duration due to other observational biases.

\subsubsection{Spectral characteristics versus redshift: searching for hints of the intrinsic spectral shapes and energy outputs}

\begin{figure}[!h]
\begin{center}
\includegraphics[width=0.8\textwidth]{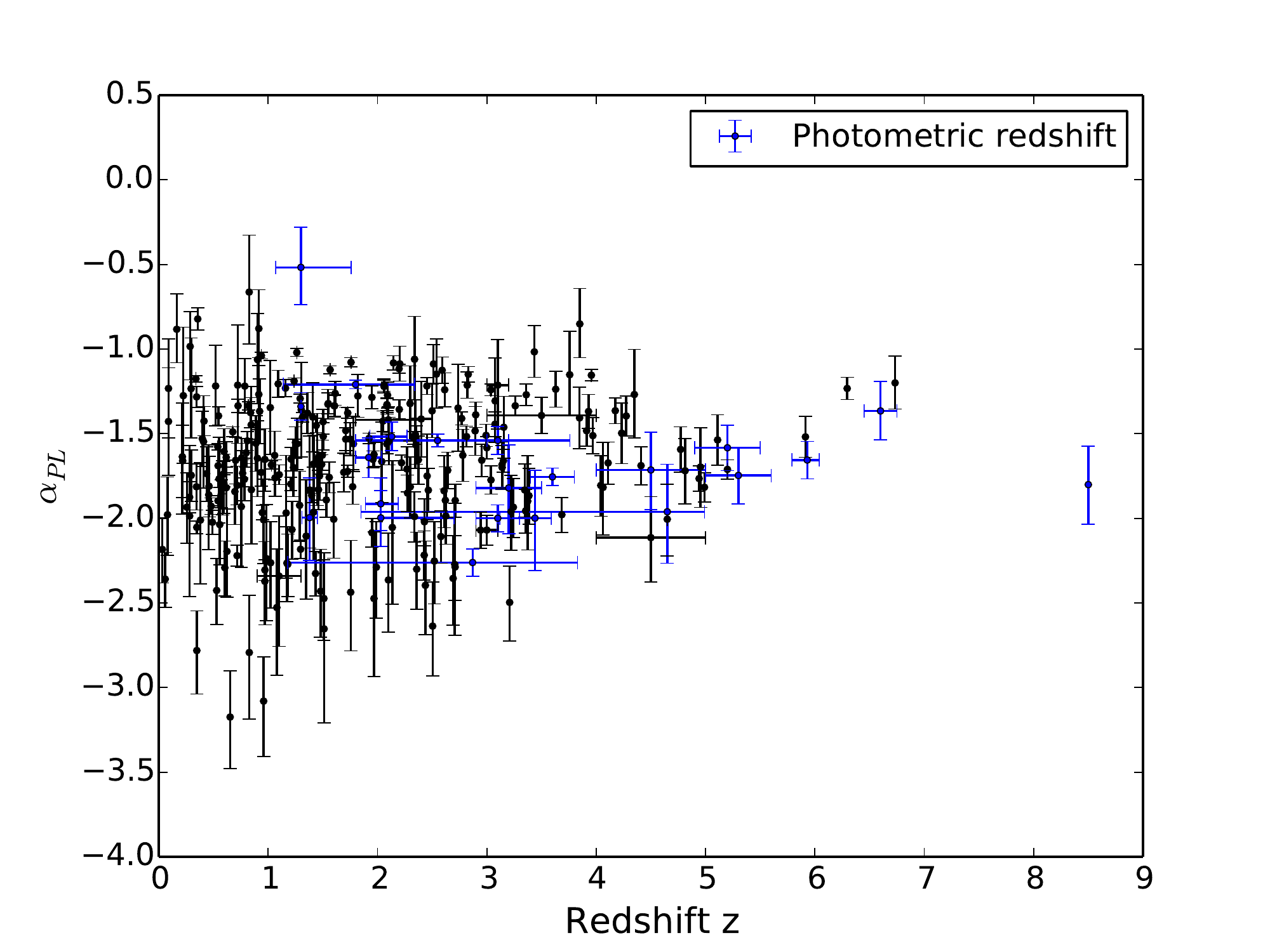}
\end{center}
\vspace{-15pt}
\caption{
The time-averaged ($T_{100}$) photon index $\alpha_{\rm PL}$ (for the bursts that are better fitted by the simple PL model) as a function of redshift $z$.
Bursts with photometric redshifts are marked in blue specifically, due to the large uncertainties in their redshift measurements.
The uncertainties for redshifts are also shown when available. However, 
note that these uncertainties are adopted from different references and might 
refer to different confidence ranges (see table \ref{tab:redshift_list} for the original sources). 
}
\label{fig:alpha_vs_z}
\end{figure}

Figure \ref{fig:alpha_vs_z} shows the time-averaged ($T_{100}$) photon index $\alpha_{\rm PL}$ (for those GRBs that are better fitted by the simple PL model) as a function of redshift $z$.
Again, the bursts with photometric redshifts are marked in blue specifically, due to the larger uncertainties in their redshift measurements.
Similar to the same figure shown in the BAT2 catalog, there is no clear correlation between $\alpha_{\rm PL}$ and redshift. However, there are probably some instrumental selection biases 
for the photon indices of the BAT-detected GRBs.
As discussed in Section~\ref{sect: BAT_selection_effect_spectrum}, 
BAT is most likely to detect bursts with a certain range of $\alpha_{\rm PL}$ that gives higher fluxes in the BAT energy range. 
Moreover, the burst needs to be bright enough to have a spectrum with uncertainties small enough to distinguish the simple PL and CPL model.
Therefore, the distribution of $\alpha_{\rm PL}$ might not represent the true intrinsic distribution.

\begin{figure}[!h]
\begin{center}
\includegraphics[width=0.8\textwidth]{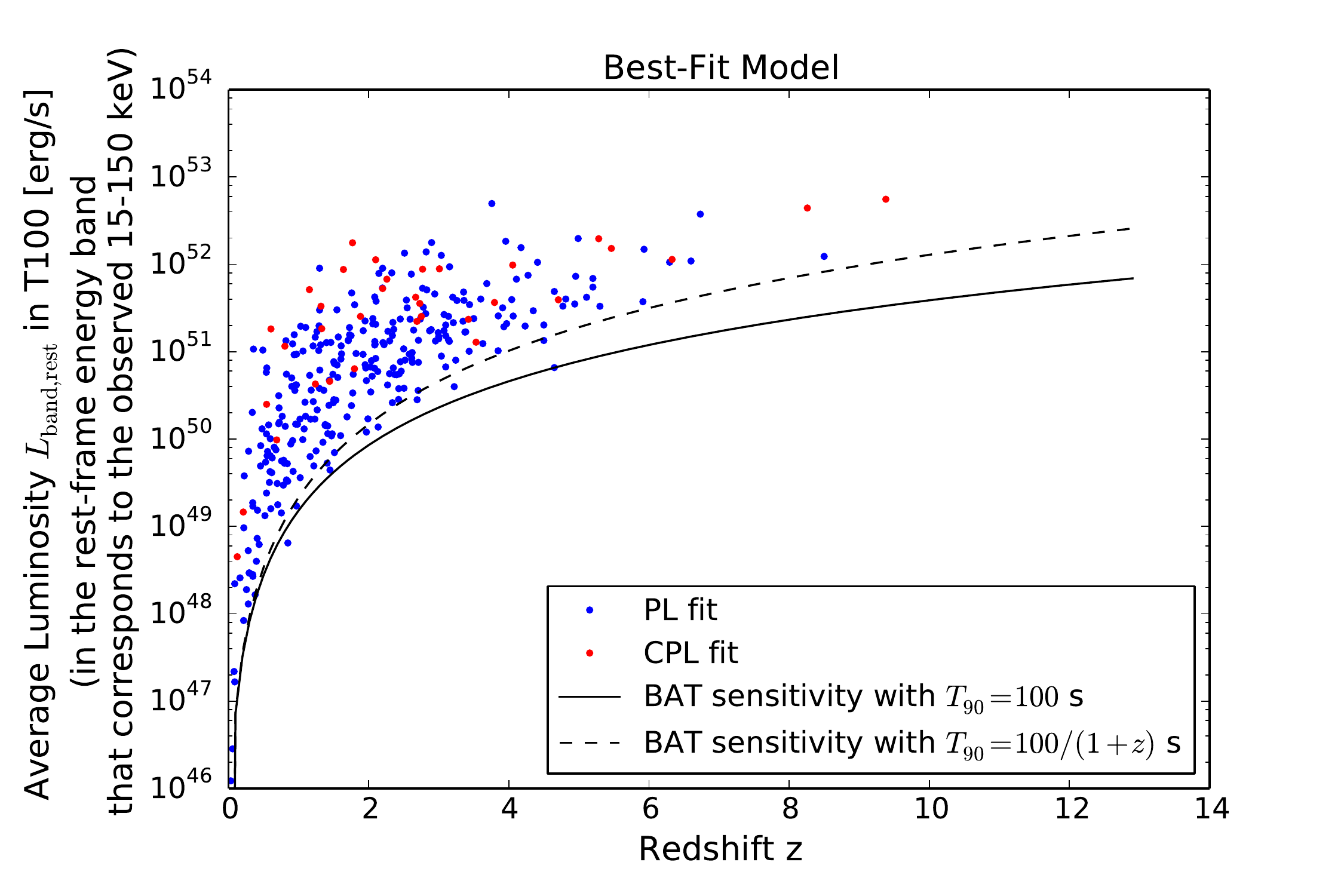}
\end{center}
\vspace{-15pt}
\caption{
GRB luminosity in the rest-frame energy band that corresponds to the observed 15-150 keV band versus redshift. The luminosity is calculated using the best-fit model (either PL or CPL) in the T100 range, with assumption that the emission is isotropic. Note that these luminosities are for different rest-frame energy bands for GRBs at different redshifts. 
}
\label{fig:lum_vs_z}
\end{figure}

Figure \ref{fig:lum_vs_z} shows the GRB luminosity in the observed 15-150 keV band as a function of redshift. 
Due to the limited energy range of the BAT, we can only constrain the energy emission within the BAT energy range.
Extrapolating the spectral fits beyond the BAT energy range is probably not a good approximation because 
the turn-over point in the spectrum (i.e., $E_{\rm peak}$) might happen somewhere above the BAT energy limit,
and thus the total luminosity calculated by extrapolating the spectral fits from the BAT spectrum can
be over-estimated.
We therefore calculate the luminosity in the observed 15-150 keV (the BAT energy range). However, this luminosity 
will correspond to a different rest-frame energy range for a GRB at a different redshift. 
Specifically, the luminosity is calculated by the following equations:
\begin{align}
L_{\rm band, rest} = \int^{E^{\rm max}_{\rm rest}}_{E^{\rm min}_{\rm rest}} L_{E,{\rm rest}} \ dE_{\rm rest} \nonumber 
         &= \int^{E^{\rm max}_{\rm rest}}_{E^{\rm min}_{\rm rest}} \frac{4 \pi D^2_L}{1+z} \ F_{E,{\rm obs}} \ dE_{\rm rest} \nonumber \\
	& =  \int^{E^{\rm max}_{\rm obs}}_{E^{\rm min}_{\rm obs}} \frac{4 \pi D^2_L}{1+z} \ F_{E, {\rm obs}} \ (1+z) dE_{\rm obs}  \nonumber \\  
	& = \int^{E^{\rm max}_{\rm obs}}_{E^{\rm min}_{\rm obs}} 4 \pi D^2_L \ F_{E,{\rm obs}} \ dE_{\rm obs} \nonumber \\
	&= 4 \pi D^2_L \ F_{\rm band, obs} 
\end{align}

where $F_{\rm band, obs} =  \int^{E^{\rm max}_{\rm obs}}_{E^{\rm min}_{\rm obs}} F_{E,{\rm obs}} \ dE_{\rm obs}$  refers to the flux in the observed energy range, and $L_{\rm band, rest}$
corresponds to the luminosity in the rest-frame energy range $E^{\rm min}_{\rm rest} = E^{\rm min}_{\rm obs}(1+z)$ to $E^{\rm max}_{\rm rest} = E^{\rm max}_{\rm obs}(1+z)$. 
More detail descriptions can be found in Appendix \ref{sect:lum_flux}.

As expected, there is a clear correlation between the minimum luminosity of detected burst with redshift, which is mainly from the Malmquist bias.
The black lines in Fig.~\ref{fig:lum_vs_z} demonstrate the effect of Malmquist bias due to the sensitivity of BAT. The solid line shows the minimum detectable luminosity 
in the observed 15-150 keV band assuming a $T_{90}$ of $100$ s. The dashed black line shows the same detectable luminosity but assuming a redshift-dependent $T_{90}$ of $100/(1+z)$ s.
Note that this $T_{90}$ becomes shorter at higher redshift, contrary to the expectation of time dilation effect. As discussed in Section~\ref{sect:T90_vs_z}, there is no clear evidence of the time-dilation effect
in the burst durations, which is likely because a larger fraction of the bursts are buried under noisy background as the burst become dimmer at higher redshift.



%
%
%
%
%

\subsection{Some statistics of observational constraints}

\begin{figure}[!h]
\begin{center}
\includegraphics[width=0.7\textwidth]{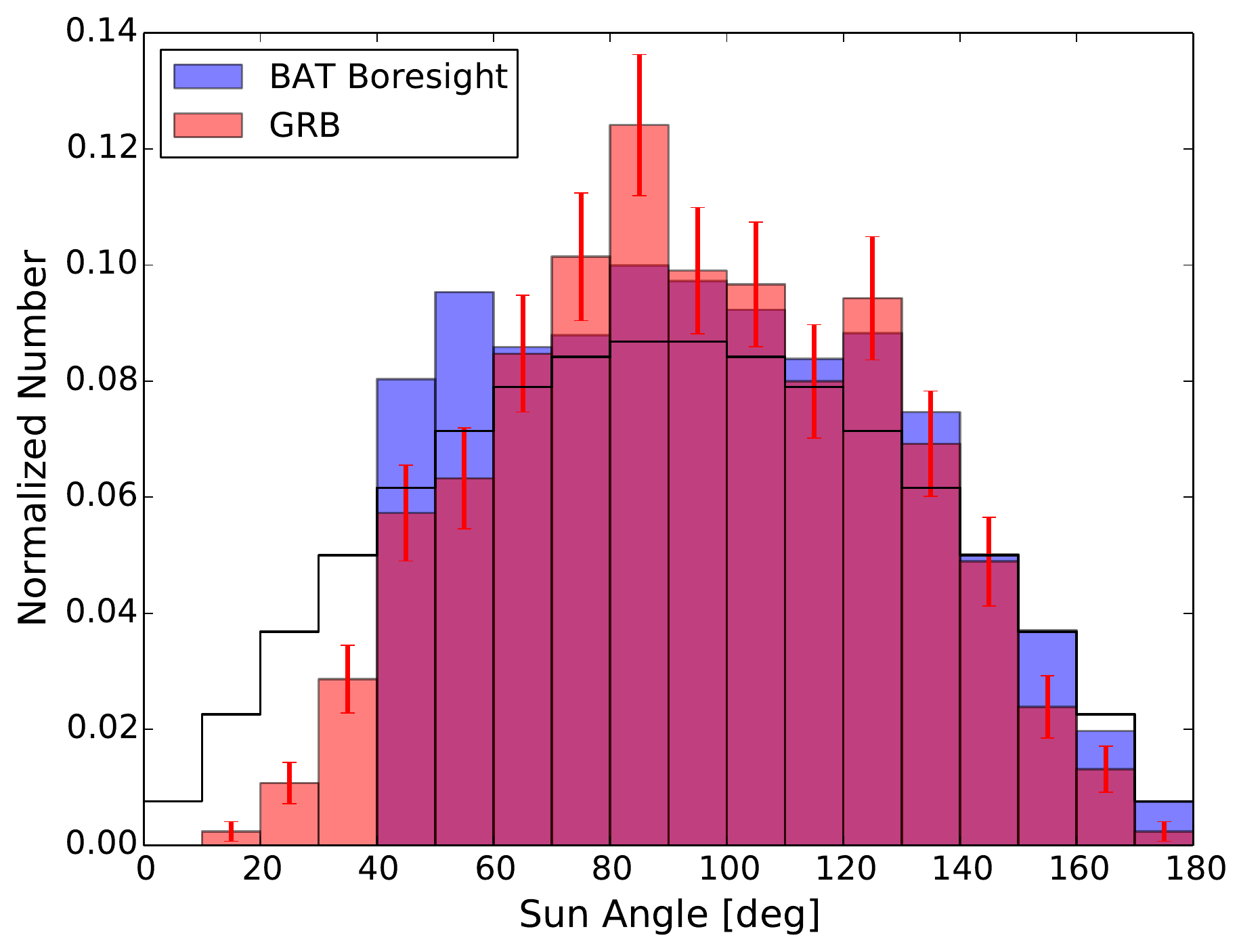}
\end{center}
\caption{
Distributions of the Sun angles of the GRBs (red bars) and the BAT boresight (blue bars).
The $1-\sigma$ poisson errors are plotted for the bins of GRB detections.
For comparison, the black lines mark the solid angle area for the Sun angle range in each bin.
}
\label{fig:GRB_sun_angle}
\end{figure}

Figure \ref{fig:GRB_sun_angle} shows the normalized distributions of both the GRB Sun angle (red bars) and the BAT boresight Sun angle (blue bars)
(i.e., the angle between BAT's pointing direction and the Sun). For the GRB Sun angles, 
a $1 \sigma$ error estimated from the Poisson distribution (i.e. $\sqrt{N}$) for each bin is also plotted.
The normalized numbers for the BAT boresight Sun angle are calculated from the Sun angle 
recorded every 5 sec (excluding the time during SAA and spacecraft slews) from 2005 to August 2015. Thus, the number for the BAT boresight Sun angle represents the fraction of time that BAT spends
at each Sun angle.

Both distributions generally follow the sin$\theta$ shape (where $\theta$ is the Sun angle), which is expected from the amount of solid angle covered by
the same angular range $\delta \theta$ (for example, there is more area covered within 80 to 90 degrees than from 170 to 180 degrees).
This relation of the solid angle versus Sun angle is shown as the black lines in the plots. 
The sharp drop off for the BAT boresight Sun angle at around 40 degree is due to the Sun constraints for the XRT, UVOT, and the star-tracker
at $\sim 44$ degrees\footnote{\label{constraint_web}http://swift.gsfc.nasa.gov/analysis/uvot\_digest/numbers.html}.
There are some GRBs detected with Sun angle less then $\sim 44$ degrees because the BAT has a large field of view. These bursts will have Sun constraints 
for prompt XRT and UVOT observations. 
This figure shows that the number of GRB detections generally follows the fraction of time that BAT spends at the location.
The GRB detection rate seem to be slightly lower than average when BAT is pointing close to the Sun. 
However, this effect is not very significant and is still consistent with the BAT pointing time (the blue bars) 
within $\sim 2-3 \sigma$ of the Poisson errors from the number of detections.
  

\begin{figure}[!h]
\begin{center}
\includegraphics[width=0.7\textwidth]{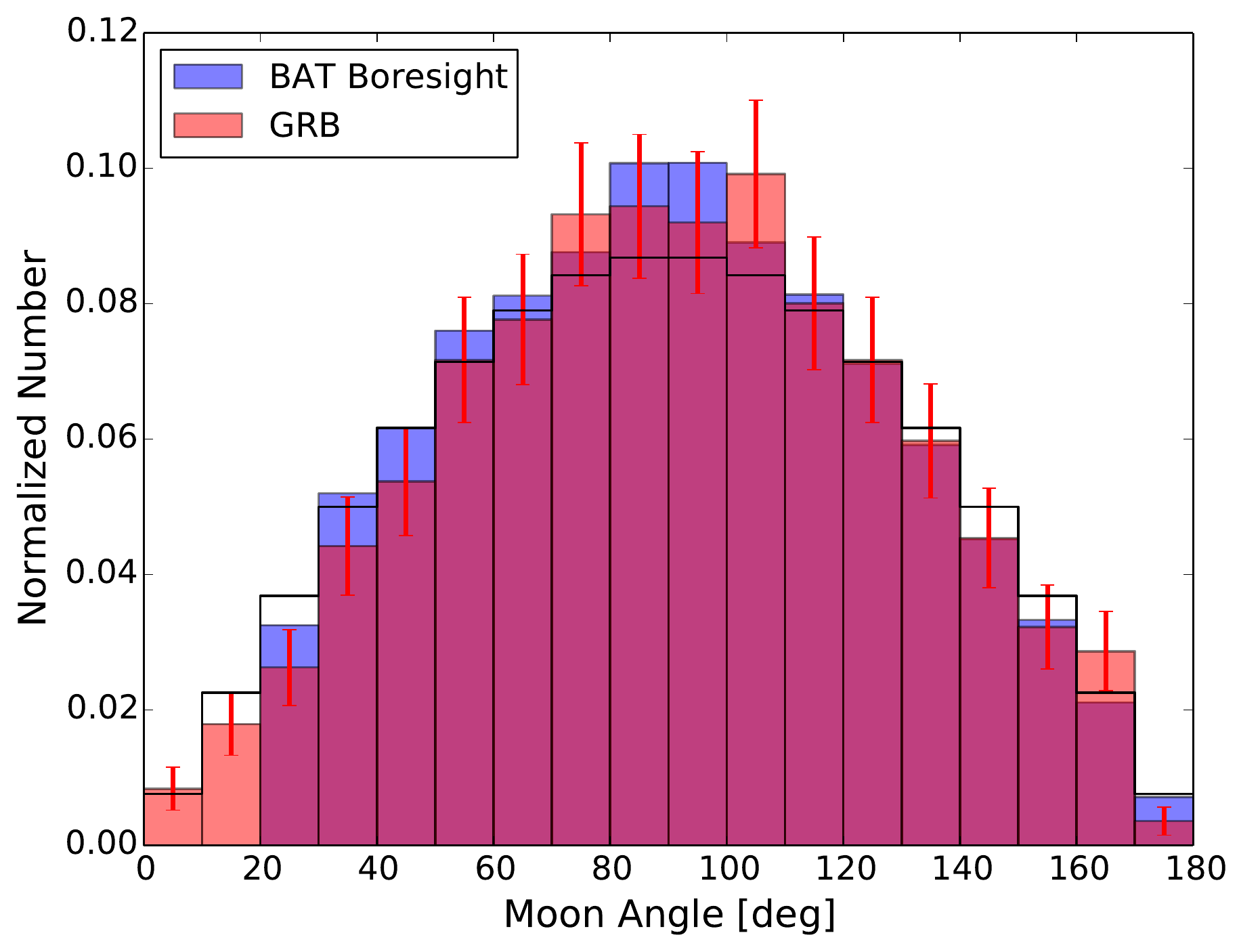}
\end{center}
\caption{
Distributions of the Moon angles of the GRBs (red bars) and the BAT boresight (blue bars).
The $1-\sigma$ poisson errors are plotted for the bins of GRB detections.
For comparison, the black lines mark the solid angle area for the Sun angle range in each bin. 
}
\label{fig:GRB_moon_angle}
\end{figure}

Similar to Fig.~\ref{fig:GRB_sun_angle}, the distributions of the Moon angles for both the GRBs (red bars) and the BAT boresights (blue bars)
are plotted in Fig.~\ref{fig:GRB_moon_angle}.
The number of GRB detections also follows the fraction of time that BAT spends at each Moon angle location, as expected.
Again, the drop off at $\sim 20$ degrees in the distribution of the BAT boresight moon angle is due to the Moon constraints for 
XRT, UVOT, and the star tracker at $\sim 19$ degrees\textsuperscript{\ref{constraint_web}}).

\vspace{-10pt}
\section{GRB analysis with the BAT survey data}
\label{sect:survey_data_analysis}

To perform further search of pre-burst or extended GRB emissions beyond 
the period of the event data, we use the results of the 104-month survey analysis, which is an extension of the previously published 70-month survey data analyses \citep{Baumgartner13}.
The survey process performs standard survey analysis using the script {\it batsurvey}\footnote{\label{note:batsurvey} https://heasarc.gsfc.nasa.gov/ftools/caldb/help/batsurvey.html},
and generates mask-weighted, cleaned images for each observations in eight energy bands (14-20, 20-24, 24-35, 35-50, 50-75, 75-100, 100-150, 150-195 keV).

We select a sub-sample of these images that have 
times close to the GRB trigger time.  In addition, we exclude the images with times overlaps roughly with the event data range when we perform the search. 
In other words, we adopt the following criteria for the search:
\begin{enumerate}
\item For GRBs occurred after GRB060319, we search in survey images with time that covers $T0-0.2$ day to $T0-50$ s, and $T0+500$ s to $T0+1$ day, where $T0$ is the BAT trigger time.
\item For GRBs occurred before (and include) GRB060319, we search in survey images with time that covers $T0-0.2$ day to $T0-50$ s, and $T0+300$ s to $T0+1$ day.
\end{enumerate}
The two different criteria is required because BAT downlinks a shorter range of event data ($\sim T0-250$ s to $\sim T0+300$ s) in early mission time, and extends
the event data range after GRB060319.
We then estimate the signal-to-noise ratio of the GRB locations in these images using the standard BAT analysis script {\it batcelldetect}\footnote{https://heasarc.gsfc.nasa.gov/ftools/caldb/help/batcelldetect.html}.
We use GRB locations estimated by the XRT, which has a resolution of $\sim$ few arcsec.
We also include a list of 21 very bright X-ray sources (e.g., Crab, Vela X-1, Cyg X-1...etc) in the input source catalog of {\it batcelldetect},
in case there is any residue of these bright sources in the background area 
that might affect detections at the GRB locations.
Furthermore, we examine the resulting detection times in survey data to make sure those are not already reported using the event data analysis.
Note that because the survey process only includes data from Dec. 2004 to Aug. 2013, 
we only search through possible GRB emissions in survey data until Aug. 2013.

\subsection{False-detection rate: searching for weak emission}
\label{sect:false_detection-rate}

We perform a study of the false-detection rate in order to find a reliable criteria to search for weak emission.
To quantify the false detection rate, we estimate the signal-to-noise ratio using 
background locations around GRBs.
We choose the background locations to be $\sim 1$ deg from the GRBs (so most of the time the background detection is from the same images as the GRB detections),
and also $\sim 1$ deg from other X-ray sources. We adopt the X-ray source list from \citet{Krimm13}.   

We quantify the false-detection rate $R_{\rm false}$ in a particular energy band with a specific signal-to-noise ratio threshold as follows,
\beq
R_{\rm false} = \frac{\rm N ({>SNR_{lim}})}{N_{\rm tot}},
\eeq
where $\rm N ({>SNR_{lim}})$ is the number of survey images with the background signal-to-noise ratio at the specific location higher than the assigned threshold.
$N_{\rm tot}$ is the total number of survey images we included in the search, that is, the subset of all survey images that are close to GRB trigger times, as described above. 
Note that because of the different event data ranges for the earlier mission time and the later mission time,
the false-detection rate study uses images that satisfy criterion 1 mentioned in previous section.
The image exposure times can vary from $\sim 300 \ \rm s$ to $\sim 2500 \ \rm s$, with the majority of the exposure time around few hundred seconds.
Ideally, one would require the observation time of each image to be identical to have a fair comparison of the signal-to-noise ratio
in each image.
However, because the survey process \citep{Baumgartner13} only produces one image for each observation (i.e., using the ``SNAPSHOT'' option in {\it batsurvey}$^{\ref{note:batsurvey}}$),
our estimation can only based on these images with different exposure times.
To produce survey images with finer time bins in each observation (i.e., using the ``DPH'' option in {\it batsurvey}$^{\ref{note:batsurvey}}$), 
one would need to re-process all survey data and the process will take 
$\sim$ months to years to finish, and hence would be beyond the scope of this paper.

We quantify the false-detection rate for a range of different signal-to-noise ratio thresholds (from $\sim$ 2.0 to $\sim$ 5.0)
in different energy bands.
The energy ranges we tried include the eight energy bands used by the survey process (14-20 keV, 20-24 keV, 24-35 keV, 35-50 keV, 50-75 keV, 100-150 keV, 
150-195 keV),
the total energy band (14-195 keV),
an energy band that combines the three soft bands (14-35 keV),
and an energy band that covers the three energy bands with the largest effective area (35-100 keV).

\begin{figure}[!h]
\begin{center}
\includegraphics[width=0.8\textwidth]{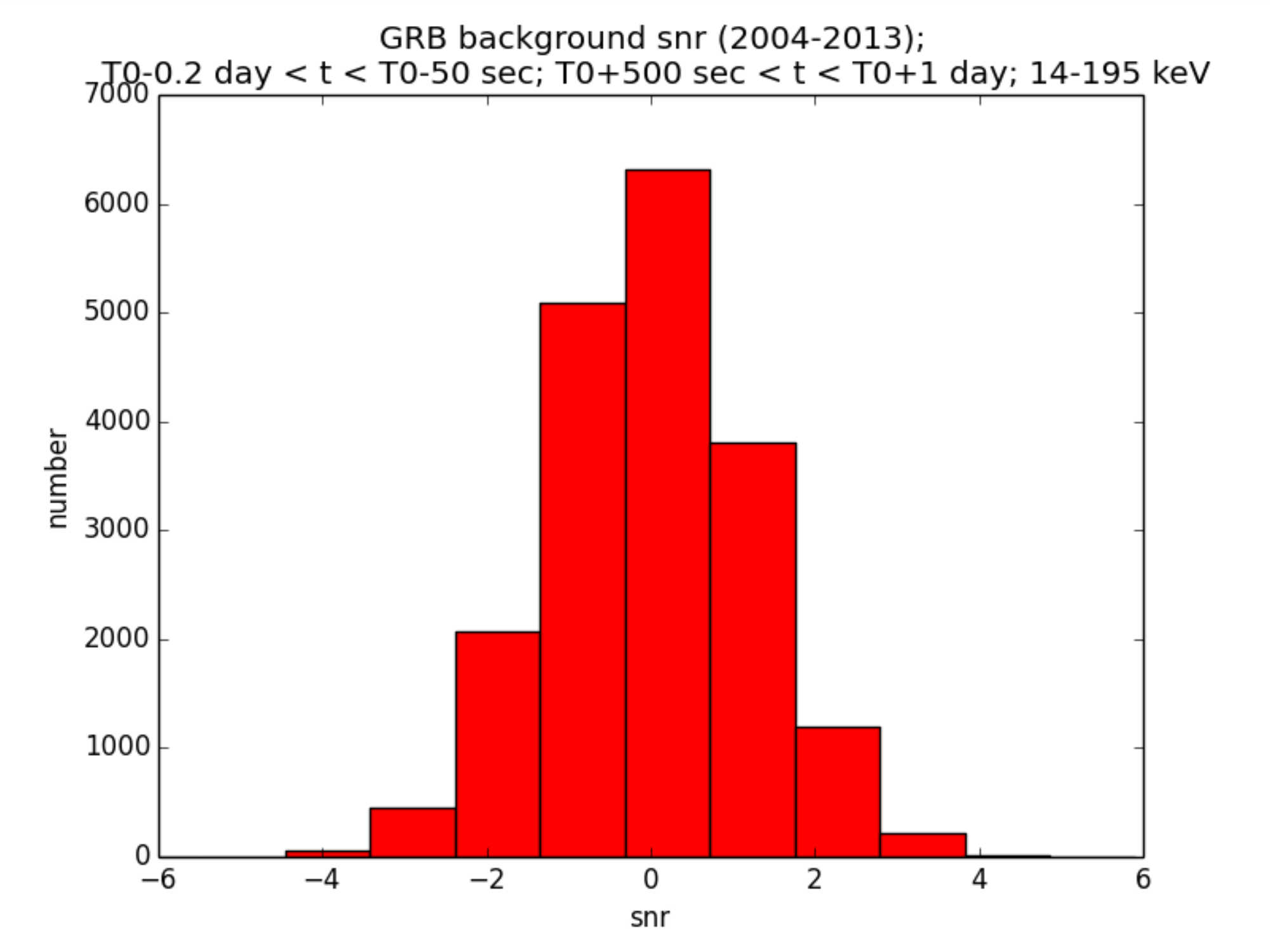}
\end{center}
\vspace{-20pt}
\caption{
The histogram of signal-to-noise ratio at the background locations in images with energy 14-195 keV, from Dec. 2004 to Aug. 2013.
}
\label{fig:snr_histogram}
\end{figure}

Figure \ref{fig:snr_histogram} shows an example of the resulting histogram of the signal-to-noise ratio at the background locations in images with energy 14-195 keV.
Table \ref{tab:snr_bgd} lists some of our calculations of false-detection rate in the most interesting ranges of signal-to-noise ratio threshold.
The numbers in parenthesis following each false-detection rate are the expected numbers of false detections out of the whole image samples
in our study (i.e., a total number of 19182 images).

We investigate the expected detection rate for each criteria, and select some potentially useful criteria to perform further tests by 
calculating the signal-to-noise ratios at the GRB locations. This gives us a total number of real detections plus false detections 
at each GRB location. 
We search through each criterion until we find one that gives the largest ratio of the number of detections at the GRB location $N_{\rm GRB\_locations}$ (i.e., number of real plus false detections) over
the number of detections at the background locations $N_{\rm bgd\_location}$ (i.e. false detections). In other words, we demand
the ratio
\beq
r_{\rm detect} \equiv \frac{N_{\rm GRB\_locations}}{N_{\rm bgd\_location}} = \frac{N(\rm real \ + \ false)}{N(\rm false)}
\eeq
to be as large as possible.

Table \ref{tab:snr_criteria} presents a list of criteria that we tried, and the results of false-detection rates, actual number of detections at the GRB locations, 
and the number of detections at the background locations. Sometimes each location can have multiple detections at different times. 
Thus, the numbers in detected locations (i.e., each location only counted once even when they are detected at different times) are shown in the parenthesis. 
The criterion using images with the energy band 14-195 keV and signal-to-noise ratio threshold above 4.3 sigma turns out to be the one that 
has the highest $r_{\rm detect}$. We thus adopt this criterion to search for possible emissions in survey data.
Results are summarized and discussed in Section~\ref{sect:survey_results}

\begin{landscape}
\setlength\LTcapwidth{\textwidth} 
\setlength\LTleft{0pt}            
\setlength\LTright{0pt}           
\begin{longtable}{@{\extracolsep{\fill}}|*{15}{c|}}
 %
 \hline
Energy & SNR $\geq$ 5.0 & SNR $\geq$ 4.9 & SNR $\geq$ 4.8 & SNR $\geq$ 4.7 & SNR $\geq$ 4.6 & SNR $\geq$ 4.5 & SNR $\geq$ 4.4 \\
 \hline
14-20 &0.0 (0)&0.0 (0)&0.0 (0)&0.0 (0)&0.0 (0)&$5.21 \times 10^{-5}$ (1)&$1.56 \times 10^{-4}$ (3) \\
20-24 &$5.21 \times 10^{-5}$ (1)&$5.21 \times 10^{-5}$ (1)&$5.21 \times 10^{-5}$ (1)&$5.21 \times 10^{-5}$ (1)&$5.21 \times 10^{-5}$ (1)&$5.21 \times 10^{-5}$ (1)&$5.21 \times 10^{-5}$ (1) \\
24-35 &0.0 (0)&0.0 (0)&0.0 (0)&0.0 (0)&0.0 (0)&$1.04 \times 10^{-4}$ (2)&$1.04 \times 10^{-4}$ (2) \\
35-50 &$1.04 \times 10^{-4}$ (2)&$1.04 \times 10^{-4}$ (2)&$1.04 \times 10^{-4}$ (2)&$1.04 \times 10^{-4}$ (2)&$1.04 \times 10^{-4}$ (2)&$1.04 \times 10^{-4}$ (2)&$1.56 \times 10^{-4}$ (3) \\
50-75 &0.0 (0)&0.0 (0)&0.0 (0)&$5.21 \times 10^{-5}$ (1)&$5.21 \times 10^{-5}$ (1)&$5.21 \times 10^{-5}$ (1)&$1.04 \times 10^{-4}$ (2) \\
75-100 &0.0 (0)&0.0 (0)&0.0 (0)&0.0 (0)&0.0 (0)&$1.56 \times 10^{-4}$ (3)&$2.08 \times 10^{-4}$ (4) \\
100-150 &$1.04 \times 10^{-4}$ (2)&$1.56 \times 10^{-4}$ (3)&$1.56 \times 10^{-4}$ (3)&$2.08 \times 10^{-4}$ (4)&$2.08 \times 10^{-4}$ (4)&$2.08 \times 10^{-4}$ (4)&$2.60 \times 10^{-4}$ (5) \\
150-195 &0.0 (0)&$5.21 \times 10^{-5}$ (1)&$5.21 \times 10^{-5}$ (1)&$5.21 \times 10^{-5}$ (1)&$5.21 \times 10^{-5}$ (1)&$5.21 \times 10^{-5}$ (1)&$5.21 \times 10^{-5}$ (1) \\
14-195 &0.0 (0)&0.0 (0)&0.0 (0)&0.0 (0)&0.0 (0)&0.0 (0)&$5.21 \times 10^{-5}$ (1) \\
14-35 &0.0 (0)&$5.21 \times 10^{-5}$ (1)&$5.21 \times 10^{-5}$ (1)&$5.21 \times 10^{-5}$ (1)&$5.21 \times 10^{-5}$ (1)&$5.21 \times 10^{-5}$ (1)&$2.08 \times 10^{-4}$ (4) \\
35-100 &$5.21 \times 10^{-5}$ (1)&$5.21 \times 10^{-5}$ (1)&$5.21 \times 10^{-5}$ (1)&$2.08 \times 10^{-4}$ (4)&$2.08 \times 10^{-4}$ (4)&$2.08 \times 10^{-4}$ (4)&$2.08 \times 10^{-4}$ (4) \\
\hline
Energy & SNR $\geq$ 4.3 & SNR $\geq$4.2 & SNR $\geq$ 4.1 & SNR $\geq$4.0 & SNR $\geq$ 3.9 & SNR $\geq$ 3.8 & SNR $\geq$ 3.7 \\ 
\hline
14-20 &$1.56 \times 10^{-4}$ (3)&$2.08 \times 10^{-4}$ (4)&$2.08 \times 10^{-4}$ (4)&$2.08 \times 10^{-4}$ (4)&$6.77 \times 10^{-4}$ (13)&$6.77 \times 10^{-4}$ (13)&$1.20 \times 10^{-3}$ (23) \\
20-24 &$5.21 \times 10^{-5}$ (1)&$2.60 \times 10^{-4}$ (5)&$2.60 \times 10^{-4}$ (5)&$3.12 \times 10^{-4}$ (6)&$6.25 \times 10^{-4}$ (12)&$6.25 \times 10^{-4}$ (12)&$1.04 \times 10^{-3}$ (20) \\
24-35 &$1.04 \times 10^{-4}$ (2)&$2.60 \times 10^{-4}$ (5)&$2.60 \times 10^{-4}$ (5)&$4.17 \times 10^{-4}$ (8)&$6.25 \times 10^{-4}$ (12)&$6.25 \times 10^{-4}$ (12)&$1.25 \times 10^{-3}$ (24) \\
35-50 &$1.56 \times 10^{-4}$ (3)&$1.56 \times 10^{-4}$ (3)&$1.56 \times 10^{-4}$ (3)&$3.12 \times 10^{-4}$ (6)&$4.69 \times 10^{-4}$ (9)&$4.69 \times 10^{-4}$ (9)&$7.29 \times 10^{-4}$ (14) \\
50-75 &$1.04 \times 10^{-4}$ (2)&$2.08 \times 10^{-4}$ (4)&$2.08 \times 10^{-4}$ (4)&$3.12 \times 10^{-4}$ (6)&$5.73 \times 10^{-4}$ (11)&$5.73 \times 10^{-4}$ (11)&$9.89 \times 10^{-4}$ (19) \\
75-100 &$2.08 \times 10^{-4}$ (4)&$3.12 \times 10^{-4}$ (6)&$3.12 \times 10^{-4}$ (6)&$3.64 \times 10^{-4}$ (7)&$5.73 \times 10^{-4}$ (11)&$5.73 \times 10^{-4}$ (11)&$8.33 \times 10^{-4}$ (16) \\
100-150 &$2.60 \times 10^{-4}$ (5)&$3.64 \times 10^{-4}$ (7)&$3.64 \times 10^{-4}$ (7)&$3.64 \times 10^{-4}$ (7)&$5.21 \times 10^{-4}$ (10)&$5.21 \times 10^{-4}$ (10)&$9.89 \times 10^{-4}$ (19) \\
150-195 &$5.21 \times 10^{-5}$ (1)&$1.56 \times 10^{-4}$ (3)&$1.56 \times 10^{-4}$ (3)&$2.60 \times 10^{-4}$ (5)&$5.73 \times 10^{-4}$ (11)&$5.73 \times 10^{-4}$ (11)&$9.37 \times 10^{-4}$ (18) \\
14-195 &$5.21 \times 10^{-5}$ (1)&$2.08 \times 10^{-4}$ (4)&$2.08 \times 10^{-4}$ (4)&$5.73 \times 10^{-4}$ (11)&$7.81 \times 10^{-4}$ (15)&$7.81 \times 10^{-4}$ (15)&$1.30 \times 10^{-3}$ (25) \\
14-35 &$2.08 \times 10^{-4}$ (4)&$4.69 \times 10^{-4}$ (9)&$4.69 \times 10^{-4}$ (9)&$5.21 \times 10^{-4}$ (10)&$1.04 \times 10^{-3}$ (20)&$1.04 \times 10^{-3}$ (20)&$1.77 \times 10^{-3}$ (34) \\
35-100 &$2.08 \times 10^{-4}$ (4)&$3.12 \times 10^{-4}$ (6)&$3.12 \times 10^{-4}$ (6)&$3.64 \times 10^{-4}$ (7)&$6.25 \times 10^{-4}$ (12)&$6.25 \times 10^{-4}$ (12)&$1.25 \times 10^{-3}$ (24) \\
\hline
\caption{The false-detection rate for different signal-to-noise ratio with different energy bands. 
The numbers in parentheses are the expected number of detections of the whole sample of 19128 images.
Energy bands listed are in units of keV.
}
\label{tab:snr_bgd}
\end{longtable}
\end{landscape}

\begin{landscape}
\setlength\LTcapwidth{\textwidth} 
\setlength\LTleft{0pt}            
\setlength\LTright{0pt}           
\begin{longtable}{||c|c||c|c|c|c|}
\hline\hline
Criterion & SNR & False-detection & At GRB locations & At background locations & $r_{\rm detect}$ \\
Energy bands & threshold  & rate & \# of detections & \# of detections &     \\
\hline
any 1 of the 8 bands & 3.0 & $3.9 \times 10^{-2}$ & 778 (479 GRBs) & 719 (338 backgrounds) &  1.0 (1.42)\\
\hline
any 1 of the 8 bands & 4.0 & $2.6 \times 10^{-3}$ & 68 (58 GRBs) & 49 (47 backgrounds)  & 1.39 (1.23) \\
\hline
any 1 of the 8 bands & 4.5 & $7.8 \times 10^{-4}$ & 24 (18 GRBs) & 14 (13 backgrounds) & 1.71 (1.38)\\
\hline
any 1 of the 8 bands & 4.8 & $3.6 \times 10^{-4}$ & 18 (14 GRBs) & 6 (6 backgrounds) & 3.0 (2.33)\\
\hline
any 2 of the 8 bands & 3.0 & $7.0 \times 10^{-3}$ & 27 (20 GRBs) & 25 (23 backgrounds) & 1.08 (0.87)\\
\hline
any 2 of the 8 bands & 3.5 & $2.8 \times 10^{-5}$ & 12 (7 GRBs) & 3 (2 backgrounds) & 4.0 (3.5) \\
\hline
any 3 of the 8 bands & 3.0 & $7.0 \times 10^{-6}$ & 11 (7 GRBs) & 0 (0 backgrounds) & N/A (N/A)\\
\hline
any 3 of the 8 bands & 2.5 & $1.9 \times 10^{-4}$ & 17 (12 GRBs) & 11 (10 backgrounds) & 1.55 (1.2)\\
\hline
any 3 of the 8 bands & 2.8 & $2.9 \times 10^{-5}$ & 12 (8 GRBs) & 2 (2 backgrounds) & 6.0 (4.0)\\
\hline
both 50-75 keV and &        &                                       &                      &                     &    \\
75-100 keV bands  & 3.0 & $2.1 \times 10^{-5}$ & 5 (3 GRBs) & 0 (0 backgrounds)    &  N/A (N/A)\\
\hline
14-35 keV band & 4.0 & $5.2 \times 10^{-4}$ & 18 (13 GRBs) & 10 (10 backgrounds) & 1.8 (1.3)\\
\hline
14-35 keV band & 4.3 & $2.1 \times 10^{-4}$ & 16 (11 GRBs) & 4 (4 backgrounds)  &  4.0 (2.75)\\
\hline
14-35 keV band & 4.5 & $5.2 \times 10^{-5}$ & 14 (9 GRBs) & 1 (1 backgrounds) & 14.0 (9.0)\\
\hline
14-195 keV band & 4.3 & $5.2 \times 10^{-5}$ & 19 (14 GRBs) & 1 (1 backgrounds) & 19.0 (14.0)\\
\hline
14-195 keV band & 4.0 & $5.7 \times 10^{-4}$ & 28 (22 GRBs) & 11 (11 backgrounds) & 2.55 (2.0)\\
\hline
35-100 keV band & 4.8 & $5.2 \times 10^{-5}$ & 11 (8 GRBs) & 1 (1 backgrounds) & 11.0 (8.0)\\
\hline
35-100 keV band & 4.3 & $2.1 \times 10^{-4}$ & 14 (11 GRBs) & 4 (4 backgrounds) & 3.5 (2.75)\\
\hline
35-100 keV band & 4.0 & $3.6 \times 10^{-4}$ & 18 (15 GRBs) & 7 (7 backgrounds) & 2.57 (2.14)\\
\hline
\caption{A summary of all the criteria we examined$^{\ref{survey_table_footnote}}$.
}
\label{tab:snr_criteria}
\end{longtable}
\end{landscape}
\footnote{\label{survey_table_footnote} Because we are using criterion 1 mentioned in Section \ref{sect:survey_data_analysis}
for selecting survey images for the false-detection rate study,
the numbers here only includes detections that occur beyond $T0+500 s$. Therefore, 
there are only 14 detected GRBs listed under the criterion ``3.4 sigma in 14-195 keV band''
(the one we used for the official search). The two GRBs listed in Table \ref{tab:survey_GRB}, GRB060218 and GRB050730, with detections before 
$T0+500$ s is not included.
}

\section{Results of BAT survey data analysis: possible ultra-long GRBs}
\label{sect:survey_results}

We find 21 detections (16 GRBs) beyond the event data range, which are summarized in Table \ref{tab:survey_GRB}.
Most of these detections happened after the BAT trigger times. However, there are two detections (GRB100316D and GRB101024A) that occurred 
before the BAT trigger times.
Within these detections, 7 GRBs are previously classified as ultra-long GRBs, which are GRB121027A, GRB111215A, 
GRB111209A, GRB101225A,  GRB100316D, GRB090417B, GRB060218 \citep[e.g.,][]{Virgili13, Gendre13, Levan14, Boer15}. 
Studies usually refer ``ultra-long GRBs'' to bursts with durations $\gtrsim$ kiloseconds \citep[e.g.,][]{Gendre13, Stratta13, Levan14, Evans14, Boer15}, however, no unified definition has been adopted. 
In this paper, an ultra-long GRB is referred to those events with observed durations longer than 1000 s in BAT energy band.
Note that the ultra-long GRB130925A \citep{Evans14, Piro14} detected by BAT is not in the list of GRBs with confirmed detection in survey data (Table \ref{tab:survey_GRB_duration}), because the currently existing survey data product required for the analysis ends before this burst.

Table \ref{tab:survey_GRB_duration} compares the total burst duration in BAT, and the $T_{90}$ estimated from the event data.
The burst duration from both the event and survey data is a rough estimation from the beginning of the $T_{100}$ (or the beginning of the event data if the burst emission starts beforehand) plus the time of middle point of the last survey detection time bin.
For those two precursor detection, the duration is estimated from the middle point of the survey detection to the end of $T_{100}$.

There are 15 GRBs in Table \ref{tab:survey_GRB_duration} with estimated duration longer than 1000 s when including the emissions in survey data,
and thus are considered as ultra-long GRBs with our definition. However, we note the potentially large uncertainty in the burst duration estimations 
due to the large time bin (i.e., long exposure time) of some survey data. Moreover, we expect one false detection in this sample, based on the 
false-detection rate study mentioned in previous section.

Seven bursts in this sample have durations exceeding the event data range (and thus only a lower limit of $T_{90}$ can be determined).
All of these seven bursts are previously recognized as ultra-long GRBs. The rest of the bursts, however, show a rather diverse $T_{90}$, ranging
from a few seconds to a few hundred seconds.
If using the lower limit for bursts without constraint $T_{90}$), the median of $T_{90}$ from event data for these ultra-long bursts is 193 s. Only $\sim 6\%$
of the total BAT GRBs with constraint burst durations have $T_{90}$ larger than 193 s.

 
Note that there are two detections (GRB060218 and GRB050730) that happen within 500 sec after the BAT trigger times.
This is because earlier in the mission, BAT downlinked a shorter range of event data that only covers until $\sim 300$ s after trigger time.
Therefore, these two detections would have been covered by event data range if the GRBs had occurred more recently.
Also, GRB080319B is the well-known naked-eye burst \citep{Racusin08}. It is possible that the extraordinary brightness of this burst is the main reason for 
the event being detectable for a long time in BAT for $\sim 1340$ sec. Thus, one needs to be cautious when exploring potential physical causes
of these late time BAT detections, particularly for brighter GRBs in our sample.

\begin{table}[h!]
\caption{\label{tab:survey_GRB}
List of GRBs detected in survey data with signal-to-noise ratio $> 4.3 \sigma$ in 14-195 keV. We expect on average $\sim 1$ false detection in
our search sample. The GRBs that are detected in all the criteria we investigated (see Section~\ref{sect:false_detection-rate}) are noted by $\star$.
}
\begin{center}
\begin{tabular}{|c|c|c|c|}
\hline\hline
GRB name & Start time of detection exposure [s] & Image exposure time [s] & SNR in 14-195 keV \\
                   &   (relative to the BAT trigger time) &                                       &   \\
\hline
GRB121027A$^\star$ & 1327.45 & 496.0 & 19.32 \\
GRB121027A$^\star$  & 5351.45 & 732.0 & 11.64 \\
GRB111215A$^\star$  &  703.0     &  840.0 &  12.27 \\
GRB111209A$^\star$  &  4814.0   & 2600.0 & 40.98 \\
GRB111209A$^\star$  & 10606.0 & 2584.0 & 14.08 \\
GRB111209A$^\star$  & 16427.0 & 2400.0 & 7.58 \\
GRB111209A$^\star$  & 565.0      & 630.0  & 92.73 \\
GRB101225A$^\star$  &  1372.0   & 300.0  & 10.28 \\
GRB101225A$^\star$  & 4936.0   & 2601.0 & 4.55 \\
GRB101024A  &  -5252.13 & 779.0 & 4.73 \\
GRB100728A  &  981.73     & 792.0 & 4.83 \\
GRB100316D$^\star$  &  -775.0  & 600.0 & 9.01 \\
GRB091127    &  5192.90    & 409.0 & 4.36 \\
GRB090417B$^\star$  &  662.0   & 1140.0 & 23.51 \\
GRB090404   &  44356.93   & 557.0 & 4.31 \\
GRB090309A   &  4075.176   & 2400.0 & 4.40 \\
GRB080319B$^\star$  &  938.1  &  799.0   & 11.26 \\
GRB070518   &  57158.83   & 1381.0 & 4.92 \\
GRB070419B  & 3724.13    & 2400.0 & 5.22 \\
GRB060218$^\star$     & 404.0   &  2327.0  & 19.20 \\
GRB050730    & 356.2   &  390.0   & 8.53 \\
\hline\hline
\end{tabular}
\end{center}
\end{table}

\begin{table}[h!]
\caption{\label{tab:survey_GRB_duration}
Comparison of the GRB duration estimated using only event data and using both event and survey data.
Again, GRBs that are detected in all the criteria we investigated (see Section~\ref{sect:false_detection-rate}) are noted by $\star$.
}
\begin{center}
\begin{tabular}{|c|c|c|c|}
\hline\hline
GRB name & Number of detections & Burst duration from & $T_{90}$  \\
                     &  in survey data            & both event and survey data & from event data \\
\hline
GRB121027A$^\star$ & 2 & $\sim 5727$ s & 80 s \\
GRB111215A$^\star$  &  1    &  $\sim 1121$ s &  $> 374$ s  \\
GRB111209A$^\star$  &  4   & $\sim 18181$ s &  $>811$ s \\
GRB101225A$^\star$  &  2  & $\sim 6416$ s &  $> 1377$ s \\
GRB101024A  &  1 & $\sim 4883$ s & 18.7 s \\
GRB100728A  &  1     & $\sim 1457$ s & 193 s \\
GRB100316D$^\star$  &  1  & $\sim 1270$ s & $> 522$ s \\
GRB091127    &  1    & $\sim 5398$ s & 7.42  \\
GRB090417B$^\star$  &  1   & $\sim 1471$ s & $> 267$ s \\
GRB090404   &  1   & $\sim 44671$ s & 82 s \\
GRB090309A   &  1   & $\sim 5276$ s & 3 s \\
GRB080319B$^\star$  &  1  &  $\sim 1341$ s   & $>125$ s\\
GRB070518   &  1   & $\sim 57851$ s & 5.5 s \\
GRB070419B  & 1    & $\sim 4934$ s & 238 s \\
GRB060218$^\star$     & 1   &  $\sim 1624$ s  & $> 602$ s \\
GRB050730    & 1   &  $\sim 608$ s  &  155 s\\
\hline\hline
\end{tabular}
\end{center}
\end{table}

Although we found relatively small number of detections in the survey data, this does not necessarily imply that 
the GRB emissions usually finish before the end of the event data.
The survey data are binned in $\gtrsim 5$ minutes, and thus it is not sensitive to late-time burst emissions
if they occur on a much shorter time scale.
In fact, throughout the process of inspecting the burst light curves created by event data by eye, we noted
52
GRBs
with incomplete burst durations, which are likely to have burst emissions beyond the event data range.
Interestingly, there are 9 GRBs (GRB121027A, GRB101024A, GRB100728A, GRB091127, GRB090404, GRB090309, GRB070518, GRB070419B, and GRB050730) 
that have detections in the survey data but are not included in the list of burst with incomplete burst durations,
which is consistent with the results shown in Table \ref{tab:survey_GRB_duration} that there is no clear relation between the prompt BAT emission
in the event data range and the late-time detections in the survey data.

\begin{figure}[!h]
\begin{center}
\includegraphics[width=0.49\textwidth]{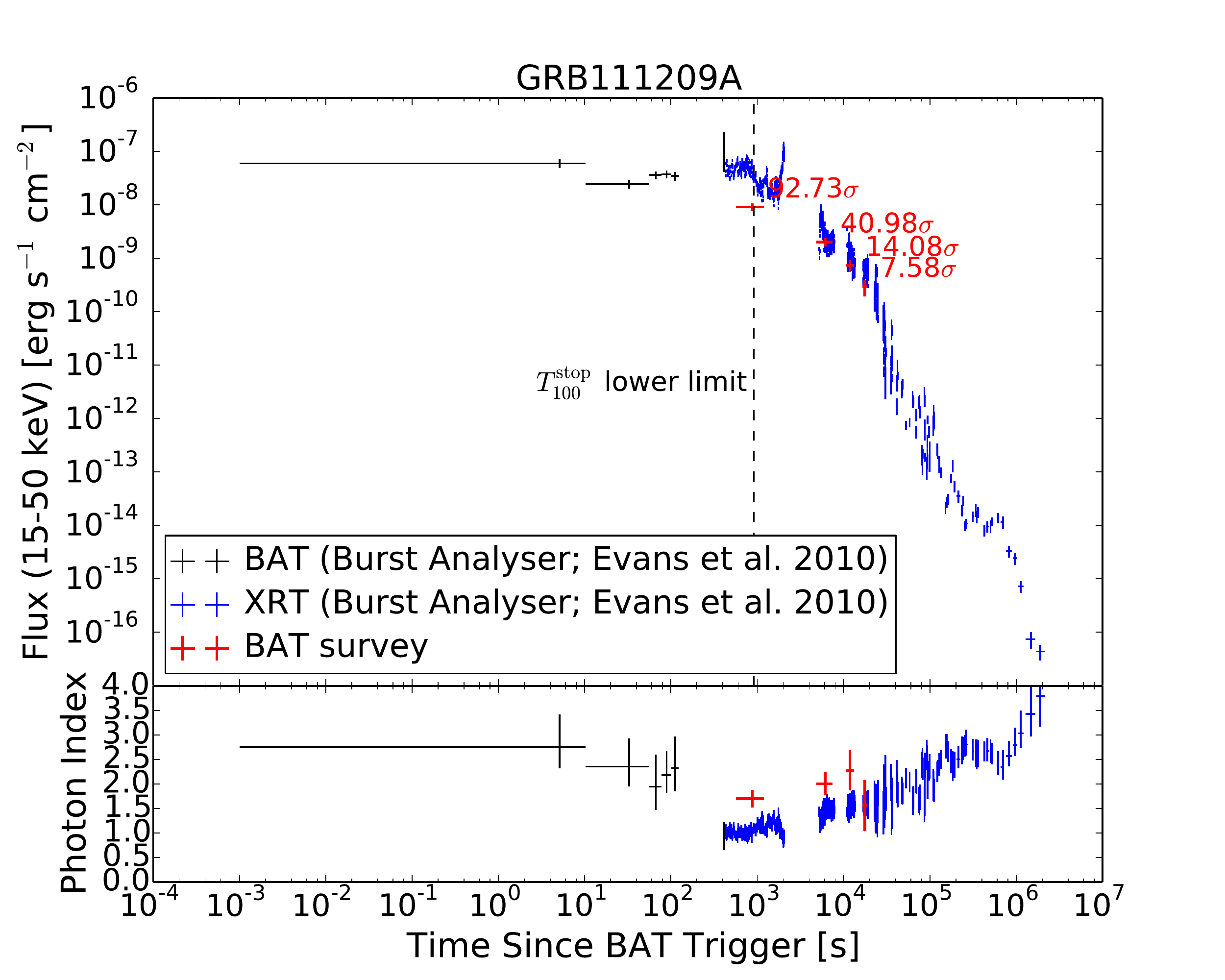}
\includegraphics[width=0.49\textwidth]{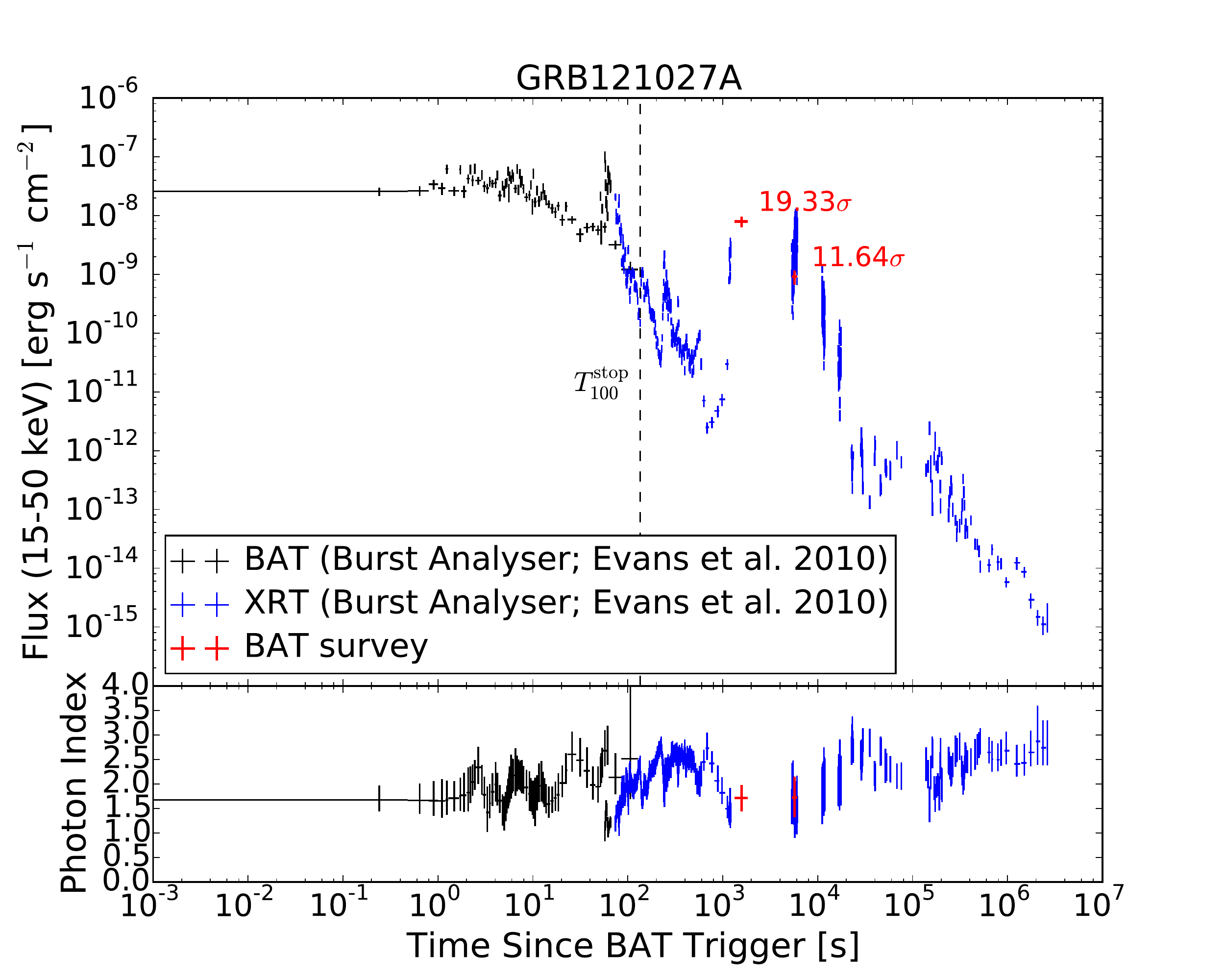}
\includegraphics[width=0.49\textwidth]{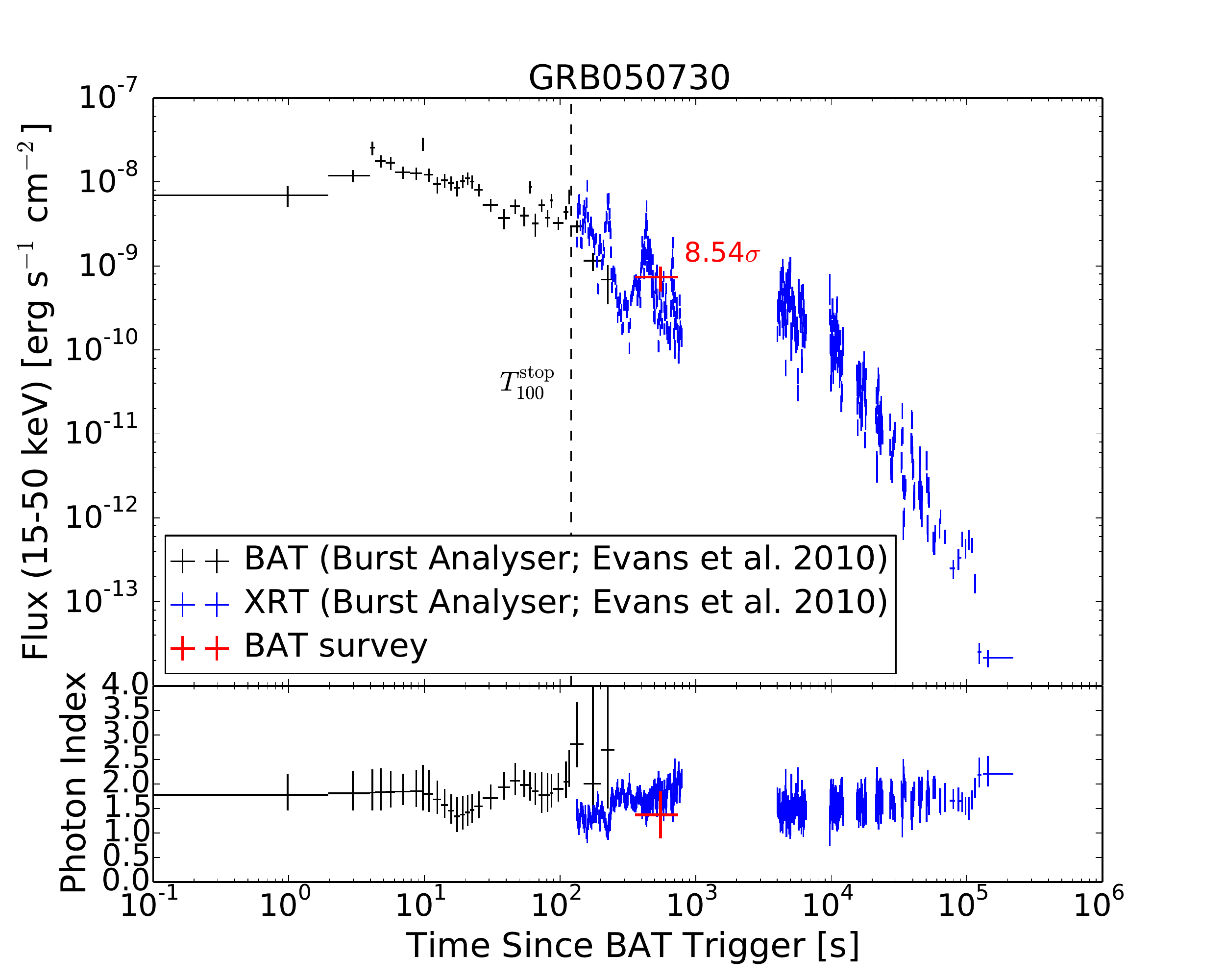}
\includegraphics[width=0.49\textwidth]{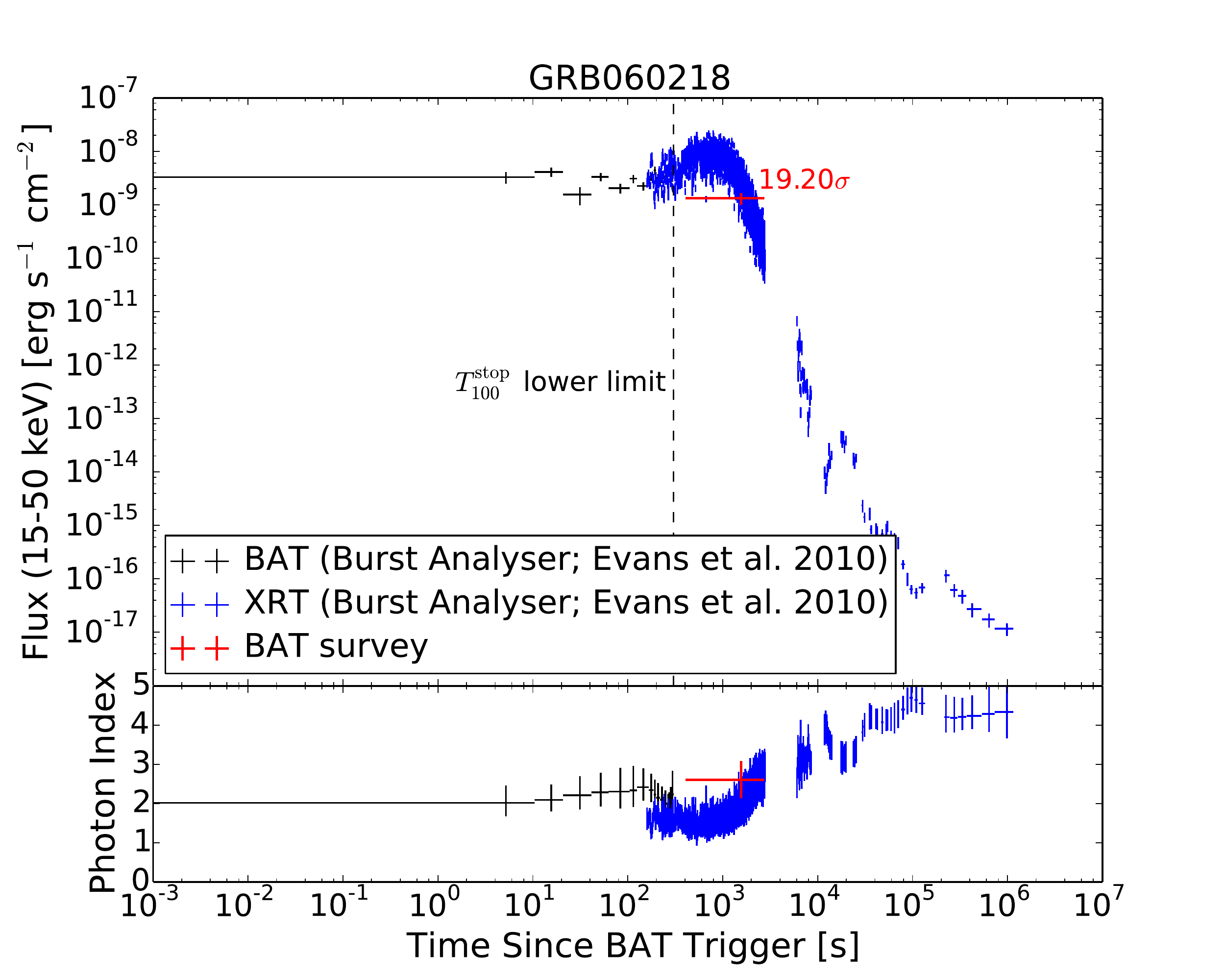}
\includegraphics[width=0.49\textwidth]{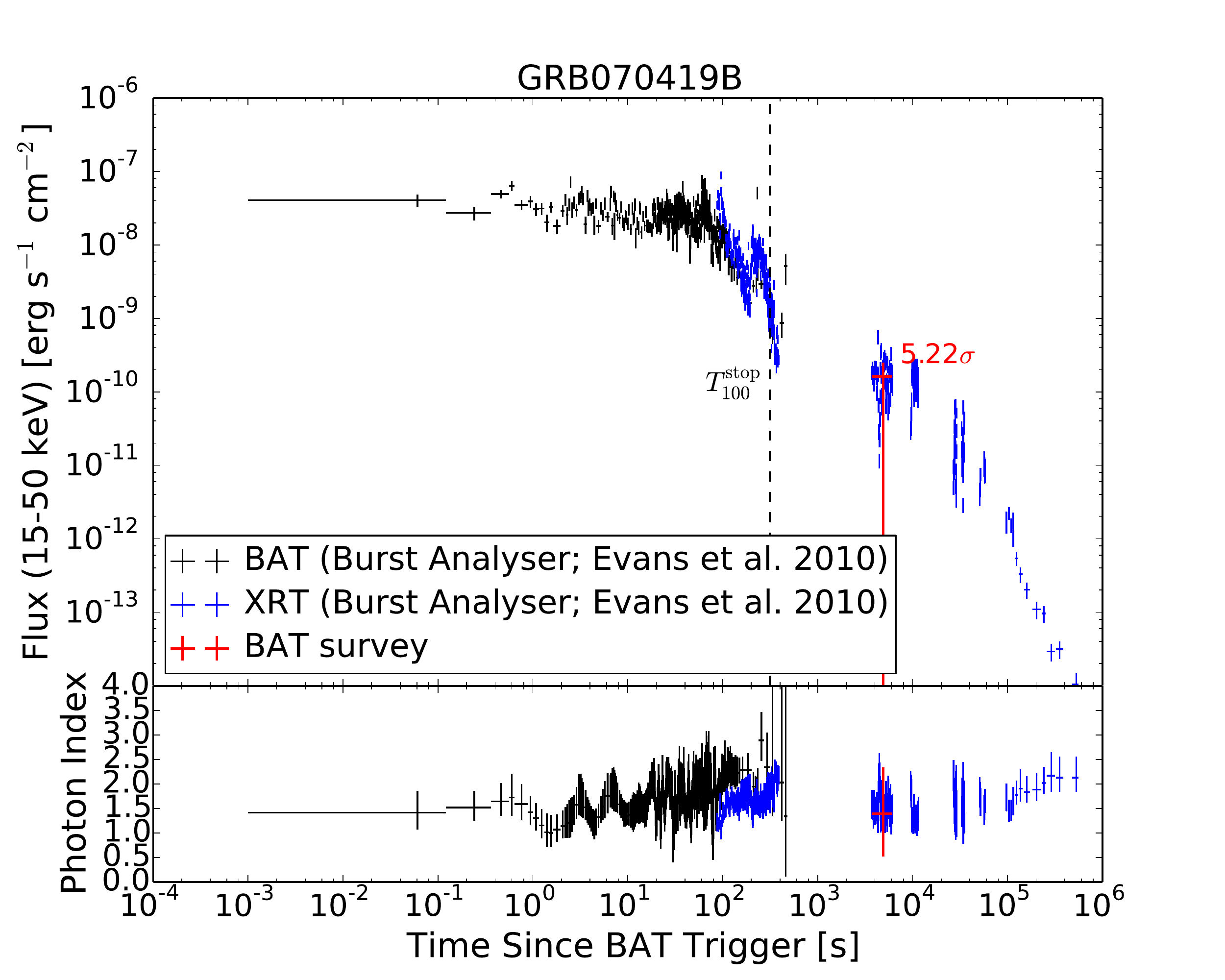}
\includegraphics[width=0.49\textwidth]{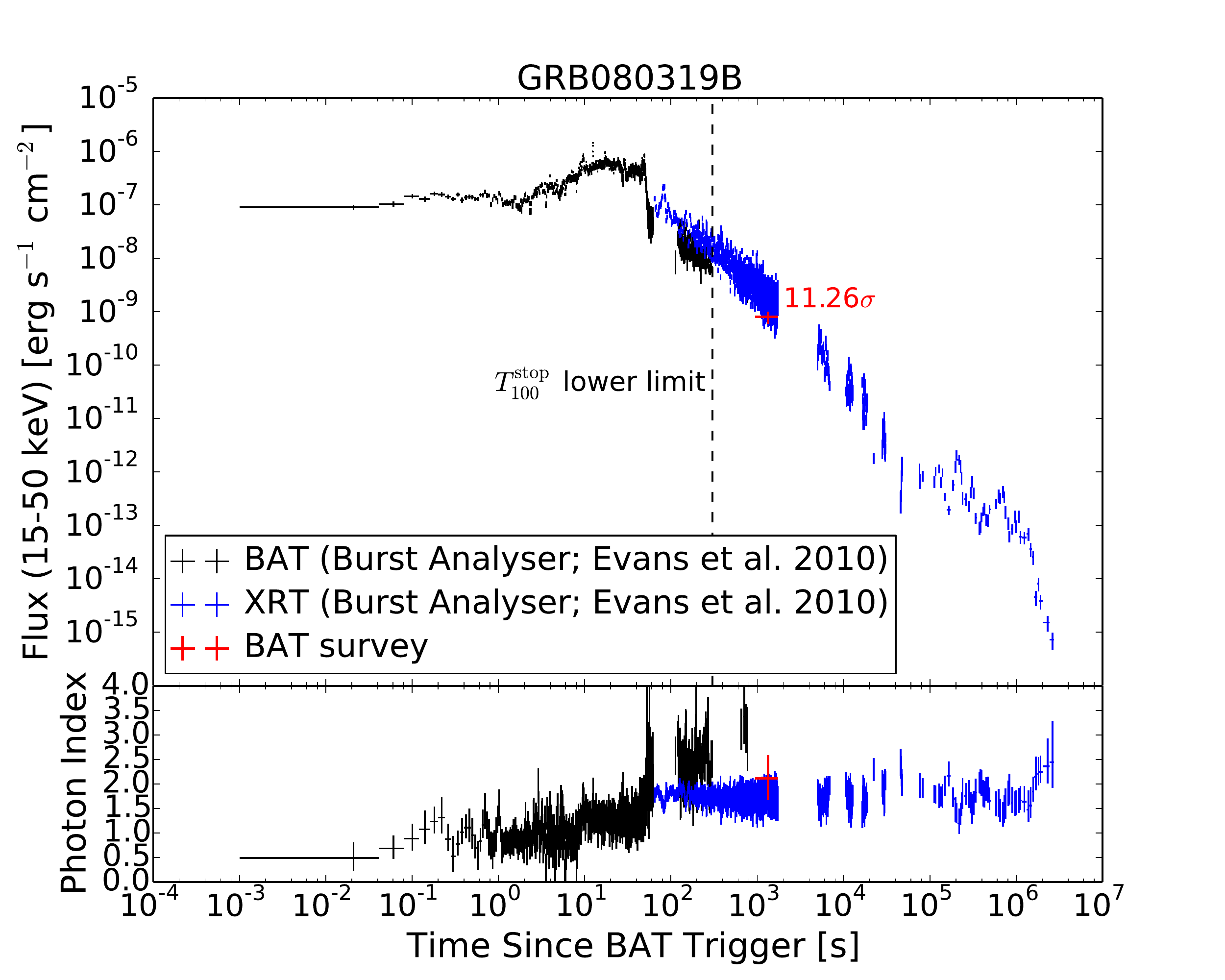}
\end{center}
\caption{
Comparisons between detections in the BAT survey data, BAT event data, and the XRT data.
The end of $T_{100}$ is marked as black-dashed line in each figure. If the burst emissions extend beyond the event data range, the black-dashed line marks the lower-limit of $T_{100}$ (i.e., the end of $T_{100}$ found by {\it battblocks}, or the end of the event data range if {\it battblocks} failed to estimate the burst durations).
}
\label{fig:lc_1}
\end{figure}

\begin{figure}[!h]
\begin{center}
\includegraphics[width=0.49\textwidth]{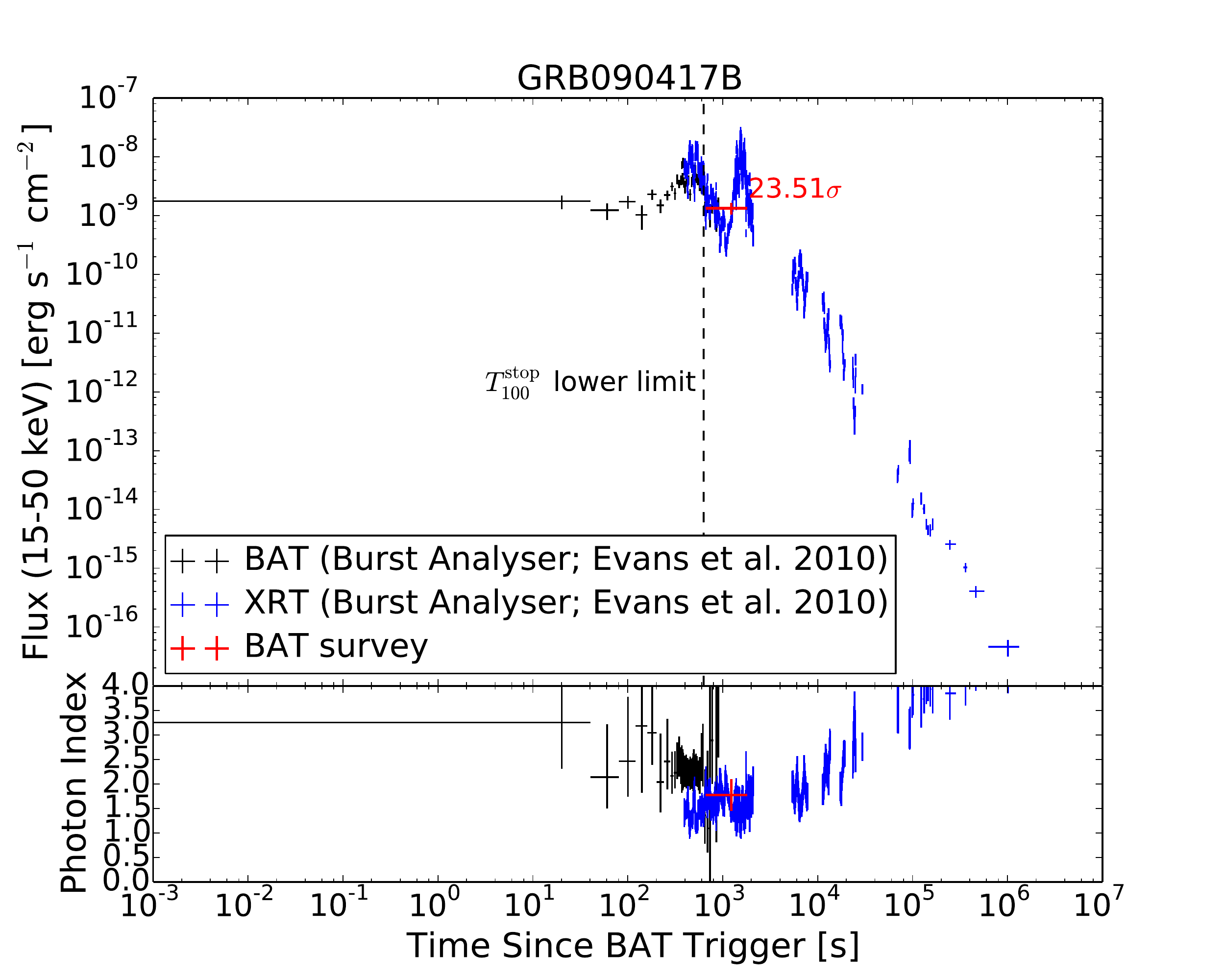}
\includegraphics[width=0.49\textwidth]{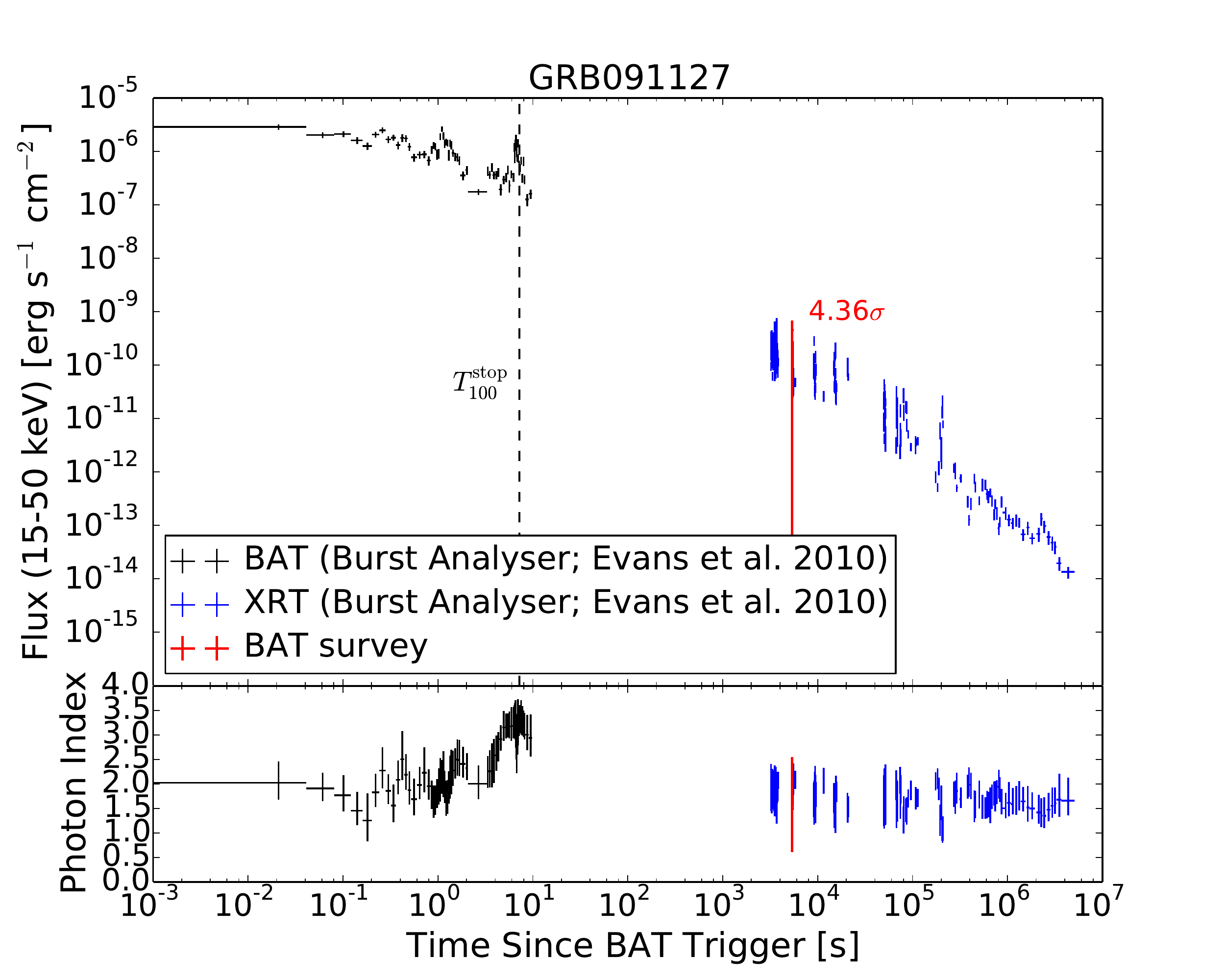}
\includegraphics[width=0.49\textwidth]{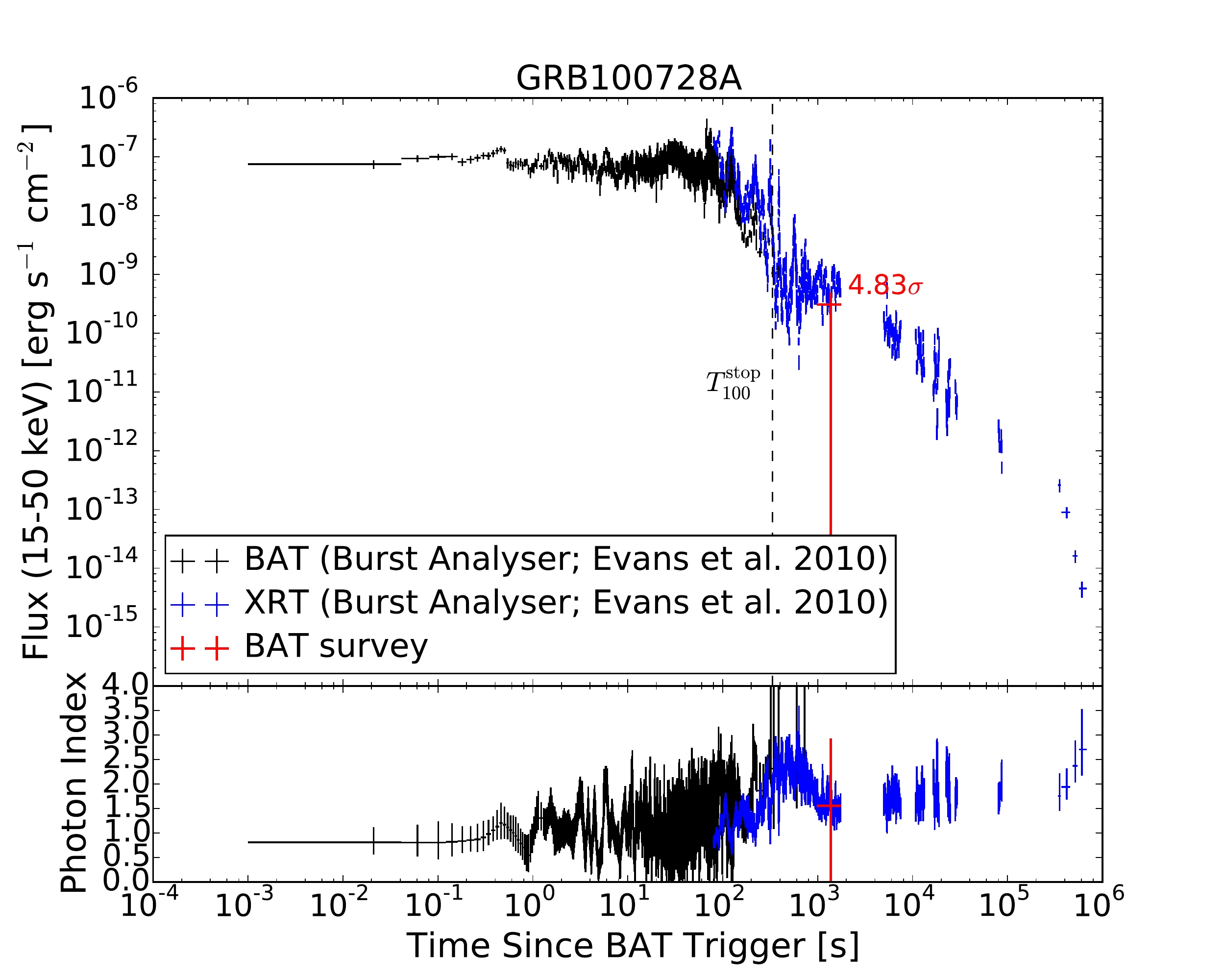}
\includegraphics[width=0.49\textwidth]{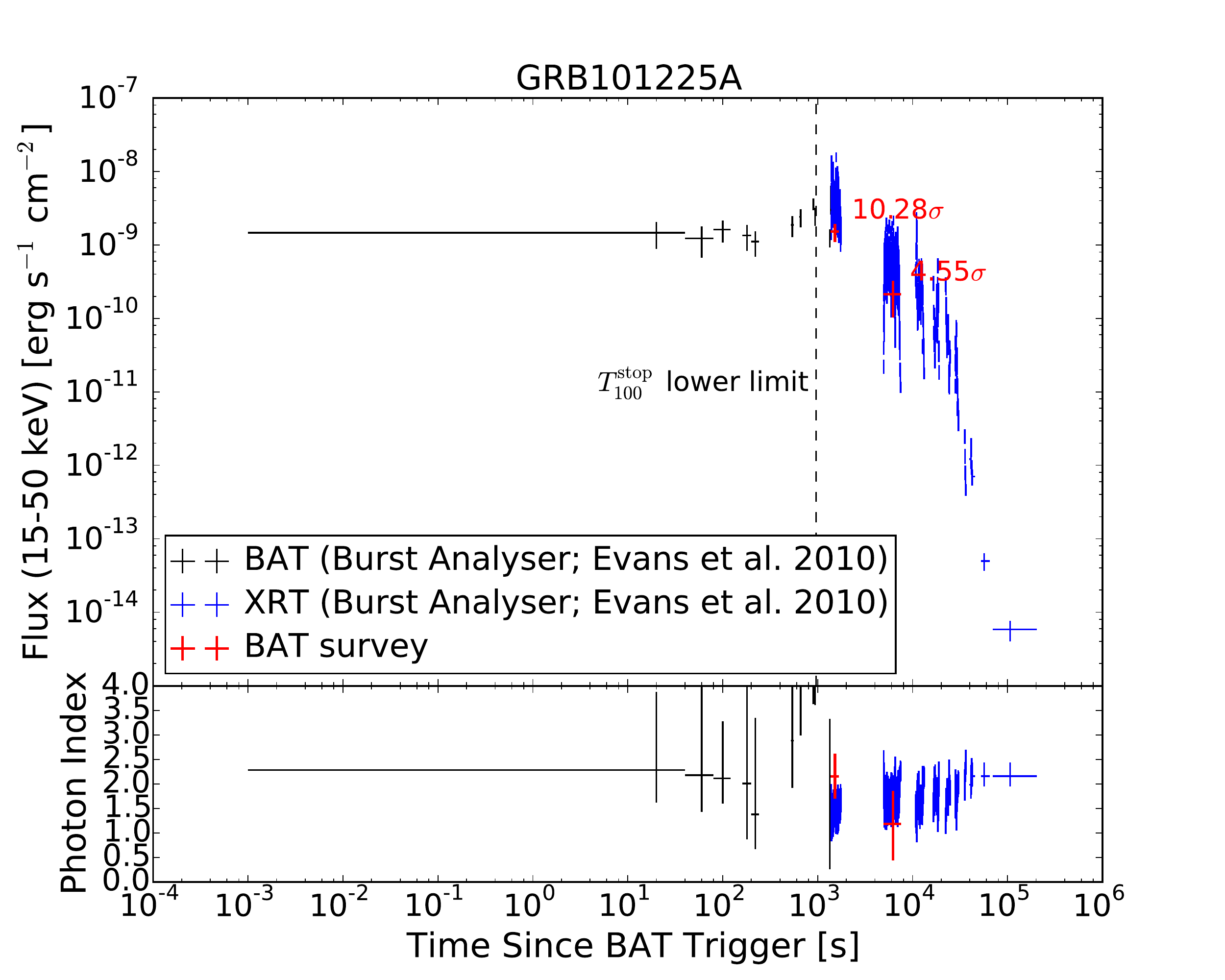}
\includegraphics[width=0.49\textwidth]{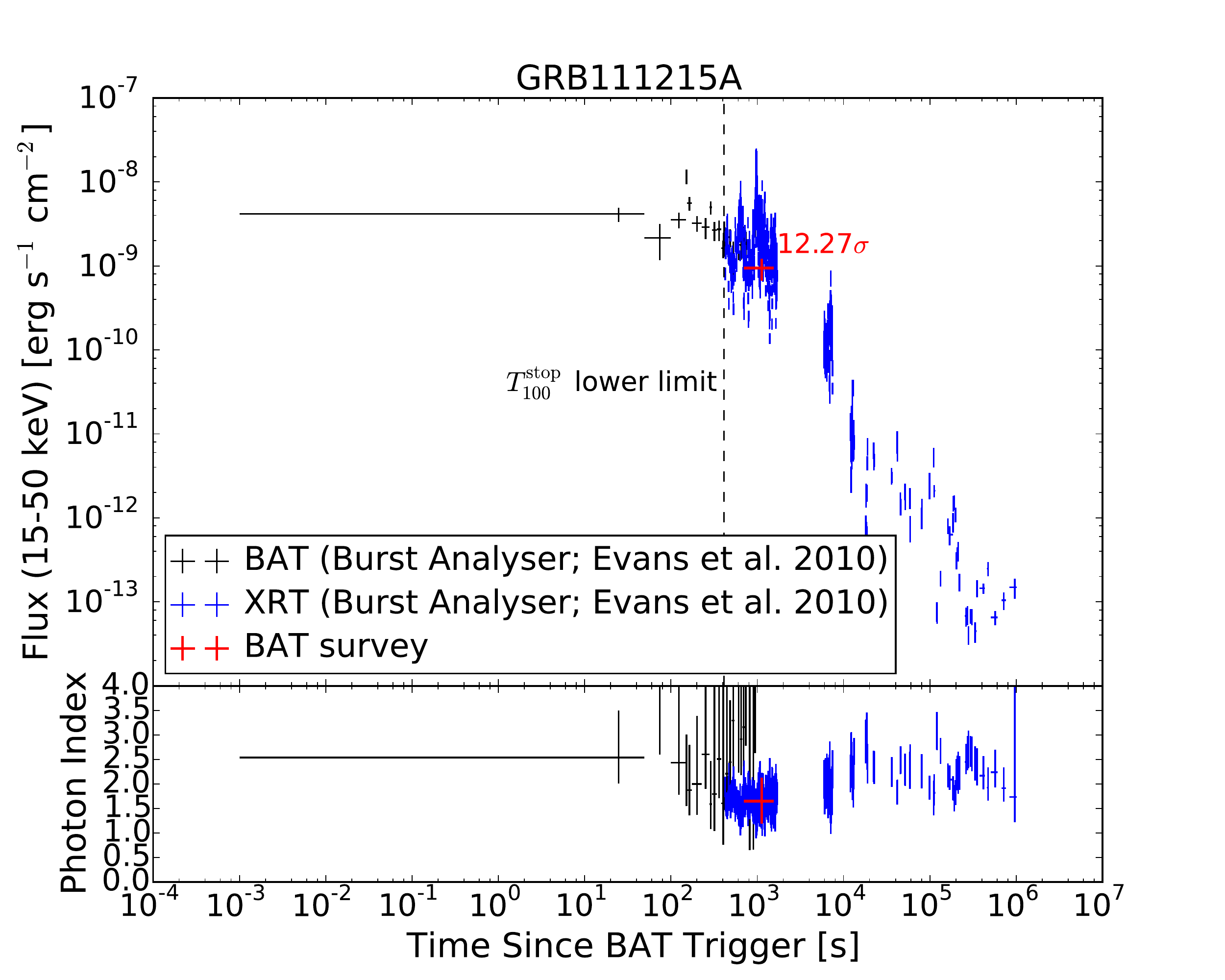}
\end{center}
\caption{
Comparisons between detections in the BAT survey data, BAT event data, and the XRT data.
The end of $T_{100}$ is marked as black-dashed line in each figure. If the burst emissions extend beyond the event data range, the black-dashed line marks the lower-limit of $T_{100}$ (i.e., the end of $T_{100}$ found by {\it battblocks}, or the end of the event data range if {\it battblocks} failed to estimate the burst durations).
}
\label{fig:lc_2}
\end{figure}

\begin{figure}[!h]
\begin{center}
\includegraphics[width=0.49\textwidth]{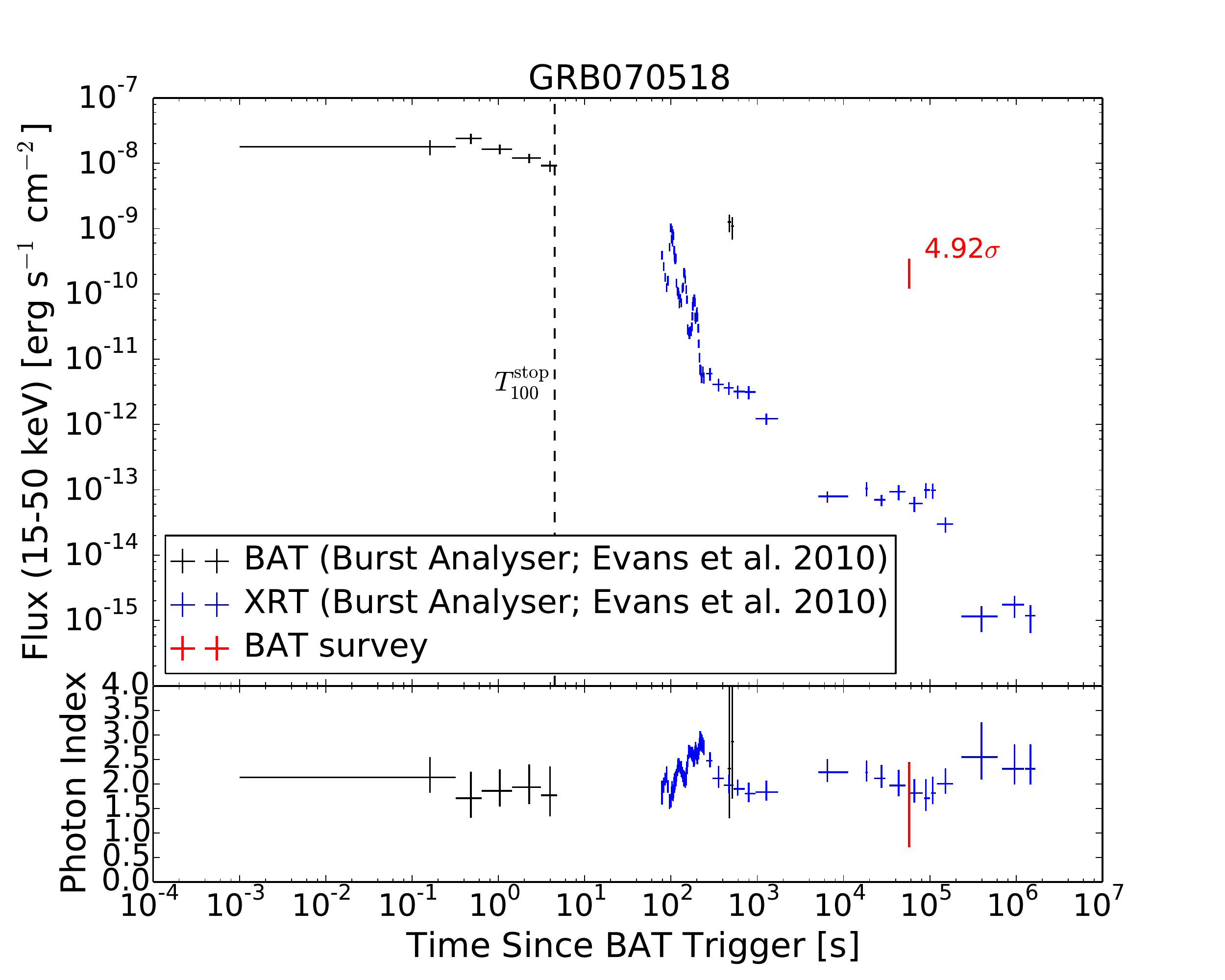}
\includegraphics[width=0.49\textwidth]{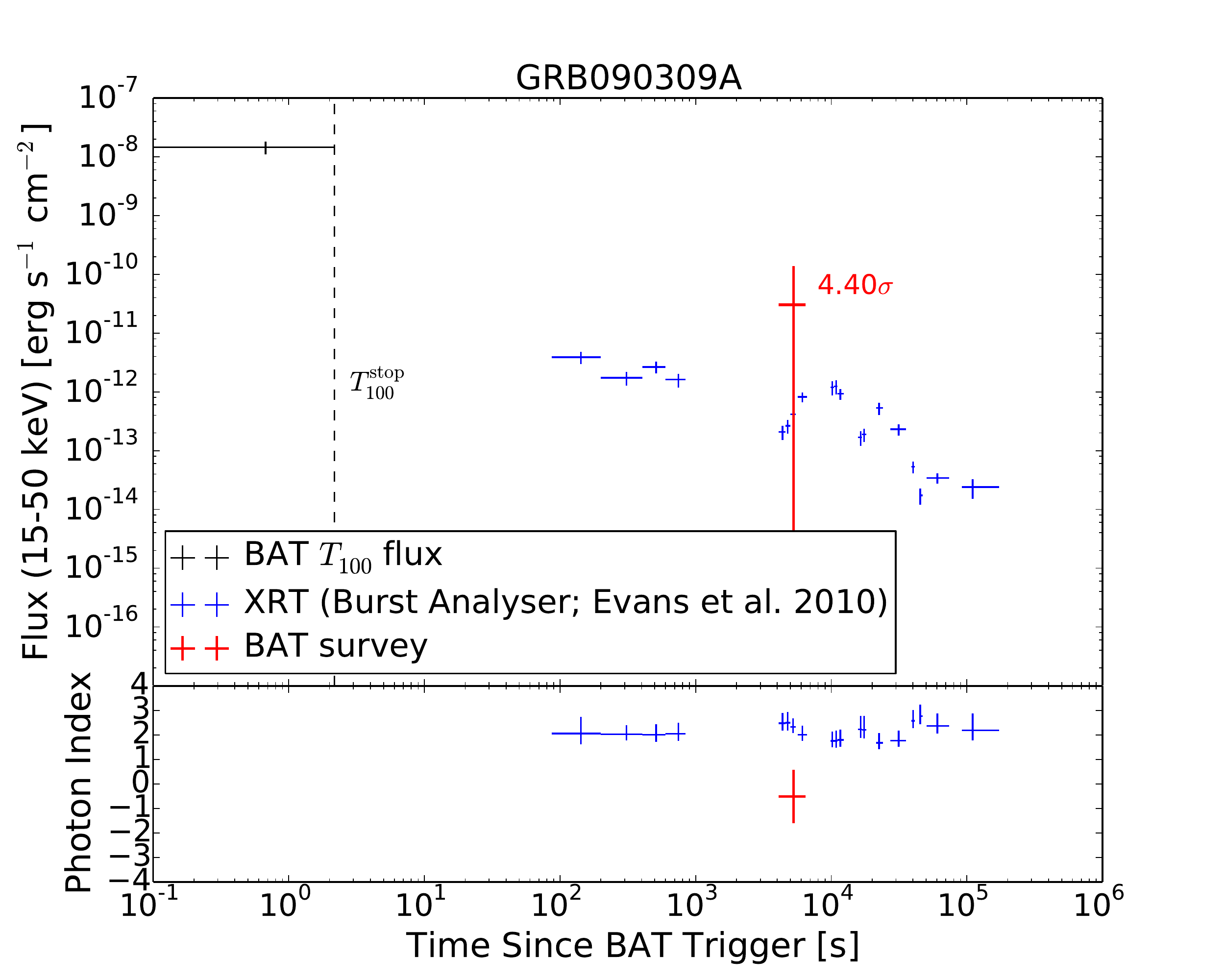}
\includegraphics[width=0.49\textwidth]{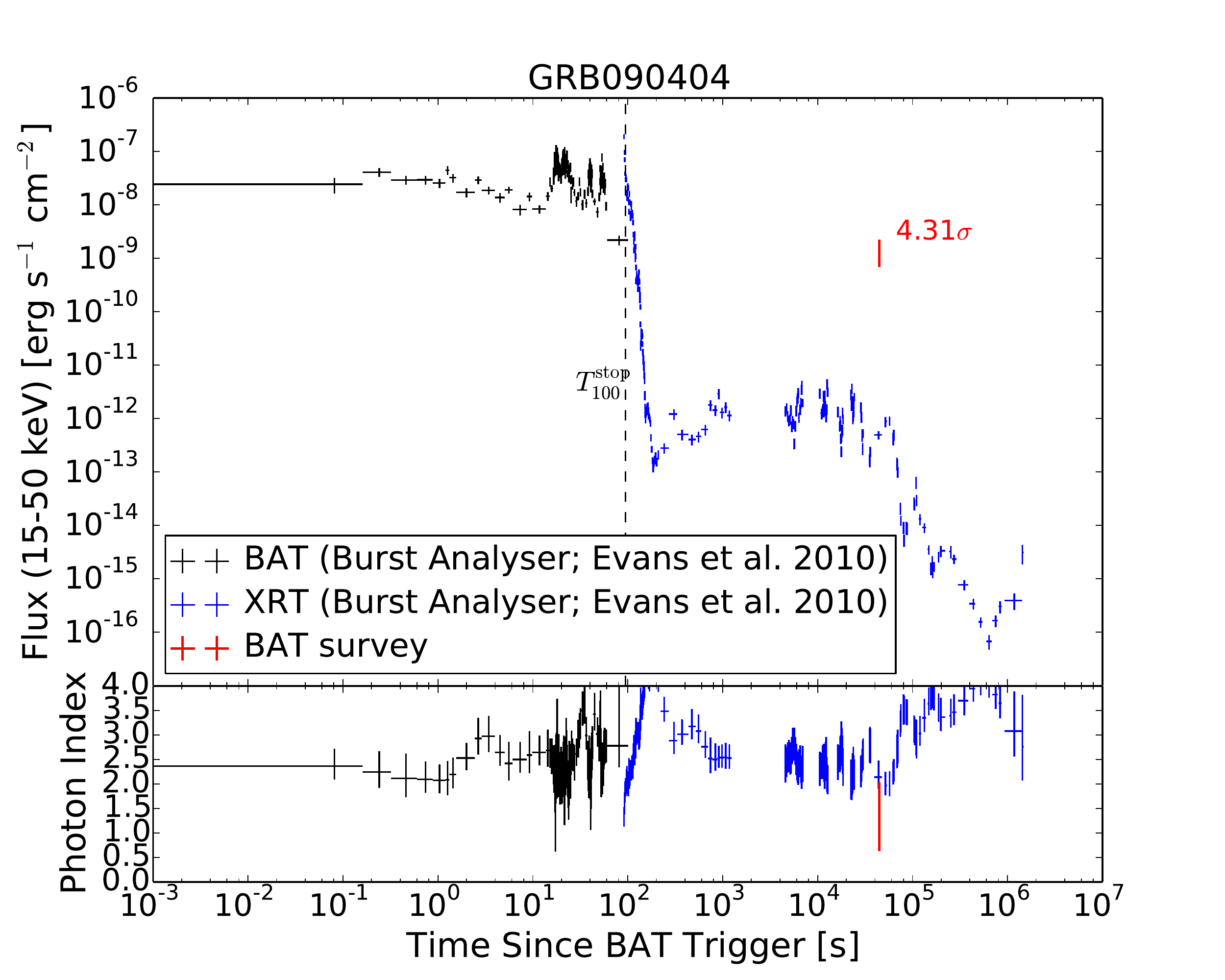}
\end{center}
\caption{
Comparisons between detections in the BAT survey data, BAT event data, and the XRT data. These GRBs
have late time BAT emissions that do not match the XRT emissions extrapolated to the same energy range.
The end of $T_{100}$ is marked as black-dashed line in each figure. If the burst emissions extend beyond the event data range, the black-dashed line marks the lower-limit of $T_{100}$ (i.e., the end of $T_{100}$ found by {\it battblocks}, or the end of the event data range if {\it battblocks} failed to estimate the burst durations).
For GRB090309A, we plot the BAT $T_{100}$ flux instead because there are no BAT fluxes produced by the Burst Analyser \citep{Evans10} for this burst.
}
\label{fig:lc_weird}
\end{figure}

\begin{figure}[!h]
\begin{center}
\includegraphics[width=0.49\textwidth]{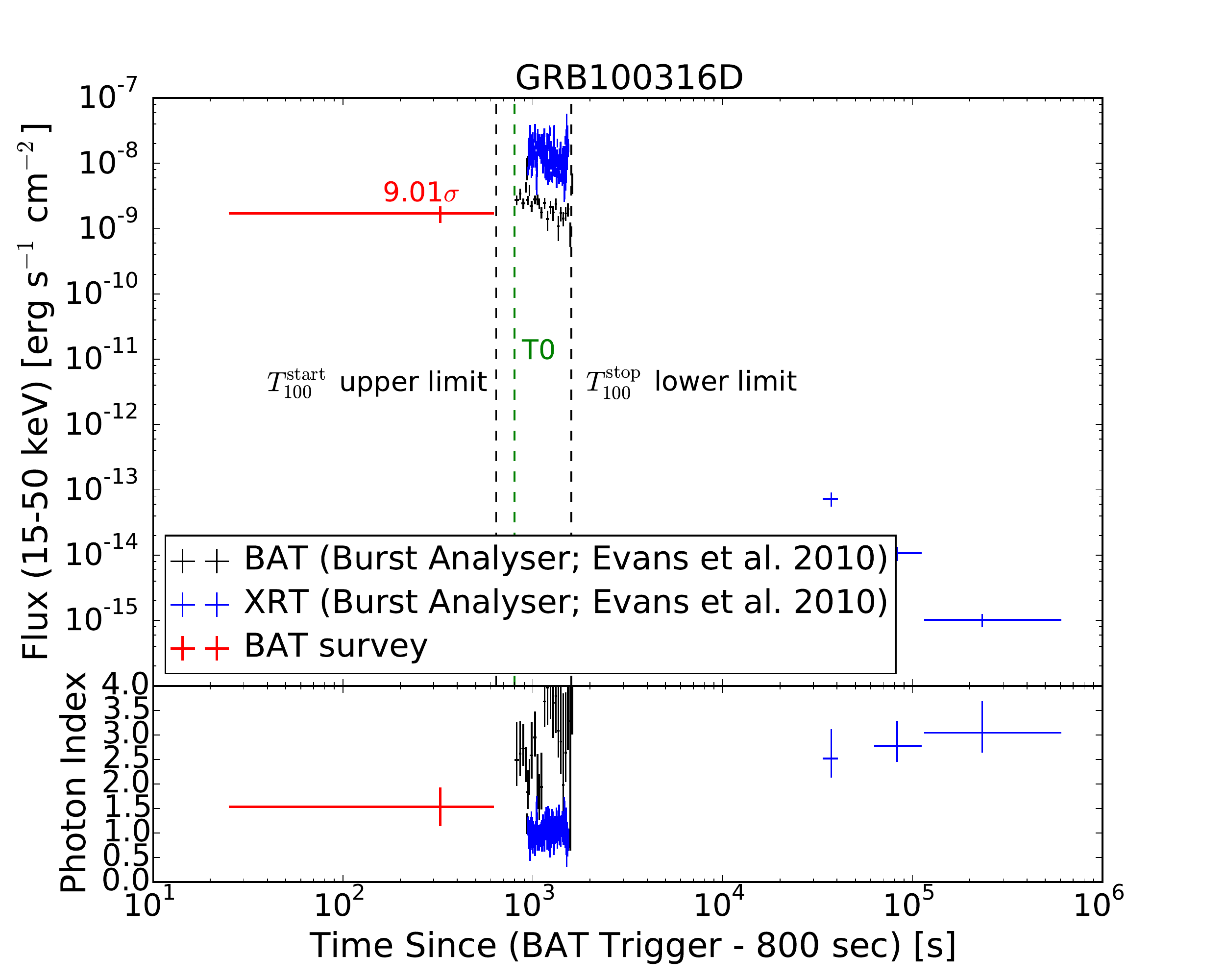}
\includegraphics[width=0.49\textwidth]{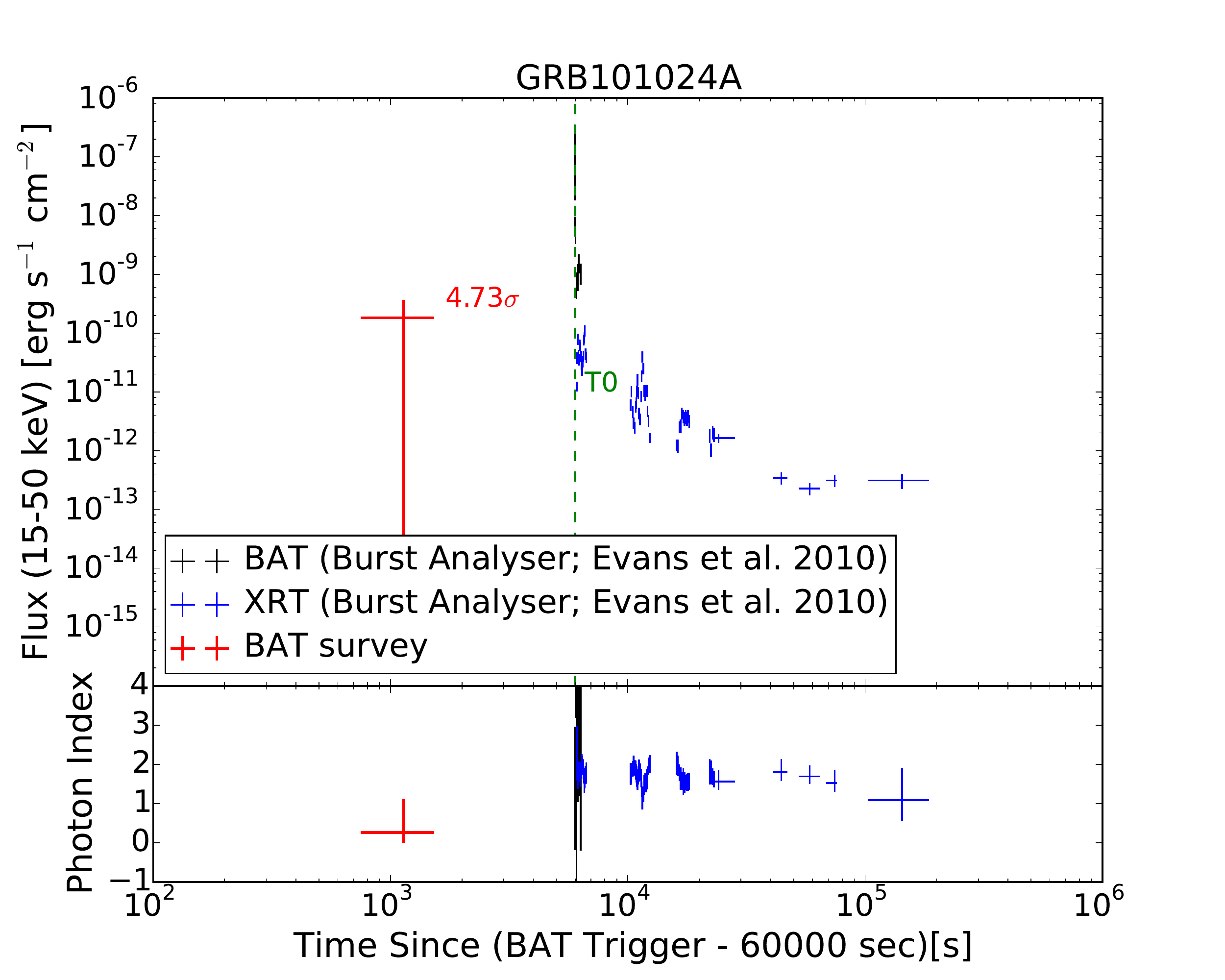}
\end{center}
\caption{
Precursor detections in the BAT survey data, overlayed on the BAT event data and XRT data for comparison.
The x-axes are shifted to avoid negative regions in the plots. The BAT trigger times $T0$ are marked as green-dashed lines.
Both the start and stop time of $T_{100}$ are marked as black-dashed line for GRB100316D.
These times are labeled as upper and lower limits because the burst emission extends beyond the event data range,
that is, the burst starts before $T^{\rm start}_{100}$ and ends after $T^{\rm stop}_{100}$.
The $T_{100}$ range (from $T0-0.25$ to $T0+20.13$s) for GRB101024A is not marked because it appears very close to the $T0$ line on the plot.
}
\label{fig:lc_precursor}
\end{figure}

We further compare these late-time emissions to the {\it Swift}/XRT light curves generated by the Burst Analyser\footnote{http://www.swift.ac.uk/burst\_analyser/} \citep{Evans10,Evans09,Evans07}.
The Burst Analyser can plot the GRB light curves from both the BAT event data and the XRT data in the 15-50 keV range. 
The equivalent XRT fluxes in the 15-50 keV range are estimated by extrapolating the XRT spectrum in 0.3-10 keV.
We also calculate the BAT fluxes in the 15-50 keV range by fitting the simple PL model to the BAT spectra generated from the survey data (eight energy bands).
Figures \ref{fig:lc_1} to \ref{fig:lc_precursor} overlay the BAT detections in the survey data on top of the observations from 
the BAT event data and the XRT data.
Results show that from most of the late-time detections in the survey data, the BAT extended emissions generally follow the behavior seen in the XRT light curves,
and the photon indices from the simple PL fit ($\alpha^{\rm PL}$) are similar to the ones derived from the XRT data.

However, there are three bursts (GRB070518, GRB090309, and GRB090404) that show late time fluxes much larger than what are measured in the XRT. The light curves of these bursts are shown 
in Fig.~\ref{fig:lc_weird}. Within these three peculiar bursts,
the spectral fits from GRB090309 is problematic with un-constrained energy flux and photon index. 
Moreover,
GRB090309 turns out to have $T_{90} = 3 \pm 1.4$ s, which makes this a possible short burst.

We found two bursts, GRB100316D and GRB101024A, with detections in survey data before the starting of event data. The light curves for these two bursts with precursor detections are
plotted in Fig.~\ref{fig:lc_precursor}. 
The time in the x-axis is shifted to prevent the problem of plotting a negative number on a log scale. Also, the spectral fits from GRB101024A 
is problematic and the flux might not be accurate.
Precursors of GRB emissions are suggested in some of the theoretical models \citep[e.g.,][]{Yamazaki09}. However, the fraction of precursor we found in this study is very low,
suggesting that the precursor, if exists, would be either weaker than the BAT's sensitivity in a regular survey image exposure time of $\sim$ few thousand seconds, or 
the emission is much shorter than the survey data exposure time and thus the signal is greatly reduced by the background noise.


\section{Summary}
\label{sect:summary}

We present the temporal and spectral analyses of the event data for GRBs detected by BAT for the past eleven years (up to 
GRB151027B). Particularly, we include analyses of the burst durations $T_{100}$, $T_{90}$, and $T_{50}$, the refined positions,
and report the spectral fits from both the simple power-law model (Eq.~\ref{eq:PL}) and the cutoff power-law model (Eq.~\ref{eq:CPL}).
We include spectral fits for the spectra made with the $T_{100}$ duration, the 1-s peak interval, the 20-ms peak duration.
Moreover, we perform spectral fits of the time-resolved spectra, with the time-resolved durations selected by the automatic pipeline {\it battblocks}.  
All the detailed numbers are summarized in tables in Appendix \ref{sect:appendix_tables}, except the time-resolved durations and spectral fits 
due to the large amount of data. The results of the time-resolved analyses can be found on the online directories at http://swift.gsfc.nasa.gov/results/batgrbcat/index.html.

Following the BAT1 and BAT2 catalog, we adopt the same criterion of $\Delta \chi^2 >6$ for determining whether the spectrum is better fitted by the CPL model
than the simple PL model. Furthermore, we introduce a few more criteria to select the acceptable spectral fits, in order to ensure
the parameters estimated from the spectral fits (e.g., photon index and flux) are reasonable. These criteria are listed in Section~\ref{sect:event_standard}.
Lists of bursts with acceptable PL and CPL fit are presented in Appendix \ref{sect:appendix_tables}.

The analysis results show that 
the general temporal and spectral distributions of the GRBs remain similar to 
those reported in the BAT2 catalog, though BAT has doubled the number of GRB detections since 2009. 
Roughly $9\%$ of the BAT GRBs are short ($T_{90} \lesssim 2$ s),
which is less than the short GRB fraction in the {\it Fermi} sample ($\sim 17\%$) and the BATSE sample ($\sim 26\%$).
The spectral analyses show that short GRBs in the BAT sample are only slightly harder than the long bursts. 
A study of the distribution of GRB partial coding fraction (which is related to the burst incident angle) suggests that short GRBs are harder to detect than long bursts
when the GRB is highly off-axis.

Some further studies of the BAT sensitivity show that the fluxes of the BAT-detected GRBs decrease as a function of $T^{-1/2}_{90}$.
That is, BAT can detect GRBs with lower flux if the burst is longer. This result is consistent with the expectation 
that the BAT sensitivity should improves with longer exposure time. Specifically, the sensitivity is inversely proportional to
the square of the exposure time, because the noise fluctuation scales as $1/\sqrt{N}$ in counting statistics, 
where N is the number of detected photons.

We construct a special list of
12
short GRBs with extended emissions in Section~\ref{sect:Short_with_EE}, and also
a list of
15
possible short GRBs with extended emissions, which either have a short pulse that is slightly longer than 2 sec,
or have some complicated factors that make it difficult to determine the reliability of the extended tail. 
The spectral analyses confirm that the photon indices of short pulses are more similar to the short GRB population,
while the spectral properties of the extended emissions resemble more closely the long burst population.

Moreover, we compile a list of GRBs with redshift measurements, and discuss how some of their properties vary
with redshift. We found that while the redshift distribution of the rate-trigger bursts are peaked at low redshift (as expected),
the redshift distribution for the image-trigger bursts are more evenly distributed throughout the redshift range,
and is composed of a higher fraction of bursts with larger redshift.

The redshift properties of BAT-detected GRBs show a hint of luminosity evolution. The number of short GRBs at high
redshift seems to decrease faster than one would expect by simply taking into account the BAT sensitivity as a function of burst duration,
with the assumption of a non-evolving upper limit of the GRB luminosity (Fig~\ref{fig:T90_obs_rest_vs_z}). 
However, due to the incompleteness of the redshift sample, we cannot rule out the possibility of we are simply missing more redshift measurements for the short bright bursts 
at large redshifts.
Furthermore, there seems to be a lack of low-flux GRBs 
at high redshift ($z \gtrsim 5$; see Fig.~\ref{fig:energy_flux_vs_z}), and a missing population of GRBs with luminosity $\gtrsim 10^{51} \rm erg \ s^{-1} \ cm^{-2}$ at low redshift ($z \lesssim$ 1; see Fig.~\ref{fig:lum_vs_z}).

In addition to the event data analyses, we search for possible emissions beyond the event data range using the survey data.
This search uses the existing survey data product, which are processed survey data and are available for events before August 2013.
After performing some studies of the false detection rate in the sample of survey data that we use, we adopted the criterion of signal-to-noise ratio $\geq 4.3$ in 
the energy band $14-195$ keV for determining a detection. We expect one false detection in our sample using this criterion.
We found 21 detections (16 GRBs) beyond the event data range, with 7 of these bursts have been previously classified as ultra-long GRBs.
We do not find an obvious relation between the detections in survey data and the $T_{90}$ estimated using the event data. 
That is, these bursts with detections in survey data are not necessarily longer in the event data range. However, we do found almost all the survey data detections
match well with the late-time XRT light curves. The detections usually happen when the burst is also bright in XRT, such as during the flares in the XRT light curves.

The overall summaries for the GRB analysis results are presented at \\
http://swift.gsfc.nasa.gov/results/batgrbcat/index.html, 
which includes summary tables of GRB properties, and a quick-look webpage and the original data product for each burst. 
This web-version summary is expected to continue to be updated with new GRBs beyond those included in this paper.

\noindent{\bf{Acknowledgments}}\\
We are grateful for valuable discussions with Sylvain Guiriec, Brett Hayes, and Antonino Cucchiara. We thank James Bubeck, Phillip Newman, and J. D. Myers for setting up the web space to host the summary results of this catalog. Moreover, we appreciate the helpful suggestions and careful number checking from the anonymous referee. The plots in the paper are made with the Python Matplotlib \citep{Hunter07}. The interactive plots on the webpages are made with the Bokeh package \citep{Bokeh}.


\appendix

\section{Appendix: Tables}
\label{sect:appendix_tables}

This section includes a list of summary tables from the analyses:

\subsection{Summary tables that includes general information of the bursts}

Table \ref{tab:summary_general}
contains some general information of the bursts. Information from these two tables are shown in one single list in the complete online version.
They are divided into two tables in the abbreviated paper version due to the limited space.

\begin{table}[h!]
\caption{
\label{tab:summary_general}
The format of the summary table that includes some general information of the bursts. 
(This table is available in its entirety in a machine-readable form in the online journal. Only the column formats are shown here for guidance regarding its form and content.)
}
\begin{center}
\small
\begin{tabular}{|c|c|c|c|}
\hline\hline
Column & Format & Unit & Description \\
\hline\hline
GRBname & A9 & -- & GRB name. \\
\hline
Trig\_ID &  I11 & -- & Special ID associated with the trigger. \\ 
& & & For GRBs found in ground analysis, \\ 
& & & the ID are associated with the trigger ID of the failed event data, \\ 
& & & or the full observation ID of the event data for the analysis.  \\
\hline
Trig\_time\_met &  F13.3 & s & The BAT trigger time shown in the Swift machine time (MET). \\
\hline
Trig\_time\_UTC & A26 & -- & The BAT trigger time shown in UTC. \\
\hline
RA\_ground  &  A12 & deg & RA of the GRB from the BAT refined position. \\
\hline
DEC\_ground   & A13 & deg & DEC of the GRB from the BAT refined position. \\
\hline
Image\_position\_err &  A12 & deg & The uncertainty of the BAT refined position. \\
\hline
Image\_SNR & A12 & -- & The signal-to-noise ratio of the detection from the BAT image. \\
\hline
T90   &  A12 & s & Burst duration $T_{90}$. \\
\hline
T90\_err   &  A12 & s & The uncertainty of $T_{90}$. \\
\hline
T50    &   A12 & s & Burst duration $T_{50}$. \\
\hline
T50\_err  & A12 & s & The uncertainty of $T_{50}$. \\
\hline
Evt\_start\_sincetrig & I4 & s & The time when the event data start, \\ 
& & & relative to the BAT trigger time. \\
\hline
Evt\_stop\_sincetrig & I4 & s & The time when the event data end, \\ 
& & &relative to the BAT trigger time. \\
\hline
pcode  & F6.4 & -- & Partial coding fraction of the burst. \\
\hline
Trigger\_method & A5 & -- & The triggering method for the burst (rate trigger, image trigger, \\
& & & ground detected, or ground detected during slew). \\
\hline
XRT\_detection    & A3 & -- & Whether or not there is an XRT detection of the burst. \\
\hline\hline
\end{tabular}
\end{center}
\end{table}

Table ~\ref{tab:burst_duration} lists the start and end time of $T_{100}$, $T_{90}$, $T_{50}$, and the 1-s peak duration, relative to the BAT trigger time.

\begin{table}[h!]
\caption{
\label{tab:burst_duration}
The format of the burst-duration table that lists the start and end time of $T_{100}$, $T_{90}$, $T_{50}$, and the 1-s peak duration, relative to the BAT trigger time.
(This table is available in its entirety in a machine-readable form in the online journal. Only the column formats are shown here for guidance regarding its form and content.)
}
\begin{center}
\small
\begin{tabular}{|c|c|c|c|}
\hline\hline
Column & Format & Unit & Description \\
\hline\hline
GRBname & A9 & -- & GRB name. \\
\hline
Trig\_ID &  I11 & -- & Special ID associated with the trigger. \\ 
& & & For GRBs found in ground analysis, \\ 
& & & the ID are associated with the trigger ID of the failed event data, \\ 
& & & or the full observation ID of the event data for the analysis.  \\
\hline
Trig\_time\_met &  F13.3 & s & The BAT trigger time shown in the Swift machine time (MET). \\
\hline
T100\_start & A11 & s & $T_{100}$ start time, \\
& & & relative to the BAT trigger time. \\ 
\hline
T100\_stop & A11 & s & $T_{100}$ end time, \\
& & & relative to the BAT trigger time. \\ 
\hline
T90\_start & A11 & s & $T_{90}$ start time, \\
& & & relative to the BAT trigger time. \\ 
\hline
T90\_stop & A11 & s & $T_{90}$ end time, \\
& & & relative to the BAT trigger time. \\ 
\hline
T50\_start & A11 & s & $T_{50}$ start time, \\
& & & relative to the BAT trigger time. \\ 
\hline
T50\_stop & A11 & s & $T_{50}$ end time, \\
& & & relative to the BAT trigger time. \\ 
\hline
1s\_peak\_start & A11 & s & 1-s peak start time, \\
& & & relative to the BAT trigger time. \\ 
\hline
1s\_peak\_stop & A11 & s & 1-s peak end time, \\
& & & relative to the BAT trigger time. \\ 
\hline\hline
\end{tabular}
\end{center}
\end{table}

\subsection{Summary tables of spectral analyses for the time-average $T_{100}$ spectra}
\label{sect:T100_spectra_tables}

Table \ref{tab:spec_T100_best_model} summarizes the best-fit model for each burst. ``PL'' refers to simple power-law model; ``CPL'' refers to cutoff power-law model. If there is no acceptable spectral fit for the burst (see Section~\ref{sect:event_standard} for the criteria), the best-fit-model column is listed as ``N/A''.

\begin{table}
\caption{
\label{tab:spec_T100_best_model}
The format of the table that summarizes the best-fit spectral model of the time-average ($T_{100}$) spectrum for each burst.
(This table is available in its entirety in a machine-readable form in the online journal. Only the column formats are shown here for guidance regarding its form and content.)
}
\begin{center}
\begin{tabular}{|c|c|c|c|}
\hline\hline
Column & Format & Unit & Description \\
\hline\hline
GRBname & A9 & -- & GRB name. \\
\hline
Trig\_ID &  I11 & -- & Special ID associated with the trigger. \\ 
& & & For GRBs found in ground analysis, \\ 
& & & the ID are associated with the trigger ID of the failed event data, \\ 
& & & or the full observation ID of the event data for the analysis.  \\
\hline
Best\_fit\_model  & A3 & -- & The best-fit model for the GRB spectrum, \\
& & & either the simple power law model (PL) \\ 
& & & or cutoff power-law model (CPL.)  \\
\hline\hline
\end{tabular}
\end{center}
\end{table}

Table \ref{tab:spec_T100_pow_parameter} shows a list of parameters from the spectral fit using the simple power-law model. 

\begin{table}
\caption{
\label{tab:spec_T100_pow_parameter}
The format of the table that presents the parameters from the PL fit for the time-averaged ($T_{100}$) spectra. Note that this table includes the fit for every GRB, regardless of whether the fit is acceptable from the criteria listed in Section~\ref{sect:event_standard}.
A list of GRBs with acceptable fits can be found in Table \ref{tab:spec_T100_best_model}.
(This table is available in its entirety in a machine-readable form in the online journal. Only the column formats are shown here for guidance regarding its form and content.)
}
\begin{center}
\footnotesize
\begin{tabular}{|c|c|c|c|}
\hline\hline
Column & Format & Unit & Description \\
\hline\hline
GRBname & A9 & -- & GRB name. \\
\hline
Trig\_ID &  I11 & -- & Special ID associated with each triggers. \\ 
& & & For GRBs found in ground analysis, \\ 
& & & the ID are associated with the trigger ID of the failed event data, \\ 
& & & or the full observation ID of the event data for the analysis.  \\
\hline
alpha & A13 & -- & $\alpha^{\rm PL}$ as defined in Eq.~\ref{eq:PL}. \\
\hline
alpha\_low & A13 & -- & The lower limit of $\alpha^{\rm PL}$. \\
\hline
alpha\_hi & A13 & -- & The upper limit of $\alpha^{\rm PL}$. \\
\hline
norm & A12 & $\rm ph \ cm^{-2} \ s^{-1} \ keV^{-1}$ & The normalization factor $K^{\rm PL}_{50}$, as defined in Eq.~\ref{eq:PL}. \\ 
\hline
norm\_low & A12 & $\rm ph \ cm^{-2} \ s^{-1} \ keV^{-1}$ & The lower limit of $K^{\rm PL}_{50}$. \\ 
\hline
norm\_hi & A12 & $\rm ph \ cm^{-2} \ s^{-1} \ keV^{-1}$ & The upper limit of $K^{\rm PL}_{50}$. \\ 
\hline
chi2 & F6.2 & -- & $\chi^2$ from the fitting, reported by XSPEC. \\  
\hline
dof & I2 & -- & degree of freedom from the fitting, reported by XSPEC. \\ 
\hline
reduced\_chi2 & F6.4 & -- & reduced $\chi^2$ (i.e., $\chi^2$ divided by the degree of freedom), \\ 
& & & reported by XSPEC. \\
\hline
null\_prob & A12 & -- & The null probability of the model, reported by XSPEC. \\ 
\hline
enorm & F7.4 & KeV & The normalization energy, which is set to 50 keV in our fits, \\ 
& & & as shown in Eq.~\ref{eq:PL}. \\ 
\hline
Exposure\_time & A6 & s & The time interval of the spectrum. \\ 
\hline
T100\_start & A11 & s & The start time of the spectrum, \\ 
& & & relative to the BAT trigger time. \\ 
\hline
T100\_stop & A11 & s & The end time of the spectrum, \\
& & & relative to the BAT trigger time.\\
\hline\hline
\end{tabular}
\end{center}
\end{table}

Table \ref{tab:spec_T100_pow_photon_flux} shows the photon fluxes in different energy ranges from the simple power-law fit. 

\begin{table}
\caption{
\label{tab:spec_T100_pow_photon_flux}
The format of the table that presents the photon flux (in unit of $\rm ph \ cm^{-2} \ s^{-1}$) from the PL fit for the time-averaged ($T_{100}$) spectra. Note that this table includes the fit for every GRB, regardless of whether the fit is acceptable from the criteria listed in Section~\ref{sect:event_standard}.
A list of GRBs with acceptable fits can be found in Table \ref{tab:spec_T100_best_model}.
(This table is available in its entirety in a machine-readable form in the online journal. Only the column formats are shown here for guidance regarding its form and content.)
}
\begin{center}
\footnotesize
\begin{tabular}{|c|c|c|c|}
\hline\hline
Column & Format & Unit & Description \\
\hline\hline
GRBname & A9 & -- & GRB name. \\
\hline
Trig\_ID &  I11 & -- & Special ID associated with the trigger. \\ 
& & & For GRBs found in ground analysis, \\ 
& & & the ID are associated with the trigger ID of the failed event data, \\ 
& & & or the full observation ID of the event data for the analysis.  \\
\hline
15\_25kev & A12 & $\rm ph \ cm^{-2} \ s^{-1}$ & Photon flux in $15-25$ keV. \\
\hline
15\_25kev\_low & A12 & $\rm ph \ cm^{-2} \ s^{-1}$ & The lower limit of photon flux in $15-25$ keV. \\ 
\hline
15\_25kev\_hi & A12 & $\rm ph \ cm^{-2} \ s^{-1}$ & The upper limit of photon flux in $15-25$ keV. \\ 
\hline
25\_50kev & A12 & $\rm ph \ cm^{-2} \ s^{-1}$ & Photon flux in $25-50$ keV. \\ 
\hline
25\_50kev\_low & A12 & $\rm ph \ cm^{-2} \ s^{-1}$ & The lower limit of photon flux in $25-50$ keV. \\ 
\hline
25\_50kev\_hi & A12 & $\rm ph \ cm^{-2} \ s^{-1}$ & The upper limit of photon flux in $25-50$ keV. \\ 
\hline
50\_100kev & A12 & $\rm ph \ cm^{-2} \ s^{-1}$ & Photon flux in $50-100$ keV. \\ 
\hline
50\_100kev\_low & A12 & $\rm ph \ cm^{-2} \ s^{-1}$ & The lower limit of photon flux in $50-100$ keV. \\ 
\hline
50\_100kev\_hi & A12 & $\rm ph \ cm^{-2} \ s^{-1}$ & The upper limit of photon flux in $50-100$ keV. \\ 
\hline
100\_150kev & A12 & $\rm ph \ cm^{-2} \ s^{-1}$ & Photon flux in $100-150$ keV. \\ 
\hline
100\_150kev\_low & A12 & $\rm ph \ cm^{-2} \ s^{-1}$ & The lower limit of photon flux in $100-150$ keV. \\ 
\hline
100\_150kev\_hi & A12 & $\rm ph \ cm^{-2} \ s^{-1}$ & The upper limit of photon flux in $100-150$ keV. \\ 
\hline
100\_350kev & A12 & $\rm ph \ cm^{-2} \ s^{-1}$ & Photon flux in $100-350$ keV. \\
\hline
100\_350kev\_low & A12 & $\rm ph \ cm^{-2} \ s^{-1}$ & The lower limit of photon flux in $100-350$ keV. \\ 
\hline
100\_350kev\_hi & A12 & $\rm ph \ cm^{-2} \ s^{-1}$ & The upper limit of photon flux in $100-350$ keV. \\
\hline
15\_150kev & A12 & $\rm ph \ cm^{-2} \ s^{-1}$ & Photon flux in $15-150$ keV. \\ 
\hline
15\_150kev\_low & A12 & $\rm ph \ cm^{-2} \ s^{-1}$ & The lower limit of photon flux in $15-150$ keV. \\
\hline
15\_150kev\_hi & A12 & $\rm ph \ cm^{-2} \ s^{-1}$ & The upper limit of photon flux in $15-150$ keV. \\ 
\hline
15\_350kev & A12 & $\rm ph \ cm^{-2} \ s^{-1}$ & Photon flux in $15-350$ keV. \\
\hline
15\_350kev\_low & A12 & $\rm ph \ cm^{-2} \ s^{-1}$ & The lower limit of photon flux in $15-350$ keV. \\
\hline
15\_350kev\_hi & A12 & $\rm ph \ cm^{-2} \ s^{-1}$ & The upper limit of photon flux in $15-350$ keV. \\
\hline
Exposure\_time & A12 & s & The time interval of the spectrum. \\
\hline
T100\_start  & A12 & s & The start time of the spectrum, \\
& & & relative to the BAT trigger time. \\
\hline
T100\_stop & A12 & s & The end time of the spectrum, \\
& & & relative to the BAT trigger time. \\
\hline\hline
\end{tabular}
\end{center}
\end{table}

Table \ref{tab:spec_T100_pow_energy_flux} shows the energy fluxes in different energy ranges from the simple power-law fit.

\begin{table}
\caption{
\label{tab:spec_T100_pow_energy_flux}
The format of the table that presents the energy flux (in unit of $\rm erg \ s^{-1} \ cm^{-2}$) from the PL fit for the time-averaged ($T_{100}$) spectra. Note that this table includes the fit for every GRB, regardless of whether the fit is acceptable from the criteria listed in Section~\ref{sect:event_standard}.
A list of GRBs with acceptable fits can be found in Table \ref{tab:spec_T100_best_model}.
(This table is available in its entirety in a machine-readable form in the online journal. Only the column formats are shown here for guidance regarding its form and content.)
}
\begin{center}
\footnotesize
\begin{tabular}{|c|c|c|c|c|}
\hline\hline
Column & Format & Unit & Description \\
\hline\hline
GRBname & A9 & -- & GRB name. \\
\hline
Trig\_ID &  I11 & -- & Special ID associated with the trigger. \\ 
& & & For GRBs found in ground analysis, \\ 
& & & the ID are associated with the trigger ID of the failed event data, \\ 
& & & or the full observation ID of the event data for the analysis.  \\
\hline
15\_25kev & A12 & $\rm erg \ cm^{-2} \ s^{-1}$ & Photon flux in $15-25$ keV. \\
\hline
15\_25kev\_low & A12 & $\rm erg \ cm^{-2} \ s^{-1}$ & The lower limit of photon flux in $15-25$ keV. \\ 
\hline
15\_25kev\_hi & A12 & $\rm erg \ cm^{-2} \ s^{-1}$ & The upper limit of photon flux in $15-25$ keV. \\ 
\hline
25\_50kev & A12 & $\rm erg \ cm^{-2} \ s^{-1}$ & Photon flux in $25-50$ keV. \\ 
\hline
25\_50kev\_low & A12 & $\rm erg \ cm^{-2} \ s^{-1}$ & The lower limit of photon flux in $25-50$ keV. \\ 
\hline
25\_50kev\_hi & A12 & $\rm erg \ cm^{-2} \ s^{-1}$ & The upper limit of photon flux in $25-50$ keV. \\ 
\hline
50\_100kev & A12 & $\rm erg \ cm^{-2} \ s^{-1}$ & Photon flux in $50-100$ keV. \\ 
\hline
50\_100kev\_low & A12 & $\rm erg \ cm^{-2} \ s^{-1}$ & The lower limit of photon flux in $50-100$ keV. \\ 
\hline
50\_100kev\_hi & A12 & $\rm erg \ cm^{-2} \ s^{-1}$ & The upper limit of photon flux in $50-100$ keV. \\ 
\hline
100\_150kev & A12 & $\rm erg \ cm^{-2} \ s^{-1}$ & Photon flux in $100-150$ keV. \\ 
\hline
100\_150kev\_low & A12 & $\rm erg \ cm^{-2} \ s^{-1}$ & The lower limit of photon flux in $100-150$ keV. \\ 
\hline
100\_150kev\_hi & A12 & $\rm erg \ cm^{-2} \ s^{-1}$ & The upper limit of photon flux in $100-150$ keV. \\ 
\hline
100\_350kev & A12 & $\rm erg \ cm^{-2} \ s^{-1}$ & Photon flux in $100-350$ keV. \\
\hline
100\_350kev\_low & A12 & $\rm erg \ cm^{-2} \ s^{-1}$ & The lower limit of photon flux in $100-350$ keV. \\ 
\hline
100\_350kev\_hi & A12 & $\rm erg \ cm^{-2} \ s^{-1}$ & The upper limit of photon flux in $100-350$ keV. \\
\hline
15\_150kev & A12 & $\rm erg \ cm^{-2} \ s^{-1}$ & Photon flux in $15-150$ keV. \\ 
\hline
15\_150kev\_low & A12 & $\rm erg \ cm^{-2} \ s^{-1}$ & The lower limit of photon flux in $15-150$ keV. \\
\hline
15\_150kev\_hi & A12 & $\rm erg \ cm^{-2} \ s^{-1}$ & The upper limit of photon flux in $15-150$ keV. \\ 
\hline
15\_350kev & A12 & $\rm erg \ cm^{-2} \ s^{-1}$ & Photon flux in $15-350$ keV. \\
\hline
15\_350kev\_low & A12 & $\rm erg \ cm^{-2} \ s^{-1}$ & The lower limit of photon flux in $15-350$ keV. \\
\hline
15\_350kev\_hi & A12 & $\rm erg \ cm^{-2} \ s^{-1}$ & The upper limit of photon flux in $15-350$ keV. \\
\hline
Exposure\_time & A12 & s & The time interval of the spectrum. \\
\hline
T100\_start  & A12 & s & The start time of the spectrum, \\
& & & relative to the BAT trigger time. \\
\hline
T100\_stop & A12 & s & The end time of the spectrum, \\
& & & relative to the BAT trigger time. \\
\hline\hline
\end{tabular}
\end{center}
\end{table}

Table \ref{tab:spec_T100_pow_energy_fluence} shows the energy fluences in different energy ranges from the simple power-law fit.
The energy fluences are calculated by multiply the flux by $T_{100}$.

\begin{table}
\caption{
\label{tab:spec_T100_pow_energy_fluence}
The format of the table that presents the energy fluences (in unit of $\rm erg \ cm^{-2}$) from the PL fit for the time-averaged ($T_{100}$) spectra. 
Note that this table includes the fit for every GRB, regardless of whether the fit is acceptable from the criteria listed in Section~\ref{sect:event_standard}.
A list of GRBs with acceptable fits can be found in Table \ref{tab:spec_T100_best_model}.
(This table is available in its entirety in a machine-readable form in the online journal. Only the column titles are shown here for guidance regarding its form and content.)
}
\begin{center}
\footnotesize
\begin{tabular}{|c|c|c|c|c|}
\hline\hline
Column & Format & Unit & Description \\
\hline\hline
GRBname & A9 & -- & GRB name. \\
\hline
Trig\_ID &  I11 & -- & Special ID associated with the trigger. \\ 
& & & For GRBs found in ground analysis, \\ 
& & & the ID are associated with the trigger ID of the failed event data, \\ 
& & & or the full observation ID of the event data for the analysis.  \\
\hline
15\_25kev & A12 & $\rm erg \ cm^{-2}$ & Photon flux in $15-25$ keV. \\
\hline
15\_25kev\_low & A12 & $\rm erg \ cm^{-2}$ & The lower limit of photon flux in $15-25$ keV. \\ 
\hline
15\_25kev\_hi & A12 & $\rm erg \ cm^{-2}$ & The upper limit of photon flux in $15-25$ keV. \\ 
\hline
25\_50kev & A12 & $\rm erg \ cm^{-2}$ & Photon flux in $25-50$ keV. \\ 
\hline
25\_50kev\_low & A12 & $\rm erg \ cm^{-2}$ & The lower limit of photon flux in $25-50$ keV. \\ 
\hline
25\_50kev\_hi & A12 & $\rm erg \ cm^{-2}$ & The upper limit of photon flux in $25-50$ keV. \\ 
\hline
50\_100kev & A12 & $\rm erg \ cm^{-2}$ & Photon flux in $50-100$ keV. \\ 
\hline
50\_100kev\_low & A12 & $\rm erg \ cm^{-2}$ & The lower limit of photon flux in $50-100$ keV. \\ 
\hline
50\_100kev\_hi & A12 & $\rm erg \ cm^{-2}$ & The upper limit of photon flux in $50-100$ keV. \\ 
\hline
100\_150kev & A12 & $\rm erg \ cm^{-2}$ & Photon flux in $100-150$ keV. \\ 
\hline
100\_150kev\_low & A12 & $\rm erg \ cm^{-2}$ & The lower limit of photon flux in $100-150$ keV. \\ 
\hline
100\_150kev\_hi & A12 & $\rm erg \ cm^{-2}$ & The upper limit of photon flux in $100-150$ keV. \\ 
\hline
100\_350kev & A12 & $\rm erg \ cm^{-2}$ & Photon flux in $100-350$ keV. \\
\hline
100\_350kev\_low & A12 & $\rm erg \ cm^{-2}$ & The lower limit of photon flux in $100-350$ keV. \\ 
\hline
100\_350kev\_hi & A12 & $\rm erg \ cm^{-2}$ & The upper limit of photon flux in $100-350$ keV. \\
\hline
15\_150kev & A12 & $\rm erg \ cm^{-2}$ & Photon flux in $15-150$ keV. \\ 
\hline
15\_150kev\_low & A12 & $\rm erg \ cm^{-2}$ & The lower limit of photon flux in $15-150$ keV. \\
\hline
15\_150kev\_hi & A12 & $\rm erg \ cm^{-2}$ & The upper limit of photon flux in $15-150$ keV. \\ 
\hline
15\_350kev & A12 & $\rm erg \ cm^{-2}$ & Photon flux in $15-350$ keV. \\
\hline
15\_350kev\_low & A12 & $\rm erg \ cm^{-2}$ & The lower limit of photon flux in $15-350$ keV. \\
\hline
15\_350kev\_hi & A12 & $\rm erg \ cm^{-2}$ & The upper limit of photon flux in $15-350$ keV. \\
\hline
Exposure\_time & A12 & s & The time interval of the spectrum. \\
\hline
T100\_start  & A12 & s & The start time of the spectrum, \\
& & & relative to the BAT trigger time. \\
\hline
T100\_stop & A12 & s & The end time of the spectrum, \\
& & & relative to the BAT trigger time. \\
\hline\hline
\end{tabular}
\end{center}
\end{table}

Similar to the tables for the simply PL fits, Table \ref{tab:spec_T100_cutpow_parameter} to \ref{tab:spec_T100_pow_energy_fluence}
present the parameters, fluxes, and fluences for the CPL fits. 

\begin{table}
\caption{
\label{tab:spec_T100_cutpow_parameter}
The format of the table that presents the parameters from the CPL fit for the time-averaged ($T_{100}$) spectra. Note that this table includes the fit for every GRB, regardless of whether the fit is acceptable from the criteria listed in Section~\ref{sect:event_standard}.
A list of GRBs with acceptable fits can be found in Table \ref{tab:spec_T100_best_model}.
(This table is available in its entirety in a machine-readable form in the online journal. Only the column formats are shown here for guidance regarding its form and content.)
}
\begin{center}
\footnotesize
\begin{tabular}{|c|c|c|c|c|}
\hline\hline
Column & Format & Unit & Description \\
\hline\hline
GRBname & A9 & -- & GRB name. \\
\hline
Trig\_ID &  I11 & -- & Special ID associated with each triggers. \\ 
& & & For GRBs found in ground analysis, \\ 
& & & the ID are associated with the trigger ID of the failed event data, \\ 
& & & or the full observation ID of the event data for the analysis.  \\
\hline
alpha & A13 & -- & $\alpha^{\rm CPL}$ as defined in Eq.~\ref{eq:CPL}. \\
\hline
alpha\_low & A13 & -- & The lower limit of $\alpha^{\rm CPL}$. \\
\hline
alpha\_hi & A13 & -- & The upper limit of $\alpha^{\rm CPL}$. \\
\hline
Epeak & A12 & keV & $E_{\rm peak}$ as defined in  Eq.~\ref{eq:CPL}. \\
\hline
Epeak\_low & A12 & keV & The lowe limit of $E_{\rm peak}$. \\
\hline
Epeak\_hi & A12 & keV & The upper limit of $E_{\rm peak}$. \\
\hline
norm & A12 & $\rm ph \ cm^{-2} \ s^{-1} \ keV^{-1}$ & The normalization factor $K^{\rm CPL}_{50}$, as defined in Eq.~\ref{eq:CPL}. \\ 
\hline
norm\_low & A12 & $\rm ph \ cm^{-2} \ s^{-1} \ keV^{-1}$ & The lower limit of $K^{\rm CPL}_{50}$. \\ 
\hline
norm\_hi & A12 & $\rm ph \ cm^{-2} \ s^{-1} \ keV^{-1}$ & The upper limit of $K^{\rm CPL}_{50}$. \\ 
\hline
chi2 & F6.2 & -- & $\chi^2$ from the fitting, reported by XSPEC. \\  
\hline
dof & I2 & -- & degree of freedom from the fitting, reported by XSPEC. \\ 
\hline
reduced\_chi2 & F6.4 & -- & reduced $\chi^2$ (i.e., $\chi^2$ divided by the degree of freedom), \\ 
& & & reported by XSPEC. \\
\hline
null\_prob & A12 & -- & The null probability of the model, reported by XSPEC. \\ 
\hline
enorm & F7.4 & KeV & The normalization energy, which is set to 50 keV in our fits, \\ 
& & & as shown in Eq.~\ref{eq:PL}. \\ 
\hline
Exposure\_time & A6 & s & The time interval of the spectrum. \\ 
\hline
T100\_start & A11 & s & The start time of the spectrum, \\ 
& & & relative to the BAT trigger time. \\ 
\hline
T100\_stop & A11 & s & The end time of the spectrum, \\
& & & relative to the BAT trigger time.\\
\hline\hline
\end{tabular}
\end{center}
\end{table}

\begin{table}
\caption{
\label{tab:spec_T100_cutpow_photon_flux}
The format of the table that presents the photon flux (in unit of $\rm ph \ s^{-1} \ cm^{-2}$) from the CPL fit for the time-averaged ($T_{100}$) spectra. Note that this table includes the fit for every GRB, regardless of whether the fit is acceptable from the criteria listed in Section~\ref{sect:event_standard}.
A list of GRBs with acceptable fits can be found in Table \ref{tab:spec_T100_best_model}.
(This table is available in its entirety in a machine-readable form in the online journal. Only the column formats are shown here for guidance regarding its form and content.)
}
\begin{center}
\footnotesize
\begin{tabular}{|c|c|c|c|c|}
\hline\hline
Column & Format & Unit & Description \\
\hline\hline
GRBname & A9 & -- & GRB name. \\
\hline
Trig\_ID &  I11 & -- & Special ID associated with the trigger. \\ 
& & & For GRBs found in ground analysis, \\ 
& & & the ID are associated with the trigger ID of the failed event data, \\ 
& & & or the full observation ID of the event data for the analysis.  \\
\hline
15\_25kev & A12 & $\rm ph \ cm^{-2} \ s^{-1}$ & Photon flux in $15-25$ keV. \\
\hline
15\_25kev\_low & A12 & $\rm ph \ cm^{-2} \ s^{-1}$ & The lower limit of photon flux in $15-25$ keV. \\ 
\hline
15\_25kev\_hi & A12 & $\rm ph \ cm^{-2} \ s^{-1}$ & The upper limit of photon flux in $15-25$ keV. \\ 
\hline
25\_50kev & A12 & $\rm ph \ cm^{-2} \ s^{-1}$ & Photon flux in $25-50$ keV. \\ 
\hline
25\_50kev\_low & A12 & $\rm ph \ cm^{-2} \ s^{-1}$ & The lower limit of photon flux in $25-50$ keV. \\ 
\hline
25\_50kev\_hi & A12 & $\rm ph \ cm^{-2} \ s^{-1}$ & The upper limit of photon flux in $25-50$ keV. \\ 
\hline
50\_100kev & A12 & $\rm ph \ cm^{-2} \ s^{-1}$ & Photon flux in $50-100$ keV. \\ 
\hline
50\_100kev\_low & A12 & $\rm ph \ cm^{-2} \ s^{-1}$ & The lower limit of photon flux in $50-100$ keV. \\ 
\hline
50\_100kev\_hi & A12 & $\rm ph \ cm^{-2} \ s^{-1}$ & The upper limit of photon flux in $50-100$ keV. \\ 
\hline
100\_150kev & A12 & $\rm ph \ cm^{-2} \ s^{-1}$ & Photon flux in $100-150$ keV. \\ 
\hline
100\_150kev\_low & A12 & $\rm ph \ cm^{-2} \ s^{-1}$ & The lower limit of photon flux in $100-150$ keV. \\ 
\hline
100\_150kev\_hi & A12 & $\rm ph \ cm^{-2} \ s^{-1}$ & The upper limit of photon flux in $100-150$ keV. \\ 
\hline
100\_350kev & A12 & $\rm ph \ cm^{-2} \ s^{-1}$ & Photon flux in $100-350$ keV. \\
\hline
100\_350kev\_low & A12 & $\rm ph \ cm^{-2} \ s^{-1}$ & The lower limit of photon flux in $100-350$ keV. \\ 
\hline
100\_350kev\_hi & A12 & $\rm ph \ cm^{-2} \ s^{-1}$ & The upper limit of photon flux in $100-350$ keV. \\
\hline
15\_150kev & A12 & $\rm ph \ cm^{-2} \ s^{-1}$ & Photon flux in $15-150$ keV. \\ 
\hline
15\_150kev\_low & A12 & $\rm ph \ cm^{-2} \ s^{-1}$ & The lower limit of photon flux in $15-150$ keV. \\
\hline
15\_150kev\_hi & A12 & $\rm ph \ cm^{-2} \ s^{-1}$ & The upper limit of photon flux in $15-150$ keV. \\ 
\hline
15\_350kev & A12 & $\rm ph \ cm^{-2} \ s^{-1}$ & Photon flux in $15-350$ keV. \\
\hline
15\_350kev\_low & A12 & $\rm ph \ cm^{-2} \ s^{-1}$ & The lower limit of photon flux in $15-350$ keV. \\
\hline
15\_350kev\_hi & A12 & $\rm ph \ cm^{-2} \ s^{-1}$ & The upper limit of photon flux in $15-350$ keV. \\
\hline
Exposure\_time & A12 & s & The time interval of the spectrum. \\
\hline
T100\_start  & A12 & s & The start time of the spectrum, \\
& & & relative to the BAT trigger time. \\
\hline
T100\_stop & A12 & s & The end time of the spectrum, \\
& & & relative to the BAT trigger time. \\\hline\hline
\end{tabular}
\end{center}
\end{table}

\begin{table}
\caption{
\label{tab:spec_T100_cutpow_energy_flux}
The format of the table that presents the energy flux (in unit of $\rm erg \ s^{-1} \ cm^{-2}$) from the CPL fit for the time-averaged ($T_{100}$) spectra. Note that this table includes the fit for every GRB, regardless of whether the fit is acceptable from the criteria listed in Section~\ref{sect:event_standard}.
A list of GRBs with acceptable fits can be found in Table \ref{tab:spec_T100_best_model}.
(This table is available in its entirety in a machine-readable form in the online journal. Only the column formats are shown here for guidance regarding its form and content.)
}
\begin{center}
\footnotesize
\begin{tabular}{|c|c|c|c|c|}
\hline\hline
Column & Format & Unit & Description \\
\hline\hline
GRBname & A9 & -- & GRB name. \\
\hline
Trig\_ID &  I11 & -- & Special ID associated with the trigger. \\ 
& & & For GRBs found in ground analysis, \\ 
& & & the ID are associated with the trigger ID of the failed event data, \\ 
& & & or the full observation ID of the event data for the analysis.  \\
\hline
15\_25kev & A12 & $\rm erg \ cm^{-2} \ s^{-1}$ & Photon flux in $15-25$ keV. \\
\hline
15\_25kev\_low & A12 & $\rm erg \ cm^{-2} \ s^{-1}$ & The lower limit of photon flux in $15-25$ keV. \\ 
\hline
15\_25kev\_hi & A12 & $\rm erg \ cm^{-2} \ s^{-1}$ & The upper limit of photon flux in $15-25$ keV. \\ 
\hline
25\_50kev & A12 & $\rm erg \ cm^{-2} \ s^{-1}$ & Photon flux in $25-50$ keV. \\ 
\hline
25\_50kev\_low & A12 & $\rm erg \ cm^{-2} \ s^{-1}$ & The lower limit of photon flux in $25-50$ keV. \\ 
\hline
25\_50kev\_hi & A12 & $\rm erg \ cm^{-2} \ s^{-1}$ & The upper limit of photon flux in $25-50$ keV. \\ 
\hline
50\_100kev & A12 & $\rm erg \ cm^{-2} \ s^{-1}$ & Photon flux in $50-100$ keV. \\ 
\hline
50\_100kev\_low & A12 & $\rm erg \ cm^{-2} \ s^{-1}$ & The lower limit of photon flux in $50-100$ keV. \\ 
\hline
50\_100kev\_hi & A12 & $\rm erg \ cm^{-2} \ s^{-1}$ & The upper limit of photon flux in $50-100$ keV. \\ 
\hline
100\_150kev & A12 & $\rm erg \ cm^{-2} \ s^{-1}$ & Photon flux in $100-150$ keV. \\ 
\hline
100\_150kev\_low & A12 & $\rm erg \ cm^{-2} \ s^{-1}$ & The lower limit of photon flux in $100-150$ keV. \\ 
\hline
100\_150kev\_hi & A12 & $\rm erg \ cm^{-2} \ s^{-1}$ & The upper limit of photon flux in $100-150$ keV. \\ 
\hline
100\_350kev & A12 & $\rm erg \ cm^{-2} \ s^{-1}$ & Photon flux in $100-350$ keV. \\
\hline
100\_350kev\_low & A12 & $\rm erg \ cm^{-2} \ s^{-1}$ & The lower limit of photon flux in $100-350$ keV. \\ 
\hline
100\_350kev\_hi & A12 & $\rm erg \ cm^{-2} \ s^{-1}$ & The upper limit of photon flux in $100-350$ keV. \\
\hline
15\_150kev & A12 & $\rm erg \ cm^{-2} \ s^{-1}$ & Photon flux in $15-150$ keV. \\ 
\hline
15\_150kev\_low & A12 & $\rm erg \ cm^{-2} \ s^{-1}$ & The lower limit of photon flux in $15-150$ keV. \\
\hline
15\_150kev\_hi & A12 & $\rm erg \ cm^{-2} \ s^{-1}$ & The upper limit of photon flux in $15-150$ keV. \\ 
\hline
15\_350kev & A12 & $\rm erg \ cm^{-2} \ s^{-1}$ & Photon flux in $15-350$ keV. \\
\hline
15\_350kev\_low & A12 & $\rm erg \ cm^{-2} \ s^{-1}$ & The lower limit of photon flux in $15-350$ keV. \\
\hline
15\_350kev\_hi & A12 & $\rm erg \ cm^{-2} \ s^{-1}$ & The upper limit of photon flux in $15-350$ keV. \\
\hline
Exposure\_time & A12 & s & The time interval of the spectrum. \\
\hline
T100\_start  & A12 & s & The start time of the spectrum, \\
& & & relative to the BAT trigger time. \\
\hline
T100\_stop & A12 & s & The end time of the spectrum, \\
& & & relative to the BAT trigger time. \\
\hline\hline
\end{tabular}
\end{center}
\end{table}

\begin{table}
\caption{
\label{tab:spec_T100_cutpow_energy_fluence}
The format of the table that presents the energy fluence in unit of $\rm erg \ s^{-1}$ from the CPL fit for the time-averaged ($T_{100}$) spectra. Note that this table includes the fit for every GRB, regardless of whether the fit is acceptable from the criteria listed in Section~\ref{sect:event_standard}.
A list of GRBs with acceptable fits can be found in Table \ref{tab:spec_T100_best_model}.
(This table is available in its entirety in a machine-readable form in the online journal. Only the column formats are shown here for guidance regarding its form and content.)
}
\begin{center}
\footnotesize
\begin{tabular}{|c|c|c|c|c|}
\hline\hline
Column & Format & Unit & Description \\
\hline\hline
GRBname & A9 & -- & GRB name. \\
\hline
Trig\_ID &  I11 & -- & Special ID associated with the trigger. \\ 
& & & For GRBs found in ground analysis, \\ 
& & & the ID are associated with the trigger ID of the failed event data, \\ 
& & & or the full observation ID of the event data for the analysis.  \\
\hline
15\_25kev & A12 & $\rm erg \ cm^{-2}$ & Photon flux in $15-25$ keV. \\
\hline
15\_25kev\_low & A12 & $\rm erg \ cm^{-2}$ & The lower limit of photon flux in $15-25$ keV. \\ 
\hline
15\_25kev\_hi & A12 & $\rm erg \ cm^{-2}$ & The upper limit of photon flux in $15-25$ keV. \\ 
\hline
25\_50kev & A12 & $\rm erg \ cm^{-2}$ & Photon flux in $25-50$ keV. \\ 
\hline
25\_50kev\_low & A12 & $\rm erg \ cm^{-2}$ & The lower limit of photon flux in $25-50$ keV. \\ 
\hline
25\_50kev\_hi & A12 & $\rm erg \ cm^{-2}$ & The upper limit of photon flux in $25-50$ keV. \\ 
\hline
50\_100kev & A12 & $\rm erg \ cm^{-2}$ & Photon flux in $50-100$ keV. \\ 
\hline
50\_100kev\_low & A12 & $\rm erg \ cm^{-2}$ & The lower limit of photon flux in $50-100$ keV. \\ 
\hline
50\_100kev\_hi & A12 & $\rm erg \ cm^{-2}$ & The upper limit of photon flux in $50-100$ keV. \\ 
\hline
100\_150kev & A12 & $\rm erg \ cm^{-2}$ & Photon flux in $100-150$ keV. \\ 
\hline
100\_150kev\_low & A12 & $\rm erg \ cm^{-2}$ & The lower limit of photon flux in $100-150$ keV. \\ 
\hline
100\_150kev\_hi & A12 & $\rm erg \ cm^{-2}$ & The upper limit of photon flux in $100-150$ keV. \\ 
\hline
100\_350kev & A12 & $\rm erg \ cm^{-2}$ & Photon flux in $100-350$ keV. \\
\hline
100\_350kev\_low & A12 & $\rm erg \ cm^{-2}$ & The lower limit of photon flux in $100-350$ keV. \\ 
\hline
100\_350kev\_hi & A12 & $\rm erg \ cm^{-2}$ & The upper limit of photon flux in $100-350$ keV. \\
\hline
15\_150kev & A12 & $\rm erg \ cm^{-2}$ & Photon flux in $15-150$ keV. \\ 
\hline
15\_150kev\_low & A12 & $\rm erg \ cm^{-2}$ & The lower limit of photon flux in $15-150$ keV. \\
\hline
15\_150kev\_hi & A12 & $\rm erg \ cm^{-2}$ & The upper limit of photon flux in $15-150$ keV. \\ 
\hline
15\_350kev & A12 & $\rm erg \ cm^{-2}$ & Photon flux in $15-350$ keV. \\
\hline
15\_350kev\_low & A12 & $\rm erg \ cm^{-2}$ & The lower limit of photon flux in $15-350$ keV. \\
\hline
15\_350kev\_hi & A12 & $\rm erg \ cm^{-2}$ & The upper limit of photon flux in $15-350$ keV. \\
\hline
Exposure\_time & A12 & s & The time interval of the spectrum. \\
\hline
T100\_start  & A12 & s & The start time of the spectrum, \\
& & & relative to the BAT trigger time. \\
\hline
T100\_stop & A12 & s & The end time of the spectrum, \\
& & & relative to the BAT trigger time. \\
\hline\hline
\end{tabular}
\end{center}
\end{table}

\subsection{Summary tables of spectral analyses for the 1-s peak spectra}

In this section, we present the tables for the 1-s peak spectra, which are the same set of the tables as those in Section~\ref{sect:T100_spectra_tables} but for the 1-s peak spectral fits.

\clearpage

\begin{table}
\caption{
\label{tab:spec_1speak_best_model}
The format of the table that summarizes the best-fit spectral model of the 1-s peak spectrum for each burst.
(This table is available in its entirety in a machine-readable form in the online journal. Only the column formats are shown here for guidance regarding its form and content.)
}
\begin{center}
\begin{tabular}{|c|c|c|c|}
\hline\hline
Column & Format & Unit & Description \\
\hline\hline
GRBname & A9 & -- & GRB name. \\
\hline
Trig\_ID &  I11 & -- & Special ID associated with the trigger. \\ 
& & & For GRBs found in ground analysis, \\ 
& & & the ID are associated with the trigger ID of the failed event data, \\ 
& & & or the full observation ID of the event data for the analysis.  \\
\hline
Best\_fit\_model  & A3 & -- & The best-fit model for the GRB spectrum, \\
& & & either the simple power law model (PL) \\ 
& & & or cutoff power-law model (CPL.)  \\
\hline\hline
\end{tabular}
\end{center}
\end{table}

\begin{table}
\caption{
\label{tab:spec_1speak_pow_parameter}
The format of the table that presents the parameters from the PL fit for the 1-s peak spectra. Note that this table includes the fit for every GRB, regardless of whether the fit is acceptable from the criteria listed in Section~\ref{sect:event_standard}.
A list of GRBs with acceptable fits can be found in Table \ref{tab:spec_1speak_best_model}.
(This table is available in its entirety in a machine-readable form in the online journal. Only the column formats are shown here for guidance regarding its form and content.)
}
\begin{center}
\footnotesize
\begin{tabular}{|c|c|c|c|}
\hline\hline
Column & Format & Unit & Description \\
\hline\hline
GRBname & A9 & -- & GRB name. \\
\hline
Trig\_ID &  I11 & -- & Special ID associated with each triggers. \\ 
& & & For GRBs found in ground analysis, \\ 
& & & the ID are associated with the trigger ID of the failed event data, \\ 
& & & or the full observation ID of the event data for the analysis.  \\
\hline
alpha & A13 & -- & $\alpha^{\rm PL}$ as defined in Eq.~\ref{eq:PL}. \\
\hline
alpha\_low & A13 & -- & The lower limit of $\alpha^{\rm PL}$. \\
\hline
alpha\_hi & A13 & -- & The upper limit of $\alpha^{\rm PL}$. \\
\hline
norm & A12 & $\rm ph \ cm^{-2} \ s^{-1} \ keV^{-1}$ & The normalization factor $K^{\rm PL}_{50}$, as defined in Eq.~\ref{eq:PL}. \\ 
\hline
norm\_low & A12 & $\rm ph \ cm^{-2} \ s^{-1} \ keV^{-1}$ & The lower limit of $K^{\rm PL}_{50}$. \\ 
\hline
norm\_hi & A12 & $\rm ph \ cm^{-2} \ s^{-1} \ keV^{-1}$ & The upper limit of $K^{\rm PL}_{50}$. \\ 
\hline
chi2 & F6.2 & -- & $\chi^2$ from the fitting, reported by XSPEC. \\  
\hline
dof & I2 & -- & degree of freedom from the fitting, reported by XSPEC. \\ 
\hline
reduced\_chi2 & F6.4 & -- & reduced $\chi^2$ (i.e., $\chi^2$ divided by the degree of freedom), \\ 
& & & reported by XSPEC. \\
\hline
null\_prob & A12 & -- & The null probability of the model, reported by XSPEC. \\ 
\hline
enorm & F7.4 & KeV & The normalization energy, which is set to 50 keV in our fits, \\ 
& & & as shown in Eq.~\ref{eq:PL}. \\ 
\hline
Exposure\_time & A6 & s & The time interval of the spectrum. \\ 
\hline
Peak\_start & A11 & s & The start time of the spectrum, \\ 
& & & relative to the BAT trigger time. \\ 
\hline
Peak\_stop & A11 & s & The end time of the spectrum, \\
& & & relative to the BAT trigger time.\\
\hline\hline
\end{tabular}
\end{center}
\end{table}

\begin{table}
\caption{
\label{tab:spec_1speak_pow_photon_flux}
The format of the table that presents the photon flux (in unit of $\rm ph \ cm^{-2} \ s^{-1}$) from the PL fit for the 1-s peak spectra. Note that this table includes the fit for every GRB, regardless of whether the fit is acceptable from the criteria listed in Section~\ref{sect:event_standard}.
A list of GRBs with acceptable fits can be found in Table \ref{tab:spec_1speak_best_model}.
(This table is available in its entirety in a machine-readable form in the online journal. Only the column formats are shown here for guidance regarding its form and content.)
}
\begin{center}
\footnotesize
\begin{tabular}{|c|c|c|c|}
\hline\hline
Column & Format & Unit & Description \\
\hline\hline
GRBname & A9 & -- & GRB name. \\
\hline
Trig\_ID &  I11 & -- & Special ID associated with the trigger. \\ 
& & & For GRBs found in ground analysis, \\ 
& & & the ID are associated with the trigger ID of the failed event data, \\ 
& & & or the full observation ID of the event data for the analysis.  \\
\hline
15\_25kev & A12 & $\rm ph \ cm^{-2} \ s^{-1}$ & Photon flux in $15-25$ keV. \\
\hline
15\_25kev\_low & A12 & $\rm ph \ cm^{-2} \ s^{-1}$ & The lower limit of photon flux in $15-25$ keV. \\ 
\hline
15\_25kev\_hi & A12 & $\rm ph \ cm^{-2} \ s^{-1}$ & The upper limit of photon flux in $15-25$ keV. \\ 
\hline
25\_50kev & A12 & $\rm ph \ cm^{-2} \ s^{-1}$ & Photon flux in $25-50$ keV. \\ 
\hline
25\_50kev\_low & A12 & $\rm ph \ cm^{-2} \ s^{-1}$ & The lower limit of photon flux in $25-50$ keV. \\ 
\hline
25\_50kev\_hi & A12 & $\rm ph \ cm^{-2} \ s^{-1}$ & The upper limit of photon flux in $25-50$ keV. \\ 
\hline
50\_100kev & A12 & $\rm ph \ cm^{-2} \ s^{-1}$ & Photon flux in $50-100$ keV. \\ 
\hline
50\_100kev\_low & A12 & $\rm ph \ cm^{-2} \ s^{-1}$ & The lower limit of photon flux in $50-100$ keV. \\ 
\hline
50\_100kev\_hi & A12 & $\rm ph \ cm^{-2} \ s^{-1}$ & The upper limit of photon flux in $50-100$ keV. \\ 
\hline
100\_150kev & A12 & $\rm ph \ cm^{-2} \ s^{-1}$ & Photon flux in $100-150$ keV. \\ 
\hline
100\_150kev\_low & A12 & $\rm ph \ cm^{-2} \ s^{-1}$ & The lower limit of photon flux in $100-150$ keV. \\ 
\hline
100\_150kev\_hi & A12 & $\rm ph \ cm^{-2} \ s^{-1}$ & The upper limit of photon flux in $100-150$ keV. \\ 
\hline
100\_350kev & A12 & $\rm ph \ cm^{-2} \ s^{-1}$ & Photon flux in $100-350$ keV. \\
\hline
100\_350kev\_low & A12 & $\rm ph \ cm^{-2} \ s^{-1}$ & The lower limit of photon flux in $100-350$ keV. \\ 
\hline
100\_350kev\_hi & A12 & $\rm ph \ cm^{-2} \ s^{-1}$ & The upper limit of photon flux in $100-350$ keV. \\
\hline
15\_150kev & A12 & $\rm ph \ cm^{-2} \ s^{-1}$ & Photon flux in $15-150$ keV. \\ 
\hline
15\_150kev\_low & A12 & $\rm ph \ cm^{-2} \ s^{-1}$ & The lower limit of photon flux in $15-150$ keV. \\
\hline
15\_150kev\_hi & A12 & $\rm ph \ cm^{-2} \ s^{-1}$ & The upper limit of photon flux in $15-150$ keV. \\ 
\hline
15\_350kev & A12 & $\rm ph \ cm^{-2} \ s^{-1}$ & Photon flux in $15-350$ keV. \\
\hline
15\_350kev\_low & A12 & $\rm ph \ cm^{-2} \ s^{-1}$ & The lower limit of photon flux in $15-350$ keV. \\
\hline
15\_350kev\_hi & A12 & $\rm ph \ cm^{-2} \ s^{-1}$ & The upper limit of photon flux in $15-350$ keV. \\
\hline
Exposure\_time & A12 & s & The time interval of the spectrum. \\
\hline
Peak\_start  & A12 & s & The start time of the spectrum, \\
& & & relative to the BAT trigger time. \\
\hline
Peak\_stop & A12 & s & The end time of the spectrum, \\
& & & relative to the BAT trigger time. \\
\hline\hline
\end{tabular}
\end{center}
\end{table}

\begin{table}
\caption{
\label{tab:spec_1speak_pow_energy_flux}
The format of the table that presents the energy flux (in unit of $\rm erg \ s^{-1} \ cm^{-2}$) from the PL fit for the 1-s peak spectra. Note that this table includes the fit for every GRB, regardless of whether the fit is acceptable from the criteria listed in Section~\ref{sect:event_standard}.
A list of GRBs with acceptable fits can be found in Table \ref{tab:spec_1speak_best_model}.
(This table is available in its entirety in a machine-readable form in the online journal. Only the column formats are shown here for guidance regarding its form and content.)
}
\begin{center}
\footnotesize
\begin{tabular}{|c|c|c|c|c|}
\hline\hline
Column & Format & Unit & Description \\
\hline\hline
GRBname & A9 & -- & GRB name. \\
\hline
Trig\_ID &  I11 & -- & Special ID associated with the trigger. \\ 
& & & For GRBs found in ground analysis, \\ 
& & & the ID are associated with the trigger ID of the failed event data, \\ 
& & & or the full observation ID of the event data for the analysis.  \\
\hline
15\_25kev & A12 & $\rm erg \ cm^{-2} \ s^{-1}$ & Photon flux in $15-25$ keV. \\
\hline
15\_25kev\_low & A12 & $\rm erg \ cm^{-2} \ s^{-1}$ & The lower limit of photon flux in $15-25$ keV. \\ 
\hline
15\_25kev\_hi & A12 & $\rm erg \ cm^{-2} \ s^{-1}$ & The upper limit of photon flux in $15-25$ keV. \\ 
\hline
25\_50kev & A12 & $\rm erg \ cm^{-2} \ s^{-1}$ & Photon flux in $25-50$ keV. \\ 
\hline
25\_50kev\_low & A12 & $\rm erg \ cm^{-2} \ s^{-1}$ & The lower limit of photon flux in $25-50$ keV. \\ 
\hline
25\_50kev\_hi & A12 & $\rm erg \ cm^{-2} \ s^{-1}$ & The upper limit of photon flux in $25-50$ keV. \\ 
\hline
50\_100kev & A12 & $\rm erg \ cm^{-2} \ s^{-1}$ & Photon flux in $50-100$ keV. \\ 
\hline
50\_100kev\_low & A12 & $\rm erg \ cm^{-2} \ s^{-1}$ & The lower limit of photon flux in $50-100$ keV. \\ 
\hline
50\_100kev\_hi & A12 & $\rm erg \ cm^{-2} \ s^{-1}$ & The upper limit of photon flux in $50-100$ keV. \\ 
\hline
100\_150kev & A12 & $\rm erg \ cm^{-2} \ s^{-1}$ & Photon flux in $100-150$ keV. \\ 
\hline
100\_150kev\_low & A12 & $\rm erg \ cm^{-2} \ s^{-1}$ & The lower limit of photon flux in $100-150$ keV. \\ 
\hline
100\_150kev\_hi & A12 & $\rm erg \ cm^{-2} \ s^{-1}$ & The upper limit of photon flux in $100-150$ keV. \\ 
\hline
100\_350kev & A12 & $\rm erg \ cm^{-2} \ s^{-1}$ & Photon flux in $100-350$ keV. \\
\hline
100\_350kev\_low & A12 & $\rm erg \ cm^{-2} \ s^{-1}$ & The lower limit of photon flux in $100-350$ keV. \\ 
\hline
100\_350kev\_hi & A12 & $\rm erg \ cm^{-2} \ s^{-1}$ & The upper limit of photon flux in $100-350$ keV. \\
\hline
15\_150kev & A12 & $\rm erg \ cm^{-2} \ s^{-1}$ & Photon flux in $15-150$ keV. \\ 
\hline
15\_150kev\_low & A12 & $\rm erg \ cm^{-2} \ s^{-1}$ & The lower limit of photon flux in $15-150$ keV. \\
\hline
15\_150kev\_hi & A12 & $\rm erg \ cm^{-2} \ s^{-1}$ & The upper limit of photon flux in $15-150$ keV. \\ 
\hline
15\_350kev & A12 & $\rm erg \ cm^{-2} \ s^{-1}$ & Photon flux in $15-350$ keV. \\
\hline
15\_350kev\_low & A12 & $\rm erg \ cm^{-2} \ s^{-1}$ & The lower limit of photon flux in $15-350$ keV. \\
\hline
15\_350kev\_hi & A12 & $\rm erg \ cm^{-2} \ s^{-1}$ & The upper limit of photon flux in $15-350$ keV. \\
\hline
Exposure\_time & A12 & s & The time interval of the spectrum. \\
\hline
Peak\_start  & A12 & s & The start time of the spectrum, \\
& & & relative to the BAT trigger time. \\
\hline
Peak\_stop & A12 & s & The end time of the spectrum, \\
& & & relative to the BAT trigger time. \\
\hline\hline
\end{tabular}
\end{center}
\end{table}

\begin{table}
\caption{
\label{tab:spec_1speak_pow_energy_fluence}
The format of the table that presents the energy fluences (in unit of $\rm erg \ cm^{-2}$) from the PL fit for the 1-s peak spectra. 
Note that this table includes the fit for every GRB, regardless of whether the fit is acceptable from the criteria listed in Section~\ref{sect:event_standard}.
A list of GRBs with acceptable fits can be found in Table \ref{tab:spec_1speak_best_model}.
(This table is available in its entirety in a machine-readable form in the online journal. Only the column titles are shown here for guidance regarding its form and content.)
}
\begin{center}
\footnotesize
\begin{tabular}{|c|c|c|c|c|}
\hline\hline
Column & Format & Unit & Description \\
\hline\hline
GRBname & A9 & -- & GRB name. \\
\hline
Trig\_ID &  I11 & -- & Special ID associated with the trigger. \\ 
& & & For GRBs found in ground analysis, \\ 
& & & the ID are associated with the trigger ID of the failed event data, \\ 
& & & or the full observation ID of the event data for the analysis.  \\
\hline
15\_25kev & A12 & $\rm erg \ cm^{-2}$ & Photon flux in $15-25$ keV. \\
\hline
15\_25kev\_low & A12 & $\rm erg \ cm^{-2}$ & The lower limit of photon flux in $15-25$ keV. \\ 
\hline
15\_25kev\_hi & A12 & $\rm erg \ cm^{-2}$ & The upper limit of photon flux in $15-25$ keV. \\ 
\hline
25\_50kev & A12 & $\rm erg \ cm^{-2}$ & Photon flux in $25-50$ keV. \\ 
\hline
25\_50kev\_low & A12 & $\rm erg \ cm^{-2}$ & The lower limit of photon flux in $25-50$ keV. \\ 
\hline
25\_50kev\_hi & A12 & $\rm erg \ cm^{-2}$ & The upper limit of photon flux in $25-50$ keV. \\ 
\hline
50\_100kev & A12 & $\rm erg \ cm^{-2}$ & Photon flux in $50-100$ keV. \\ 
\hline
50\_100kev\_low & A12 & $\rm erg \ cm^{-2}$ & The lower limit of photon flux in $50-100$ keV. \\ 
\hline
50\_100kev\_hi & A12 & $\rm erg \ cm^{-2}$ & The upper limit of photon flux in $50-100$ keV. \\ 
\hline
100\_150kev & A12 & $\rm erg \ cm^{-2}$ & Photon flux in $100-150$ keV. \\ 
\hline
100\_150kev\_low & A12 & $\rm erg \ cm^{-2}$ & The lower limit of photon flux in $100-150$ keV. \\ 
\hline
100\_150kev\_hi & A12 & $\rm erg \ cm^{-2}$ & The upper limit of photon flux in $100-150$ keV. \\ 
\hline
100\_350kev & A12 & $\rm erg \ cm^{-2}$ & Photon flux in $100-350$ keV. \\
\hline
100\_350kev\_low & A12 & $\rm erg \ cm^{-2}$ & The lower limit of photon flux in $100-350$ keV. \\ 
\hline
100\_350kev\_hi & A12 & $\rm erg \ cm^{-2}$ & The upper limit of photon flux in $100-350$ keV. \\
\hline
15\_150kev & A12 & $\rm erg \ cm^{-2}$ & Photon flux in $15-150$ keV. \\ 
\hline
15\_150kev\_low & A12 & $\rm erg \ cm^{-2}$ & The lower limit of photon flux in $15-150$ keV. \\
\hline
15\_150kev\_hi & A12 & $\rm erg \ cm^{-2}$ & The upper limit of photon flux in $15-150$ keV. \\ 
\hline
15\_350kev & A12 & $\rm erg \ cm^{-2}$ & Photon flux in $15-350$ keV. \\
\hline
15\_350kev\_low & A12 & $\rm erg \ cm^{-2}$ & The lower limit of photon flux in $15-350$ keV. \\
\hline
15\_350kev\_hi & A12 & $\rm erg \ cm^{-2}$ & The upper limit of photon flux in $15-350$ keV. \\
\hline
Exposure\_time & A12 & s & The time interval of the spectrum. \\
\hline
Peak\_start  & A12 & s & The start time of the spectrum, \\
& & & relative to the BAT trigger time. \\
\hline
Peak\_stop & A12 & s & The end time of the spectrum, \\
& & & relative to the BAT trigger time. \\
\hline\hline
\end{tabular}
\end{center}
\end{table}

\begin{table}
\caption{
\label{tab:spec_1speak_cutpow_parameter}
The format of the table that presents the parameters from the CPL fit for the 1-s peak spectra. Note that this table includes the fit for every GRB, regardless of whether the fit is acceptable from the criteria listed in Section~\ref{sect:event_standard}.
A list of GRBs with acceptable fits can be found in Table \ref{tab:spec_1speak_best_model}.
(This table is available in its entirety in a machine-readable form in the online journal. Only the column formats are shown here for guidance regarding its form and content.)
}
\begin{center}
\footnotesize
\begin{tabular}{|c|c|c|c|c|}
\hline\hline
Column & Format & Unit & Description \\
\hline\hline
GRBname & A9 & -- & GRB name. \\
\hline
Trig\_ID &  I11 & -- & Special ID associated with each triggers. \\ 
& & & For GRBs found in ground analysis, \\ 
& & & the ID are associated with the trigger ID of the failed event data, \\ 
& & & or the full observation ID of the event data for the analysis.  \\
\hline
alpha & A13 & -- & $\alpha^{\rm CPL}$ as defined in Eq.~\ref{eq:CPL}. \\
\hline
alpha\_low & A13 & -- & The lower limit of $\alpha^{\rm CPL}$. \\
\hline
alpha\_hi & A13 & -- & The upper limit of $\alpha^{\rm CPL}$. \\
\hline
Epeak & A12 & keV & $E_{\rm peak}$ as defined in  Eq.~\ref{eq:CPL}. \\
\hline
Epeak\_low & A12 & keV & The lowe limit of $E_{\rm peak}$. \\
\hline
Epeak\_hi & A12 & keV & The upper limit of $E_{\rm peak}$. \\
\hline
norm & A12 & $\rm ph \ cm^{-2} \ s^{-1} \ keV^{-1}$ & The normalization factor $K^{\rm CPL}_{50}$, as defined in Eq.~\ref{eq:CPL}. \\ 
\hline
norm\_low & A12 & $\rm ph \ cm^{-2} \ s^{-1} \ keV^{-1}$ & The lower limit of $K^{\rm CPL}_{50}$. \\ 
\hline
norm\_hi & A12 & $\rm ph \ cm^{-2} \ s^{-1} \ keV^{-1}$ & The upper limit of $K^{\rm CPL}_{50}$. \\ 
\hline
chi2 & F6.2 & -- & $\chi^2$ from the fitting, reported by XSPEC. \\  
\hline
dof & I2 & -- & degree of freedom from the fitting, reported by XSPEC. \\ 
\hline
reduced\_chi2 & F6.4 & -- & reduced $\chi^2$ (i.e., $\chi^2$ divided by the degree of freedom), \\ 
& & & reported by XSPEC. \\
\hline
null\_prob & A12 & -- & The null probability of the model, reported by XSPEC. \\ 
\hline
enorm & F7.4 & KeV & The normalization energy, which is set to 50 keV in our fits, \\ 
& & & as shown in Eq.~\ref{eq:PL}. \\ 
\hline
Exposure\_time & A6 & s & The time interval of the spectrum. \\ 
\hline
Peak\_start & A11 & s & The start time of the spectrum, \\ 
& & & relative to the BAT trigger time. \\ 
\hline
Peak\_stop & A11 & s & The end time of the spectrum, \\
& & & relative to the BAT trigger time.\\
\hline\hline
\end{tabular}
\end{center}
\end{table}

\begin{table}
\caption{
\label{tab:spec_1speak_cutpow_photon_flux}
The format of the table that presents the photon flux (in unit of $\rm ph \ s^{-1} \ cm^{-2}$) from the CPL fit for the 1-s peak spectra. Note that this table includes the fit for every GRB, regardless of whether the fit is acceptable from the criteria listed in Section~\ref{sect:event_standard}.
A list of GRBs with acceptable fits can be found in Table \ref{tab:spec_1speak_best_model}.
(This table is available in its entirety in a machine-readable form in the online journal. Only the column formats are shown here for guidance regarding its form and content.)
}
\begin{center}
\footnotesize
\begin{tabular}{|c|c|c|c|c|}
\hline\hline
Column & Format & Unit & Description \\
\hline\hline
GRBname & A9 & -- & GRB name. \\
\hline
Trig\_ID &  I11 & -- & Special ID associated with the trigger. \\ 
& & & For GRBs found in ground analysis, \\ 
& & & the ID are associated with the trigger ID of the failed event data, \\ 
& & & or the full observation ID of the event data for the analysis.  \\
\hline
15\_25kev & A12 & $\rm ph \ cm^{-2} \ s^{-1}$ & Photon flux in $15-25$ keV. \\
\hline
15\_25kev\_low & A12 & $\rm ph \ cm^{-2} \ s^{-1}$ & The lower limit of photon flux in $15-25$ keV. \\ 
\hline
15\_25kev\_hi & A12 & $\rm ph \ cm^{-2} \ s^{-1}$ & The upper limit of photon flux in $15-25$ keV. \\ 
\hline
25\_50kev & A12 & $\rm ph \ cm^{-2} \ s^{-1}$ & Photon flux in $25-50$ keV. \\ 
\hline
25\_50kev\_low & A12 & $\rm ph \ cm^{-2} \ s^{-1}$ & The lower limit of photon flux in $25-50$ keV. \\ 
\hline
25\_50kev\_hi & A12 & $\rm ph \ cm^{-2} \ s^{-1}$ & The upper limit of photon flux in $25-50$ keV. \\ 
\hline
50\_100kev & A12 & $\rm ph \ cm^{-2} \ s^{-1}$ & Photon flux in $50-100$ keV. \\ 
\hline
50\_100kev\_low & A12 & $\rm ph \ cm^{-2} \ s^{-1}$ & The lower limit of photon flux in $50-100$ keV. \\ 
\hline
50\_100kev\_hi & A12 & $\rm ph \ cm^{-2} \ s^{-1}$ & The upper limit of photon flux in $50-100$ keV. \\ 
\hline
100\_150kev & A12 & $\rm ph \ cm^{-2} \ s^{-1}$ & Photon flux in $100-150$ keV. \\ 
\hline
100\_150kev\_low & A12 & $\rm ph \ cm^{-2} \ s^{-1}$ & The lower limit of photon flux in $100-150$ keV. \\ 
\hline
100\_150kev\_hi & A12 & $\rm ph \ cm^{-2} \ s^{-1}$ & The upper limit of photon flux in $100-150$ keV. \\ 
\hline
100\_350kev & A12 & $\rm ph \ cm^{-2} \ s^{-1}$ & Photon flux in $100-350$ keV. \\
\hline
100\_350kev\_low & A12 & $\rm ph \ cm^{-2} \ s^{-1}$ & The lower limit of photon flux in $100-350$ keV. \\ 
\hline
100\_350kev\_hi & A12 & $\rm ph \ cm^{-2} \ s^{-1}$ & The upper limit of photon flux in $100-350$ keV. \\
\hline
15\_150kev & A12 & $\rm ph \ cm^{-2} \ s^{-1}$ & Photon flux in $15-150$ keV. \\ 
\hline
15\_150kev\_low & A12 & $\rm ph \ cm^{-2} \ s^{-1}$ & The lower limit of photon flux in $15-150$ keV. \\
\hline
15\_150kev\_hi & A12 & $\rm ph \ cm^{-2} \ s^{-1}$ & The upper limit of photon flux in $15-150$ keV. \\ 
\hline
15\_350kev & A12 & $\rm ph \ cm^{-2} \ s^{-1}$ & Photon flux in $15-350$ keV. \\
\hline
15\_350kev\_low & A12 & $\rm ph \ cm^{-2} \ s^{-1}$ & The lower limit of photon flux in $15-350$ keV. \\
\hline
15\_350kev\_hi & A12 & $\rm ph \ cm^{-2} \ s^{-1}$ & The upper limit of photon flux in $15-350$ keV. \\
\hline
Exposure\_time & A12 & s & The time interval of the spectrum. \\
\hline
Peak\_start  & A12 & s & The start time of the spectrum, \\
& & & relative to the BAT trigger time. \\
\hline
Peak\_stop & A12 & s & The end time of the spectrum, \\
& & & relative to the BAT trigger time. \\\hline\hline
\end{tabular}
\end{center}
\end{table}

\begin{table}
\caption{
\label{tab:spec_1speak_cutpow_energy_flux}
The format of the table that presents the energy flux (in unit of $\rm erg \ s^{-1} \ cm^{-2}$) from the CPL fit for the 1-s peak spectra. Note that this table includes the fit for every GRB, regardless of whether the fit is acceptable from the criteria listed in Section~\ref{sect:event_standard}.
A list of GRBs with acceptable fits can be found in Table \ref{tab:spec_1speak_best_model}.
(This table is available in its entirety in a machine-readable form in the online journal. Only the column formats are shown here for guidance regarding its form and content.)
}
\begin{center}
\footnotesize
\begin{tabular}{|c|c|c|c|c|}
\hline\hline
Column & Format & Unit & Description \\
\hline\hline
GRBname & A9 & -- & GRB name. \\
\hline
Trig\_ID &  I11 & -- & Special ID associated with the trigger. \\ 
& & & For GRBs found in ground analysis, \\ 
& & & the ID are associated with the trigger ID of the failed event data, \\ 
& & & or the full observation ID of the event data for the analysis.  \\
\hline
15\_25kev & A12 & $\rm erg \ cm^{-2} \ s^{-1}$ & Photon flux in $15-25$ keV. \\
\hline
15\_25kev\_low & A12 & $\rm erg \ cm^{-2} \ s^{-1}$ & The lower limit of photon flux in $15-25$ keV. \\ 
\hline
15\_25kev\_hi & A12 & $\rm erg \ cm^{-2} \ s^{-1}$ & The upper limit of photon flux in $15-25$ keV. \\ 
\hline
25\_50kev & A12 & $\rm erg \ cm^{-2} \ s^{-1}$ & Photon flux in $25-50$ keV. \\ 
\hline
25\_50kev\_low & A12 & $\rm erg \ cm^{-2} \ s^{-1}$ & The lower limit of photon flux in $25-50$ keV. \\ 
\hline
25\_50kev\_hi & A12 & $\rm erg \ cm^{-2} \ s^{-1}$ & The upper limit of photon flux in $25-50$ keV. \\ 
\hline
50\_100kev & A12 & $\rm erg \ cm^{-2} \ s^{-1}$ & Photon flux in $50-100$ keV. \\ 
\hline
50\_100kev\_low & A12 & $\rm erg \ cm^{-2} \ s^{-1}$ & The lower limit of photon flux in $50-100$ keV. \\ 
\hline
50\_100kev\_hi & A12 & $\rm erg \ cm^{-2} \ s^{-1}$ & The upper limit of photon flux in $50-100$ keV. \\ 
\hline
100\_150kev & A12 & $\rm erg \ cm^{-2} \ s^{-1}$ & Photon flux in $100-150$ keV. \\ 
\hline
100\_150kev\_low & A12 & $\rm erg \ cm^{-2} \ s^{-1}$ & The lower limit of photon flux in $100-150$ keV. \\ 
\hline
100\_150kev\_hi & A12 & $\rm erg \ cm^{-2} \ s^{-1}$ & The upper limit of photon flux in $100-150$ keV. \\ 
\hline
100\_350kev & A12 & $\rm erg \ cm^{-2} \ s^{-1}$ & Photon flux in $100-350$ keV. \\
\hline
100\_350kev\_low & A12 & $\rm erg \ cm^{-2} \ s^{-1}$ & The lower limit of photon flux in $100-350$ keV. \\ 
\hline
100\_350kev\_hi & A12 & $\rm erg \ cm^{-2} \ s^{-1}$ & The upper limit of photon flux in $100-350$ keV. \\
\hline
15\_150kev & A12 & $\rm erg \ cm^{-2} \ s^{-1}$ & Photon flux in $15-150$ keV. \\ 
\hline
15\_150kev\_low & A12 & $\rm erg \ cm^{-2} \ s^{-1}$ & The lower limit of photon flux in $15-150$ keV. \\
\hline
15\_150kev\_hi & A12 & $\rm erg \ cm^{-2} \ s^{-1}$ & The upper limit of photon flux in $15-150$ keV. \\ 
\hline
15\_350kev & A12 & $\rm erg \ cm^{-2} \ s^{-1}$ & Photon flux in $15-350$ keV. \\
\hline
15\_350kev\_low & A12 & $\rm erg \ cm^{-2} \ s^{-1}$ & The lower limit of photon flux in $15-350$ keV. \\
\hline
15\_350kev\_hi & A12 & $\rm erg \ cm^{-2} \ s^{-1}$ & The upper limit of photon flux in $15-350$ keV. \\
\hline
Exposure\_time & A12 & s & The time interval of the spectrum. \\
\hline
Peak\_start  & A12 & s & The start time of the spectrum, \\
& & & relative to the BAT trigger time. \\
\hline
Peak\_stop & A12 & s & The end time of the spectrum, \\
& & & relative to the BAT trigger time. \\
\hline\hline
\end{tabular}
\end{center}
\end{table}

\begin{table}
\caption{
\label{tab:spec_1speak_cutpow_energy_fluence}
The format of the table that presents the energy fluence in unit of $\rm erg \ s^{-1}$ from the CPL fit for the 1-s peak spectra. Note that this table includes the fit for every GRB, regardless of whether the fit is acceptable from the criteria listed in Section~\ref{sect:event_standard}.
A list of GRBs with acceptable fits can be found in Table \ref{tab:spec_1speak_best_model}.
(This table is available in its entirety in a machine-readable form in the online journal. Only the column formats are shown here for guidance regarding its form and content.)
}
\begin{center}
\footnotesize
\begin{tabular}{|c|c|c|c|c|}
\hline\hline
Column & Format & Unit & Description \\
\hline\hline
GRBname & A9 & -- & GRB name. \\
\hline
Trig\_ID &  I11 & -- & Special ID associated with the trigger. \\ 
& & & For GRBs found in ground analysis, \\ 
& & & the ID are associated with the trigger ID of the failed event data, \\ 
& & & or the full observation ID of the event data for the analysis.  \\
\hline
15\_25kev & A12 & $\rm erg \ cm^{-2}$ & Photon flux in $15-25$ keV. \\
\hline
15\_25kev\_low & A12 & $\rm erg \ cm^{-2}$ & The lower limit of photon flux in $15-25$ keV. \\ 
\hline
15\_25kev\_hi & A12 & $\rm erg \ cm^{-2}$ & The upper limit of photon flux in $15-25$ keV. \\ 
\hline
25\_50kev & A12 & $\rm erg \ cm^{-2}$ & Photon flux in $25-50$ keV. \\ 
\hline
25\_50kev\_low & A12 & $\rm erg \ cm^{-2}$ & The lower limit of photon flux in $25-50$ keV. \\ 
\hline
25\_50kev\_hi & A12 & $\rm erg \ cm^{-2}$ & The upper limit of photon flux in $25-50$ keV. \\ 
\hline
50\_100kev & A12 & $\rm erg \ cm^{-2}$ & Photon flux in $50-100$ keV. \\ 
\hline
50\_100kev\_low & A12 & $\rm erg \ cm^{-2}$ & The lower limit of photon flux in $50-100$ keV. \\ 
\hline
50\_100kev\_hi & A12 & $\rm erg \ cm^{-2}$ & The upper limit of photon flux in $50-100$ keV. \\ 
\hline
100\_150kev & A12 & $\rm erg \ cm^{-2}$ & Photon flux in $100-150$ keV. \\ 
\hline
100\_150kev\_low & A12 & $\rm erg \ cm^{-2}$ & The lower limit of photon flux in $100-150$ keV. \\ 
\hline
100\_150kev\_hi & A12 & $\rm erg \ cm^{-2}$ & The upper limit of photon flux in $100-150$ keV. \\ 
\hline
100\_350kev & A12 & $\rm erg \ cm^{-2}$ & Photon flux in $100-350$ keV. \\
\hline
100\_350kev\_low & A12 & $\rm erg \ cm^{-2}$ & The lower limit of photon flux in $100-350$ keV. \\ 
\hline
100\_350kev\_hi & A12 & $\rm erg \ cm^{-2}$ & The upper limit of photon flux in $100-350$ keV. \\
\hline
15\_150kev & A12 & $\rm erg \ cm^{-2}$ & Photon flux in $15-150$ keV. \\ 
\hline
15\_150kev\_low & A12 & $\rm erg \ cm^{-2}$ & The lower limit of photon flux in $15-150$ keV. \\
\hline
15\_150kev\_hi & A12 & $\rm erg \ cm^{-2}$ & The upper limit of photon flux in $15-150$ keV. \\ 
\hline
15\_350kev & A12 & $\rm erg \ cm^{-2}$ & Photon flux in $15-350$ keV. \\
\hline
15\_350kev\_low & A12 & $\rm erg \ cm^{-2}$ & The lower limit of photon flux in $15-350$ keV. \\
\hline
15\_350kev\_hi & A12 & $\rm erg \ cm^{-2}$ & The upper limit of photon flux in $15-350$ keV. \\
\hline
Exposure\_time & A12 & s & The time interval of the spectrum. \\
\hline
Peak\_start  & A12 & s & The start time of the spectrum, \\
& & & relative to the BAT trigger time. \\
\hline
Peak\_stop & A12 & s & The end time of the spectrum, \\
& & & relative to the BAT trigger time. \\
\hline\hline
\end{tabular}
\end{center}
\end{table}

\subsection{Summary tables of spectral analyses for the 20-ms peak spectra}
\label{sect:20ms_tables}

In this section, we present the tables for the 20-ms peak spectra, which are the same set of the tables as those in Section~\ref{sect:T100_spectra_tables}.

\clearpage

\begin{table}
\caption{
\label{tab:spec_20mspeak_best_model}
The format of the table that summarizes the best-fit spectral model of the 20-ms peak spectrum for each burst.
(This table is available in its entirety in a machine-readable form in the online journal. Only the column formats are shown here for guidance regarding its form and content.)
}
\begin{center}
\begin{tabular}{|c|c|c|c|}
\hline\hline
Column & Format & Unit & Description \\
\hline\hline
GRBname & A9 & -- & GRB name. \\
\hline
Trig\_ID &  I11 & -- & Special ID associated with the trigger. \\ 
& & & For GRBs found in ground analysis, \\ 
& & & the ID are associated with the trigger ID of the failed event data, \\ 
& & & or the full observation ID of the event data for the analysis.  \\
\hline
Best\_fit\_model  & A3 & -- & The best-fit model for the GRB spectrum, \\
& & & either the simple power law model (PL) \\ 
& & & or cutoff power-law model (CPL.)  \\
\hline\hline
\end{tabular}
\end{center}
\end{table}

\begin{table}
\caption{
\label{tab:spec_20mspeak_pow_parameter}
The format of the table that presents the parameters from the PL fit for the 20-ms peak spectra. Note that this table includes the fit for every GRB, regardless of whether the fit is acceptable from the criteria listed in Section~\ref{sect:event_standard}.
A list of GRBs with acceptable fits can be found in Table \ref{tab:spec_20mspeak_best_model}.
(This table is available in its entirety in a machine-readable form in the online journal. Only the column formats are shown here for guidance regarding its form and content.)
}
\begin{center}
\footnotesize
\begin{tabular}{|c|c|c|c|}
\hline\hline
Column & Format & Unit & Description \\
\hline\hline
GRBname & A9 & -- & GRB name. \\
\hline
Trig\_ID &  I11 & -- & Special ID associated with each triggers. \\ 
& & & For GRBs found in ground analysis, \\ 
& & & the ID are associated with the trigger ID of the failed event data, \\ 
& & & or the full observation ID of the event data for the analysis.  \\
\hline
alpha & A13 & -- & $\alpha^{\rm PL}$ as defined in Eq.~\ref{eq:PL}. \\
\hline
alpha\_low & A13 & -- & The lower limit of $\alpha^{\rm PL}$. \\
\hline
alpha\_hi & A13 & -- & The upper limit of $\alpha^{\rm PL}$. \\
\hline
norm & A12 & $\rm ph \ cm^{-2} \ s^{-1} \ keV^{-1}$ & The normalization factor $K^{\rm PL}_{50}$, as defined in Eq.~\ref{eq:PL}. \\ 
\hline
norm\_low & A12 & $\rm ph \ cm^{-2} \ s^{-1} \ keV^{-1}$ & The lower limit of $K^{\rm PL}_{50}$. \\ 
\hline
norm\_hi & A12 & $\rm ph \ cm^{-2} \ s^{-1} \ keV^{-1}$ & The upper limit of $K^{\rm PL}_{50}$. \\ 
\hline
chi2 & F6.2 & -- & $\chi^2$ from the fitting, reported by XSPEC. \\  
\hline
dof & I2 & -- & degree of freedom from the fitting, reported by XSPEC. \\ 
\hline
reduced\_chi2 & F6.4 & -- & reduced $\chi^2$ (i.e., $\chi^2$ divided by the degree of freedom), \\ 
& & & reported by XSPEC. \\
\hline
null\_prob & A12 & -- & The null probability of the model, reported by XSPEC. \\ 
\hline
enorm & F7.4 & KeV & The normalization energy, which is set to 50 keV in our fits, \\ 
& & & as shown in Eq.~\ref{eq:PL}. \\ 
\hline
Exposure\_time & A6 & s & The time interval of the spectrum. \\ 
\hline
Peak\_start & A11 & s & The start time of the spectrum, \\ 
& & & relative to the BAT trigger time. \\ 
\hline
Peak\_stop & A11 & s & The end time of the spectrum, \\
& & & relative to the BAT trigger time.\\
\hline\hline
\end{tabular}
\end{center}
\end{table}

\begin{table}
\caption{
\label{tab:spec_20mspeak_pow_photon_flux}
The format of the table that presents the photon flux (in unit of $\rm ph \ cm^{-2} \ s^{-1}$) from the PL fit for the 20-ms peak spectra. Note that this table includes the fit for every GRB, regardless of whether the fit is acceptable from the criteria listed in Section~\ref{sect:event_standard}.
A list of GRBs with acceptable fits can be found in Table \ref{tab:spec_20mspeak_best_model}.
(This table is available in its entirety in a machine-readable form in the online journal. Only the column formats are shown here for guidance regarding its form and content.)
}
\begin{center}
\footnotesize
\begin{tabular}{|c|c|c|c|}
\hline\hline
Column & Format & Unit & Description \\
\hline\hline
GRBname & A9 & -- & GRB name. \\
\hline
Trig\_ID &  I11 & -- & Special ID associated with the trigger. \\ 
& & & For GRBs found in ground analysis, \\ 
& & & the ID are associated with the trigger ID of the failed event data, \\ 
& & & or the full observation ID of the event data for the analysis.  \\
\hline
15\_25kev & A12 & $\rm ph \ cm^{-2} \ s^{-1}$ & Photon flux in $15-25$ keV. \\
\hline
15\_25kev\_low & A12 & $\rm ph \ cm^{-2} \ s^{-1}$ & The lower limit of photon flux in $15-25$ keV. \\ 
\hline
15\_25kev\_hi & A12 & $\rm ph \ cm^{-2} \ s^{-1}$ & The upper limit of photon flux in $15-25$ keV. \\ 
\hline
25\_50kev & A12 & $\rm ph \ cm^{-2} \ s^{-1}$ & Photon flux in $25-50$ keV. \\ 
\hline
25\_50kev\_low & A12 & $\rm ph \ cm^{-2} \ s^{-1}$ & The lower limit of photon flux in $25-50$ keV. \\ 
\hline
25\_50kev\_hi & A12 & $\rm ph \ cm^{-2} \ s^{-1}$ & The upper limit of photon flux in $25-50$ keV. \\ 
\hline
50\_100kev & A12 & $\rm ph \ cm^{-2} \ s^{-1}$ & Photon flux in $50-100$ keV. \\ 
\hline
50\_100kev\_low & A12 & $\rm ph \ cm^{-2} \ s^{-1}$ & The lower limit of photon flux in $50-100$ keV. \\ 
\hline
50\_100kev\_hi & A12 & $\rm ph \ cm^{-2} \ s^{-1}$ & The upper limit of photon flux in $50-100$ keV. \\ 
\hline
100\_150kev & A12 & $\rm ph \ cm^{-2} \ s^{-1}$ & Photon flux in $100-150$ keV. \\ 
\hline
100\_150kev\_low & A12 & $\rm ph \ cm^{-2} \ s^{-1}$ & The lower limit of photon flux in $100-150$ keV. \\ 
\hline
100\_150kev\_hi & A12 & $\rm ph \ cm^{-2} \ s^{-1}$ & The upper limit of photon flux in $100-150$ keV. \\ 
\hline
100\_350kev & A12 & $\rm ph \ cm^{-2} \ s^{-1}$ & Photon flux in $100-350$ keV. \\
\hline
100\_350kev\_low & A12 & $\rm ph \ cm^{-2} \ s^{-1}$ & The lower limit of photon flux in $100-350$ keV. \\ 
\hline
100\_350kev\_hi & A12 & $\rm ph \ cm^{-2} \ s^{-1}$ & The upper limit of photon flux in $100-350$ keV. \\
\hline
15\_150kev & A12 & $\rm ph \ cm^{-2} \ s^{-1}$ & Photon flux in $15-150$ keV. \\ 
\hline
15\_150kev\_low & A12 & $\rm ph \ cm^{-2} \ s^{-1}$ & The lower limit of photon flux in $15-150$ keV. \\
\hline
15\_150kev\_hi & A12 & $\rm ph \ cm^{-2} \ s^{-1}$ & The upper limit of photon flux in $15-150$ keV. \\ 
\hline
15\_350kev & A12 & $\rm ph \ cm^{-2} \ s^{-1}$ & Photon flux in $15-350$ keV. \\
\hline
15\_350kev\_low & A12 & $\rm ph \ cm^{-2} \ s^{-1}$ & The lower limit of photon flux in $15-350$ keV. \\
\hline
15\_350kev\_hi & A12 & $\rm ph \ cm^{-2} \ s^{-1}$ & The upper limit of photon flux in $15-350$ keV. \\
\hline
Exposure\_time & A12 & s & The time interval of the spectrum. \\
\hline
Peak\_start  & A12 & s & The start time of the spectrum, \\
& & & relative to the BAT trigger time. \\
\hline
Peak\_stop & A12 & s & The end time of the spectrum, \\
& & & relative to the BAT trigger time. \\
\hline\hline
\end{tabular}
\end{center}
\end{table}

\begin{table}
\caption{
\label{tab:spec_20mspeak_pow_energy_flux}
The format of the table that presents the energy flux (in unit of $\rm erg \ s^{-1} \ cm^{-2}$) from the PL fit for the 20-ms peak spectra. Note that this table includes the fit for every GRB, regardless of whether the fit is acceptable from the criteria listed in Section~\ref{sect:event_standard}.
A list of GRBs with acceptable fits can be found in Table \ref{tab:spec_20mspeak_best_model}.
(This table is available in its entirety in a machine-readable form in the online journal. Only the column formats are shown here for guidance regarding its form and content.)
}
\begin{center}
\footnotesize
\begin{tabular}{|c|c|c|c|c|}
\hline\hline
Column & Format & Unit & Description \\
\hline\hline
GRBname & A9 & -- & GRB name. \\
\hline
Trig\_ID &  I11 & -- & Special ID associated with the trigger. \\ 
& & & For GRBs found in ground analysis, \\ 
& & & the ID are associated with the trigger ID of the failed event data, \\ 
& & & or the full observation ID of the event data for the analysis.  \\
\hline
15\_25kev & A12 & $\rm erg \ cm^{-2} \ s^{-1}$ & Photon flux in $15-25$ keV. \\
\hline
15\_25kev\_low & A12 & $\rm erg \ cm^{-2} \ s^{-1}$ & The lower limit of photon flux in $15-25$ keV. \\ 
\hline
15\_25kev\_hi & A12 & $\rm erg \ cm^{-2} \ s^{-1}$ & The upper limit of photon flux in $15-25$ keV. \\ 
\hline
25\_50kev & A12 & $\rm erg \ cm^{-2} \ s^{-1}$ & Photon flux in $25-50$ keV. \\ 
\hline
25\_50kev\_low & A12 & $\rm erg \ cm^{-2} \ s^{-1}$ & The lower limit of photon flux in $25-50$ keV. \\ 
\hline
25\_50kev\_hi & A12 & $\rm erg \ cm^{-2} \ s^{-1}$ & The upper limit of photon flux in $25-50$ keV. \\ 
\hline
50\_100kev & A12 & $\rm erg \ cm^{-2} \ s^{-1}$ & Photon flux in $50-100$ keV. \\ 
\hline
50\_100kev\_low & A12 & $\rm erg \ cm^{-2} \ s^{-1}$ & The lower limit of photon flux in $50-100$ keV. \\ 
\hline
50\_100kev\_hi & A12 & $\rm erg \ cm^{-2} \ s^{-1}$ & The upper limit of photon flux in $50-100$ keV. \\ 
\hline
100\_150kev & A12 & $\rm erg \ cm^{-2} \ s^{-1}$ & Photon flux in $100-150$ keV. \\ 
\hline
100\_150kev\_low & A12 & $\rm erg \ cm^{-2} \ s^{-1}$ & The lower limit of photon flux in $100-150$ keV. \\ 
\hline
100\_150kev\_hi & A12 & $\rm erg \ cm^{-2} \ s^{-1}$ & The upper limit of photon flux in $100-150$ keV. \\ 
\hline
100\_350kev & A12 & $\rm erg \ cm^{-2} \ s^{-1}$ & Photon flux in $100-350$ keV. \\
\hline
100\_350kev\_low & A12 & $\rm erg \ cm^{-2} \ s^{-1}$ & The lower limit of photon flux in $100-350$ keV. \\ 
\hline
100\_350kev\_hi & A12 & $\rm erg \ cm^{-2} \ s^{-1}$ & The upper limit of photon flux in $100-350$ keV. \\
\hline
15\_150kev & A12 & $\rm erg \ cm^{-2} \ s^{-1}$ & Photon flux in $15-150$ keV. \\ 
\hline
15\_150kev\_low & A12 & $\rm erg \ cm^{-2} \ s^{-1}$ & The lower limit of photon flux in $15-150$ keV. \\
\hline
15\_150kev\_hi & A12 & $\rm erg \ cm^{-2} \ s^{-1}$ & The upper limit of photon flux in $15-150$ keV. \\ 
\hline
15\_350kev & A12 & $\rm erg \ cm^{-2} \ s^{-1}$ & Photon flux in $15-350$ keV. \\
\hline
15\_350kev\_low & A12 & $\rm erg \ cm^{-2} \ s^{-1}$ & The lower limit of photon flux in $15-350$ keV. \\
\hline
15\_350kev\_hi & A12 & $\rm erg \ cm^{-2} \ s^{-1}$ & The upper limit of photon flux in $15-350$ keV. \\
\hline
Exposure\_time & A12 & s & The time interval of the spectrum. \\
\hline
Peak\_start  & A12 & s & The start time of the spectrum, \\
& & & relative to the BAT trigger time. \\
\hline
Peak\_stop & A12 & s & The end time of the spectrum, \\
& & & relative to the BAT trigger time. \\
\hline\hline
\end{tabular}
\end{center}
\end{table}

\begin{table}
\caption{
\label{tab:spec_20mspeak_pow_energy_fluence}
The format of the table that presents the energy fluences (in unit of $\rm erg \ cm^{-2}$) from the PL fit for the 20-ms peak spectra. 
Note that this table includes the fit for every GRB, regardless of whether the fit is acceptable from the criteria listed in Section~\ref{sect:event_standard}.
A list of GRBs with acceptable fits can be found in Table \ref{tab:spec_20mspeak_best_model}.
(This table is available in its entirety in a machine-readable form in the online journal. Only the column titles are shown here for guidance regarding its form and content.)
}
\begin{center}
\footnotesize
\begin{tabular}{|c|c|c|c|c|}
\hline\hline
Column & Format & Unit & Description \\
\hline\hline
GRBname & A9 & -- & GRB name. \\
\hline
Trig\_ID &  I11 & -- & Special ID associated with the trigger. \\ 
& & & For GRBs found in ground analysis, \\ 
& & & the ID are associated with the trigger ID of the failed event data, \\ 
& & & or the full observation ID of the event data for the analysis.  \\
\hline
15\_25kev & A12 & $\rm erg \ cm^{-2}$ & Photon flux in $15-25$ keV. \\
\hline
15\_25kev\_low & A12 & $\rm erg \ cm^{-2}$ & The lower limit of photon flux in $15-25$ keV. \\ 
\hline
15\_25kev\_hi & A12 & $\rm erg \ cm^{-2}$ & The upper limit of photon flux in $15-25$ keV. \\ 
\hline
25\_50kev & A12 & $\rm erg \ cm^{-2}$ & Photon flux in $25-50$ keV. \\ 
\hline
25\_50kev\_low & A12 & $\rm erg \ cm^{-2}$ & The lower limit of photon flux in $25-50$ keV. \\ 
\hline
25\_50kev\_hi & A12 & $\rm erg \ cm^{-2}$ & The upper limit of photon flux in $25-50$ keV. \\ 
\hline
50\_100kev & A12 & $\rm erg \ cm^{-2}$ & Photon flux in $50-100$ keV. \\ 
\hline
50\_100kev\_low & A12 & $\rm erg \ cm^{-2}$ & The lower limit of photon flux in $50-100$ keV. \\ 
\hline
50\_100kev\_hi & A12 & $\rm erg \ cm^{-2}$ & The upper limit of photon flux in $50-100$ keV. \\ 
\hline
100\_150kev & A12 & $\rm erg \ cm^{-2}$ & Photon flux in $100-150$ keV. \\ 
\hline
100\_150kev\_low & A12 & $\rm erg \ cm^{-2}$ & The lower limit of photon flux in $100-150$ keV. \\ 
\hline
100\_150kev\_hi & A12 & $\rm erg \ cm^{-2}$ & The upper limit of photon flux in $100-150$ keV. \\ 
\hline
100\_350kev & A12 & $\rm erg \ cm^{-2}$ & Photon flux in $100-350$ keV. \\
\hline
100\_350kev\_low & A12 & $\rm erg \ cm^{-2}$ & The lower limit of photon flux in $100-350$ keV. \\ 
\hline
100\_350kev\_hi & A12 & $\rm erg \ cm^{-2}$ & The upper limit of photon flux in $100-350$ keV. \\
\hline
15\_150kev & A12 & $\rm erg \ cm^{-2}$ & Photon flux in $15-150$ keV. \\ 
\hline
15\_150kev\_low & A12 & $\rm erg \ cm^{-2}$ & The lower limit of photon flux in $15-150$ keV. \\
\hline
15\_150kev\_hi & A12 & $\rm erg \ cm^{-2}$ & The upper limit of photon flux in $15-150$ keV. \\ 
\hline
15\_350kev & A12 & $\rm erg \ cm^{-2}$ & Photon flux in $15-350$ keV. \\
\hline
15\_350kev\_low & A12 & $\rm erg \ cm^{-2}$ & The lower limit of photon flux in $15-350$ keV. \\
\hline
15\_350kev\_hi & A12 & $\rm erg \ cm^{-2}$ & The upper limit of photon flux in $15-350$ keV. \\
\hline
Exposure\_time & A12 & s & The time interval of the spectrum. \\
\hline
Peak\_start  & A12 & s & The start time of the spectrum, \\
& & & relative to the BAT trigger time. \\
\hline
Peak\_stop & A12 & s & The end time of the spectrum, \\
& & & relative to the BAT trigger time. \\
\hline\hline
\end{tabular}
\end{center}
\end{table}

\begin{table}
\caption{
\label{tab:spec_20mspeak_cutpow_parameter}
The format of the table that presents the parameters from the CPL fit for the 20-ms peak spectra. Note that this table includes the fit for every GRB, regardless of whether the fit is acceptable from the criteria listed in Section~\ref{sect:event_standard}.
A list of GRBs with acceptable fits can be found in Table \ref{tab:spec_20mspeak_best_model}.
(This table is available in its entirety in a machine-readable form in the online journal. Only the column formats are shown here for guidance regarding its form and content.)
}
\begin{center}
\footnotesize
\begin{tabular}{|c|c|c|c|c|}
\hline\hline
Column & Format & Unit & Description \\
\hline\hline
GRBname & A9 & -- & GRB name. \\
\hline
Trig\_ID &  I11 & -- & Special ID associated with each triggers. \\ 
& & & For GRBs found in ground analysis, \\ 
& & & the ID are associated with the trigger ID of the failed event data, \\ 
& & & or the full observation ID of the event data for the analysis.  \\
\hline
alpha & A13 & -- & $\alpha^{\rm CPL}$ as defined in Eq.~\ref{eq:CPL}. \\
\hline
alpha\_low & A13 & -- & The lower limit of $\alpha^{\rm CPL}$. \\
\hline
alpha\_hi & A13 & -- & The upper limit of $\alpha^{\rm CPL}$. \\
\hline
Epeak & A12 & keV & $E_{\rm peak}$ as defined in  Eq.~\ref{eq:CPL}. \\
\hline
Epeak\_low & A12 & keV & The lowe limit of $E_{\rm peak}$. \\
\hline
Epeak\_hi & A12 & keV & The upper limit of $E_{\rm peak}$. \\
\hline
norm & A12 & $\rm ph \ cm^{-2} \ s^{-1} \ keV^{-1}$ & The normalization factor $K^{\rm CPL}_{50}$, as defined in Eq.~\ref{eq:CPL}. \\ 
\hline
norm\_low & A12 & $\rm ph \ cm^{-2} \ s^{-1} \ keV^{-1}$ & The lower limit of $K^{\rm CPL}_{50}$. \\ 
\hline
norm\_hi & A12 & $\rm ph \ cm^{-2} \ s^{-1} \ keV^{-1}$ & The upper limit of $K^{\rm CPL}_{50}$. \\ 
\hline
chi2 & F6.2 & -- & $\chi^2$ from the fitting, reported by XSPEC. \\  
\hline
dof & I2 & -- & degree of freedom from the fitting, reported by XSPEC. \\ 
\hline
reduced\_chi2 & F6.4 & -- & reduced $\chi^2$ (i.e., $\chi^2$ divided by the degree of freedom), \\ 
& & & reported by XSPEC. \\
\hline
null\_prob & A12 & -- & The null probability of the model, reported by XSPEC. \\ 
\hline
enorm & F7.4 & KeV & The normalization energy, which is set to 50 keV in our fits, \\ 
& & & as shown in Eq.~\ref{eq:PL}. \\ 
\hline
Exposure\_time & A6 & s & The time interval of the spectrum. \\ 
\hline
Peak\_start & A11 & s & The start time of the spectrum, \\ 
& & & relative to the BAT trigger time. \\ 
\hline
Peak\_stop & A11 & s & The end time of the spectrum, \\
& & & relative to the BAT trigger time.\\
\hline\hline
\end{tabular}
\end{center}
\end{table}

\begin{table}
\caption{
\label{tab:spec_20mspeak_cutpow_photon_flux}
The format of the table that presents the photon flux (in unit of $\rm ph \ s^{-1} \ cm^{-2}$) from the CPL fit for the 20-ms peak spectra. Note that this table includes the fit for every GRB, regardless of whether the fit is acceptable from the criteria listed in Section~\ref{sect:event_standard}.
A list of GRBs with acceptable fits can be found in Table \ref{tab:spec_20mspeak_best_model}.
(This table is available in its entirety in a machine-readable form in the online journal. Only the column formats are shown here for guidance regarding its form and content.)
}
\begin{center}
\footnotesize
\begin{tabular}{|c|c|c|c|c|}
\hline\hline
Column & Format & Unit & Description \\
\hline\hline
GRBname & A9 & -- & GRB name. \\
\hline
Trig\_ID &  I11 & -- & Special ID associated with the trigger. \\ 
& & & For GRBs found in ground analysis, \\ 
& & & the ID are associated with the trigger ID of the failed event data, \\ 
& & & or the full observation ID of the event data for the analysis.  \\
\hline
15\_25kev & A12 & $\rm ph \ cm^{-2} \ s^{-1}$ & Photon flux in $15-25$ keV. \\
\hline
15\_25kev\_low & A12 & $\rm ph \ cm^{-2} \ s^{-1}$ & The lower limit of photon flux in $15-25$ keV. \\ 
\hline
15\_25kev\_hi & A12 & $\rm ph \ cm^{-2} \ s^{-1}$ & The upper limit of photon flux in $15-25$ keV. \\ 
\hline
25\_50kev & A12 & $\rm ph \ cm^{-2} \ s^{-1}$ & Photon flux in $25-50$ keV. \\ 
\hline
25\_50kev\_low & A12 & $\rm ph \ cm^{-2} \ s^{-1}$ & The lower limit of photon flux in $25-50$ keV. \\ 
\hline
25\_50kev\_hi & A12 & $\rm ph \ cm^{-2} \ s^{-1}$ & The upper limit of photon flux in $25-50$ keV. \\ 
\hline
50\_100kev & A12 & $\rm ph \ cm^{-2} \ s^{-1}$ & Photon flux in $50-100$ keV. \\ 
\hline
50\_100kev\_low & A12 & $\rm ph \ cm^{-2} \ s^{-1}$ & The lower limit of photon flux in $50-100$ keV. \\ 
\hline
50\_100kev\_hi & A12 & $\rm ph \ cm^{-2} \ s^{-1}$ & The upper limit of photon flux in $50-100$ keV. \\ 
\hline
100\_150kev & A12 & $\rm ph \ cm^{-2} \ s^{-1}$ & Photon flux in $100-150$ keV. \\ 
\hline
100\_150kev\_low & A12 & $\rm ph \ cm^{-2} \ s^{-1}$ & The lower limit of photon flux in $100-150$ keV. \\ 
\hline
100\_150kev\_hi & A12 & $\rm ph \ cm^{-2} \ s^{-1}$ & The upper limit of photon flux in $100-150$ keV. \\ 
\hline
100\_350kev & A12 & $\rm ph \ cm^{-2} \ s^{-1}$ & Photon flux in $100-350$ keV. \\
\hline
100\_350kev\_low & A12 & $\rm ph \ cm^{-2} \ s^{-1}$ & The lower limit of photon flux in $100-350$ keV. \\ 
\hline
100\_350kev\_hi & A12 & $\rm ph \ cm^{-2} \ s^{-1}$ & The upper limit of photon flux in $100-350$ keV. \\
\hline
15\_150kev & A12 & $\rm ph \ cm^{-2} \ s^{-1}$ & Photon flux in $15-150$ keV. \\ 
\hline
15\_150kev\_low & A12 & $\rm ph \ cm^{-2} \ s^{-1}$ & The lower limit of photon flux in $15-150$ keV. \\
\hline
15\_150kev\_hi & A12 & $\rm ph \ cm^{-2} \ s^{-1}$ & The upper limit of photon flux in $15-150$ keV. \\ 
\hline
15\_350kev & A12 & $\rm ph \ cm^{-2} \ s^{-1}$ & Photon flux in $15-350$ keV. \\
\hline
15\_350kev\_low & A12 & $\rm ph \ cm^{-2} \ s^{-1}$ & The lower limit of photon flux in $15-350$ keV. \\
\hline
15\_350kev\_hi & A12 & $\rm ph \ cm^{-2} \ s^{-1}$ & The upper limit of photon flux in $15-350$ keV. \\
\hline
Exposure\_time & A12 & s & The time interval of the spectrum. \\
\hline
Peak\_start  & A12 & s & The start time of the spectrum, \\
& & & relative to the BAT trigger time. \\
\hline
Peak\_stop & A12 & s & The end time of the spectrum, \\
& & & relative to the BAT trigger time. \\\hline\hline
\end{tabular}
\end{center}
\end{table}

\begin{table}
\caption{
\label{tab:spec_20mspeak_cutpow_energy_flux}
The format of the table that presents the energy flux (in unit of $\rm erg \ s^{-1} \ cm^{-2}$) from the CPL fit for the 20-ms peak spectra. Note that this table includes the fit for every GRB, regardless of whether the fit is acceptable from the criteria listed in Section~\ref{sect:event_standard}.
A list of GRBs with acceptable fits can be found in Table \ref{tab:spec_20mspeak_best_model}.
(This table is available in its entirety in a machine-readable form in the online journal. Only the column formats are shown here for guidance regarding its form and content.)
}
\begin{center}
\footnotesize
\begin{tabular}{|c|c|c|c|c|}
\hline\hline
Column & Format & Unit & Description \\
\hline\hline
GRBname & A9 & -- & GRB name. \\
\hline
Trig\_ID &  I11 & -- & Special ID associated with the trigger. \\ 
& & & For GRBs found in ground analysis, \\ 
& & & the ID are associated with the trigger ID of the failed event data, \\ 
& & & or the full observation ID of the event data for the analysis.  \\
\hline
15\_25kev & A12 & $\rm erg \ cm^{-2} \ s^{-1}$ & Photon flux in $15-25$ keV. \\
\hline
15\_25kev\_low & A12 & $\rm erg \ cm^{-2} \ s^{-1}$ & The lower limit of photon flux in $15-25$ keV. \\ 
\hline
15\_25kev\_hi & A12 & $\rm erg \ cm^{-2} \ s^{-1}$ & The upper limit of photon flux in $15-25$ keV. \\ 
\hline
25\_50kev & A12 & $\rm erg \ cm^{-2} \ s^{-1}$ & Photon flux in $25-50$ keV. \\ 
\hline
25\_50kev\_low & A12 & $\rm erg \ cm^{-2} \ s^{-1}$ & The lower limit of photon flux in $25-50$ keV. \\ 
\hline
25\_50kev\_hi & A12 & $\rm erg \ cm^{-2} \ s^{-1}$ & The upper limit of photon flux in $25-50$ keV. \\ 
\hline
50\_100kev & A12 & $\rm erg \ cm^{-2} \ s^{-1}$ & Photon flux in $50-100$ keV. \\ 
\hline
50\_100kev\_low & A12 & $\rm erg \ cm^{-2} \ s^{-1}$ & The lower limit of photon flux in $50-100$ keV. \\ 
\hline
50\_100kev\_hi & A12 & $\rm erg \ cm^{-2} \ s^{-1}$ & The upper limit of photon flux in $50-100$ keV. \\ 
\hline
100\_150kev & A12 & $\rm erg \ cm^{-2} \ s^{-1}$ & Photon flux in $100-150$ keV. \\ 
\hline
100\_150kev\_low & A12 & $\rm erg \ cm^{-2} \ s^{-1}$ & The lower limit of photon flux in $100-150$ keV. \\ 
\hline
100\_150kev\_hi & A12 & $\rm erg \ cm^{-2} \ s^{-1}$ & The upper limit of photon flux in $100-150$ keV. \\ 
\hline
100\_350kev & A12 & $\rm erg \ cm^{-2} \ s^{-1}$ & Photon flux in $100-350$ keV. \\
\hline
100\_350kev\_low & A12 & $\rm erg \ cm^{-2} \ s^{-1}$ & The lower limit of photon flux in $100-350$ keV. \\ 
\hline
100\_350kev\_hi & A12 & $\rm erg \ cm^{-2} \ s^{-1}$ & The upper limit of photon flux in $100-350$ keV. \\
\hline
15\_150kev & A12 & $\rm erg \ cm^{-2} \ s^{-1}$ & Photon flux in $15-150$ keV. \\ 
\hline
15\_150kev\_low & A12 & $\rm erg \ cm^{-2} \ s^{-1}$ & The lower limit of photon flux in $15-150$ keV. \\
\hline
15\_150kev\_hi & A12 & $\rm erg \ cm^{-2} \ s^{-1}$ & The upper limit of photon flux in $15-150$ keV. \\ 
\hline
15\_350kev & A12 & $\rm erg \ cm^{-2} \ s^{-1}$ & Photon flux in $15-350$ keV. \\
\hline
15\_350kev\_low & A12 & $\rm erg \ cm^{-2} \ s^{-1}$ & The lower limit of photon flux in $15-350$ keV. \\
\hline
15\_350kev\_hi & A12 & $\rm erg \ cm^{-2} \ s^{-1}$ & The upper limit of photon flux in $15-350$ keV. \\
\hline
Exposure\_time & A12 & s & The time interval of the spectrum. \\
\hline
Peak\_start  & A12 & s & The start time of the spectrum, \\
& & & relative to the BAT trigger time. \\
\hline
Peak\_stop & A12 & s & The end time of the spectrum, \\
& & & relative to the BAT trigger time. \\
\hline\hline
\end{tabular}
\end{center}
\end{table}

\begin{table}
\caption{
\label{tab:spec_20mspeak_cutpow_energy_fluence}
The format of the table that presents the energy fluence in unit of $\rm erg \ s^{-1}$ from the CPL fit for the 20-ms peak spectra. Note that this table includes the fit for every GRB, regardless of whether the fit is acceptable from the criteria listed in Section~\ref{sect:event_standard}.
A list of GRBs with acceptable fits can be found in Table \ref{tab:spec_20mspeak_best_model}.
(This table is available in its entirety in a machine-readable form in the online journal. Only the column formats are shown here for guidance regarding its form and content.)
}
\begin{center}
\footnotesize
\begin{tabular}{|c|c|c|c|c|}
\hline\hline
Column & Format & Unit & Description \\
\hline\hline
GRBname & A9 & -- & GRB name. \\
\hline
Trig\_ID &  I11 & -- & Special ID associated with the trigger. \\ 
& & & For GRBs found in ground analysis, \\ 
& & & the ID are associated with the trigger ID of the failed event data, \\ 
& & & or the full observation ID of the event data for the analysis.  \\
\hline
15\_25kev & A12 & $\rm erg \ cm^{-2}$ & Photon flux in $15-25$ keV. \\
\hline
15\_25kev\_low & A12 & $\rm erg \ cm^{-2}$ & The lower limit of photon flux in $15-25$ keV. \\ 
\hline
15\_25kev\_hi & A12 & $\rm erg \ cm^{-2}$ & The upper limit of photon flux in $15-25$ keV. \\ 
\hline
25\_50kev & A12 & $\rm erg \ cm^{-2}$ & Photon flux in $25-50$ keV. \\ 
\hline
25\_50kev\_low & A12 & $\rm erg \ cm^{-2}$ & The lower limit of photon flux in $25-50$ keV. \\ 
\hline
25\_50kev\_hi & A12 & $\rm erg \ cm^{-2}$ & The upper limit of photon flux in $25-50$ keV. \\ 
\hline
50\_100kev & A12 & $\rm erg \ cm^{-2}$ & Photon flux in $50-100$ keV. \\ 
\hline
50\_100kev\_low & A12 & $\rm erg \ cm^{-2}$ & The lower limit of photon flux in $50-100$ keV. \\ 
\hline
50\_100kev\_hi & A12 & $\rm erg \ cm^{-2}$ & The upper limit of photon flux in $50-100$ keV. \\ 
\hline
100\_150kev & A12 & $\rm erg \ cm^{-2}$ & Photon flux in $100-150$ keV. \\ 
\hline
100\_150kev\_low & A12 & $\rm erg \ cm^{-2}$ & The lower limit of photon flux in $100-150$ keV. \\ 
\hline
100\_150kev\_hi & A12 & $\rm erg \ cm^{-2}$ & The upper limit of photon flux in $100-150$ keV. \\ 
\hline
100\_350kev & A12 & $\rm erg \ cm^{-2}$ & Photon flux in $100-350$ keV. \\
\hline
100\_350kev\_low & A12 & $\rm erg \ cm^{-2}$ & The lower limit of photon flux in $100-350$ keV. \\ 
\hline
100\_350kev\_hi & A12 & $\rm erg \ cm^{-2}$ & The upper limit of photon flux in $100-350$ keV. \\
\hline
15\_150kev & A12 & $\rm erg \ cm^{-2}$ & Photon flux in $15-150$ keV. \\ 
\hline
15\_150kev\_low & A12 & $\rm erg \ cm^{-2}$ & The lower limit of photon flux in $15-150$ keV. \\
\hline
15\_150kev\_hi & A12 & $\rm erg \ cm^{-2}$ & The upper limit of photon flux in $15-150$ keV. \\ 
\hline
15\_350kev & A12 & $\rm erg \ cm^{-2}$ & Photon flux in $15-350$ keV. \\
\hline
15\_350kev\_low & A12 & $\rm erg \ cm^{-2}$ & The lower limit of photon flux in $15-350$ keV. \\
\hline
15\_350kev\_hi & A12 & $\rm erg \ cm^{-2}$ & The upper limit of photon flux in $15-350$ keV. \\
\hline
Exposure\_time & A12 & s & The time interval of the spectrum. \\
\hline
Peak\_start  & A12 & s & The start time of the spectrum, \\
& & & relative to the BAT trigger time. \\
\hline
Peak\_stop & A12 & s & The end time of the spectrum, \\
& & & relative to the BAT trigger time. \\
\hline\hline
\end{tabular}
\end{center}
\end{table}

\subsection{A list of GRBs with bright X-ray sources in the same field of view}

\begin{table}
\caption{
\label{tab:brigh_sources}
The format of the table that presents a list of GRBs with bright X-ray sources in the same field of view (see Section~\ref{sect:bright_src} for detailed discussions).
(This table is available in its entirety in a machine-readable form in the online journal. Only the column format is shown here for guidance regarding its form and content.)
}
\begin{center}
\begin{tabular}{|c|c|c|c|}
\hline\hline
Column & Format & Unit & Description \\
\hline\hline
GRBname & A9 & -- & GRB name. \\
\hline\hline
\end{tabular}
\end{center}
\end{table}

\subsection{Redshift list}

Table \ref{tab:redshift_list} presents a list of GRB redshifts with the references. The methods of redshift measurements are noted with the following symbols:
``a'': spectroscopic measurement from absorption lines; ``h'': spectroscopic measurement from host galaxy; ``p'': photometric redshift; ``hp'': photometric redshift from host galaxy. The redshift uncertainty for photometric redshift is presented when available. Note that the redshift uncertainties is adopted from the reference, and thus might represent different confidence level.
Moreover, some of the redshift, especially the one from host galaxies and photometric redshifts, can be less robust due to the uncertainty in their measurements. For example, sometime the reference reported two solution for the photometric redshifts, with one of them stated as a more likely solution. We thus record the more likely one in this table. However, the reader is strongly recommended to refer to the original reference for more information.

\begin{table}
\caption{
\label{tab:redshift_list}
A list of GRB redshifts. The measurement methods are also listed: ``ba'': spectroscopic measurement from burst afterglow absorption lines; ``he'': spectroscopic measurement from host galaxy emission lines; ``bp'': photometric redshift from burst afterglow; ``hp'': photometric redshift from host galaxy.
(This table is available in its entirety in a machine-readable form in the online journal. Only the column titles are shown here for guidance regarding its form and content.)
}
\begin{center}
\begin{tabular}{|c|c|c|c|}
\hline\hline
Column & Format & Unit & Description \\
\hline\hline
GRBname & A9 & -- & GRB name. \\
z & A6 & -- & GRB redshift. \\ 
Method & A2 & -- & The method of how the redshift is determined: \\
& & & ``ba'': spectroscopic measurement from \\
& & & the burst afterglow absorption lines. \\
& & & ``he'': spectroscopic measurement \\
& & &  from host galaxy emission lines. \\ 
& & & ``bp'': photometric redshift from burst afterglow. \\
& & & ``hp'': photometric redshift from host galaxy.  \\
Uncertainty (when available) &  A11 & -- & The redshift uncertainty (when available). \\ 
Ref. & A100 & -- & References of the GRB redshift measurements. \\
& & & The reference marked with ``$\star$'' is the one  \\ 
& & & related to the value presented in this table.\\
\hline\hline
\end{tabular}
\end{center}
\end{table}

\section{Appendix: The relation between the rest-frame luminosity and the observed flux}
\label{sect:lum_flux}

Calculating the rest-frame luminosity from the observed flux involves dealing with the redshifting effects of both the time and energy, which can be confusing. 
We thus write out detail derivation of the equation used to calculate the rest-frame luminosity in the rest-frame energy band in Section \ref{sect:T90_vs_z}.
In addition, we also include cross-checks of our equations with equations used for luminosity/energy calculations in other papers.

The luminosity distance $D_L$ is defined by the following equation,
\beq
D_L = \bigg( \frac{L_{\rm bol, rest}}{4 \pi F_{\rm bol, obs}} \bigg)^{1/2}
\eeq
such that the relation between the bolometric luminosity $L_{\rm bol, rest}$ and the bolometric flux $F_{\rm bol, obs}$ is
\beq
L_{\rm bol, rest} = 4 \pi D^2_L F_{\rm bol, obs}
\eeq
which is in the same form as the usual one in the local frame. All the cosmological redshifting effects between the rest-frame bolometric luminosity $L_{\rm bol, rest}$ and the observer-frame bolometric flux $F_{\rm bol, obs}$
are handled by $D_L$.

To be more specific, the rest-frame bolometric luminosity is defined as
\beq
L_{\rm bol, rest} = \frac{dE_{\rm rest}}{dt_{\rm rest}}
\eeq
(i.e., the energy output in the rest frame divided by the time measured in the rest frame),
while the observer-frame bolometric flux is defined as
\beq
F_{\rm bol, obs} = \frac{dE_{\rm obs}}{dA \ dt_{\rm obs}}
\eeq
where dA is the surface area. Energy and time in the rest frame and the observer frame are related by
\beq
E_{\rm rest} = (1+z) E_{\rm obs}
\eeq
\beq
t_{\rm rest} = \frac{t_{\rm obs}}{1+z}
\eeq

In the real astronomy cases, it is common that the bolometric flux and luminosity cannot be measured directly due to the limited bandpass of the instrument.
The relationship between the luminosity and flux in certain bandpass can be derived from the luminosity density (i.e., the differential luminosity) and the flux density (i.e., the differential flux).
The luminosity density and flux density can be defined as either (1) luminosity per photon frequency ($dL_{\rm bol, rest}/d\nu_{\rm rest}$) and flux per photon frequency ($dF_{\rm bol, obs}/d\nu_{\rm obs}$), or (2) luminosity per photon energy ($dL_{\rm bol, rest}/dE_{\rm rest}$) and flux per photon energy ($dF_{\rm bol, obs}/dE_{\rm obs}$). The photon energy E can be simply related to the photon frequency $\nu$ by $E = h \nu$, where $h$ is the Planck constant. 
The high energy astrophysics community commonly refers to photon energy rather than photon frequency. We thus adopt the ``per-energy'' definition for the 
luminosity density and flux density hereafter. In other words, the luminosity density $L_E$ and the flux density $F_E$ are defined as follows:
\beq
L_{E, \rm rest} = \frac{dL_{\rm bol, rest}}{dE_{\rm rest}} = \frac{dE_{\rm rest}}{dt_{\rm rest} \ dE_{\rm rest}}
\eeq
\beq
F_{E, \rm obs} = \frac{dF}{dE_{\rm obs}} =  \frac{dE_{\rm obs}}{dt_{\rm obs} \ dA \ dE_{\rm obs}}
\eeq
The energy in the denominator does not cancel with the energy in the nominator, because the one in the nominator refers to the energy output, and the one in the denominator 
refers to the photon energy, despite they have the same unit.

The relation between the luminosity density $L_E$ in the rest frame and the flux density $F_E$ in the observer frame can be derived as follows,
\begin{align}
L_{E, \rm rest} = \frac{dL_{\rm bol, rest}}{dE_{\rm rest}} \nonumber &= \frac{d}{dE_{\rm rest}} ( 4 \pi D^2_L F_{\rm bol, obs}) \nonumber \\
	&= 4 \pi D^2_L \frac{dF_{\rm bol, obs}}{dE_{\rm rest}} \nonumber \\
	&= 4 \pi D^2_L \frac{dF_{\rm bol, obs}}{dE_{\rm obs}} \frac{dE_{\rm obs}}{dE_{\rm rest}} \nonumber \\ 
	&= 4 \pi D^2_L \frac{dF_{\rm bol, obs}}{dE_{\rm obs}} \ \frac{1}{(1+z)}  \nonumber \\
	&= \frac{4 \pi D^2_L \ F_{E, \rm obs}}{(1+z)}
\end{align}

The luminosity at a certain energy band $L_{\rm band, rest}$ can be calculated by integrating $L_{E, \rm rest}$,
\begin{align}
\label{eq:lum_flux}
L_{\rm band, rest} = \int^{E_{2, \rm rest}}_{E_{1, \rm rest}} L_{E, \rm rest} \ dE_{\rm rest} \nonumber &= \int^{E_{2, \rm rest}}_{E_{1, \rm rest}}  \frac{dL_{\rm bol, rest}}{dE_{\rm rest}} \ dE_{\rm rest} \nonumber \\
         &= \int^{E_{2, \rm rest}}_{E_{1, \rm rest}} \frac{4 \pi D^2_L}{1+z} \ F_{E, \rm obs}  \ dE_{\rm rest} \nonumber \\
	& =  \int^{E_{2, \rm obs}}_{E_{1, \rm obs}} \frac{4 \pi D^2_L}{1+z} \ F_{E, \rm obs} \ (1+z) dE_{\rm obs}  \nonumber \\  
	& \ \ \ \rm (Change \ of \ variables \ with \ dE_{\rm rest} = (1+z) dE_{\rm obs}) \nonumber \\
	& = \int^{E_{2, \rm obs}}_{E_{1, \rm obs}} 4 \pi D^2_L \ F_{E, \rm obs} \ dE_{\rm obs} \nonumber \\
	&=  4 \pi D^2_L \ \int^{E_{2, \rm obs}}_{E_{1, \rm obs}} \ F_{E, \rm obs} \ dE_{\rm obs} \nonumber \\
	&= 4 \pi D^2_L \ F_{\rm band, obs} 
\end{align}
Where $F_{\rm band, obs} = \int^{E_{2, \rm obs}}_{E_{1, \rm obs}} \ F_{E, \rm obs} \ dE_{\rm obs}$ is the observed flux in a the observed energy band, $E_{1, \rm obs}$ to $E_{2, \rm obs}$,
which correspond to the rest-frame energy band, $E_{2, \rm rest} = (1+z) \ E_{1, \rm obs}$ to $E_{2, \rm rest} = (1+z) \ E_{2, \rm obs}$. This is the equation we use to calculate the ``luminosity''
in Section \ref{sect:T90_vs_z}. Since we use the flux measured in the BAT energy band 15-150 keV, the ``luminosity'' we presented in this paper corresponds to different rest-frame energy for GRBs at different redshifts.

We cross-check our derivation with the equations mentioned in previous papers. \cite{Bloom01} mentioned the relation between the energy $E_{\rm band, rest}$ in some comoving bandpass $[E_{1, \rm rest}, E_{2, \rm rest}]$ and the observed fluence $S_{\rm band, obs}$ in the corresponded-redshifted bandpass of $[E_{1, \rm rest}/(1+z), E_{2, \rm rest}/(1+z)]$ to be
\beq
\label{eq:energy_fluence}
E_{\rm band, rest} =  \frac{4 \pi D^2_L}{(1+z)} \ S_{\rm band, obs}
\eeq
This relationship can be derived from Eq.~\ref{eq:lum_flux} by integrating the luminosity from time $t_{1, \rm rest}$ to $t_{2, \rm rest}$ in the rest-frame,
\begin{align}
E_{\rm band, rest} &= \int^{t2, \rm rest}_{t1,\rm rest} L_{\rm band, rest} \ dt_{\rm rest}  \nonumber \\
	 		     &= \int^{t2, \rm rest}_{t1,\rm rest} 4 \pi D^2_L \ F_{\rm band, obs} \ dt_{\rm rest}  \nonumber \\
			     &=  \int^{t2, \rm obs}_{t1,\rm obs} 4 \pi D^2_L \ F_{\rm band, obs} \ \frac{dt_{\rm obs}}{(1+z)}  \nonumber \\
			     & \ \ \ \rm (Change \ of \ variables \ with \ dt_{\rm rest} = \frac{dt_{\rm obs}}{(1+z)}  \nonumber \\
			     &= \frac{4 \pi D^2_L}{(1+z)} \int^{t2, \rm obs}_{t1,\rm obs} F_{\rm band, obs} \ dt_{\rm obs}  \nonumber \\
			     &= \frac{4 \pi D^2_L}{(1+z)} \ S_{E, \rm band, obs}
\end{align}

\citet{Amati02} uses a slightly different approach by first blue-shifting the GRB spectrum back to the rest frame to get the total fluences in the rest frame, which can be expressed as follows,
\begin{align}
S_{\rm band, rest} &= \int^{E_{2, \rm rest}}_{E_{1, \rm rest}} \ \bigg( \int^{t2, \rm rest}_{t1,\rm rest}  F_{\rm band, rest} \ dt_{\rm rest} \bigg) \ dE_{\rm rest}  \nonumber \\
		             &=  \int^{E_{2, \rm rest}}_{E_{1, \rm rest}} \ \bigg( \int^{t2, \rm rest}_{t1,\rm rest} \frac{dF}{dE_{\rm rest}} \ dt_{\rm rest} \bigg) \ dE_{\rm rest}  \nonumber \\
		              &= \int^{E_{2, \rm rest}}_{E_{1, \rm rest}} \ \bigg( \int^{t2, \rm rest}_{t1,\rm rest} \ \frac{dE_{\rm rest}}{dt_{\rm rest} \ dA \ dE_{\rm rest}} \ dt_{\rm rest} \bigg) \ dE_{\rm rest}
\end{align}
$S_{\rm band, rest}$ corresponds to the $\int^{10000}_1 \ EN(E,\alpha,E_0,\beta,A)dE$ in the original equation in \citet{Amati02}.
Using Eq.~\ref{eq:energy_fluence} and converting the observer-frame fluence to the rest-frame fluence, one can get the equation used in \citet{Amati02},
\begin{align}
E_{\rm band, rest} &=  \frac{4 \pi D^2_L}{(1+z)} \ S_{\rm band, obs}  \nonumber \\
		             &= \frac{4 \pi D^2_L}{(1+z)} \ \int^{E_{2, \rm obs}}_{E_{1, \rm obs}} \ \bigg( \int^{t2, \rm obs}_{t1,\rm obs} \ \frac{dE_{\rm obs}}{dt_{\rm obs} \ dA \ dE_{\rm obs}} \ dt_{\rm obs} \bigg) \ dE_{\rm obs}  \nonumber \\
		             &= \frac{4 \pi D^2_L}{(1+z)} \ \int^{E_{2, \rm rest}}_{E_{1, \rm rest}} \ \bigg( \int^{t2, \rm rest}_{t1,\rm rest} \ \frac{dE_{\rm rest}/(1+z)}{dt_{\rm rest}(1+z) \ dA \ dE_{\rm rest}/(1+z)} \ dt_{\rm rest}(1+z) \bigg) \ \frac{dE_{\rm rest}}{(1+z)}  \nonumber \\
		             &= \frac{4 \pi D^2_L}{(1+z)^2} \ \int^{E_{2, \rm rest}}_{E_{1, \rm rest}} \ \bigg( \int^{t2, \rm rest}_{t1,\rm rest} \ \frac{dE_{\rm rest}}{dt_{\rm rest} \ dA \ dE_{\rm rest}} \ dt_{\rm rest} \bigg) \ dE_{\rm rest}  \nonumber \\
		             &=  \frac{4 \pi D^2_L}{(1+z)^2} \ S_{\rm band, rest}
\end{align}

Therefore, the equation we use for calculating the rest-frame luminosity from the observed flux is consistent with the equations used in \citet{Bloom01} and \citet{Amati02}.

In summary, as mentioned in \citet{Bloom01}, if one would like to calculate the luminosity or energy in the rest-frame bandpass that matches perfectly with the redshifted band in the observer frame, one can use Eq.~\ref{eq:lum_flux} and Eq.~\ref{eq:energy_fluence} without extra factor of k-correction.
However, if one would like calculate the energy or luminosity in other energy bands, an extra factor of ``k-correction'' based on the source spectrum would be required to adjust the flux difference between the original energy range ($E_{1, \rm rest}$
to $E_{2, \rm rest}$) and the desired energy range \citep[for more details of the k-correction factor, see][]{Bloom01}.

\bibliographystyle{apj}
\bibliography{ref}

\end{document}